\documentclass[
12pt, % The default document font size, options: 10pt, 11pt, 12pt
english, % ngerman for German
onehalfspacing, % Single line spacing, alternatives: onehalfspacing or 
nolistspacing, % If the document is onehalfspacing or doublespacing, uncomment 
liststotoc, % Uncomment to add the list of figures/tables/etc to the table of contents
headsepline, % Uncomment to get a line under the header
]{MastersDoctoralThesis} % The class file specifying the document structure
\usepackage{amsmath}
\usepackage{amssymb}
\usepackage{amsfonts}
\usepackage{amsbsy}
\usepackage{mathrsfs}
\usepackage{bm}
\usepackage{enumitem}
\usepackage{soul}
\usepackage[page,toc,titletoc,title]{appendix}

\usepackage[T1]{fontenc} % Output font encoding for international characters
\usepackage{mathpazo} % Use the Palatino font by default

\RequirePackage[numbers,sort&compress]{natbib}
\AtBeginDocument{\hypersetup{breaklinks=true,citecolor=darkred,linkcolor=darkblue,linktoc=page}}
\usepackage[autostyle=true]{csquotes} 
\usepackage{setspace}
\usepackage{graphicx}
\usepackage{subfig}
\usepackage{wasysym}
\usepackage{comment}
\usepackage{textgreek}
\usepackage{adjustbox}
\usepackage{multirow}
\usepackage{makecell}

\usepackage{chngcntr} 
\counterwithout{footnote}{chapter}
\usepackage[flushmargin]{footmisc}
\usepackage{relsize}
\usepackage{wrapfig}
\usepackage{nomencl}
\usepackage{environ}

\newcounter{numln}
\newlist{numitemise}{itemize}{2}
\setlist[numitemise]{wide}%
\setlist[numitemise, 1]{labelindent=0pt,labelwidth=2em, label=\stepcounter{numln}\makebox[2em]{\thenumln.\hfill}, leftmargin=\dimexpr\labelwidth+\labelsep\relax}%
\setlist[numitemise, 2]{labelindent=\dimexpr -2em-\labelsep\relax, labelwidth=\dimexpr 2em+\labelsep\relax, label=\stepcounter{numln}\makebox[\dimexpr\labelwidth + \labelsep\relax]{\thenumln.\hfill\textbullet}, leftmargin=\dimexpr\leftmargin+2\labelsep\relax}%

\NewEnviron{myequation}{%
\begin{equation}
\scalebox{0.9}{$\BODY$}
\end{equation}

}

\definecolor{darkblue}{rgb}{0.0, 0.0, 0.62}
\definecolor{deepmagenta}{rgb}{0.8, 0.0, 0.7}
\definecolor{darkred}{rgb}{0.55, 0.0, 0.0}

\RequirePackage{mymacro}
\renewcommand{\thechapter}{\arabic{chapter}}
\renewcommand{\mdtChapapp}{}{}
\renewcommand{\chapteralign}{\raggedleft}
\renewcommand{\chapterfont}{\color{black}\huge\MakeUppercase}
\renewcommand{\chapterprefixfont}{\fontsize{50pt}{0pt}\selectfont\color{mdtRed}}
\makenomenclature
\makeatletter
\def\blfootnote{\gdef\@thefnmark{}\@footnotetext}
\makeatother

\newlength{\tocindent}

 \makeatletter
\def\@chapter[#1]#2{\ifnum \c@secnumdepth >\m@ne
                         \refstepcounter{chapter}%
                         \typeout{\@chapapp\space\thechapter.}%
                         \settowidth{\tocindent}{\normalsize\bfseries \@chapapp\space}% complute additonal indentation
                         \addtocontents{toc}{\protect\tocindent=\the\tocindent}% store in TOC
                         \addcontentsline{toc}{chapter}%
                                   {\@chapapp~\protect\numberline{\thechapter}#1}%
                    \else
                      \addcontentsline{toc}{chapter}{#1}%
                    \fi
                    \chaptermark{#1}%
                    \addtocontents{lof}{\protect\addvspace{10\p@}}%
                    \addtocontents{lot}{\protect\addvspace{10\p@}}%
                    \if@twocolumn
                      \@topnewpage[\@makechapterhead{#2}]%
                    \else
                      \@makechapterhead{#2}%
                      \@afterheading
                    \fi}

\renewcommand*\l@chapter[2]{%
  \ifnum \c@tocdepth >\m@ne
    \addpenalty{-\@highpenalty}%
    \vskip 1.0em \@plus\p@
    \setlength\@tempdima{1.5em}%
    \begingroup
      \parindent \z@ \rightskip \@pnumwidth
      \parfillskip -\@pnumwidth
      \leavevmode \bfseries
      \advance\leftskip\@tempdima
      \advance\leftskip\tocindent% only change
      \hskip -\leftskip
      #1\nobreak\hfil
      \nobreak\hb@xt@\@pnumwidth{\hss #2%
                                 \kern-\p@\kern\p@}\par
      \penalty\@highpenalty
    \endgroup
  \fi}

 \def\ps@plain{% first page of chapter
      \def\@oddhead{\normalfont\hfil\thepage}%
      \let\@evenhead\@oddhead
      \let\@oddfoot\@empty
      \let\@evenfoot\@empty}
 \makeatother

\thesistitle{Perturbations In Some Dark Energy Models} \iffalse( Your thesis title, 
\supervisor{Prof.\ Narayan Banerjee} \iffalse( Your supervisor's name, this is used in the title 
\examiner{} \iffalse( Your examiner's name, this is not currently used anywhere 
\degree{Doctor of Philosophy} \iffalse( Your degree name, this is used in the 
\author{Srijita Sinha} \iffalse( Your name, this is used in the title page and 
\addresses{Mohanpur 741246, West Bengal, India} % Your address, this is not currently used anywhere in the template, print it elsewhere with \addressname

\subject{Physical Sciences} % Your subject area, this is not currently used anywhere in the template, print it elsewhere with \subjectname
 % Keywords for your thesis, this is not currently used anywhere in the template, print it elsewhere with \keywordnames
\registration{13IP012} % Keywords for your thesis, this is not currently used anywhere in the template, print it elsewhere with \rollno
\university{Indian Institute of Science Education and Research Kolkata} % Your university's name and URL, this is used in the title page and abstract, print it elsewhere with \univname
\department{Department of Physical Sciences} % Your department's name and URL, this is used in the title page and abstract, print it elsewhere with \deptname
\def\DateSub{September 20, 2021}

\AtBeginDocument{
\hypersetup{pdftitle=\ttitle} % Set the PDF's title to your title
\hypersetup{pdfauthor=\authorname} % Set the PDF's author to your name
\hypersetup{pdfkeywords=\keywordnames} % Set the PDF's keywords to your keywords
}

\begin{document}

%{
%  \hypersetup{hidelinks}
%  \tableofcontents
%}

\frontmatter \iffalse( Use roman page numbering style (i, ii, iii, iv...) for 
the pre-content pages )\fi

\pagestyle{plain} \iffalse( Default to the plain heading style until the thesis 
style is called for the body content )\fi

\pagestyle{empty}

%-------------------------------------------------------------------------------
%	TITLE PAGE
%-------------------------------------------------------------------------------

\begin{titlepage}
\begin{center}
\vspace*{-0.05\textheight}
%{\scshape\LARGE \univname\par}\vspace{1.5cm} % University name
\textsc{\Large Doctoral Thesis}\\[0.55cm] % Thesis type
%
%\HRule \\[0.4cm] % Horizontal line
%{\fontsize{26pt}{1pt} \bfseries \ttitle \par}\vspace{0.3cm} % Thesis title
{\Huge \bfseries \ttitle \par}\vspace{0.35cm} % Thesis title
%\color{mdtRed}
\HRule \\[1.cm] % Horizontal line

\begin{center}
	\Large{By\\
    \textbf{\authorname} \\
	Roll No.: \rollno}\\
\vspace{0.05\linewidth}
    \large{\deptname\\\vspace{0.0cm}
    \univname\vspace{0.0cm}}
\end{center}
\vspace{0.05\linewidth}
\begin{center} 
	\large{\textbf{Supervisor:} \supname}
\end{center}
\vspace{1.2cm}
\centering
	\includegraphics[width=0.3\linewidth]{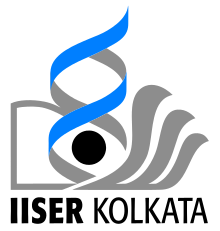}
	
\vspace{1.5cm}
\textit{ A thesis submitted in fulfilment of the requirements for 
the degree of\\ \degreename \ % University requirement text
in the \deptname \ at the\\ \univname}

\vspace{0.7cm}
{\large September, 2021}\\[3cm] % Date
\vfill
\end{center}
\end{titlepage}

%-------------------------------------------------------------------------------
%	DECLARATION PAGE
%-------------------------------------------------------------------------------
%
%%
\sloppy
\begin{declaration}
%\addchaptertocentry{\authorshipname} % Add the declaration to the table of contents
\begin{flushright}
Date: {\DateSub} \\
\end{flushright}
\noindent
I, Ms.\ \textbf{\authorname} Registration No.\ \textbf{\rollno} dated 
25/07/2013, a student of \deptname\ of the Integrated PhD Programme of the \univname\ (IISER Kolkata), hereby declare that this thesis is my own work and, to the best of my knowledge, it neither contains materials previously published or written by any other person, nor it has been submitted for any degree/diploma or any other academic award anywhere before.
\par 
I also declare that all copyrighted material incorporated into this thesis is in compliance with the Indian Copyright Act, 1957 (amended in 2012) and that I have received written permission from the copyright owners for my use of their work.
%I also declare that all copyrighted material incorporated into this thesis is in compliance with the Indian Copyright (Amendment) Act, 2012 and that I have received written permission from the copyright owners for my use of their work.
\par 
I hereby grant permission to IISER Kolkata to store the thesis in a database which can be accessed by others.

\vspace{2.5cm}
\begin{flushleft}
	\setul{-0.5ex}{1.pt}
	\ul{\mbox{\hspace{5.5cm}}}\\[0.5ex]
%	-----------------------------------\\
	{\bf \authorname}\\
	\deptname\\
	\univname\\
	\addressname
\end{flushleft}

\vspace{2.3cm}

\end{declaration}
\cleardoublepage

%\addcontentsline{toc}{chapter}{Certificate from the Supervisor}
\thispagestyle{plain}
\null\vfil
{\noindent\huge\bfseries Certificate from the Supervisor \par\vspace{10pt}}

\begin{flushright}
	Date: {\DateSub} \\
\end{flushright}
\noindent
\par
%\vspace{1.5cm}
{\doublespacing This is to certify that the thesis entitled \textbf{``\ttitle''} submitted by \textbf{\authorname} Registration No.\ \textbf{\rollno} dated 25/07/2013, a 
student of \deptname\ of the Integrated PhD 
Programme of \univname\ (IISER Kolkata), is based upon her own research work under my 
supervision. This is also to certify that neither the thesis nor any part of it 
has been submitted for any degree/diploma or any other academic award anywhere 
before. In my opinion, the thesis fulfils the requirement for the award of the 
degree of \degreename.}

\vspace{2.5cm}

\begin{flushleft}
\setul{-0.5ex}{1.pt}
\ul{\mbox{\hspace{5.5cm}}}\\[0.5ex]
%-----------------------------------\\
	{\bf \supname} \\
	Professor\\
	\deptname \\
	\univname\\
	\addressname
\end{flushleft}
\vspace{2.3cm}

\cleardoublepage

%-------------------------------------------------------------------------------
%	ACKNOWLEDGEMENTS
%-------------------------------------------------------------------------------

\sloppy
\doublespacing
%%
%%\iffalse
%%\begin{acknowledgements}
%%\addchaptertocentry{\acknowledgementname} 
%%% Add the acknowledgements to the table of contents
%%\end{acknowledgements}
%%\fi
%
%\addcontentsline{toc}{chapter}{Acknowledgements}
\thispagestyle{plain}
%\null\vfil
{\noindent\huge\bfseries Acknowledgements \par\vspace{15pt}}
First and foremost, I wish to express my sincere and deepest gratitude to my supervisor, Prof.\ Narayan Banerjee, for his immense patience, kindness, continuous support, enthusiastic guidance and encouragement throughout my research. He has always guided me with his vast knowledge, experience, and expertise and has constantly stimulated my interests while allowing me to pursue my ideas with complete freedom. I am indebted to him for his time, patience and effort and for believing in me even when I could not. It was a great pleasure to work under his supervision with countless interesting academic and non-academic discussions. I could not have wished for a better \emph{guru}.

I am grateful to Dr.\ Golam Mortuza Hossain, Prof.\ Rajesh Kumble Nayak, Prof.\ Dibyendu Nandi and Dr.\ Ananda Dasgupta for their support and valuable advice. 
I would also like to express my deepest gratitude towards Prof.\ Satyabrata Raj for his valuable and unconditional help and support regarding every hurdle I faced while using the computational facility during my research.

I would like to convey my sincere gratitude to Dr.\ Tuhin Ghosh (National Institute of Science Education and Research, Bhubaneswar, Odisha, India) for his invaluable help in learning \camb and \cosmomc and for his welcoming discussions and suggestions on cosmological data analysis. I am also grateful to my seniors Ankan Mukherjee and Supriya Pan for their valuable suggestions, patient and inexhaustible discussions on cosmological data analysis and Nandan Roy for his comprehensible explanations on dynamical systems analysis.

I want to thank my seniors Soumya Chakrabarti, Anushree Datta, 
Biswarup Ash,
Chiranjeeb Singha, 
Gopal Sardar, 
Nivedita Bhadra, 
Rafiqul Rahman, 
Santanu Tripathi, 
Souvik Pramanik 
Subhajit Barman and
Swati Sen 
for their help and advice. I am thankful to my friends and colleagues 
Ankit,
Anurag,
Sachin, 
Sajal, 
Sayak, 
Soumik, 
Souraj, 
Purba,
Shibendu and
Tanima 
for making my stay at IISER Kolkata a remarkable experience. 

I owe a debt of gratitude to my dear old friend and colleague, Avijit Chowdhury, for being there unwaveringly with his kindness, understanding, support, motivation, and encouragement through thick and thin of my life.

I would also like to extend my deepest thanks to the Department of Physical Sciences office, IISER Kolkata, and Assistant Librarian, IISER-K Library, Dr Siladitya Jana, for their kind help and cooperation. I would like to thank IISER Kolkata for the use of ``Dirac Supercomputing Facility'' and for the use of the software \texttt{Mathematica}.

Lastly, my deep and sincere gratitude to my parents for never losing confidence in me and always being there as the last resort I will run to at any time of my life's journey. Without their love, support and encouragement, this thesis would not have been possible. I want to thank my `little' brother Amitariddhi Sinha, who is no longer little and has always been there for me with his love and support as an elder brother.

%-------------------------------------------------------------------------------
%	DEDICATION
%-------------------------------------------------------------------------------

\dedicatory{To my parents.}

%-------------------------------------------------------------------------------
%	ABSTRACT PAGE
%-------------------------------------------------------------------------------
\sloppy
\cleardoublepage
%\addchaptertocentry{List of Publications}
\thispagestyle{plain}
%\null\vfil
{\noindent\huge\bfseries List of Publications \par \vspace{15pt}}

\onehalfspacing

\paragraph*{Research works included in the thesis:}
  \begin{numitemise}%{A}
  
        \item {\textbf{Srijita Sinha} and Narayan Banerjee, ``Density perturbation in the models reconstructed from jerk parameter'', Gen.\ Rel.\ Grav.,  \textbf{50} 67 (2018), arXiv:1805.02824 (Chapter \ref{chap3:grg}).}
        
        \item {\textbf{Srijita Sinha} and Narayan Banerjee, ``Density perturbation in an interacting holographic dark energy model'', Eur. Phys. J. Plus, \textbf{135} 779 (2020), arXiv:1911.06520 (Chapter \ref{chap4:epjp}).}

     	\item {\textbf{Srijita Sinha} and Narayan Banerjee, ``Perturbations in a scalar field model with virtues of $\Lambda$CDM'', JCAP \textbf{04} 060 (2021), arXiv:2010.02651 (Chapter \ref{chap5:jcap}).}

        \item {\textbf{Srijita Sinha}, ``Differentiating dark interactions with perturbation'', Phys.\ Rev.\ D \textbf{103}, 123547 (2021), arXiv:2101.08959 (Chapter \ref{chap6:prd}).}
  	
\end{numitemise}

%\paragraph*{Publications not included in this thesis:}
%
%\begin{numitemise}
%
%    \item {\textbf{Avijit Chowdhury} and Narayan Banerjee, “Echoes from a Singularity”, Phys.\ Rev.\ D \textbf{102}, 124051 (2020); arXiv:2006.16522 [gr-qc] }
%
%
%\end{numitemise}

\doublespacing
\cleardoublepage
%\addcontentsline{toc}{chapter}{Abstract}
\thispagestyle{plain}
%\null\vfil
{\noindent\huge\bfseries Abstract \par\vspace{15pt}}
Dark energy is the candidate that can produce effective negative pressure and make the galaxies and galaxy clusters move away from each other in an accelerated way. The structures of the Universe have evolved from some initial primordial fluctuations and depend on the background dynamics of different components of the Universe like dark matter, dark energy and others. The motivation of this thesis is to investigate how some of the dark energy models manifest themselves in the formation of the structures in the Universe. 

In this work, four different types of dark energy models are discussed in Chapter \ref{chap3:grg} to Chapter \ref{chap6:prd}. The dark energy models considered in Chapter \ref{chap3:grg} are reconstructed from a kinematical quantity jerk parameter, which is the third-order derivative of the scale factor. The reconstructed models are such that dark energy has an energy transfer with dark matter in the first case, and they conserve together, while in the second case, both dark matter and dark energy are conserved independently. Here the evolution of dark matter perturbation has been studied, and it has been found that non-interacting models perform better in the context of structure formation.

Holographic dark energy with future event horizon as the characteristic Infra-Red cut-off is another model considered in the present thesis. In this case, dark energy interacts with dark matter. Here perturbations of both dark matter and dark energy are studied, and it is found that the evolution of perturbation is consistent with the requirement of structure formation. Furthermore, the effective sound speed of dark energy perturbation is considered to be a parameter, and it has been found that dark energy can cluster similar to dark matter in the absence of effective sound speed.

A scalar field with a suitable potential is considered as the next model. The potential is constructed such that at the early epoch, the scalar field tracks the dominant background component and at the later epoch, the scalar field behaves like a cosmological constant ($\Lambda$) and drives the recent cosmic acceleration. This cosmological constant like behaviour is ensured for any values of the model parameters. Perturbations in dark matter and dark energy have been studied, and it is concluded that the evolution of dark matter perturbations is similar to the concordance \lcdm model.
 
Lastly, an interacting dark matter and dark energy model in which the coupling between them has an ``evolving'' parameter is considered. This coupling parameter is so chosen that the interaction is either dominant in the early epoch or at a later time. It is found that an early interaction describes the evolution of the perturbations better than a late interaction. In this case, the models are also constrained with recent observational datasets, and Bayesian evidence is computed.

\cleardoublepage

%-------------------------------------------------------------------------------
%	LIST OF CONTENTS/FIGURES/TABLES PAGES
%-------------------------------------------------------------------------------
\sloppy
\tableofcontents % Prints the main table of contents

\listoffigures % Prints the list of figures

\listoftables % Prints the list of tables

%-------------------------------------------------------------------------------
%	ABBREVIATIONS
%-------------------------------------------------------------------------------

%%%
\sloppy
\printnomenclature %[5cm]
\addcontentsline{toc}{chapter}{List of Abbreviations}

%-------------------------------------------------------------------------------
%	PHYSICAL CONSTANTS/OTHER DEFINITIONS
%-------------------------------------------------------------------------------
%
%\begin{constants}{lr@{${}={}$}l} % The list of physical constants is a three column table
%
%% The \SI{}{} command is provided by the siunitx package, see its documentation for instructions on how to use it
%
%Speed of Light & $c_{0}$ & \SI{2.99792458e8}{\meter\per\second} (exact)\\
%%Constant Name & $Symbol$ & $Constant Value$ with units\\
%
%\end{constants}

%-------------------------------------------------------------------------------
%	SYMBOLS
%-------------------------------------------------------------------------------

%\begin{symbols}{lll} % Include a list of Symbols (a three column table)
%
%$a$ & distance & \si{\meter} \\
%$P$ & power & \si{\watt} (\si{\joule\per\second}) \\
%%Symbol & Name & Unit \\
%
%\addlinespace % Gap to separate the Roman symbols from the Greek
%
%$\omega$ & angular frequency & \si{\radian} \\
%
%\end{symbols}

%-------------------------------------------------------------------------------
%	THESIS CONTENT - CHAPTERS
%-------------------------------------------------------------------------------

\mainmatter % Begin numeric (1,2,3...) page numbering
\pagestyle{thesis} % Return the page headers back to the "thesis" style
\sloppy
% Chapter 1
\chapter{Introduction } % Main chapter titleto Cosmology
\label{chapter1}
\chaptermark{Introduction}
For the last two decades, the most bewildering observation in cosmology is that the galaxies and galaxy clusters are moving away from one other in an accelerated way. Explaining this bizarre behaviour needs an agent that leads to an effective repulsive gravity strong enough to overcome the gravitational attraction of normal matter and make matter move away from each other at an increasing speed. An enormous amount of cosmological models have been put forward to explain the repulsive nature of gravity including modification of General Relativity as the theory of gravity, but arguably the most widely accepted one is the presence of an exotic component named \emph{dark energy}. However, one of the essential aspects of any cosmological model is to produce large scale structures like the galaxies, galaxy clusters in the Universe as we see today. These large scale structures have grown from some initial density fluctuations and depends on the background evolution. Different cosmological models will have different imprints on the large scale structures. Thus structure formation helps in breaking the degeneracy between different cosmological models and can identify a more suitable one. The motivation of the thesis is to investigate if some dark energy models can provide a congenial environment for structure formation.

In the remainder of this chapter, we will discuss the background dynamics of the standard cosmological model.

%%%% Einstein's general theory of relativity %%%%%
\section{Cosmology: A Brief History}\label{sec1:cos}  
The human mind, has always been fascinated by the night sky. The inquiring mind could not help but wonder what surrounds us and started the quest to know the unknown --- the \emph{cosmos}. The study of the cosmos or Universe from its origin to the future, its evolution, is called \emph{Cosmology}. Though the study of the Universe dates back to the 16th century BC to 12th century BC, and then Copernicus, Galileo, Kepler, Newton in the 16th - 17th century AD, the dawn of modern cosmology with a mathematical model was marked by Einstein's development of the General Theory of Relativity in 1915~\cite{kragh2017cosmology}. The General Theory of Relativity\footnote{English translations are available at: \href{https://einsteinpapers.press.princeton.edu/vol6-trans/}{https://einsteinpapers.press.princeton.edu/vol6-trans/}}~\cite{einstein1914grformalintro, einstein1915gr1, einstein1915gr2, einstein1915grfieldeq} is the theory of gravity that unites matter distribution with the geometry of spacetime. The gravitational attraction among the matter distribution is determined by curving the spacetime around the matter distribution and is described by the famous Einsteins field equations. In 1917, Einstein himself provided the first exact cosmological solution to the Einsteins field equations after modifying the equations for a static Universe~\cite{einstein1917cosmology}. In the same year, de Sitter provided another exact solution for an empty Universe~\cite{desitter1917mnras3}.

In 1922, Friedmann presented new solutions to the Einsteins field equations, and for the first time, predicted the possibility of an expanding (or contracting) Universe~\cite{friedman1922, friedman1991grg}. In 1927, Lema\^itre independently derived the expanding solution and pointed out that galaxies (``\emph{nubel\ae}'' then) would be moving away from one another due to cosmic expansion. At that time, the dataset measuring the galaxy recession velocities by Slipher was the only dataset available~\cite{slipher1913}. It was Robertson who, in 1928, attributed the galaxy velocities to overall cosmic expansion and predicted the linear velocity distance relationship. A year later, in 1929, Hubble observed the same linear relation combining the extragalactic distance measurements from his previous work and the measurements from Slipher~\cite{slipher1917,hubble1929,hubble1931}. The same year saw the homogenous and isotropic solution for expanding Universe by Robertson and Walker. The solution presented by Friedmann-Lema\^itre-Robertson-Walker (FLRW) \nomenclature{FLRW}{Friedmann-Lema\^itre-Robertson-Walker} laid the cornerstone of the Standard Cosmological Model.

The following decades witnessed significant developments in both the theoretical and observational fronts. Many important ideas were introduced and discussed, of which the Hot Big Bang model supported by Lema\^itre, Gamow~\cite{lemaitre1931nature, eddington1931nature, gamow1946pr, gamow1948nature} and the Steady State model supported by Bondi, Gold, Hoyle, Narlikar~\cite{bondi1948mnras, hoyle1948mnras, hoyle1964rspa} were the competing models. The observation of Hubble that the Universe was expanding indicated that it started from \emph{a singularity} called a ``Big Bang''\footnote{The term was coined later by Hoyle in 1950}, and all matter was created at thereafter. In 1933, Milne formulated the \emph{perfect cosmological principle} --- that the Universe would look identical at all place and at all time~\cite{milne1934qjm}, which eventually led to the development of the Steady State model. The one-time singularity of the FLRW model was unaesthetic for many favouring the Steady State model, while the success story of the cosmic nucleosynthesis~\cite{gamow1946pr, alpher1948pr1, alpher1948nature, alpher1948pr2} clearly favoured the Hot Big Bang model creating a caustic dispute on which model describes the Universe best. The discovery of cosmic microwave background (CMB) radiation by Penzias and Wilson in 1965 settled the debate once and for all, announcing the Hot Big Bang model as the winner~\cite{penzias1965apj,peebles1999apj}. The cosmic microwave background radiation is the \emph{relic of the radiation} after the hot and dense Universe cooled as it expanded~\cite{peebles1965apj1}. The following year, 1966, sealed the triumph of the Hot Big Bang model by Peebles with the prediction of currect Helium abundance, produced as a result of early Universe Big Bang Nucleosynthesis~\cite{peebles1966prl,peebles1966apj} and by Hawking and Ellis with the \emph{singularity theorem}, stating that a singularity is inevitable in the Universe within a finite past~\cite{hawking1966prl}. Meanwhile, in 1933 astronomer Fritz Zwicky inferred the existence of non-luminous, non-baryonic \emph{dark matter} from the rotation curves of the Coma Cluster~\cite{andernach2017zwiky}.  

The detection of the relic radiation presented the most outstanding puzzle of that era --- the origin of galaxies. The remarkable uniform temperature of the CMB across the sky indicated that the early Universe was smooth, whereas the Universe being dotted with galaxies would appear clumpy at a smaller scale. Long before the discovery of CMB and the establishment of Hot Big Bang model, James Jeans, in 1902, put forth the Newtonian theory of \emph{gravitational instability} in a non-expanding incompressible fluid medium~\cite{jeans1902ptrs} and Gamow and Teller in 1939, using Jeans length in the context of the expanding Universe made early efforts to explain the formation of galaxies~\cite{gamow1939pr}. Later, in 1946, Lifshitz formulated the relativistic treatment of gravitational instabilities in an expanding Universe and set the groundwork for the formation of cosmic structures~\cite{lifshitz1946, lifshitz2017grg}. With his work on Big Bang Nucleosynthesis (BBN)\nomenclature{BBN}{Big Bang Nucleosynthesis}, Gamow showed  in 1948, that radiation density should be less than matter density for ``gravitational instability'' to take over~\cite{gamow1948pr}. The year of detection of the CMB radiation, 1965,  witnessed another pioneering development --- Peebles explained how the blackbody spectrum of the remnant \emph{fireball} radiation would affect the formation of galaxies~\cite{peebles1965apj2}. In 1967, Sachs and Wolfe first predicted the presence of anisotropies in the background radiation and showed that cosmic structures were formed from those primordial inhomogeneities~\cite{sachs1967apj}. The following years focused on understanding how the primordial anisotropies lead to the large scale structures of the Universe. Notable contributions came from Doroshkevich, Zel'dovich, Novikov, Sakharov, Silk, Sunyaev, Peebles and Yu~\cite{doroshkevich1965zetf, sakharov1966sjetp, doroshkevich1967az, silk1968apj, zeldovich1970aa, sunyaev1970ass, peebles1970apj}. By the end of that decade in 1970, the existence of dark matter was confirmed by Rubin, Ford and Freeman~\cite{rubin1970apj, freeman1970apj}.

Though the Big Bang model was gaining success in connection with structure formation, its success was challenged once again when it could not address puzzles like why the CMB temperature in the sky is so evenly distributed in all directions even for causally disconnected regions, why the Universe is spatially almost flat, why there is no abundance of magnetic monopoles, where the primordial fluctuations came from.  In the late 1970s and early 80s, the theory of inflation came to its rescue and became a landmark in the history of cosmology. The cosmic \emph{inflation} or exponential expansion of the Universe at a very early epoch came with the revolutionary works of Starobinsky, Sato, Guth, Linde and Steinhardt~\cite{starobinskii1979zpr, starobinskii1979jetpl, guth1981prd, sato1981plb, sato1981mnras, linde1982plb, albrecht1982prl, starobinsky1982plb}. However, it was Guth who employed the inflationary models to solve the ``horizon problem'', ``flatness problem'' and ``monopole problem'' in 1981~\cite{guth1981prd}. Inflation also provided quantum fluctuations as the seeds for the growth of large scale structures in the Universe.

The subsequent years underwent a journey of remarkable theoretical and observational advancements. Though the study of cosmic structures using simulations dates back several decades~\cite{gott1977araa}, the cosmological simulations with high resolution became a reality with the Particle-Particle-Particle-Mesh ($P^{3}M$) simulation code by Efstathiou, Davis, Frenk and White~\cite{davis1985apj, white1987apj, white1987nature}. Further developments on the N-body simulations of structure formation opened avenues for studying more realistic models of the Universe. Mapping of the Universe with different surveys dominated the period. Redshift surveys like the Center for Astrophysics (CfA)\nomenclature{CfA}{Center for Astrophysics} Redshift Survey~\cite{huchra1983apj, falco1999pasp} comprising a ``slice of the sky'' passing through the Coma cluster being one of them. The slice named \emph{de Lapparent Slice} revealed clusters of galaxies, filaments and voids on an enormous scale ($\sim 1100$ galaxies in a $6^{\circ}$ wide and $130^{\circ}$ long strip on the sky)~\cite{lapparent1986apjl}. The primary filament of size $60 \times 170\,\mpc$, observed in the slice is called the \emph{Great Wall}~\cite{geller1989science}. Galaxy catalogue surveys like the Automatic Plate Measuring (APM) survey\nomenclature{APM}{Automatic Plate Measuring}~\cite{maddox1990mnras} catalogued millions of galaxies to unprecedented depths over a large area of the sky, correlation function consistent with previous measurements. The Cosmic Background Explorer (COBE)\nomenclature{COBE}{Cosmic Background Explorer} satellite launched by NASA in 1989 not only presented concrete evidence of the perfect blackbody spectrum~\cite{smoot1992apj} of the CMB of temperature $2.726\pm 0.01$ K but made a breakthrough discovery of the anisotropies of $10^{-5}$ in the CMB temperature~\cite{mather1999apj} as predicted by Sachs and Wolfe and measured its amplitude accurately by 1992. Another ambitious redshift survey of the age was the Two-degree-Field Galaxy Redshift Survey (2dFGRS)\nomenclature{2dFGRS}{Two-degree-Field Galaxy Redshift Survey} that mapped more than $200,000$ galaxies to a depth of around 2.5 billion light-years and became the largest redshift survey around 1998. In 1998, two groups, Supernova Cosmology Project and High-Z Supernova Search Team, measuring the luminosity distance of the Type Ia supernova, observed the dimming of the supernovae. The observation that the distant supernovae, objects of fixed intrinsic brightness, appear to be fainter than expected in an empty Universe led to the conclusion that the light sources are moving away from each other at a faster rate, possibly with an \emph{acceleration}~\cite{riess1998anj,schmidt1998apj,perlmutter1999apj, scolnic2018apj}. The discovery that the Universe is expanding with an acceleration brought back the long lost idea of cosmological constant and has set a new milestone in the history of cosmology. Subsequently, with the onset of the millennium commenced the era of precision cosmology, with a flurry of high precision observational data from surveys like Sloan Digital Sky Survey (SDSS)\nomenclature{SDSS}{Sloan Digital Sky Survey}, Wilkinson Microwave Anisotropy Probe (WMAP)\nomenclature{WMAP}{Wilkinson Microwave Anisotropy Probe}, \Planck, Dark Energy Survey (DES)\nomenclature{DES}{Dark Energy Survey}~\cite{eisenstein1998apj, planck2018cp, partphys2018prd, reid2010mnras, alam2017sdss3, descol2018prd1, descol2018prd2, descol2019prl, alam2020sdss4} and the notion of accelerated expansion of the Universe is quite firmly established.
%
%
%%%%%%%%%%%%%%%%%%%%%%%%%%%%%%%%%%%%%%%%%%%%%%%%%%%%%

\section{The Standard Model}
The standard model or the Big Bang model is based on the \emph{cosmological principle}, which states that the Universe is spatially homogeneous and isotropic on sufficiently large scales. The cosmological principle gives up the homogeneity in time, unlike the perfect cosmological principle. Large-scale structures like galaxies and galaxy clusters indicate that the Universe is inhomogeneous and anisotropic at smaller scales. The Universe is inhomogeneous at length scales smaller than $100$ Mpc to $300$ Mpc~\cite{yadav2005mnras, clowes2013mnras}\footnote{1 pc = $3.0857 \times 10^{16}$ m = 3.26 lightyears}, while it is anisotropic to one part in $10^{-5}$ as observed from the cosmic microwave background radiation in the sky~\cite{hu2002araa}.

On large scales, the Universe is described by a homogeneous and isotropic Friedmann-Lema\^itre-Robertson-Walker (FLRW) metric, given as
\begin{equation}\label{eq1:metric}
ds^2= \gmn d x^{\mu} dx^{\nu} = -d t^{2} + a^2(t) \left[\frac{dr^{2}}{1-K r^{2}}+ r^{2}\paren*{d \theta^{2}+ \sin^{2}\theta\,d \phi^{2}}\right],
\end{equation}
where $\gmn$ is the metric tensor describing the space-time geometry. Hereafter, the Greek indices $\mu , \nu \dotsc $ denote the space-time coordinates or 4-coordinates while the Latin indices $i ,j \dotsc$ denote the coordinates in the spatial hypersurface or 3-coordinates and speed of light $c$ is considered to be unity. In Eq.\ (\ref{eq1:metric}), the coordinate $t$ is \emph{cosmic time} or \emph{proper time} as measured by a comoving observer, $(r, \theta, \phi)$ are comoving spherical polar coordinates and $a(t)$ is the \emph{scale factor}. The scale factor accounts for the time evolution of the physical distance, $D$, between two comoving spatial coordinates as, $D= a\Delta x^{i}$. The constant $K$ is the \emph{curvature parameter} that determines curvature of the spatial geometry. If $K =1$, the spatial geometry is a 3-sphere (closed), while for $K = 0$, the spatial geometry is an infinite plane (flat) and for $K =-1$, the spatial geometry is a 3-hyperboloid (open). Recent observations like WMAP and \Planck satellite suggest that the space section is very close to flat~\cite{komatsu2011apjs, bennett2013apjs, planck2015cp} and hence, for the rest of the work, the Universe is considered to spatially flat, homogeneous and isotropic, given by,
\begin{equation}
ds^2=-d t^{2} + a^2(t) \paren*{dx^{2}+ d y^{2} + d z^{2}},\label{eq1:metric-cart}
\end{equation}
in cartesian coordinates.

Another crucial assumption of the standard model is that the Universe in all length scale is described by Einstein's Theory of General Relativity defined by the Einstein-Hilbert action
\begin{equation}
\mathcal{S} = \int \sqrt{-g}\paren*{\frac{1}{16\pi G_{N}} R + \mathcal{L}_{m}} d^{4}x,\label{eq1:action}
\end{equation}
where $G_N$ is the Newtonian gravitational constant, $g \equiv \mbox{det}\paren*{\gmn}$ is the determinant of the space-time metric tensor and the Lagrangian density $\mathcal{L}$ of the gravitational sector is the Ricci scalar $R = R^{\mu\nu}\gmn$ obtained by contracting the Ricci tensor $R^{\mu\nu}$ while $\mathcal{L}_{m}$ is the Lagrangian density of matter sector.

The Einsteins equations, connecting the geometry and the matter content of the Universe, is arrived at by varying the action $\mathcal{S}$ (Eq.\ (\ref{eq1:action})) with respect to $\gmn$ as
\begin{equation}\label{eq1:Enstein1}
G^{\mu}{}_{\nu} \equiv R^{\mu}{}_{\nu} -\frac{1}{2} R g^{\mu}{}_{\nu} =  8 \pi G_{N} T^{\mu}{}_{\nu},
\end{equation}
where $G^{\mu}{}_{\nu}$ is the Einstein tensor and $T^{\mu}{}_{\nu}$ is the total energy-momentum tensor of the matter and is defined as 
\begin{equation}
\Tmn = -2 \frac{\partial \mathcal{L}_{m}}{\partial \gmn} + \delta^{\mu}{}_{\nu} \mathcal{L}_{m}.
\end{equation}
It is considered that the Universe is filled with different \emph{fluid} components labelled as fluid `A', each with an energy-momentum tensor $T^\mu_{\left(A\right)\nu}$ such that $\Tmn= \sum_A T^\mu_{\left(A\right)\nu}$. The energy-momentum tensor for fluid `A' is given by,
\begin{equation}\label{eq1:stress-imperf}
T^\mu_{\left(A\right) \nu} = \paren*{\rA + \pA}u^{\mu}_{\paren{A}}u_{\nu\paren{A}} + \pA g^{\mu}{}_{\nu} + \pi^{\mu}{}_{\nu\paren{A}}+q_{\mu\paren{A}}u^{\nu}_{\left(A\right)}+q_{\nu\paren{A}}u^{\mu}_{\left(A\right)},
\end{equation}
where $\rA$ is the \emph{energy density}, $\pA$ is the \emph{pressure}, $u_{\mu\paren{A}}$ is the \emph{comoving 4-velocity}, $\pi^{\mu}{}_{\nu\paren{A}}$ is the \emph{anisotropic stress} (with $\pi^{\mu}{}_{\mu\paren{A}} = 0$) and $q_{\mu\paren{A}}$ is the \emph{momentum density} or the spacelike heat flux vector such that $q_{\mu\paren{A}}u^{\mu}_{\paren{A}} = 0$. In absence of any  anisotropic stress and heat dissipation, the fluid is said to be a homogeneous, isotropic \emph{perfect fluid} described by the stress-energy tensor,
\begin{equation}
T^\mu_{\left(A\right) \nu}= \left(\rA + \pA \right)u^{\mu}_{\paren{A}}u_{\nu\paren{A}} +  p_A  \delta^\mu{}_{\nu}.\label{eq1:stress-perf}
\end{equation}

Using the metric given by Eq.\ (\ref{eq1:metric-cart}) and the stress-energy tensor given by Eq.\ (\ref{eq1:stress-perf}), the field equations (\ref{eq1:Enstein1}) are written as
\begin{eqnarray}
3 H^2 &=& -8 \pi G_N\,\sum_{A}\rA \label{eq1:fd1},\\ 
2H^2+ 3\dot{H} &=& 8 \pi G_N\,\sum_{A} \pA, \label{eq1:fd2} 
\end{eqnarray}
where, $H(t)= \frac{\dot{a}}{a}$ is the \emph{Hubble parameter} and an overhead dot $(\dot{})$ denotes the derivative with respect to time. The Hubble parameter relates how fast two distant points are moving away from each other to their distance through \emph{Hubble's law},
\begin{equation}
V \simeq H D.\label{eq1:Hubbles-law}
\end{equation}
In Hubble's law given by Eq.\ (\ref{eq1:Hubbles-law}), $V = \dot{a}\Delta x^{i}$, is the recession velocity of distant galaxies. A spatially flat geometry demands the Universe to have some critical amount of energy density, called the \emph{critical density}, $\rho_{\scriptsize\mbox{crit}} = 3\,H^2 /8 \pi G_N$. If the energy density the Universe is greater than the critical value, the geometry is \emph{closed} ($K=-1$), whereas a smaller density leads to an \emph{open} geometry ($K=+1$).

In absence of any non-gravitational interaction among the different components, from the \emph{contracted Bianchi identity}, $G^\mu{}_{\nu;\,\mu} = 0$, the energy conservation equation for the fluid `A' follows as 
\begin{equation}\label{eq1:energy-cons}
\dot{\rho}_{A}+ 3 H \paren*{\rA + \pA} = 0.
\end{equation}
It must be noted that for a perfect fluid there is no momentum transfer. 

Equation (\ref{eq1:energy-cons}) can be obtained from Eqs.\ (\ref{eq1:fd1}) and (\ref{eq1:fd2}) as a consequence of Bianchi identities, thus does not add to the number of independent equations.

For a barotropic fluid the pressure, $\pA$ is related to the energy density $\rA$ by an equation of state as
\begin{equation}
\pA = \wA \rA.\label{eq1:eos}
\end{equation}
In the equation (\ref{eq1:eos}), $\wA$ is called the \emph{equation of state} (EoS) parameter of the fluid. For any EoS parameter, $\wA\paren*{a}$, Eq.\ (\ref{eq1:energy-cons}) can be integrated as
\begin{equation}
\rA = \rho_{A0}\, e^{-3\int\left[1+\wA\paren*{a}\right] \frac{da}{a}}.\label{eq1:rhoA}
\end{equation}
Here, $\rho_{A0}$ is the value of the energy density at the present epoch, $a=a_{0}$. It is always convenient to express the results in terms of dimensionless quantities. Hence, two important dimensionless quantities, namely \emph{density parameter} and \emph{cosmological redshift} are defined below.
\paragraph*{Density Parameter :} Density parameter is defined as the fractional energy density of the Universe as $\Omega_{A} = \frac{\rho_{A}}{3\,H^2 /8 \pi G_N}$, such that for a spatially flat Universe $\sum_{A} \Omega_{A} = 1$.
%
%\subsection*{Cosmological Redshift}
\paragraph*{Cosmological Redshift :} Cosmological redshift or redshift is the fractional change in wavelength of a distant luminous source due to the expansion of the spacetime. If $\lambda_{e}$ is the emitted wavelength and $\lambda_{0}$ is the observed wavelength at the present epoch, then redshift in wavelength as observed by an observer at the present epoch is given as,
\begin{equation}
z = \frac{\lambda_{0} -\lambda_{e}}{\lambda_{e}}.
\end{equation}
Since, the wavelength emitted by the source is stretched as $\lambda_{e} \propto a$, redshift, $z$ is related to the scale factor $a$ at the epoch of emission as,
\begin{equation} 
1+z = \frac{a_{0}}{a}.
\end{equation}
It should be noted that in all the subsequent discussion, the scale factor, $a$, is scaled such that its present value, $a_{0}=1$.
%%%%%%%%%%%%%%%%%%%%%%%%%%%%%%%%%%%%%%%%%%%%%%%%%%%%%

\section{Need For An Exotic Component}\label{sec1:acc}
The contents of the Universe can be broadly divided into \emph{non-relativistic} and \emph{relativistic} fluids. For non-relativistic fluid, pressure due to the thermal energy is negligible compared to the rest-mass energy, and the EoS parameter can be assumed to be $\wA = 0$. For relativistic fluid, the pressure due to the thermal energy dominates the energy density, and the EoS is written as $\wA = 1/3$. Hence, non-relativistic fluid corresponds to pressure-less (cold) matter or \emph{dust}, while relativistic fluid corresponds to radiation. Thus for constant EoS, Eq.\ (\ref{eq1:rhoA}) can be readily written as
\begin{equation}
\rA = \rho_{A0}\,\paren*{\frac{a}{a_{0}}}^{-3\paren*{1+\wA}}.\label{eq1:rhoA-const}
\end{equation}
Pressure-less matter ($m$) in the Universe constitutes of baryons ($b$), cold dark matter ($c$) with $w_{b} = w_{c} = 0$, whereas radiation ($r$) constitutes of photons ($\gamma$) and neutrinos ($\nu$) with $w_{\gamma}=w_{\nu}=1/3$. For a matter dominated Universe, the matter density evolves as
\begin{equation}
\rdm = \rho_{m0}\,\paren*{\frac{a}{a_{0}}}^{-3},\label{eq1:rhom}
\end{equation}
while for a radiation dominated Universe, the radiation energy density evolves as
\begin{equation}
\rho_{r} = \rho_{r0}\,\paren*{\frac{a}{a_{0}}}^{-4}.\label{eq1:rhor}
\end{equation}

The dimming of the high redshift supernovae~\cite{perlmutter1999apj, riess1998anj, schmidt1998apj,scolnic2018apj} predicts that the Universe is expanding with an \emph{acceleration} which immediately suggests the existence of an exotic matter. Rewriting the Friedmann \emph{acceleration equation} (\ref{eq1:fd1}) and (\ref{eq1:fd2}) as,
\begin{equation}
\frac{\ddot{a}}{a} = -\frac{4 \pi G_N}{3} \sum_{A}\paren*{\rA + 3\pA},\label{eq1:acce-eq}
\end{equation}
 it can be directly seen that for the Universe to expand with acceleration ($\ddot{a}>0$); the \emph{strong energy condition} ($\sum_{A}\paren*{\rA + 3\pA} > 0$) must be violated. Gravity being an attractive force, cannot make matter move away from each other in accelerated way; hence there must be an \emph{exotic component} called ``dark energy'' (DE) that enables the Universe to overcome the attractive nature of gravity and make matter move away from each other at a faster rate. Mathematically, the pressure ($\pA$) of the exotic component must be sufficiently negative, making its ratio with the energy density ($\rA$) at least less than $-\frac{1}{3}$ ($\pA/\rA = \wA < -1/3$). A non-zero \emph{cosmological constant}, $\Lambda$, is undoubtedly one of the preferred choices~\cite{sahni2000ijmpd, carroll2001lrr, paddy2003pr, peebles2003rmp, frieman2008araa, amendola2010prl, mehrabi2018prd, planck2018cp}. For a very recent review on accelerated expansion, we refer to the work of Haridasu \etal~\cite{haridasu} and also Rubin and Hayden~\cite{rubin2016apj}.  The story so far is summarised in a recent work by Brax~\cite{brax2018rpp}.
% 
%%%%%%%%%%%%%%%%%%%%%%%%%%%%%%%%%%%%%%%%%%%%%%%%%%%%%

\section{Cosmological Constant}
The \emph{cosmological constant} ($\Lambda$), introduced by Einstein for a static cosmological solution in 1917, was discarded for expanding cosmological solution and Hubble's discovery in 1929. Later in 1981, it returned to explain the exponential expansion in the context of inflation, and finally came on-stage to explain the repulsive gravity in 1998. 

Theoretically, a cosmological constant ($\Lambda = 8\pi G_{N} \rho_{\scriptsize\mbox{vac}}$) is predicted to arise from the zero-point vacuum fluctuations of quantum fields, and the energy-momentum tensor of the cosmological constant is equivalent to that of the vacuum energy. The vacuum energy-momentum tensor is given as,
\begin{equation}
T^\mu_{\left(\scriptsize\mbox{vac}\right) \nu} = -\rho_{\scriptsize\mbox{vac}}\,g^{\mu}{}_{\nu},\label{eq1:em-vac}
\end{equation}
which represents a perfect fluid with a negative pressure such that,
\begin{equation}
p_{\scriptsize\mbox{vac}} =- \rho_{\scriptsize\mbox{vac}},
\end{equation}
or equivalently,
\begin{equation}
p_{\Lambda} = -\rho_{\Lambda} \equiv -\frac{\Lambda}{8\pi G_{N}},
\end{equation}
with an EoS parameter, $w_{\Lambda} = -1$. Hence a positive cosmological constant ($\Lambda>0$) with $\paren*{\rho_{\Lambda} + 3 p_{\Lambda}} =- 2 \rho_{\Lambda}$ can lead to repulsive gravity. In an expanding Universe with a decreasing $\rdm$ and a constant $\rho_{\Lambda}$, matter will accelerate away from each other when $\rdm<2\rho_{\Lambda}$, in agreement with the observation from the Type Ia Supernovae (SNe Ia) measurements~\cite{riess1998anj, schmidt1998apj,perlmutter1999apj,scolnic2018apj}. 

%%%%%%%%%%%%%%%%%%%%%%%%%%%%%%%%%%%%%%%%%%%%%%%%%%%%%
\subsection{Shortcomings Of Cosmological Constant}
%
%\%\%\%\%\%\%\%\%\%\%\%\%\%\%\%\%\%\%\%\%\%\%\%\%\%\%\%\%\%\%\%\%\\%
The cosmological constant is plagued with quite a few problems. Recent observational data~\cite{planck2018cp} predicts the value of $\Lambda$ to be 123 orders of magnitude smaller than the value estimated from Quantum Field Theory calculation, in different theories. This huge discrepancy between the observationally required value and the theoretically predicted one is called the \emph{cosmological constant problem}. The discovery of the recent accelerated expansion confirmed the existence of a small cosmological constant and has worsened the problem. To match the observational value of $\Lambda$ with the theoretical one, extreme fine-tuning is required.

Recent observations confirm that dark matter constitutes 27\% of the energy budget in the Universe, while cosmological constant as dark energy constitutes 68\% of the energy budget. When the Universe was small, dark matter dominated the energy content, but as the Universe expands, the energy density of matter decreases while the energy density of the cosmological constant remains constant throughout the evolution. The \emph{coincidence problem} is why both dark matter and dark energy have comparable energy densities precisely at the present epoch? Had it been before, the Universe would have accelerated before the formation of galaxies, whereas if it were to happen in future, observations would not have shown accelerated expansion at the current epoch. For detailed review on the shortcomings of the cosmological constant we refer to~\cite{weinberg1989rmp, sahni2000ijmpd, carroll2001lrr, sahni2002cqg, paddy2003pr, martin2012crp, steinhardt2003jstor, velten2014epjc}.
%
%%%%%%%%%%%%%%%%%%%%%%%%%%%%%%%%%%%%%%%%%%%%%%%%%%%%%
\section{Possible Alternatives}
In the era of precision cosmology, the cosmological constant with cold dark matter ($\Lambda$CDM) gives quite an accurate description of the evolution of the Universe, but it is still plagued with the fine-tuning problem and the cosmic coincidence problem. These problems in the $\Lambda$CDM model have forced one to look for other candidates that can drive the acceleration or even go beyond General Relativity at large scales. 

\subsection{Quintessence Models}\label{sec1:sfm}

A scalar field rolling down a slowly varying potential, introduced by Ratra \& Peebles~\cite{ratra1988prd}, Peebles \& Ratra~\cite{peebles1988apjl} and by Wetterich~\cite{wetterich1988npb}, not only gives rise to acceleration but also alleviates the cosmological coincidence problem. Such a scalar field dubbed ``quintessence'' has been studied extensively in the literature~\cite{frieman1995prl,ferreira1997prl, ferreira1998prd, carroll1998prl, caldwell1998prl, copeland1998prd, brax2000prd2, sahni2002cqg, sen2002plb, banerjee2006mpla, banerjee2006ass, das2006grg, banerjee2005grg, roy2014epjp, banerjee2015grg, durrive2018prd} and many more.

The ``tracking'' model was first introduced by Ratra \& Peebles~\cite{ratra1988prd} and Peebles \& Ratra~\cite{peebles1988apjl}. The idea was to resolve the fine-tuning problem, long before the discovery of the present accelerated expansion of the Universe. A lot of work followed from there for various purpose~\cite{peebles1988apjl, ratra1988prd, ferreira1997prl, ferreira1998prd}. Some ``scaling'' models, which alleviates the fine-tuning problem, were also discussed in~\cite{ferreira1997prl, ferreira1998prd, copeland1998prd, amendola2006prd}, which do not however drive the present acceleration. Zlatev \etal~\cite{zlatev1999prl} and Steinhardt \etal~\cite{steinhardt1999prd} later utilised a modified version to incorporate the accelerated expansion. 

The exponential potential, $\Vp = V_{0}\, e^{- \lambda \vphi} $ introduced by Ratra \& Peebles~\cite{ratra1988prd} as a ``tracking'' model has two attractor solutions,
\begin{itemize}
\item[(a)] where the scalar field follows the evolution of the dominant background fluid with $\wphi = w_{D}$ and $\Ophi = 3\paren*{1+w_{D}}/\lambda^{2}$ with the condition $\lambda^{2} > 3\paren*{1+w_{D}}$, $w_{D}$  being the EoS parameter of the background fluid and $\Ophi$ is the energy density parameter. This solution is called the \emph{scaling solution}~\cite{ferreira1997prl, ferreira1998prd,copeland1998prd} and
\item[(b)] where the scalar field acts as the dominant energy component with $\wphi = -1+\lambda^{2}/3$ and $\Ophi = 1$ with the condition $\lambda^{2} < 3\paren*{1+w_{D}}$. 
\end{itemize}
The attractor (a) allows the scalar field energy density to maintain a constant ratio with the background component starting from any initial condition. The attractor (b) allows cosmic acceleration for $\lambda^{2}<2$. As $\lambda$ is a constant, the scalar field cannot exit the scaling regime (a) and approach (b) to give an accelerated expansion. Solution to this problem was provided by Sahni and Wang~\cite{sahni2000prd} with the potential $\Vp = \paren*{\cosh \vphi-1}^{\alpha}$ and Barreiro \etal~\cite{barreiro2000prd} with the double exponential potential, $\Vp = V_{0}\, \paren*{e^{\,\alpha \kappa \vphi} + e^{\, \beta \kappa \vphi}}$. 

Another possible solution to the coincidence problem was introduced by Griest~\cite{griest2002prd} as ``thawing'' model, where the EoS parameter $w \approx -1$ at the early epoch and increases with time. On the other hand, the tracking scenario represents a ``freezing'' model, where EoS parameter freezes at close to $-1$ at late time, starting from any other value. The potential of the pseudo-Nambu-Goldstone boson (PNGB) $\Vp = V_{0}\, \left[1 + \cos \paren*{\vphi / f}\right]$ proposed by Kim~\cite{kim1999jhep} representing the thawing model. On the other hand, Albrecht and Skordis~\cite{albrecht2000prl} have developed an interesting freezing model with potential $\Vp = V_{0}\, \left[\paren*{\vphi - B}^\alpha + A\right] e^{-\lambda\vphi}$ from string theory where the scalar field enters a regime of damped oscillations with $w \rightarrow  -1$ leading to an acceleration. Some other scalar field potentials that have been studied in literature in the context of quintessence models  are given in table (\ref{table:pot}).
\begin{table}[!h]
\begin{center}
\caption{Some scalar field potentials that have been studied in the literature.}\label{table:pot}
\begin{adjustbox}{width=0.7\textwidth}
\begin{tabular}{cc}
\hline \hline
\rule[-1ex]{0pt}{2.5ex}Potential  & \hspace{24ex}References \\ 
\hline \hline \vspace*{-2.5mm}\\
\rule[-1ex]{0pt}{2.5ex}$V_{0}\, \vphi^{-\alpha}\,, ~\alpha >0 $ & \hspace{24ex}\cite{ratra1988prd}  \\ \vspace*{-2.mm} \\
\rule[-1ex]{0pt}{2.5ex}$m^{2} \vphi^{2}, V_{0}\, \vphi^{4}$  & \hspace{24ex}\cite{frieman1995prl}  \\ \vspace*{-2.mm} \\
\rule[-1ex]{0pt}{2.5ex}$V_{0}\, \paren*{e^{\,\, M_{p}/\vphi}-1}$  & \hspace{24ex}\cite{steinhardt1999prd} \\ \vspace*{-2.mm} \\
\rule[-1ex]{0pt}{2.5ex}$V_{0}\, e^{\, \, \lambda \vphi^{2}}/\vphi^{\alpha}\, , ~\alpha>0$  & \hspace{24ex}\cite{brax2000prd1,brax1999plb}\\ \vspace*{-2.mm} \\
\rule[-1ex]{0pt}{2.5ex}$V_{0}\, \sinh^{-\alpha}\paren*{\lambda \vphi} $ & \hspace{24ex}\cite{sahni2000ijmpd, urena2000prd} \\ \vspace*{-2.mm} \\ 
\rule[-1ex]{0pt}{2.5ex}$V_{0}\, e^{- \lambda \vphi} \paren*{1 + A \sin \nu \vphi}$ & \hspace{24ex}\cite{dodelson2000prl} \\ 
\hline \hline
\end{tabular}
\end{adjustbox}
\end{center}
\end{table}
A comprehensive study on theoretical and observational aspects of different scalar field dark energy models can be found in~\cite{carvalho2006prl, scherrer2008prd1, scherrer2008prd2, chiba2009prd, chiba2013prd, roy2014epjp, banerjee2015grg, pantazis2016prd} whereas excellent reviews on quintessence models are found in~\cite{brax2000prd2, sahni2002cqg, martin2008mpla, tsujikawa2013cqg}.

The Lagrangian density for a scalar field $\vphi$ with a potential $\Vp$ is written as
\begin{equation}
\mathcal{L}_{\vphi} = -\frac{1}{2} g^{\mu\nu}\vphi_{;\mu}\vphi_{;\nu} - \Vp,\label{eq1:lag}
\end{equation}
and the energy-momentum tensor is obtained as
\begin{equation}
\Tmn = g^{\mu\alpha}\vphi_{;\alpha}\vphi_{;\nu} - \delta^{\mu}{}_{\nu} \paren*{\frac{1}{2} g^{\alpha\beta}\vphi_{;\alpha}\vphi_{;\beta} +\Vp}.
\end{equation}
The energy density and pressure of the scalar field are $\rphi = \frac{1}{2}\dot{\vphi}^{2} +\Vp$ and $\pphi = \frac{1}{2}\dot{\vphi}^{2} -\Vp$ respectively. The equation of state (EoS) parameter is given as 
\begin{eqnarray}
\wphi = \frac{\pphi}{\rphi}~=~ \frac{\frac{1}{2}\dot{\vphi}^{2} -\Vp}{\frac{1}{2}\dot{\vphi}^{2} +\Vp}~=~ 1-\frac{2 \,\Vp}{\rphi}.\label{eq1:wphi}
\end{eqnarray}
The Klein-Gordon equation or the \emph{equation of motion} of the scalar field can be obtained as a consequence of the Bianchi identities as
\begin{equation}\label{eq1:kg1}
\ddot{\vphi} + 3 H \dot{\vphi} + \frac{d V}{d \vphi}= 0.
\end{equation}
It is clear from the expression (\ref{eq1:wphi}) that $\wphi$ has an evolution and ranges between $-1 \leq \wphi \leq1$ for a real scalar field and a positive definite $\Vp$. When the kinetic energy ($E_{K} = \frac{1}{2}\dot{\vphi}^{2}$) is dominant with a negligible potential energy ($E_{P} = \Vp$), the scalar field behaves as a stiff fluid with $\wphi = 1$, and when $E_{P}$ dominates with a negligible $E_{K}$, it gives rise to a cosmological constant with $\wphi = -1$. For the recent accelerated expansion of the Universe, the scalar field at late time should roll sufficiently slowly along the potential such that $E_{K} \ll E_{P}$. 

Among the quintessence models, some may evolve in a way such that the EoS parameter of the dark energy attains a value less than $-1$ at the present epoch or in a finite future called the ``phantom'' model~\cite{chiba2000prd2, caldwell2002plb, carroll2003prd, caldwell2005prl}. In such cases, the Universe has a future singularity where the scale factor $a$ and Hubble parameter $H$ attain infinitely large values. The scalar field models in which the EoS parameter evolve to mimic the phantom fluid are called ``quintom'' models~\cite{feng2005plb,cai2007plb1, cai2007plb2, cai2008cqg}. 
%%%%%%%%%%%%%%%%%%%%%%%%%%%%%%%%%%%%%%%%%%%%%%%%%%%%%
\subsection{Interacting Models}
A cosmological model with an energy transfer between the dark matter (DM) and dark energy (DE) can give rise to comparable energy densities at the present epoch. Wetterich~\cite{wetterich1995aa} introduced coupling between the dark matter and scalar field to settle the coincidence problem, and later Amendola~\cite{amendola2000prd} used it in the context of recent acceleration. If there is an energy transfer only in the dark sector of the Universe, that the conservation equations are
\begin{eqnarray}
\dot{\rho}_c+ 3 H \paren*{1+w_{c}} \rho_c &=& -Q\,,\label{eq1:con1}\\
\dot{\rho}_{de}+ 3 H \paren*{1+\wde} \rde&=& Q. \label{eq1:con2}
\end{eqnarray}
The other three fluids --- photons ($\gamma$), neutrinos ($\nu$) and baryons ($b$) conserve independently and have no energy transfer among themselves. Their conservation equations are written as
\begin{equation}
\dot{\rho}_{A}+ 3 H \paren*{1+w_{A}} \rA = 0\, \label{eq1:con3},
\end{equation}
where $w_{A} = \pA/\rA$ is the equation of state parameter (EoS) of the $A$-th fluid and $A = \gamma, \nu, b$. For photons and neutrinos, the EoS parameter is $w_{\gamma} = w_{\nu} = 1/3$, for baryons, the EoS parameter is $w_{b} = 0$. In Eqs.\ (\ref{eq1:con1}) and (\ref{eq1:con2}), the EoS parameter of cold dark matter is $w_{c} = 0$  and that of dark energy is $\wde = \pde/\rde$.  It is clear from equations (\ref{eq1:con1}) and (\ref{eq1:con2}) that the total of dark matter and dark energy is conserved. 

In Eqs.\ (\ref{eq1:con1}) and (\ref{eq1:con2}), $Q$ gives the rate of energy transfer between the two fluids. If $Q<0$, energy is transferred from dark energy to dark matter (DE $\rightarrow$ DM) and if $Q>0$, energy is transferred from dark matter to dark energy (DM $\rightarrow$ DE). When $Q>0$, dark matter redshifts faster than $a^{-3}$ and when $Q<0$, dark matter redshifts slower than $a^{-3}$. The dark energy evolution depends on the difference $\wde-\frac{a Q}{3 \cH \rde}$. Thus, the interaction manifests itself by changing the scale factor dependence of the dark matter as well as dark energy. 
There are different forms of the choice of the phenomenological interaction term $Q$, the models with $Q$ proportional to either $\rdc$ or $\rde$ or any combination of them are among the more popular choices,~\cite{bohmer2008prd,clemson2012prd,acosta2014prd,yang2014prd1,yang2018prd} to mention a few. It must be mentioned here that there is no particular theoretical compulsion for any of these choices.

%%%%%%%%%%%%%%%%%%%%%%%%%%%%%%%%%%%%%%%%%%%%%%%%%%%%%
\subsection{Other Dark Energy Models}
\paragraph*{Tachyon field model:} A tachyon field produced at the time of decay of D-brane with a negative squared mass and EoS parameter ($\wphi$) varying between $-1$ and $0$ can successfully produce the late time acceleration~\cite{sen1999jhep, bergshoeff2000jhep, kluson2000prd, sen2002jhep1, gibbons2002plb, paddy2002prd, bagla2003prd, garousi2003jhep, abramo2003plb, jassal2004pramana, aguirregabiria2004prd, guo2004jcap, copeland2005prd, singh2019jcap, singh2020jcap, rajvanshi2021ax}. A tachyon field ($\vphi$) rests on the maximum of the potential ($\Vp$) and, when subjected to a perturbation, rolls down to the real mass. The Lagrangian density is given as
\begin{equation}
\mathcal{L} = \Vp \sqrt{\mbox{det}\paren*{\gmn+\partial_{\mu}\vphi\partial_{\nu}\vphi}}.
\end{equation}
The energy density and pressure are given as
\begin{equation}
\rphi = \frac{\Vp}{\sqrt{1-\dot{\vphi}^{2}}} \hspace{0.5cm} \mbox{and} \hspace{0.5cm} \pphi = -\Vp \sqrt{1-\dot{\vphi}^{2}},
\end{equation}       
respectively in the context of a spatially isotropic and homogeneous cosmological model. Hence the EoS parameter becomes $\wphi = \frac{\pphi}{\rphi}~=~\paren*{\dot{\vphi}^{2}-1}$, where an over-dot indicates derivative with respect to cosmic time, $t$. Thus it can be seen that a tachyon field can generate the recent acceleration when $\dot{\vphi}^{2}<\frac{2}{3}$.

\paragraph*{Chaplygin gas:} A generalised Chaplygin gas is defined with an EoS parameter $w=\frac{p}{\rho}=-\frac{A}{\rho^{\alpha+1}}$, where $A$ is a positive constant and $\alpha$ is a parameter ($0<\alpha\le 1$). A generalised Chaplygin gas can not only generate the recent acceleration but also provide an unification of dark matter and dark energy~\cite{kamenshchik2001plb, bilic2002plb, bento2002prd, padmanabhan2002prd, bento2003prd, chimento2011prd}. The CMB measurement constraints the allowed region of the parameter $\alpha$ to be $0\le\alpha\le 0.2$~\cite{amendola2003jcap}.

\paragraph*{Holographic Dark Energy:} Holographic dark energy (HDE) is based on the \emph{holographic principle} in quantum gravity theory~\cite{susskind}. The holographic principle, following the 't Hooft conjecture~\cite{thooft}, states that \emph{the information contained in a volume can be ascertained with the knowledge about the degrees of freedom residing on its boundary}. This principle actually stems from Bekenstein's idea that the entropy of a black hole is related to its area~\cite{beken7}. The quantum zero-point energy of a system with size $L$ should not exceed the mass of a black hole with the same size and this relates the short distance or ultraviolet (UV) cut-off to the long distance or infrared (IR)\nomenclature{IR}{Infra-Red} cut-off~\cite{cohen}. If $\rho_{vac}$ is quantum zero-point energy caused by a UV cut-off then the inequality, $L^{3}\rho_{vac} \le LM_{P}^2$, specifies the maximum allowed size of the system or the IR cut-off. For largest allowed value $L$ of IR cut-off, the HDE is given as
\begin{equation}\label{eq2:def}
\rho_{HDE}=3 C^2 M_{P}^2 L^2, 
\end{equation}
where $M_{P} = \paren*{8\pi G_{N}}^{-2}$ is the reduced Planck mass. For a detailed study on different HDE models we refer to~\cite{li2004plb, pavon2006aip, zimdahl2007cqg, elizalde2004prd, nojiri2006grg, zhang2012jcap, chimento2012prd, akhlaghi2018mnras}.

There are excellent reviews that summarise the list of candidates and their strength and weakness~\cite{copeland2006ijmpd, sahni2006ijmpd, sami2016ijmpd}. The list of candidates as dark energy is ever increasing in the absence of a universally accepted one.

%%%%%%%%%%%%%%%%%%%%%%%%%%%%%%%%%%%%%%%%%%%%%%%%%%%%%
\subsection{Modified Gravity Models}
The presence of dark energy in the contents of the Universe modifies the matter contribution to Einstein field equations. However, the late time acceleration can also be achieved by suitably modifying the contribution from the gravity sector. \emph{Modified theories of gravity} also provide possible solution to the cosmological constant problem. The gravity sector needs to be modified so that the effect of such modifications is suppressed at scales where Einstein gravity or General Relativity has been well tested. The stringent Solar System tests such as the bending of light rays~\cite{shapiro2004prl} and time delation~\cite{bertotti2003nature} by the Sun's gravitational field provide strong constraints on the modified gravity models. Moreover, the modified gravity models must also survive the latest cosmological observational tests like the EoS parameter of DE as $\wde=-1.03\pm 0.03$ within the $68\%$ confidence level~\cite{planck2018cp}. Different screening mechanisms have been developed to conceal or ``screen'' the effects of modified gravity on small scales~\cite{khoury2004prd, khoury2004prl}. 

One possible modification to General Relativity is in the form of the \emph{scalar-tensor theories} of gravity, where a scalar field is non-minimally coupled to geometry. The simplest example of scalar-tensor theory where a scalar field is coupled to gravity is the \emph{Brans-Dicke} theory~\cite{brans1961pr}. The scalar-tensor theories are some of the most established theories of gravity and some examples are in~\cite{bergmann1968ijtp, wagoner1970prd, nordtved1970apj, horndeski1974ijtp, uzan1999prd, amendola1999prd, chiba1999prd, nicolis2009prd, deffayet2009prd1}. In the context of late time cosmic acceleration, some of the notables investigations can be found in~\cite{bartolo1999prd, banerjee2001cqg, sen2001prd, banerjee2001prd, nb1, sudipta2008prd}.

In General Relativity, field equations derived from the Einstein-Hilbert action (Eq.\ (\ref{eq1:action})) in four dimensions are at most second-order derivatives of the metric tensor, $\gmn$~\cite{lovelock1971jmp}. Thus, one way to extend General Relativity is to allow the field equations to be higher than the second-order derivative, and the models are classified as \emph{higher derivative theories}. One of the most simplistic options is replacing the Ricci scalar $R$ in the Einstein-Hilbert Lagrangian density with some non-linear function, $f(R)$ and are popularly known as \emph{$f(R)$ gravity} models. Some examples of $f(R)$ gravity models can be found in~\cite{starobinsky1980plb, mena2006prl, felice2006jcap, hu2007prd, appleby2007plb, nojiri2007plb, starobinsky2007jetpl}. In the context of late time cosmic acceleration, $f(R)$ gravity models have been utilised in~\cite{capozziello2002ijmpd, capozziello2003ijmpd, nojiri2003prd, carroll2004prd, das2006cqg, vollick2003prd}. Examples on cosmological viability can be found in~\cite{amendola2007prl, amendola2007prd1}. The Lagrangian density can also be a function of any other scalar quantities constructed from the contraction of the Ricci or Riemann tensors as $R_{\mu\nu}R^{\mu\nu}$ and $R_{\mu\nu\alpha\beta}R^{\mu\nu\alpha\beta}$~\cite{carroll2005prd, nojiri2005plb}. Other modified theories include the \emph{braneworld} gravity in the form of the Dvali-Gabadadze-Porrati (DGP) model~\cite{dvali2000plb}, $f(T)$ gravity~\cite{cai2016rpp} and $f(T,T_{G})$ gravity~\cite{kofinas2014prd} models. For detailed reviews of different modified gravity theories we refer to~\cite{sotiriou2010rmp, defelice2010lrr, maartens2010lrr, tsujikawa2010, clifton2012pr, joyce2015pr, koyama2016rpp}.

%%%%%%%%%%%%%%%%%%%%%%%%%%%%%%%%%%%%%%%%%%%%%%%%%%%%%
\section{Outline Of The Thesis}
The Universe on larger scales is spatially homogeneous and isotropic. As we start zooming in, at some smaller scales, we start seeing the inhomogeneities ie.\ the structures like galaxies, galaxy clusters and so on. These structures have grown from some primordial fluctuations generated at the time of inflation through \emph{gravitational instability}. The principal idea is that any small overdensity will accrete matter from its surrounding area and will grow in time, eventually collapsing under self-gravity. However, due to increased matter in a small region, the pressure due to random thermal motion tends to counter the increasing gravitational attraction and cease the growth. Thus if the pressure is small, density fluctuation grows, and if the pressure is large, fluctuations oscillate with time. The typical length scale of fluctuation $\lambda_{J} \simeq c_{a}\paren*{G_{N}\rho}^{-1/2}$ ($c_{a}=\sqrt{\frac{p}{\rho}}$ is the sound speed) above which fluctuations can grow is called the \emph{Jeans length}. The primordial fluctuations evolve through different stages of the evolutionary history of the Universe and a detailed discussion is required. As already mentioned, this thesis investigates the possibility of the growth of matter perturbation in some dark energy models. The perturbation theory is the mainstay of the present work, so it will be discussed in detail in chapter \ref{chap2:gi-pert}.
%%%%%%%%%%%%%%%%%%%%%%%%%%%%%%%%%%%%%%%%%%%%%%%%%%%%%

In chapter \ref{chap3:grg}, we considered density perturbation in dark energy models reconstructed from the kinematical quantity, the \emph{jerk parameter}, $j = - \frac{1}{a^3} \frac{d^3 a}{dt^3}$. The idea behind the reconstruction from the \emph{kinematical quantity} is that one ignores the theory of gravity and takes an ansatz of the kinematical quantity and then attempts to develop the model from observation. The \emph{Hubble parameter}, $H = \frac{\dot{a}}{a}$ is the oldest known observational quantity in cosmology and was found to evolve with time. So the natural choice for the reconstruction has been the next higher order derivative, the \emph{deceleration parameter}, $q = - \frac{a\ddot{a}}{{\dot{a}}^2}$. From recent observational data it is found that $q$ is also evolving with time. Hence the next higher order derivative, $j$ plays a significant role in the game of reconstruction through kinematical quantities. The models used to study the growth of dark matter perturbations in chapter \ref{chap3:grg} are discussed in detail in~\cite{mukherjee2016prd, mukherjee2017cqg}. The reconstructed models are such that in one of them there is an interaction in the dark sector while in the other one the constituents of the dark sector conserve individually. We have showed that the models allowing interaction in the dark sector mostly fail to yield the large scale structures as the perturbations decay during the late time. The density fluctuation for the non-interacting models has growing modes during the later time. So the non-interacting models appear to be favoured for structure formation.
%%%%%%%%%%%%%%%%%%%%%%%%%%%%%%%%%%%%%%%%%%%%%%%%%%%%%

Chapter \ref{chap4:epjp} deals with the evolution of the density contrasts for a cosmological model where along with the standard cold dark matter (CDM)\nomenclature{CDM}{Cold Dark Matter}, the present Universe also contains holographic dark energy (HDE)\nomenclature{HDE}{Holographic Dark Energy}. The HDE is allowed to interact with the CDM. The characteristic IR cut-off is considered to be the future event horizon, following~\cite{li2004plb}. An inclusion of interaction between components of dark sector prevents the future ``big-rip'' singularity, which is the ripping apart of the Universe due to an accelerated expansion with an effective EoS parameter of the Universe, $w<-1$. The equations for the density contrasts of both dark matter and dark energy are integrated numerically. It is found that irrespective of the presence of an interaction, the matter perturbation has growing modes. The HDE is also found to have a growth of perturbation in the absence of any effective sound speed ($c_{s,de}^2\equiv \frac{\delta p_{de}}{\delta \rde}=0$), hence it can be said that HDE can also cluster. 
%The interesting point to note is that the density contrast corresponding to HDE has a peak at a recent past and is decaying at the present epoch.
%%%%%%%%%%%%%%%%%%%%%%%%%%%%%%%%%%%%%%%%%%%%%%%%%%%%%

In chapter \ref{chap5:jcap}, we investigate the perturbations in a scalar field model with a potential. The potential of scalar field is constructed such that the scalar field drives the recent acceleration in a similar fashion that the cosmological constant does and has the dark energy (DE)\nomenclature{DE}{Dark Energy} density comparable to the dark matter (DM)\nomenclature{DM}{Dark Matter} energy density at the recent epoch starting from arbitrary initial conditions. Thus the scalar field model is free from the initial condition problem. We have considered perturbation of both dark matter and scalar field and studied their evolution. The perturbations show that this model, though it keeps the virtues of a \lcdm model, has a distinctive qualitative feature --- it reduces the amplitude of the matter power spectrum on a scale of $8\,h^{-1}\, \mpc$, $\se$ at the present epoch. 
%%%%%%%%%%%%%%%%%%%%%%%%%%%%%%%%%%%%%%%%%%%%%%%%%%%%%

Dark matter and dark energy are evolving together from the early epoch and an interaction between the two cannot be ruled out \emph{a priori}. This naturally raises the question when is the interaction significant in the evolutionary history of the Universe --- if it was there from the early epoch and stays through the evolution or it is a more recent phenomenon or it was entirely an early phenomenon. A simple modification of the interaction term with \emph{evolving coupling parameter} may answer this question. The motivation of the work in chapter \ref{chap6:prd} is to look for any preferable stage of evolution when the interaction is significant. Chapter \ref{chap6:prd} deals with the perturbation analysis, parameter estimation and Bayesian evidence calculation of interacting models with dynamical coupling parameter that determines the strength of the interaction. We have considered two cases, where the interaction is a more recent phenomenon and where the interaction is a phenomenon in the distant past. Moreover, we have considered the quintessence DE equation of state with Chevallier-Polarski-Linder (CPL)\nomenclature{CPL}{Chevallier-Polarski-Linder} parametrisation and energy flow from DM to DE. Using the current observational datasets like the cosmic microwave background (CMB)\nomenclature{CMB}{Cosmic Microwave Background}, baryon acoustic oscillation (BAO)\nomenclature{BAO}{Baryon Acoustic Oscillation}, Type Ia Supernovae (SNe Ia)\nomenclature{SNe Ia}{Type Ia Supernovae} and redshift-space distortions (RSD)\nomenclature{RSD}{Redshift-Space Distortions}, we have estimated the mean values of the parameters. Using the perturbation analysis and Bayesian evidence calculation, we have shown that interaction present as a brief early phenomenon is preferred over its being a recent phenomenon.

Finally, in chapter \ref{chap7}, we conclude with a brief summary and relevant discussions regarding the work presented in this thesis. 
%%%%%%%%%%%%%%%%%%%%%%%%%%%%%%%%%%%%%%%%%%%%%%%%%%%%%

%%%%%%%%%%%%%%%%%%%%%%%%%%%%%%%%%%%%%%%%%%%%%%%%%%%%%

%%%%%%%%%%%%%%%%%%%%%%%%%%%%%%%%%%%%%%%%%%%%%%%%%%%%%

%%%% Gauge invariant perturbation theory %%%%%
\chapter{Cosmological Perturbation Theory}\label{chap2:gi-pert}
\chaptermark{Cosmological Perturbation Theory}

\section{Introduction}\label{sec2:metric-pert}
The large scale structures like the galaxies and cluster of galaxies that we observe today grew from the primordial density fluctuations with very small amplitudes. On scales greater than the Hubble horizon, $H^{-1}$, the amplitude of the perturbations in geometrical quantities are comparable or larger than the perturbations in density. To understand the growth of density fluctuations, in a homogeneous and isotropic background on super-horizon scales, gauge-invariant perturbation theory is useful.
%%%%%%%%%%%%%%%%%%%%%%%%%%%%%%%%%%%%%%%%%%%%%%%%%%%%%
%\section{Scalar-Vector-Tensor Decomposition}\label{sec2:svt}  
\section{Metric Perturbation}\label{sec2:svt}  
The Friedmann-Lema\^itre-Robertson-Walker (FLRW) metric in an unperturbed Universe is written as
\begin{equation}\label{eq2:metric}
ds^2= \bar{g}_{\mu\nu}\paren*{x^{\alpha}} d x^{\mu} dx^{\nu} = a^2(\tau)\paren*{- d \tau ^2+\gamma_{ij}\,d x^i dx^j},
\end{equation}
where $a(\tau)$ is the scale factor and the conformal time $\tau$ is related to the cosmic time $t$ as $a^2 d\tau^2 = dt^2$. In Eq.\ (\ref{eq2:metric}), $\bar{g}_{\mu\nu}$ is the unperturbed 4-metric tensor and $\gamma_{ij} = \delta_{ij}$ is the 3-metric tensor of the spatially flat, constant-$\tau$ hypersurface. Small deviations from the spatially homogeneous and isotropic spacetime in the form of first-order perturbations, $\delta \gmn$ are considered such that the metric tensor $\gmn$ can be spilt as 
\begin{equation}\label{eq2:gmn-split}
\gmn\paren*{x^{\alpha}} = \bar{g}_{\mu\nu}\paren*{x^{\alpha}} + \delta \gmn\paren*{x^{\alpha}}
\end{equation}
As a symmetric, $4 \times 4$ matrix, the metric tensor, $\gmn$ has $10$ independent components, hence $10$ degrees of freedom.  For a linear theory of perturbation, the metric can be decomposed into scalar, vector and tensor perturbations depending on their transformation properties on the spatial hypersurfaces. The \emph{scalar-vector-tensor} (SVT) decomposition of the metric was presented by Lifshitz~\cite{lifshitz2017grg}, Lifshitz and Khalatnikov~\cite{lifshitz1963ap} in 1946. After that, SVT decomposition in cosmological perturbations were studied in detail by Peebles~\cite{peebles1980}, Bardeen~\cite{bardeen1980prd}, Kodama \& Sasaki~\cite{kodama1984ptps} to name a few. Later, Stewart~\cite{stewart1990cqg} had given a covariant description of tensor decomposition. This decomposition of the metric is based on the $\paren*{3 + 1}$ Arnowitt, Deser and Misner (ADM) formalism on slicing of the spacetime~\cite{adm1960pr,mtw1973}. The $\paren*{3+1}$-decomposition is a ``slicing'' of spacetime into a series of spatial hypersurfaces each of which is labelled by a coordinate time $\tau$.

The different components of the metric perturbations in general are written as follows,
\begin{equation}\label{eq2:met-pert}
\delta g_{00} = -a^{2} 2\phi, \hspace{0.6cm} \delta g_{0i} = a^{2}w_{i}, \hspace{0.6cm} \delta g_{ij} = a^{2}2 \paren*{\psi \gamma_{ij} + h_{ij}} \hspace{0.15cm} \mbox{with} \hspace{0.15cm} \gamma^{ij}h_{ij} = 0 .
\end{equation}
Here, two 3-scalar fields $\phi\paren*{x^{i},\tau}$ and $\psi\paren*{x^{i},\tau}$, one 3-vector field $B_{i}\paren*{x^{i},\tau}$ and one symmetric traceless second-rank 3-tensor field $h_{ij}\paren*{x^{i},\tau}$ are introduced. The trace part of $h_{ij}$ has been absorbed into $\psi$ without any loss of generality, leaving $h_{ij}$ with $5$ independent components. Thus, $10$ independent fields are introduced; same as the number of independent components of the perturbed metric. The inverse 3-metric, $\gamma^{ij} \paren*{\gamma^{ik}\gamma_{kj} = \delta^{i}_{j}}$, is used to raise indices of 3-vectors and 3-tensors on the spatial hypersurfaces. 

Any 3-vector on the 3-hypersurface can be decomposed as the sum of longitudinal and transverse components as
\begin{equation}\label{eq2:vec-dec}
w_{i} = w_{i}^{\parallel} + w_{i}^{\perp}.
\end{equation}
The decomposition is such that the longitudinal part, $w_{i}^{\parallel}$ is irrotational (curl-free) while the transverse part, $w_{i}^{\perp} = -S_{i}$ is solenoidal (divergence-free) and $w_{i}^{\parallel} = \nabla_{i} B$, for some scalar field $B$. The longitudinal component that can be derived from the covariant derivative of a scalar field contributes to the \emph{scalar} part of the metric perturbation, while the transverse component, which is not derived from any scalar, contributes to the \emph{vector} components of the perturbation. This decomposition of a vector field into longitudinal and transverse parts in Euclidean space comes from the Helmholtz's theorem.

Similarly, any symmetric, traceless, second-rank 3-tensor on the spatial hypersurface can be decomposed as the sum of doubly longitudinal, singly longitudinal and doubly transverse components as
\begin{equation}\label{eq2:ten-dec}
h_{ij} = h_{ij}^{\parallel} + h_{ij}^{\perp} + h_{ij}^{TT} \hspace{0.15cm} \mbox{with} \hspace{0.15cm} \gamma^{ik}\nabla_{i} h_{kj}^{TT} = 0.
\end{equation}
The doubly longitudinal (or longitudinal) and singly longitudinal (or solenoidal) parts can be derived from the gradients of a scalar field, $E$ and a transverse vector field $F_{k}$ respectively as
\begin{equation}\label{eq2:ten-dec-con}
h_{ij}^{\parallel} = \paren*{\nabla_{i}\nabla_{j} - \frac{1}{3}\gamma_{ij}\nabla^{2}}E, \hspace{0.6cm} h_{ij}^{\perp} = \frac{1}{2}\paren*{\nabla_{i} F_{j} + \nabla_{j} F_{i}}.
\end{equation}
The longitudinal part, $h_{ij}^{\parallel}$ and the solenoidal part, $h_{ij}^{\perp}$ contribute to the \emph{scalar} and \emph{vector} components of the perturbation respectively and the divergence-free, traceless transverse part, $h_{ij}^{TT}$  contributes to the \emph{tensor} components of the perturbation.

Thus, the linearly perturbed line element in a flat spacetime can be written in the general form as\footnote{The factor $- \frac{1}{3}\gamma_{ij}\nabla^{2}E$ is absorbed $2\psi$ and the factor of 2 is absorbed in $h_{ij}^{TT}$, so that the resulting form matches with the standard notation~\cite{mukhanov1992pr,malik2001thesis,brandenberger2004}}
\begin{equation}\label{eq2:svt-metric2}
\begin{split}
ds^2=  a^2(\tau)&\Big\{-\paren*{1+2\phi} d \tau ^2 + 2 \paren*{\partial_{i}B - S_{i}} d\tau d x^{i} \\
&+\left[\paren*{1-2\psi}\delta_{ij}+ 2 \partial_{i}\partial_{j}E + 2\partial_{i} F_{j} + h_{ij}^{TT}\right] d x^i dx^j\Big\},
\end{split}
\end{equation}
consisting of four scalars ($\phi, \psi, B, E$) each having $1$ degree of freedom, two divergence-less vectors ($S_{i}, F_{i}$) each with $2$ degrees of freedom and a symmetric, traceless transverse tensor ($h_{ij}^{TT}$) having $2$ degrees of freedom, totalling $10$ degrees of freedom~\cite{mukhanov1992pr,bertschinger1993cosdy,bertschinger2001,brandenberger2004}. Out of these $10$ degrees of freedom, only six are physical degrees of freedom as the four coordinates can be transformed without affecting the physical quantities, making $4$ degrees of freedom coordinate dependent. The $6$ physical degrees of freedom come from two scalar fields, two vector fields and two tensor fields as the tensor components are gauge-invariant. The $4$ coordinate or ``gauge'' degrees of freedom correspond to two of the scalar fields, and one transverse vector~\cite{mukhanov1992pr,bertschinger1993cosdy,bertschinger2001}. The different coordinate choices can eliminate these $4$ coordinate degrees of freedom and is discussed later in Sections \ref{sec2:gt} and \ref{sec2:gaugechoices}.

Physically, the scalar perturbations of the metric coupled with matter perturbations, lead to the growth of inhomogeneities, resulting in the formation of the galaxies and galaxy clusters in the Universe. The scalar perturbations are the relativistic modifications of Newtonian gravity; the $2$ physical scalar degrees of freedom correspond to the Newtonian gravitational potential and the relativistic correction. The vector perturbations exhibit no instability and decay in expanding background. The $2$ physical degrees of freedom of vector perturbations correspond to gravitomagnetism. On the other hand, the tensor perturbations produce gravitational waves and do not couple to matter perturbations in the first order. The $2$ tensor degrees of freedom are the two polarization states of gravitational waves. In the linear perturbation theory, the scalar, vector and tensor perturbations evolve independently. As the scalar perturbation contributes to the clustering of matter this will only be considered in the present work. The scalar perturbation is described by the line element
\begin{equation}\label{eq2:sc-metric2}
ds^2=  a^2(\tau)\Big\{-\paren*{1+2\phi} d \tau ^2 + 2\partial_{i}B d\tau d x^{i} +\left[\paren*{1-2\psi}\delta_{ij}+ 2 \partial_{i}\partial_{j}E \right] d x^i dx^j\Big\}.
\end{equation}
The scalar metric perturbation will be discussed in the subsequent sections and chapters. In the $\paren*{3 + 1}$-formalism, the term $\phi$ called the \emph{lapse function}, is the perturbation in the lapse of proper-time between any two neighbouring constant time hypersurfaces, the term $\partial_{i}B$ called the \emph{shift vector}, represents the perturbation in the rate of deviation of a constant space-coordinate line from a line normal to a constant time hypersurface. The term $\psi$ is the spatial curvature perturbation and $E$ is the anisotropic perturbation of each constant time hypersurface (off-diagonal spatial perturbation). 
%
%%%%%%%%%%%%%%%%%%%%%%%%%%%%%%%%%%%%%%%%%%%%%%%%%%%%%
\section{Gauge Transformation}\label{sec2:gt}
General relativity is a theory which is covariant under general changes of coordinates (or equivalently a theory about differential manifolds with no preferred coordinate charts)~\cite{stewart1990cqg}. In simple terms, all the physical quantities that can be calculated are independent of any particular coordinate choice. However, an infinitesimal change in the coordinates will introduce coordinate induced (fictitious) fluctuations in the physical quantities like the density perturbations. These fluctuations in density are not physical inhomogeneities that can grow in time to produce the large scale structures. This infinitesimal change in the spacetime coordinates is called \emph{gauge transformation}. The introduction of these ``spurious'' fluctuations due to gauge transformation is called the \emph{gauge problem} while the freedom to choose the coordinate system is called \emph{gauge freedom}. 

It is necessary to explain the meaning of gauge transformation to study the physical perturbations further. Gauge transformation can be realised both from a coordinate and covariant point of view. The coordinate approach was first developed by Lifshitz~\cite{lifshitz2017grg} in 1946, used in the gauge-invariant description of cosmological perturbations by Bardeen~\cite{bardeen1980prd} in 1980 and was further developed by Kodama \& Sasaki~\cite{kodama1984ptps}, Mukhanov~\cite{mukhanov1992pr}, Ma \& Bertschinger~\cite{ma1995apj}. The second approach, known as the covariant approach, was first developed by Sachs~\cite{sachs1964} in 1964 and used for relativistic fluids by Hawking~\cite{hawking1966apj} in 1966. Thereafter, the coordinate-independent approach was further developed by Stewart and Lemma~\cite{stewart1974prsa}, Stewart~\cite{stewart1990cqg}, Ellis~\cite{ellis1971grg,ellis1989prd,ellis1999} and Bruni~\cite{bruni1992apj}. 

\begin{figure}[!h]
  \centering
   \includegraphics[width=0.7\linewidth]{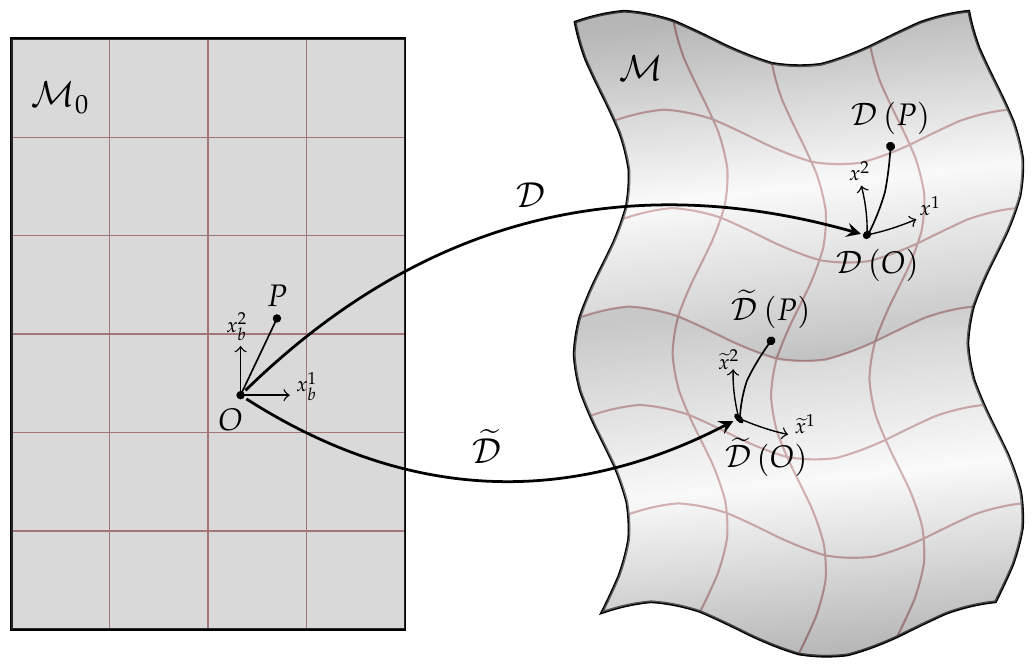}
\caption{Two dimensional representation of two diffeomorphisms $\cD$ and $\widetilde{\cD}$ between the two manifolds $\cM_{0}$ (background Universe) and $\cM$ (physical perturbed Universe). The coordinate system $x_{b}^{\mu}$ on $\cM_{0}$ is mapped to two different coordinate systems $x^{\mu}$ and $\ctx^{\mu}$ on $\cM$ by the maps $\cD$ and $\widetilde{\cD}$ respectively. The point, $P$ corresponds to the same point in different coordinate systems.}\label{im2:mani}
\end{figure}
In a coordinate dependent approach~\cite{mukhanov1992pr,brandenberger2004}, an ideal manifold $\cM_{0}$, describing the homogeneous and isotropic Universe given by FLRW metric and a physical manifold $\cM$, describing the physical perturbed spacetime are considered. The coordinate system in $\cM_{0}$ is represented as $x_{b}^{\mu}$, ($b$ stands for `background') while that in $\cM$ is represented as $x^{\mu}$. Any diffeomorphism\footnote{For any two manifolds $\cM$ and $\mathcal{N}$, a map $f\mathbin{:} \cM \rightarrow \mathcal{N}$ is called \emph{diffeomorphism}, if $f$ is bijective, differentiable and its inverse, $f^{-1}\mathbin{:} \mathcal{N} \rightarrow \mathcal{M}$ is also differentiable.} $\cD \mathbin{:}\cM_{0} \rightarrow \cM$, maps the points $x_{b}^{\mu}$ into $x^{\mu}$ on $\cM$. Another diffeomorphism $\widetilde{\cD}\mathbin{:}\cM_{0} \rightarrow \cM$ maps the same points $x_{b}^{\mu}$ on $\cM_{0}$, into different points $\ctx^{\mu}$ on $\cM$, as shown in Fig.\ (\ref{im2:mani}). Now a physical quantity $\cQ^{(0)}$ on $\cM_{0}$ and the corresponding physical quantity $\cQ$ on $\cM$ are considered. For the given diffeomorphism $\cD$, the perturbation $\delta \cQ$ of the quantity $\cQ$ at any point, $P \in \cM$ associated with coordinates $x^{\mu}\paren*{P}$ is defined as
\begin{equation}\label{eq2:map1}
\delta \cQ\paren*{P} = \cQ\paren*{P} - \cQ^{(0)}\paren*{\cD^{-1}\paren*{P}}.
\end{equation}
Similarly, for the other diffeomorphism $\widetilde{\cD}$, the perturbation $\widetilde{\delta \cQ}$ of the same quantity $\widetilde{\cQ}$ at the same point, $P$ associated with coordinates $\widetilde{x}^{\mu}\paren*{P}$ is defined as
\begin{equation}\label{eq2:map2}
\widetilde{\delta \cQ}\paren*{P} = \widetilde{\cQ}\paren*{P} - \cQ^{(0)}\paren*{\widetilde{\cD}^{-1}\paren*{P}}.
\end{equation}
The transformation $\delta \cQ\paren*{P} \rightarrow \widetilde{\delta \cQ}\paren*{P}$ is called the \emph{gauge transformation}, generated by the change of \emph{correspondence}, $\cD \rightarrow \widetilde{\cD}$ between the manifolds $\cM_{0}$ and $\cM$. This change in correspondence can be associated with the change in coordinates, $x^{\mu} \rightarrow \ctx^{\mu}$ induced on $\cM$. The difference between the two choices
\begin{equation}\label{eq2:lie}
\Delta \delta \cQ = \widetilde{\delta \cQ} - \delta \cQ
\end{equation}
generates to the fictitious fluctuations and are called \emph{gauge modes}.

For an infinitesimal coordinate transformation,
\begin{equation}\label{eq2:coor-in}
\ctx^{\mu} = x^{\mu} + \xi^{\mu},
\end{equation}
described by the 4-vector $\xi^{\mu}$, the difference is given by
\begin{equation}\label{eq2:diff-lie}
\Delta \delta \cQ = \widetilde{\delta \cQ} - \delta \cQ = \mathcal{L}_{\xi} \cQ
\end{equation}
where $\mathcal{L}_{\xi} \cQ$ is the Lie derivative of $\cQ$ in the direction of the vector $\xi$. The quantity $\cQ$ is called \emph{guage independent} only if either $\cQ=0$ on $\cM_{0}$, or $\cQ$ is a constant scalar field on $\cM_{0}$ or $\cQ$ is a constant linear combination of products of $\delta^{\mu}_{\nu}$~\cite{stewart1974prsa}. Thus any physical quantity like, density perturbation will be gauge invariant only in a non-expanding Universe.

For convenience, the spatial parts of the 4-vector $\xi^{\mu}$ have been split into a gradient of a scalar, $\xi$ and a divergence-free 3-vector, $\hat{\xi}^{i}$ components.
\begin{equation}\label{eq2:coor-ch}
\ctt = \tau + \xi^{0}\paren*{\tau, x^{k}}, \hspace{0.6cm} \ctx^{i} = x^{i}+\gamma^{ij}\partial_{j}\xi\paren*{\tau, x^{k}} + \hat{\xi}^{i}\paren*{\tau, x^{k}}
\end{equation}
Here, the scalar function $\xi^{0}$ determines the constant $\tau$-hypersurface while $\xi$ and $\hat{\xi}^{i}$ determines the coordinates within the hypersurface. These infinitesimal coordinate transformations must also change the perturbation fields ($\phi, \psi, w_{i}, h_{ij}$) so that the line element $ds^{2}$ remains invariant. Under the transformation (\ref{eq2:coor-in}), the metric tensor transforms as
\begin{eqnarray}
\widetilde{g}_{\mu\nu}\paren*{x^{\sigma}} & = & \frac{\partial x^{\alpha}}{\partial \ctx^{\mu}}\frac{\partial x^{\beta}}{\partial \ctx^{\nu}} g_{\alpha \beta}\paren*{x^{\sigma}-\xi^{\sigma}},\nonumber\\
& = & \gmn\paren*{x^{\sigma}} + g_{\alpha \nu} \partial_{\mu} \xi^{\alpha} + g_{\alpha \mu} \partial_{\nu} \xi^{\alpha} - \xi^{\lambda} \partial_{\lambda} \gmn + \mathcal{O}\paren*{\xi^{2}}\label{eq2:metric-trans1}.
\end{eqnarray}
It must be noted that the transformation is evaluated at the same coordinate point, $x^{\sigma}$ of the two different gauges. Since the background quantities are same for both the guages, to first order the perturbation in the metric tensor transforms as
\begin{equation}\label{eq2:metric-trans2}
\widetilde{\delta g}_{\mu\nu}\paren*{x^{\sigma}} = \delta \gmn\paren*{x^{\sigma}} + g_{\alpha \nu} \partial_{\mu} \xi^{\alpha} + g_{\alpha \mu} \partial_{\nu} \xi^{\alpha} - \xi^{\lambda} \partial_{\lambda} \gmn,
\end{equation}
and the scale factor $a$, transforms as 
\begin{equation}\label{eq2:scale-trans}
a\paren*{\ctt}  = a\paren*{\tau} \paren*{1+\xi^{0} \cH}.
\end{equation}
Using Eqns.\ (\ref{eq2:metric-trans2}) and (\ref{eq2:scale-trans}) to the Eqn.\ (\ref{eq2:sc-metric2}), gives the transformation of the scalar metric perturbations as
\begin{subequations}
\begin{eqnarray}
\widetilde{\phi} &=& \phi -\cH \xi^{0} - \xi^{0\,\prime},\label{eq2:metric-pert-trans1}\\
\widetilde{\psi} &=& \psi + \cH \xi^{0},\label{eq2:metric-pert-trans2}\\
\widetilde{B} &=& B + \xi^{0} - \xi^{\prime},\label{eq2:metric-pert-trans3}\\
\widetilde{E} &=& E - \xi \label{eq2:metric-pert-trans4}.
\end{eqnarray}
\end{subequations}
Here $\cH(\tau)= \frac{\,\,a^\prime}{a}$ is the conformal Hubble parameter and prime $\paren*{^{\prime}}$ denotes a derivative with respect to conformal time $\tau$. Though only scalar perturbation will be discussed further, for completeness the vector transformations are given. Similarly, using Eqns.\ (\ref{eq2:metric-trans2}) and (\ref{eq2:scale-trans}) to the vector metric components in Eqn.\ (\ref{eq2:svt-metric2}), leads the vector metric perturbations to transform as
\begin{subequations}
\begin{eqnarray}
\widetilde{F}_{i} &=& F_{i} - \hat{\xi}_{i},\\
\widetilde{S}_{i} &=& S_{i} + \hat{\xi}_{\prime\,i},
\end{eqnarray}
\end{subequations}
and the tensor metric perturbation $h_{ij}^{TT}$ is gauge invariant.

The presence of gauge transformation variables, $\xi^{0}$ and $\xi$ make the perturbations gauge-dependent and, only gauge-independent quantities are physically relevant~\cite{bardeen1980prd}. The simplest gauge-invariant linear combinations constructed from $\psi$, $\phi$, $B$ and $E$ being
\begin{eqnarray}\label{eq2:bardeen-pot}
\Phi &\equiv& \phi + \cH \paren*{B-E^{\prime}} + \paren*{B-E^{\prime}}^{\prime},\\
\Psi &\equiv& \psi - \cH \paren*{B-E^{\prime}}.
\end{eqnarray}
These gauge-invariant variables, $\Phi$ and $\Psi$ were introduced by Bardeen~\cite{bardeen1980prd} in 1980 and are known as \emph{Bardeen potentials}. The significance of these two variables will be discussed later. Needless to say, there are infinite numbers of gauge-invariant variables, since any combination of gauge-invariant variables will also be gauge invariant.
%
%
%%%%%%%%%%%%%%%%%%%%%%%%%%%%%%%%%%%%%%%%%%%%%%%%%%%%%
\section{Matter Perturbation}\label{sec2:matter-pert}
The energy momentum tensor of the perfect fluid in an unperturbed Universe is given by
\begin{equation}\label{eq2:stress-energy1}
\bTmn = \paren*{\brho+\bp}\bar{u}^{\mu}\bar{u}_{\nu} + \bp \delta^{\mu}{}_{\nu},
\end{equation}
where $\brho$ is the background energy density, $\bp$ is the background isotropic pressure  and $\bar{u}_{\mu}$ is the background energy frame 4-velocity of the fluid as measured by a comoving observer at rest with respect to the fluid at any instant of time. For the unperturbed fluid, the momentum flux relative to $\bar{u}_{\mu}$ is zero such that $\bTmn\bar{u}^{\nu} = \brho \bar{u}^{\mu}$. From the normalisation condition, $\bar{u}^{\mu}\bar{u}_{\nu} = -1$, the comoving 4-velocity can be written as
\begin{equation}\label{eq2:vel-1}
\bar{u}^{\mu} = a^{-1}\paren*{1, \vec{0}}, \hspace{0.6cm} \bar{u}_{\nu} = a\paren*{-1,\vec{0}}.
\end{equation}
Small deviations from the background isotropic fluid in the form of first-order perturbations are considered such that the stress-energy tensor of the perturbed fluid, $\Tmn$, can be spilt as
\begin{equation}\label{eq2:tot-stress}
\Tmn\paren*{x^{\alpha}} = \bTmn\paren*{x^{\alpha}} + \delta \Tmn\paren*{x^{\alpha}},
\end{equation} 
with the perturbations in energy density, pressure and 4-velocity written respectively as
\begin{equation}\label{eq2:pert-def}
\rho  \equiv \brho + \delta \rho, \hspace{0.6cm} p  \equiv \bp + \delta p \hspace{0.6cm} \mbox{and} \hspace{0.6cm}u^{\mu}  \equiv \bar{u}^{\mu} + \delta u^{\mu}.
\end{equation}
The normalisation condition $u^{\mu} u_{\nu} = -1$, gives the 4-velocity, $u^{\mu}$ as
\begin{equation}\label{eq2:vel-2}
u^{\mu} = a^{-1}\paren*{1-\phi, v_{i}}, \hspace{0.6cm} u_{\nu} = a\paren*{-1-\phi, v_{i}-w_{i}} 
\end{equation}
where $v_{i}$ is the coordinate velocity of the fluid and $w_{i}$ comes from metric perturbation in Eqn.\ (\ref{eq2:met-pert}). Using Eqns.\ (\ref{eq2:pert-def}) and (\ref{eq2:vel-2}), the perturbed components of the stress-energy tensor are written as
\begin{subequations}
\begin{eqnarray}
T^{0}{}_{0} &=& -\paren*{\brho +\delta \rho},\\
T^{0}{}_{j} &=& \paren*{\brho+\bp}\paren*{v_{i}+w_{i}},\\
T^{i}{}_{0} &=&  -\paren*{\brho+\bp}v^{i},\\
T^{i}{}_{j} &=& \paren*{\bp+\delta p}\delta^{i}{}_{j} + \Pi^{i}{}_{j}\,, \hspace{0.6cm} \Pi^{i}{}_{i} = 0.
\end{eqnarray}
\label{eq2:stress-energy2}
\end{subequations}
Here, $\Pi^{i}{}_{j}$ is the traceless anisotropic stress tensor present in the perturbed fluid. Like the metric, the stress-energy tensor is also symmetric, has 10 independent components ---  two 4-scalar fields $\delta \rho$ and $\delta p$, one 3-vector field $v_{i}$ and one symmetric, traceless second-rank 3-tensor field $\Pi_{ij}$. These fields are gauge-dependent and are divided into scalar, vector and tensor (SVT) parts, for first order perturbations. Similar to the metric perturbations, on the 3-hypersurface, the 3-vector can be decomposed into scalar part $v$ and divergence-free vector part $V_{i}$ as
\begin{equation}
v_{i} = v_{i}^{\parallel}+ v_{i}^{\perp} = \nabla_{i} v + V_{i},
\end{equation}
and the symmetric, traceless second-rank 3-tensor can be decomposed into into a scalar part $\pi$, transverse vector part $\Sigma_{i}$, and transverse, traceless tensor part $\Pi_{ij}^{TT}$, expressed as
\begin{equation}
\Pi_{ij} = \Pi_{ij}^{\parallel} + \Pi_{ij}^{\perp} + \Pi_{ij}^{TT},
\end{equation}
\begin{equation}
\mbox{where} \hspace{0.6cm} \Pi^{\parallel}_{ij} = \paren*{\nabla_{i}\nabla_{j} - \frac{1}{3}\gamma_{ij}\nabla^{2}}\pi, \hspace{0.6cm} \Pi_{ij}^{\perp} = \frac{1}{2}\paren*{\nabla_{i} \Sigma_{j} + \nabla_{j} \Sigma_{i}}, 
\end{equation}
\begin{equation}
\mbox{and} \hspace{0.6cm}\gamma^{ik}\nabla_{i} \Pi_{kj}^{TT} = 0.
\end{equation}
Thus, the stress-energy tensor has 4 scalar ($\delta \rho,\delta p,v, \pi$) degrees of freedom, 4 vector ($V_{i}, \Sigma_{i}$) degrees of freedom and 2 tensor ($\Pi_{ij}^{TT}$) degrees of freedom, of which 6 are physical and 4 are gauge degrees of freedom. Under the transformation rule defined in Eqn.\ (\ref{eq2:coor-ch}), the perturbed energy-momentum tensor transforms, to first order, 
\begin{equation}\label{eq2:stress-trans2}
\widetilde{\delta T}^{\mu}{}_{\nu}\paren*{x^{\sigma}} = \delta \Tmn\paren*{x^{\sigma}} - T^{\alpha}{}_{\nu} \partial_{\alpha} \xi^{\mu} + T^{\mu}{}_{\alpha} \partial_{\nu} \xi^{\alpha} - \xi^{\lambda} \partial_{\lambda} \Tmn.
\end{equation}
Using Eqn.\ (\ref{eq2:stress-energy2}) with Eqns.\ (\ref{eq2:scale-trans}) and (\ref{eq2:stress-trans2}), the gauge transformation of the scalar perturbations of the physical quantities are given as
\begin{subequations}
\begin{eqnarray}
\widetilde{\delta \rho} &=& \delta \rho - \brho^{\prime}\xi^{0},\\
\widetilde{\delta p} &=& \delta p - \bp^{\prime}\xi^{0},\\
\widetilde{v} &=& v + \xi^{\prime},\\
\widetilde{\pi} &=& \pi.
\end{eqnarray}
\end{subequations}
The physical scalar fields $\delta \rho$ and $\delta p$ are dependent only on the choice of the temporal gauge,$\xi^{0}$, and are independent of the spatial gauge $\xi^{i}$ on the 3-hypersurface. It must be mentioned that like the metric counterpart, the traceless, transverse tensor perturbation $\Pi_{ij}^{TT}$ is gauge-invariant. 

%%%%%%%%%%%%%%%%%%%%%%%%%%%%%%%%%%%%%%%%%%%%%%%%%%%%%
\section{Conservation Of Multi-Component Fluid}\label{sec2:multi-fluid}
The Universe is filled with a mixture of different fluid components like photons ($\gamma$), neutrinos ($\nu$), baryons ($b$), cold dark matter ($c$) and dark energy ($de$), which dominate at different epochs of the evolutionary history. In a multi-component fluid mixture, the different components may transfer energy and momenta among themselves and are not conserved independently. The non-relativistic fluid description is valid for the baryons and the cold dark matter. In this Section, an interaction only between cold dark matter and dark energy is considered. From the Bianchi identity, it follows that the total matter content of the Universe must be conserved and consequently
\begin{equation}\label{eq2:conserve1}
T^\mu{}_{\nu;\,\mu} = 0,
\end{equation}
where $\Tmn$ is the energy-momentum tensor of the total fluid mixture and $\Tmn= \sum_A T^\mu_{\left(A\right)\nu}$. Here, $T^\mu_{\left(A\right)\nu}$ is the energy-momentum tensor of the individual fluid `A'; `A' can be cold dark matter and dark energy, if dark energy is assumed to be represented by a perfect fluid. The properties of the energy-momentum tensor discussed in the previous Section \ref{sec2:matter-pert} are valid for the individual components as well for the combination. as the total fluid, The total energy density is $\rho = \sum_{A} \rho_{A}$, the total pressure is $p = \sum_{A} p_{A}$ and the total 4-velocity $u^{\mu}$ is the average for all fluids\footnote{Individual fluid components will be labelled with subscript `A' and total fluid will be without label}. Thus, for the total energy frame with zero momentum flux, the coordinate velocity of the total fluid $v^{i}$ is related to the coordinate velocity, $v^{i}_{A}$ of fluid `A' as
\begin{equation}
\paren*{\brho + \bp} v^{i} = \Sum_{A} \paren*{\brho_{A} + \bp_{A}} v^{i}_{A}.
\end{equation}

When there is an energy transfer among the different components, the divergence of the energy-momentum tensor of each component has a general source term, $\bar{Q}_{\left(A\right)\nu}$. Thus, the covariant form of the conservation equation for fluid `A', follows from Eqn.\ (\ref{eq2:conserve1}) as,
\begin{equation} \label{eq2:condition}
\bar{T}^\mu_{\paren*{A} \nu;\,\mu} = \bar{Q}_{\left(A\right)\nu} ~, \hspace{0.2cm} \mbox{where}  \hspace{0.2cm} \Sum_A \bar{Q}_{\left(A\right)\nu} =0.
\end{equation}
It must be mentioned here that energy-momentum is transferred between fluid `A' and other fluid so that $\bar{Q}_{(A)}^{\mu}$ is source for fluid `A' and sink for other fluids, which leads to $\Sum_A \bar{Q}_{\left(A\right)\nu} =0$. The source term for the interaction is a $4$-vector and has the form
\begin{equation} \label{eq2:Q-def}
\bar{Q}^\mu_{\paren*{A}}=\frac{1}{a} \left(\bar{Q}_{A}, \vec{0}\right),
\end{equation}
where $\bar{Q}_{A}= \bar{Q}_{(A)}^{0}$, the time component of the four vector $\bar{Q}_{(A)}^{\mu}$. It is assumed that there is no momentum transfer in the background Universe. Thus, in a homogeneous and isotropic background, the energy balance equation for the fluid `A'  is written as
\begin{equation}\label{eq2:balaneA}
\brhoA^{\prime}+ 3 \cH \left(1+\bar{w}_{A}\right) \brhoA= a \bar{Q}_A,
\end{equation}
where $\bar{w}_{A} \paren*{= \bpA/\brhoA}$ is the equation of state (EoS) parameter of fluid `A'. The non-interacting scenario can be recovered by setting $\bar{Q}_A = 0$.

To study the effect of interaction in the interaction, it is convenient to decompose the source term relative to the total 4-velocity as,
\begin{subequations}
\begin{eqnarray}
&Q^{\,\mu}_{\paren*{A}} = Q_{A} u^{\,\mu} + F^{\,\mu}_{\paren*{A}}\,, \hspace{1cm} &Q^{\,\mu}_{\paren*{A}} = \bar{Q}^{\,\mu}_{\paren*{A}} + \delta Q^{\,\mu}_{\paren*{A}},\label{eq2:Q-pert-def}\\
&u_{\mu} F^{\,\mu}_{\paren*{A}} = 0\,, \hspace{1cm} &F^{\,\mu}_{\paren*{A}} = a^{-1}\, \paren*{0,\partial^{i}\, f_A}\label{eq2:F-pert-def}.
\end{eqnarray}
\label{eq2:pert-def}
\end{subequations}
Here, $\delta Q^{\,\mu}_{\paren*{A}}$ is the perturbation in energy transfer rate, $F^{\,\mu}_{\paren*{A}}$ is the perturbation in the momentum density transfer rate and $f_{A}$ is the momentum transfer potential. From Eqn.\ (\ref{eq2:condition}), it is clear that the perturbations, $\delta Q_{A}$ and $f_{A}$, will vanish when the total fluid mixture is considered,
\begin{equation}
\Sum_A \delta Q_A =\Sum_A f_A =0.\label{eq2:pert-condition}
\end{equation}
In the linear perturbation theory, the 4-vector, $Q_{\left(A\right)\nu}$ can also be decomposed into scalar, vector and tensor parts. Using Eqns.\ (\ref{eq2:vel-2}) and (\ref{eq2:pert-def}) the temporal and spatial components of the source term are obtained as,
\begin{subequations}
\begin{eqnarray}
Q^{0}_{\paren*{A}} &=& a^{-1}\,\left[\bar{Q}_A(1-\phi)+\delta Q_A\right],\\
Q^{i}_{\paren*{A}} &=& a^{-1}\,\partial^{i}\paren*{\bar{Q}_A\,v+ f_A}.
\end{eqnarray}
\label{eq2:inter-trans1}
\end{subequations}
Under the transformation rule defined in Eqn.\ (\ref{eq2:coor-ch}) to the first order, the perturbed source vector transforms as, 
\begin{equation}\label{eq2:inter-trans2}
\widetilde{\delta Q}^{\mu}_{\paren*{A}}\paren*{x^{\sigma}} = \delta Q^{\mu}_{\paren*{A}}\paren*{x^{\sigma}} -Q^{\alpha}_{\paren*{A}} \partial_{\alpha} \xi^{\mu} - \xi^{\lambda} \partial_{\lambda} Q^{\mu}_{\paren*{A}}.
\end{equation}
Using Eqn.\ (\ref{eq2:inter-trans1}) with Eqns.\ (\ref{eq2:scale-trans}) and (\ref{eq2:inter-trans2}), the gauge transformation of the scalar perturbations of the energy transfer rate and the momentum transfer potential are given as,
\begin{eqnarray}
\widetilde{\delta Q}_{\paren*{A}} &=& \delta Q_{\paren*{A}} -Q^{\prime}_{\paren*{A}} \xi^{0},\\
\widetilde{f}_{\paren*{A}} &=&  f_{\paren*{A}}.
\end{eqnarray}
Thus, the momentum transfer potential is invariant under gauge transformation. 

In the perturbed spacetime, the conservation of the energy-momentum of the fluid `A' are governed by the \emph{energy conservation} and the \emph{momentum conservation} equations respectively, as
\begin{eqnarray}
\begin{split}
\delta \rA^\prime + 3 \cH \,\paren*{\delta \rA+\delta \pA}-3\paren*{\brhoA+\bpA}\psi^{\prime}+&\nabla^{2}\paren*{\brhoA+\bpA}\paren*{\vA+E^{\prime}} \\
 =&~a\bar{Q}_A\phi +a \delta Q_A , \label{eq2:e1}
\end{split}\\
\begin{split}
\left[\paren*{\brhoA+\bpA}\paren*{\vA+B}\right]'+4\cH \paren*{\brhoA+\bpA}\paren*{\vA+B}&+\paren*{\brhoA+\bpA}\phi+\delta \pA \\
 +\frac{2}{3 a^{2}}\nabla^{2} \pi &=a\bar{Q}_A\paren*{v+B}+a f_A. \label{eq2:m1}
\end{split}
\end{eqnarray}
The conservation of the energy density, Eqn.\ (\ref{eq2:e1}), depends on the background expansion rate, evolution of the metric perturbation, the spatial evolution of the velocity field and the energy transfer with other components. Similarly, the conservation of the momentum, Eqn.\ (\ref{eq2:m1}), depends on the background expansion rate, the metric perturbation, the spatial evolution of the anisotropic pressure field and the momentum transfer with other components. It must be noted again, there is no momentum transfer in the unperturbed background, but there is a momentum transfer in the perturbed Universe. 
%
%%%%%%%%%%%%%%%%%%%%%%%%%%%%%%%%%%%%%%%%%%%%%%%%%%%%%
\subsection{Pressure Perturbation And Sound Velocity}
In an arbitrary gauge, the pressure perturbation, $\delta \pA$, depends on the density perturbation, $\delta \rA$, as~\cite{waynehu1998apj, valiviita2008jcap, bean2004prd} 
\begin{equation}\label{eq2:pert-p1}
\delta \pA = c_{a,A}^{2} \delta \rA + \paren*{c_{s,A}^{2}-c_{a,A}^{2}}\left[ \delta \rA+ \brhoA^{\prime}\paren*{\vA+B}\right],
\end{equation}
where the first term in the right hand side represents \emph{adiabatic} pressure perturbation and the second term represents the \emph{non-adiabatic} or \emph{entropic} pressure perturbation. The quantity
\begin{equation}\label{eq2:ca2}
c_{a,A}^2=\frac{\bpA^\prime}{\brhoA^\prime}= \bar{w}_A+\frac{\bar{w}_A^\prime}{\brhoA^\prime/\bpA^\prime}
\end{equation}
is the square of adiabatic speed of sound in the fluid `A'and $c_{s,A}^2$ is the square of effective speed of sound in the fluid `A', defined as
\begin{equation}\label{eq2:cs2}
c_{s,A}^2=\frac{\delta p_{A}}{\delta \rho_{A}} \bigg \rvert_{rest,A}
\end{equation}
is the ratio of pressure fluctuation to density fluctuation in the rest frame of fluid `A'. In presence of an interaction, the dependance on the interaction term, $\bar{Q}_A$, manifests itself through the non-adiabatic pressure. Thus, from Eqns.\ (\ref{eq2:balaneA}) and (\ref{eq2:pert-p1}), the pressure perturbation can be expressed as
\begin{equation}
\delta \pA=c_{s,A}^2 \delta \rA+\paren*{c_{s,A}^2-c_{a,A}^2}\left[-3 \cH\paren*{1+\bar{w}_A}\brhoA +a \bar{Q}_A\right]\paren*{\vA+B}.
\end{equation}
It deserves mention that, in general, the effective sound speed $c_{s,A}^2$, in Eqn.\ (\ref{eq2:cs2}), is different from the adiabatic sound speed $c_{a,A}^{2}$, in Eqn.\ (\ref{eq2:ca2}). For a fluid with $\pA \propto \rA$, the two sound speeds are equal, $c_{a,A}^{2} = c_{s,A}^2$. For a dark energy model with dynamical EoS parameter, $w_{de}$, the effective sound speed is set as, $c_{s,de}^2 = 1$, to avoid instabilities in the dark energy perturbations. As shown in~\cite{mehrabi2015mnras}, $c_{s,de}^2$ plays a significant role in DE clustering and hence DM clustering. When $c_{s,de}^2 \simeq 1$, the pressure perturbation should suppress any growth in DE perturbation whereas when $c_{s,de}^2 \ll 1$, DE perturbation should grow like that of DM.  It is shown in~\cite{mehrabi2015mnras, batista2013jcap, batista2017jcap} that DE can cluster like DM when $c_{s,de}^2=0$.

%%%%%%%%%%%%%%%%%%%%%%%%%%%%%%%%%%%%%%%%%%%%%%%%%%%%%
\section{Gauge-dependent Field Equations}\label{sec2:field-eq2}
The Einstein equation, Eqn.\ (\ref{eq1:Enstein1}), in an inhomogeneous and anisotropic Universe can be split into background and perturbation equations. The perturbed Einstein equations to first order are written as
\begin{equation}\label{eq2:Enstein2}
\delta G^{\mu}{}_{\nu} =  8 \pi G_{N} \delta T^{\mu}{}_{\nu}.
\end{equation}
For the scalar metric perturbation given by the line element in Eqn.\ (\ref{eq2:sc-metric2}) in an arbitrary gauge, the $\paren*{0\mhyphen 0}$, $\paren*{0\mhyphen i}$ and $\paren*{i\mhyphen j}$ components of the perturbed field equations~\cite{mukhanov1992pr} are, respectively, 
\begin{equation}
\left[-3 \cH \paren*{\cH \phi +\psi^{\prime}} + \nabla^{2} \lbrace \psi -\cH \paren*{B- E^{\prime}}\rbrace\right] = - 4\piGa \delta \rho,\label{eq2:g002}
\end{equation}
\begin{equation}
\partial_{i}\left[\cH\phi +\psi^\prime -\cH \paren*{B- E^{\prime}}\right] = 4\piGa \paren*{\brho+\bp} \partial_{i} v,\label{eq2:g0j2}
\end{equation}
\begin{equation}\label{eq2:gij2}
\begin{split}
\left[\psi^{\prime \prime} + \paren*{2 \cH^{\prime} + \cH^{2}} \phi + \cH\paren*{\phi+2\psi}^{\prime} +\frac{1}{2} \nabla^{2} \mathcal{D} \right]\delta^{i}{}_{j} -\frac{1}{2}\gamma^{i k}&\partial_{k}\partial_{j} \mathcal{D}\\
= 4\piGa \left[\delta p \delta^{i}{}_{j} + \bp \paren*{\gamma^{i k}\,\partial_{k}\partial_{j} - \frac{1}{3}\gamma^{i}{}_{j}\nabla^{2}}\pi\right]&,\\
\mbox{with} \hspace{0.6cm}\mathcal{D} = \paren*{\phi-\psi} +2 \cH \paren*{B- E^{\prime}} + \paren*{B- E^{\prime}}^{\prime}.~~~~&
\end{split}
\end{equation}
The $\paren*{i\mhyphen j}$ component, Eqn.\ (\ref{eq2:gij2}), can be separated into trace and traceless parts as
\begin{subequations}
\begin{eqnarray}
&\psi^{\prime \prime} + \paren*{2 \cH^{\prime} + \cH^{2}} \phi + \cH\paren*{\phi+2\psi}^{\prime} +\frac{1}{3} \nabla^{2} \mathcal{D} = 4 \piGa \delta p,&\\
&\paren*{\gamma^{i k}\,\partial_{k}\partial_{j} - \frac{1}{3}\gamma^{i}{}_{j}\nabla^{2}} \mathcal{D} = -8 \piGa \bp \paren*{\gamma^{i k}\,\partial_{k}\partial_{j} - \frac{1}{3}\gamma^{i}{}_{j}\nabla^{2}}\pi.&\label{eq2:gij22b}
\end{eqnarray}\label{eq2:gij22}
\end{subequations}
The Eqns.\ (\ref{eq2:g002}), (\ref{eq2:g0j2}) and (\ref{eq2:gij22}), are gauge-dependent linear perturbation equations. For a set of linearised equations, it is convenient to solve them in the Fourier space~\cite{liddle2000}. Any perturbation,  $f\paren*{\tau, \vec{x}}$ can be expanded in Fourier modes as
\begin{equation}\label{eq2:fourier}
f\paren*{\tau, \vec{x}} = \int \frac{d^{3} k}{\paren*{2 \pi}^{3}} f_{k} \paren*{\tau} e^{\im\vec{k}\cdot\vec{x}}, \hspace{0.6cm} \im \equiv \sqrt{-1}
\end{equation}
where $\vec{k}$ is the comoving wave-vector and $\vec{x} = x^{i}$ are the spatial coordinates. The comoving wavenumber, $k \equiv \abs{\vec{k}}$, is related to the physical wavenumber, $k_{p}$ as $k = a\,k_{p}$, where $a$ is the scale factor. Since each Fourier mode evolves independently for first-order perturbation, they can be studied separately. So, for brevity, the subscript $k$ is omitted henceforth. In Fourier space the perturbed field Eqns.\ (\ref{eq2:g002}), (\ref{eq2:g0j2}) and (\ref{eq2:gij22}) are expressed as
\begin{equation}
\left[-3 \cH \paren*{\cH \phi +\psi^{\prime}} -k^{2} \lbrace \psi -\cH \paren*{B- E^{\prime}}\rbrace\right] = - 4\piGa \delta \rho,\label{eq2:g003}
\end{equation}
\begin{equation}
\left[\cH\phi +\psi^\prime -\cH \paren*{B- E^{\prime}}\right] = 4\piGa \paren*{\brho+\bp} v, \label{eq2:g0j3}
\end{equation}
\begin{equation}
\psi^{\prime \prime} + \paren*{2 \cH^{\prime} + \cH^{2}} \phi + \cH\paren*{\phi+2\psi}^{\prime} -\frac{1}{3} k^{2} \mathcal{D} = 4\piGa \delta p,\label{eq2:gij31}
\end{equation}
\begin{equation}
k^{2} \mathcal{D} = -8\piGa \bp \pi,\label{eq2:gij32}
\end{equation}
\begin{equation*}
\mbox{with} \hspace{0.6cm}\mathcal{D} = \paren*{\phi-\psi} +2 \cH \paren*{B- E^{\prime}} + \paren*{B- E^{\prime}}^{\prime}.
\end{equation*}
Equations (\ref{eq2:g0j3}) and (\ref{eq2:gij31}) give the evolution of the metric perturbations whereas Eqns.\ (\ref{eq2:g003}) and (\ref{eq2:gij32}) provide the necessary constraint on the metric perturbations. The evolution equations with the constraint are solved using specific gauges.

%%%%%%%%%%%%%%%%%%%%%%%%%%%%%%%%%%%%%%%%%%%%%%%%%%%%%
\section{Choice of Gauges}\label{sec2:gaugechoices}
To specify a gauge, two conditions on the gauge-dependent variables must be imposed --- one for the time-coordinate and one for the space-coordinate. The fixing of time-coordinate is the choice of time \emph{slicing} of the perturbed spacetime while that of the scape-coordinate is the \emph{threading} of the perturbed spacetime. Thus, $\xi^{0}$ changes the slicing and $\xi^{i}$ changes the threading. In the following, only two particular choices of slicing and threading will be considered.
\paragraph*{Longitudinal gauge :}
Longitudinal gauge conditions can be implemented by setting $\widetilde{B}=0$ and $\widetilde{E}=0$~\cite{mukhanov1992pr} in Eqns.\ (\ref{eq2:metric-pert-trans3}) and (\ref{eq2:metric-pert-trans4}) respectively such that
\begin{subequations}
\begin{eqnarray}
\xi^{0} &=& -\paren*{B-E^{\prime}},\label{eq2:long-gauge1}\\
\xi &=& E.\label{eq2:long-gauge2}
\end{eqnarray}
\end{subequations}
Thus, the condition $E = 0$, fixes $\xi$ uniquely while the condition $B = 0$ which means the threads are orthogonal to the slices, fixes $\xi^{0}$ uniquely. Hence, the longitudinal gauge has no residual gauge modes. With the conditions (\ref{eq2:long-gauge1}) and (\ref{eq2:long-gauge2}), the metric perturbation variables transform  as
\begin{subequations}
\begin{eqnarray}
\phi^{L} &=& \Phi, \hspace{0.6cm} \psi^{L} = \Psi,\\
B^{L} &=& 0, \hspace{0.6cm} E^{L} = 0,
\end{eqnarray}
\end{subequations}
and the matter perturbation variables transform as
\begin{subequations}
\begin{eqnarray}
\delta \rho^{L} &=& \delta\rho+ \brho^{\prime}\paren*{B-E^{\prime}},\\
\delta p^{L} &=& \delta p + \bp^{\prime}\paren*{B-E^{\prime}},\\
v^{L} &=& v - E^{\prime},\\
\pi^{L} &=& \pi.
\end{eqnarray}
\end{subequations}
The superscript `L' in the above expressions stands for \emph{longitudinal} gauge. In longitudinal gauge, the gauge-invariant variables, $\Phi$ and $\Psi$ denote the amplitude of metric perturbations and the perturbed metric takes the form
\begin{equation}\label{eq2:long-metric2}
ds^{2}= a^{2}\paren*{\tau}\left[-\paren*{1+2\Phi}d \tau^2 +\paren*{1-2\Psi}dx^{i} dx^{j}\right].
\end{equation}
From Eqn.\ (\ref{eq2:gij22}), in absence of anisotropic stress, $\phi = \psi$ or $\Phi = \Psi$, the metric perturbation depends on the generalised Newtonian potential, $\Phi$ only. For this reason, the \emph{longitudinal} gauge is also called the \emph{conformal-Newtonian} gauge or simply \emph{Newtonian} gauge. This choice of gauge will be used in Chapter \ref{chap4:epjp}.
\paragraph*{Synchronous gauge :}
Synchronous gauge conditions can be arrived at from any arbitrary initial coordinate system by setting $\widetilde{\phi} = 0$ and $\widetilde{B} = 0$~\cite{lifshitz2017grg, mukhanov1992pr} in Eqns.\  (\ref{eq2:metric-pert-trans2}) and (\ref{eq2:metric-pert-trans3}) respectively and solving the differential equations for $\xi^{0}$ and $\xi$ as,
\begin{subequations}
\begin{eqnarray}
\xi^{0\,\prime} &=& \phi - \cH \xi^{0},\label{eq2:sync-gauge2}\\
\xi^{\prime}  &=& -\xi^{0} - B.\label{eq2:sync-gauge2}
\end{eqnarray}
\end{subequations}
The condition $\phi = 0$ implies that for the two adjacent hypersurfaces, proper-time distance coincides with the coordinate-time distance along the normal vector defining these hypersurfaces, while $B=0$, means that the space coordinates are specified by the condition that the threads are orthogonal to the slices. Synchronous gauge conditions do not specify the initial hypersurface and the coordinate system, and hence synchronous gauge conditions do not eliminate all gauge freedom. It can be assumed that there exists a set of non-trivial comoving observers, free-falling along the constant space-coordinate lines or threads ($x^{i} = $ constant) called the ``fundamental observers''. These fundamental observers densely populate the spacetime. The synchronous gauge spacetime coordinates are defined by the conformal time read by the observer's clock and the constant spatial coordinate labelled by their positions. The residual gauge freedom arises from the freedom to choose the initial clock settings and the spatial coordinate labels of the observers.

In synchronous gauge only the spatial hypersurface is perturbed and the metric takes the form
\begin{equation} \label{eq2:sync-metric2}
\begin{split}
ds^{2}=a^{2}\paren*{\tau} & \left\{ -d\tau^2 +\left[\paren*{1-2\psi}\delta_{ij}+2\partial_i\partial_jE\right]dx^idx^j \right\},
\end{split}
\end{equation}
Inspite of having residual gauge freedom, synchronous gauge is used in the publicly available Boltzmann's solvers like \camb\footnote{Available at \href{https://camb.info}{https://camb.info}}\nomenclature{CAMB}{Code for Anisotropies in the Microwave Background}~\cite{lewis1999bs} and will be used in the Chapters \ref{chap5:jcap} and \ref{chap6:prd}.

In synchronous gauge, $\phi^{S}=B^{S}=0$ and the perturbation is only in the spatial hypersurface,
\begin{equation}
h_{ij} = -2 \psi\delta_{ij}+ 2\partial_{i}\partial_{j}E.
\end{equation}
Using the notation of~\cite{ma1995apj}, two functions $\eta\paren{\tau,\vec{k}}$ and $\msh\paren{\tau,\vec{k}}$ in the Fourier space are defined as $\eta\equiv\psi$ and $\msh \equiv h^{i}_{i}$, such that spatial perturbation $h_{ij}$ in Fourier space is
\begin{equation}\label{eq2:h-sync}
h_{ij} = -2\eta\delta_{ij} + 2\partial_{i}\partial_{j}E.
\end{equation}
In synchronous gauge, the metric perturbation is discussed in terms of the variables $\eta$ and $\msh$ in Fourier space~\cite{ma1995apj}. Rearranging the trace of Eqn.\ (\ref{eq2:h-sync}) as, 
\begin{equation}\label{eq2:syn-con}
2\nabla^{2}E = \msh +6\eta,
\end{equation}
yields the relation between $E$ and $\eta$ in Fourier space, as
\begin{equation}
k^{2}E=-\msh/2-3\eta.
\end{equation}

%%%%%%%%%%%%%%%%%%%%%%%%%%%%%%%%%%%%%%%%%%%%%%%%%%%%%
\section{Perturbation In A Scalar Field Model}
For the scalar field $\vphi$ with a potential $\Vp$, energy density and pressure are given as $\rphi = \frac{1}{2 a^{2}}\vphi^{\prime \,2} +\Vp$ and $\pphi = \frac{1}{2 a^{2}}\vphi^{\prime \,2} -\Vp$ respectively and the EoS parameter is given by 
\begin{eqnarray}
\wphi = \frac{\pphi}{\rphi}~=~ \frac{\frac{1}{2 a^{2}}\vphi^{\prime \,2} -\Vp}{\frac{1}{2 a^{2}}\vphi^{\prime \,2} +\Vp}~=~ 1-\frac{2 \,\Vp}{\rphi}~. \label{eq5:wphi}
\end{eqnarray}
The Klein-Gordon equation in conformal time can be obtained as
\begin{equation} \label{eq5:kg1}
\vphi^{\prime \prime} + 2 \cH \vphi^{\prime} + a^{2} \frac{d V}{d \vphi}= 0~.
\end{equation}

The perturbation $\delta \vphi$ in the scalar field has the equation of motion
\begin{equation} \label{eq5:kg2}
\delta \vphi^{\prime \prime} + 2 \cH \delta \vphi^{\prime} - \nabla^{2}\delta \vphi +a^{2}\,\frac{d^{2} V}{d \vphi^{2}} \delta \vphi + \frac{1}{2} \vphi^{\prime} \msh^{\prime}= 0,
\end{equation}
\begin{equation}
\delta \vphi^{\prime \prime} + 2 \cH \delta \vphi^{\prime} - \nabla^{2}\delta \vphi +a^{2}\,\frac{d^{2} V}{d \vphi^{2}} \delta \vphi+ 2a^{2}\,\frac{d V}{d \vphi} \phi - \vphi^{\prime}\phi^{\prime} -3 \vphi^{\prime}\psi^{\prime} = 0,
\end{equation}
in the Fourier space with wavenumber $k$. The perturbation in energy density $\delta \rphi$ and pressure $\delta \pphi$ are given as
\begin{eqnarray}
\delta \rphi &=& -\delta T^{0}{}_{0 \paren*{\vphi}}~=~ \frac{\vphi^{\prime} \delta \vphi^{\prime}}{a^{2}}+\delta \vphi \frac{d V}{d \vphi} , \label{eq5:pe2}\\
\delta T^{0}{}_{j\paren*{\vphi}} &=& - \frac{\vphi^{\prime}}{a^{2}}\partial_{j}\delta \vphi, \label{eq5:pv2}\\
\delta \pphi \delta^{i}{}_{j}&=& \delta T^{i}{}_{j\paren*{\vphi}}~=~ \paren*{\frac{\vphi^{\prime} \delta \vphi^{\prime}}{a^{2}}-\delta \vphi \frac{d V}{d \vphi}}\delta^{i}{}_{j}\label{eq5:pp2},
\end{eqnarray}
when expanded in the Fourier space. Here, $\delta T^{\mu}{}_{\nu\paren*{\vphi}}$ is the perturbed stress-energy tensor of the scalar field. For the evolution of perturbations in different scalar field models we refer to~\cite{jassal2008prd, jassal2009prd, jassal2010prd, jassal2012prd}.\\
%
%%%%%%%%%%%%%%%%%%%%%%%%%%%%%%%%%%%%%%%%%%%%%%%%%%%%%
\section{Evolution Equations For Radiation}\label{sec2:rad-evol}
The evolution equations for the relativistic components like photons and neutrino are obtained by the first order \emph{Boltzmann equation}. Schematically, the evolution of the distribution function $f$, in absence of any collision is written as
\begin{equation}\label{eq2:coll-less}
\frac{d f}{dt} = 0.
\end{equation}
This is known as the \emph{collisionless Boltzmann equation} and expresses the \emph{Liouville's theorem} which states that \emph{the number of particles in a given element of phase space does not change with time.} The distribution function, $f$ is a function of spacetime $x^{\mu}$ and on the momentum vector in comoving frame as $P^{\mu} \equiv \frac{d x^{\mu}}{d \lambda}$, where $\lambda$ is the parameter giving the particle's path. 

In presence of all the possible collision terms $C\left[f\right]$, the \emph{collisional Boltzmann equation} is schematically written as
\begin{equation}\label{eq2:coll}
\frac{d f}{dt} = C\left[f\right].
\end{equation}
The collision term includes the Thomson scattering of photons with the electrons. The electrons on the other hand are tightly coupled to the baryons and can be considered as a single electron-baryon fluid.

%%%%%%%%%%%%%%%%%%%%%%%%%%%%%%%%%%%%%%%%%%
\subsection{Boltzmann Equation For Photons}
%, $c=1$ is the speed of light 
In an inhomogeneous spacetime, the photon distribution function $f$ depends on cosmic time $t$, position $x^{i}$, magnitude of generalised momentum $p=E$, $E$ being the energy of photon and direction vector $\hat{p}^{i}$ which is normalised as $\delta_{ij}\hat{p}^{i}\hat{p}^{j} = 1$. The left hand side of Eqn.\ (\ref{eq2:coll}) can be expanded as
\begin{equation}\label{eq2:f-expand1}
\frac{d f}{d t} = \frac{\partial f}{\partial t} + \frac{\partial f}{\partial x^{i}} \cdot \frac{d x^{i}}{d t} + \frac{\partial f}{\partial p} \frac{d p}{d t} + \underbrace{\frac{\partial f}{\partial \hat{p}^{i}} \cdot \frac{d \hat{p}^{i}}{d t} }_{\mathcal{O}\paren*{2}}.
\end{equation}
The photon distribution in the background Universe depends only on the magnitude of momentum, $p$ and not on the direction $\hat{p}^{i}$ and the photon travels in straight line. So, both $\frac{\partial f}{\partial \hat{p}^{i}}$ and $\frac{d \hat{p}^{i}}{d t}$ are first order in perturbation and are neglected as a product. The coefficients $\frac{d x^{i}}{d t}$ and $\frac{d p}{d t}$ are computed in cosmic time $t$ following~\cite{dodelson2003}. Due to the masslessness of photon
\begin{equation}
P^{2} \equiv \gmn P^{\mu} P^{\nu} = g_{00}\paren{P^{0}}^{2}+g_{ij}P^{i}P^{j} =0.
\end{equation}
Defining, $g_{ij}P^{i}P^{j} = p^{2}$ and using the longitudinal gauge metric components in (\ref{eq2:long-metric2}) in cosmic time, the time component of $P^{\mu}$ is obtained as
\begin{equation}\label{eq2:p0}
P^{0} = p \paren*{1-\Phi}.
\end{equation}
Using the definition of the comoving momentum vector, the coefficient $\frac{d x^{i}}{d t}$ is written as
\begin{equation}\label{eq2:dxi1}
\frac{d x^{i}}{d t} = \frac{d x^{i}}{d \lambda} \frac{d \lambda}{d t} = \frac{P^{i}}{P^{0}},
\end{equation}
where $P^{0} \equiv \frac{dt}{d\lambda}$ is given by Eqn.\ (\ref{eq2:p0}) and $P^{i} \equiv \frac{dx^{i}}{d\lambda}$ is obtained as $\hat{p}^{i} p\paren*{1-\Psi}/a$, following~\cite{ma1995apj, dodelson2003}. Combining, $P^{0}$ and $P^{i}$, Eqn.\ (\ref{eq2:dxi1}) is written as 
\begin{equation}\label{eq2:dxi2}
\frac{d x^{i}}{d t} = \frac{\hat{p}^{i}}{a} \paren*{1+\Phi-\Psi}.
\end{equation}
In an overdense region, $\Phi<0$ and $\Psi>0$, making the term in the parentheses less than one. Thus, Eqn.\ (\ref{eq2:dxi2}) says that the gravitational force of the overdense region slows down the photon ($\frac{d x^{i}}{d t}$ becomes smaller).

Following~\cite{dodelson2003}, the coefficient $\frac{d p}{d t}$ is obtained as
\begin{equation}\label{eq2:dp1}
\frac{1}{p} \frac{dp}{dt} = -H - \frac{\partial \Psi}{\partial t} - \frac{\hat{p}^{i}}{a} \frac{\partial \Phi}{\partial x^{i}}. 
\end{equation}
Eqn.\ (\ref{eq2:dp1}) gives the change in momentum as the photon moves through the perturbed Universe. The first term on the right hand side denotes the loss of momentum due to the Hubble expansion. The second term says in an overdense region, the photon loses more energy to emerge from a deepening gravitational well ($\frac{\partial \Psi}{\partial t}>0$). The third term says that a photon travelling into a well ($\hat{p}^{i}\frac{\partial \Phi}{\partial x^{i}} <0$) gains energy as it is being pulled into the centre. Plugging expressions (\ref{eq2:dxi2}) and (\ref{eq2:dp1}) in Eqn.\ (\ref{eq2:f-expand1}) and neglecting all second order terms, the total derivative of $f$ is written as
\begin{equation}\label{eq2:f-expand2}
\frac{d f}{d t} = \frac{\partial f}{\partial t} +\frac{\hat{p}^{i}}{a} \frac{\partial f}{\partial x^{i}} - p \frac{\partial f}{\partial p} \left[H +\frac{\partial \Psi}{\partial t} +\frac{\hat{p}^{i}}{a} \frac{\partial \Phi}{\partial x^{i}} \right].
\end{equation}
In Eqn.\ (\ref{eq2:f-expand2}), $ax^{i}$ is the \emph{physical distance}. 

In an unperturbed Universe, the photon distribution is identified as the \emph{Bose-Einstein distribution} with zero chemical potential and is expressed as
\begin{equation}
f^{\paren*{0}} = \frac{1}{\exp\paren*{p/T}-1},
\end{equation}
where $T\paren*{t}$ is the temperature of the photon distribution and is obtained to be $2.7255$ K as the microwave background temperature~\cite{fixsen2009apj}. Adding a perturbation $\delta T$ to the temperature $T$ such that the temperature anisotropy is $\Theta \equiv \delta T/T$, the distribution function is written as
\begin{equation}
f\paren*{t, \vec{x}, p, \hat{p}} = \frac{1}{\exp\lbrace p/T\left[1+\Theta\paren*{t, \vec{x},\hat{p}}\right]\rbrace-1}.
\end{equation}
Expanding upto first order, $f$ is written as
\begin{equation}\label{eq2:f-expand3}
f \simeq f^{\paren*{0}} + T \frac{\partial f^{\paren*{0}}}{\partial T} \Theta = f^{\paren*{0}} - p \frac{\partial f^{\paren*{0}}}{\partial p} \Theta.
\end{equation}
Putting Eqn.\ (\ref{eq2:f-expand3}) into Eqn.\ (\ref{eq2:f-expand2}) and collecting terms first order in perturbation, the first order Boltzmann equation results into%
\begin{equation}\label{eq2:f-order1}
\frac{d f^{\paren*{1}}}{d t} = - p\frac{\partial f^{\paren*{0}}}{\partial p} \left[\frac{\partial \Theta}{\partial t} + \frac{\hat{p}^{i}}{a}\frac{\partial \Theta}{\partial x^{i}} + \frac{\partial \Psi}{\partial t} + \frac{\hat{p}^{i}}{a}\frac{\partial \Phi}{\partial x^{i}}\right].
\end{equation}
The first two terms on the right hand side of Eqn.\ (\ref{eq2:f-order1}), accounts for \emph{free streaming} which denotes anisotropies on increasingly small scales as the Universe evolves. The last two terms account for the effect of gravity.

The collision term $C\left[f\paren*{\vec{p}}\right]$ on the right hand side of Eqn.\ (\ref{eq2:coll}) for the photon is obtained following~\cite{dodelson2003} as
\begin{equation}\label{eq2:c-order1}
C\left[f\paren*{\vec{p}}\right] = - p\frac{\partial f^{\paren*{0}}}{\partial p} n_{e} \sigma_{T} \left[\Theta_{0} - \Theta\paren*{\hat{p}} + \hat{p}\cdot \vec{v}_{b} \right],
\end{equation}
where $n_{e}$ is the electron density, $\sigma_{T}$ is the Thomson scattering cross-section and $\vec{v}_{b}$ is the velocity of the electrons, which are tightly coupled to the baryons. Here $\Theta_{0}$ is the \emph{monopole} part of the perturbation and is defined as an integral of the perturbation over all the directions as
\begin{equation}
\Theta_{0}\paren*{t, \vec{x}} = \frac{1}{4\pi}\int \Theta\paren*{t, \vec{x}, \tilde{\hat{p}}} d \widetilde{\Omega}.
\end{equation}
The monopole term $\Theta_{0}$ denotes the perturbation at a given point in space over its average in all space.

Finally, combining Eqns.\ (\ref{eq2:f-order1}) and (\ref{eq2:c-order1}), the collisional Boltzmann equation at first order for photon is written as%
\begin{equation}
\frac{\partial \Theta}{\partial t} + \frac{\hat{p}^{i}}{a}\frac{\partial \Theta}{\partial x^{i}} + \frac{\partial \Psi}{\partial t} + \frac{\hat{p}^{i}}{a}\frac{\partial \Phi}{\partial x^{i}} = n_{e} \sigma_{T} \left[\Theta_{0} - \Theta + \hat{p}\cdot \vec{v}_{b} \right].
\end{equation}
In terms of conformal time $\tau$, the Boltzmann equation becomes
\begin{equation}\label{eq2:col-final}
\Theta^{\prime}+ \hat{p}^{i}\frac{\partial \Theta}{\partial x^{i}} + \Psi^{\prime} + \hat{p}^{i}\frac{\partial \Phi}{\partial x^{i}} = a n_{e} \sigma_{T} \left[\Theta_{0} - \Theta + \hat{p}\cdot \vec{v}_{b} \right].
\end{equation}
Using Eqn.\ (\ref{eq2:fourier}), in Fourier mode, Eqn.\ (\ref{eq2:col-final}) can be written as
\begin{equation}\label{eq2:col-final2}
\Theta^{\prime}+ \im\,k\,\mu\,\Theta + \Psi^{\prime} + \im\,k\,\mu\,\Phi =-\tau_{d}^{\prime}\left[\Theta_{0} - \Theta + \mu v_{b} \right],
\end{equation}
where $\mu\equiv \frac{\vec{k}\cdot \hat{p}}{k}$ is the cosine of the angle between the wavevector $\vec{k}$ and photon direction $\hat{p}$, electron velocity is irrotational ie.\ $\hat{p}\cdot \vec{v}_{b}=\mu v_{b}$ and $\tau_{d}$ is the optical depth defined as
\begin{equation}\label{eq2:op-dep}
\tau_{d}\paren*{\tau} \equiv \int_{\tau}^{\tau_{0}} n_{e}\sigma_{T}a d\tilde{\tau}.
\end{equation}
Thus the evolution equation of photon in longitudinal gauge is given by Eqn.\ (\ref{eq2:col-final2}). For the corresponding equations in synchronous gauge we refer to~\cite{ma1995apj}.
%%%%%%%%%%%%%%%%%%%%%%%%%%%%%%%%%%%%%%%%%%%%%%%%%%%%%

\subsection{Boltzmann Equation For Neutrinos}
For massless neutrino distribution, the collisionless Boltzmann equation similar to photon distribution is obtained as
\begin{equation}
\mathcal{N}^{\prime}+ \im\,k\,\mu\,\mathcal{N} + \Psi^{\prime} + \im\,k\,\mu\,\Phi = 0,
\end{equation}
where $\mathcal{N}$ is the neutrino temperature distribution. For discussion on massive neutrinos we refer to~\cite{ma1995apj}.
%%%%%%%%%%%%%%%%%%%%%%%%%%%%%%%%%%%%%%%%%%%%%%%%%%%%%

\section{Perturbation Equations For Baryons}
Protons and electrons couple tightly via Coulomb scattering to form the \emph{baryonic matter}. The coupling is so strong that the overdensities of the both components attain a common value. Following~\cite{dodelson2003}, perturbation equations for baryons in longitudinal gauge are obtained from the Boltzmann Eq.\ (\ref{eq2:coll}). For baryons, energy $E$ is related to mass $m$ and momentum $p$ as
\begin{equation}
E = \sqrt{p^{2}+m^{2}},
\end{equation}
such that 
\begin{equation}
P^{2} \equiv \gmn P^{\mu} P^{\nu} = -m^{2}.
\end{equation}
Momentum $p$ is defined as $p^{2} = g_{ij} P^{i} P^{j}$. Following in the similar fashion as for photon, the comoving 4-momentum of a particle of mass $m$ is obtained as
\begin{equation}
P^{\mu} = \left[E\paren*{1-\Phi}, \hat{p}^{i} p\frac{\paren*{1-\Psi}}{a}\right].
\end{equation}
Defining the common density fluctuation as $\delta_{b} = \delta \rho_{b}/\rho_{b}$ and 3-velocity as $v_{b}$, the perturbation equations are written as
\begin{eqnarray}\label{eq2:baryon-eq}
&\delta_{b}^{\prime} + \im\,k\,v_{b} + 3 \Psi^{\prime} = 0,\\
&v_{b}^{\prime} + \cH v_{b} +\im\,k\,\Phi= \tau_{d}^{\prime}\,\frac{4 \rho_{\gamma}}{3 \rho_{b}}\left[3\im\Theta_{1}+v_{b}\right].  \paren*{\theta_{\gamma}-\theta_{b}} .
\end{eqnarray}
Here $\tau_{d}$ is defined via Eqn.\ (\ref{eq2:op-dep}) and comes into Eqn.\ (\ref{eq2:baryon-eq}) from the Thomson scattering of electrons and photons. For the corresponding equations in synchronous gauge, we refer to~\cite{ma1995apj}.
%%%%%%%%%%%%%%%%%%%%%%%%%%%%%%%%%%%%%%%%%%%%%%%%%%%%%
\section{Density Fluctuations}
A simple way to analyse the matter distribution is through the \emph{matter power spectrum}. In Fourier space, the fluctuations are considered as uncorrelated and Gaussian random fields. Defining the density fluctuations as $\delta = \delta \rho/\rho$, the \emph{two-point} function of $\delta$ in Fourier mode called the \emph{matter power spectrum} $P\paren*{k}$, and is defined as 
\begin{equation}\label{eq2:corr-k1}
\langle\delta\paren{\vec{k}_{1}}\delta^{\ast}\paren{\vec{k}_{2}}\rangle = \paren*{2\pi}^{3} P\paren*{k} \delta_{D}^{3} \paren*{\vec{k}_{1}-\vec{k}_{2}},
\end{equation}
where angular brackets indicate ensemble average and $\delta^{3}_{D}$ is the Dirac delta function. In Eqn.\ (\ref{eq2:corr-k1}), the perturbations are considered real, so $\delta^{\ast}\paren{\vec{k}}= \delta\paren{-\vec{k}}$ and $k = \abs{\vec{k}}$. The power spectrum is the spread in the matter distribution implying the power spectrum will be small if the matter distribution is smooth while it will be large if there are a lot of over and under-densities. Following~\cite{dodelson2003}, the power spectrum is given as
\begin{equation} \label{eq2:power}
P\left(k,a\right)= A_{s} k^{n_s} T^2\left(k\right) D^2\left(a\right),
\end{equation}
where $A_{s}$ is the normalizing constant, $n_s$ is the spectral index, $T\left(k\right)$ is the\emph{ matter transfer function} which relates the primordial fluctuations with those at some later epoch and $D\left(z\right)=\frac{\delta_m\left(a\right)}{\delta_m\left(a=1\right)}$ is the normalized density contrast. 

Another important statistical quantity is the root-mean-square (rms) mass fluctuation in sphere with radius $8\,h^{-1}\,\mpc$ called $\se$. The mean-square or variance of the matter density fluctuation within the sphere of radius $R$ is given by
\begin{equation} \label{eq6:sigma2}
\sigma^2\left(R,a\right)=\frac{1}{2 \pi^2} \int k^3 P\left(k,a\right) W\left( kR\right)^2 \frac{d k}{k},
\end{equation}
where $P\left(k,a\right)$ is the matter power spectrum given in Eqn.\ (\ref{eq2:power}) and $W\left( kR\right)$ is the top-hat window function given by
\begin{equation} \label{eq6:window}
W\left( kR\right)=3\left[\frac{\sin\left(k R\right)}{\left(k R\right)^3}-\frac{\cos\left(k R\right)}{\left(k R\right)^2}\right].
\end{equation}
When the size of the filter is $R = 8\,h^{-1}\,\mpc$, $\sigma^{2}\paren*{R,a} \equiv \sigma_{8}^{2}\paren*{a}$.

%%%%%%%%%%%%%%%%%%%%%%%%%%%%%%%%%%%%%%%%%%%%%%%%%%%%%
\section{Temperature Anisotropies}
The anisotropies observed in the CMB radiation is the fluctuations in the photon temperature coming at the present epoch from different directions.
The temperature fluctuation $\Theta$ is expanded in spherical harmonics $Y_{\ell m}$ with coefficients $a_{\ell m}$ as
\begin{equation}
\Theta\paren{\tau, \vec{x},\hat{p}} = \SSum_{\ell=0}^{\infty} \SSum_{m=-\ell}^{\ell} a_{\ell m}\paren*{\tau, \vec{x}} Y_{\ell m}\paren{\hat{p}}.
\end{equation}
Here, $\ell$ is the multipole index and is related to the angular size of the sky as $\ell\sim 180^{\circ}/\theta$ while $m$ is the phase. Considering initial Gaussian fluctuations, mean of all $a_{\ell m}$ vanishes and the variance of $a_{\ell m}$ is obtained as%
\begin{equation}
\langle a_{\ell m}a^{\ast}_{\ell^{\prime} m^{\prime}} \rangle = \delta_{\ell \ell^{\prime}}\delta_{m m^{\prime}}C_{\ell}, 
\end{equation}
where $C_{\ell}$ is the \emph{angular power spectrum} of the temperature field. The angular power spectrum is related to the statistical correlation between temperature fluctuations in two directions $\hat{p}$ and $\hat{q}$ averaged over the entire sky via
\begin{equation}
\langle\Theta\paren{\hat{p}}\Theta^{\ast}\paren{\hat{q}}\rangle = \frac{1}{4\pi}\Sum_{\ell}\paren*{2\ell+1}C_{\ell}P_{\ell}\paren*{\hat{p}\cdot\hat{q}},
\end{equation}
where, $\hat{p}\cdot\hat{q}\equiv\cos\theta$, $\theta$ being the angle between the two directions and $P_{\ell}$ are Legendre polynomials. The temperature fluctuations in the Fourier space is now expanded in terms of  Legendre polynomials as
\begin{equation}
\Theta\paren{\tau,\vec{k},\mu} \equiv \Sum_{l=0}^{\infty} \paren{-\im}^{\ell} \Theta_{\ell}\paren{\tau,\vec{k}}P_{\ell}\paren*{\mu},
\end{equation}
where $\Theta_{\ell}$ are the \emph{multipole moments} of the distribution. The \emph{temperature transfer function} is defined as $\Theta_{\ell}\paren*{k}\equiv \frac{\Theta_{\ell}\paren{\tau,\vec{k}}}{\mathcal{R}\paren*{\vec{k}}}$, where $\mathcal{R}$ is the curvature perturbation. The transfer function relates the primordial power spectrum $\Delta_{\mathcal{R}}\paren*{k}$ to the angular power spectrum $C_{\ell}$ through
\begin{equation}
C_{\ell} = \frac{4\pi}{\paren*{2\ell+1}^{2}} \int d \ln k \,\Delta_{\mathcal{R}}^{2}\paren*{k} \Theta_{\ell}^{2}\paren*{k}.
\end{equation}
The angular transfer function is calculated by solving the Boltzmann equation for photons (\ref{eq2:col-final2}) for different multipole index $\ell$. For a detailed analysis on the CMB spectrum one may refer to~\cite{hu1995apj,seljak1996apj,dodelson2003}. 
%%%%%%%%%%%%%%%%%%%%%%%%%%%%%%%%%%%%%%%%%%%%%%%%%%%%%

% Chapter 3
\chapter{Density Perturbation In The Models Reconstructed From Jerk Parameter}\blfootnote{The work presented in this chapter is based on ``Density Perturbation In The Models Reconstructed From Jerk Parameter'', \textbf{\authorname} and Narayan Banerjee, Gen.\ Rel.\ Grav.,  \textbf{50}, 67 (2018)} 
\label{chap3:grg}
\chaptermark{Density Perturbation In Reconstructed Models}

%%%%%%%%%%%%%%%%%%%%%%%%%%%%%%%%%%%%%%%%%%%%%%%%%%%%%
\section{Introduction}\label{sec3:intro}
In the absence of a universally accepted model for the accelerated expansion, attempts have also been there to ``reconstruct'' a dark energy model. The idea is to guess a model of evolution that explains the observations and find the distribution of matter that can give rise to that~\cite{ellis1991cqg}. As there is no dearth of observational data now, various improvisations of the methods of reconstruction have been suggested, making use of the data efficiently. For a scalar field model, Starobinsky showed that one can exploit the data on density perturbation to reconstruct the scalar potential~\cite{staro}. Huterer and Turner, on the other hand, utilized the distance measurement data for the same purpose~\cite{huterer1, huterer2}.

Reconstruction of dark energy models normally start with some physical quantity, like the potential of the quintessence field, or the equation of state parameter $w$ which is the ratio of the pressure and the energy density of the corresponding matter~\cite{sahni2000ijmpd, saini2000prl, gerke2002mnras,gong2007prd,holsclaw2010prd, chevallier2001ijmpd, linder2003prl, scherrer2008prd1, scherrer2008prd2, hu, slp1, slp2, czp}. 

There is a new trend in reconstruction of dark energy models where one ignores the physical quantities and bank on kinematical parameters. With the assumption of spatial isotropy and homogeneity, the spacetime geometry is determined by the scale factor $a$. The derivatives of $a$ with respect to the cosmic time of various orders yield the kinematical quantities. The first few of them are the Hubble parameter $H = \frac{\dot{a}}{a}$, the deceleration parameter $q = - \frac{a\ddot{a}}{{\dot{a}}^2}$, the jerk parameter $j = - \frac{1}{a^3} \frac{d^3 a}{dt^3}$ and so on. The basic idea is to write an ansatz for some kinematical quantity which involves some parameters and then an estimation of the parameters using observational data. In a way, this is an attempt to construct the model through cosmography, where one builds up the model from observables rather than modelling from a theory. Cosmography may be related to a particular data set. For example, the baryon acoustic oscillation (BAO)~\cite{alcaniz}, Supernovae data~\cite{cattoen}, observations at high redshift~\cite{vitagliano} to name a few. Cosmographic methods without using standard candles and standard rulers are discussed by Xia {\it et al.}~\cite{xia}. Cosmography, using a Markov Chain method, has been discussed by by Capozziello, Lazcoz and Salzano~\cite{salzano}. For a very recent account of cosmography, one can see~\cite{li}.

Hubble parameter is the oldest observational quantity in cosmology and it was found to have an evolution. So the natural choice for a reconstruction through a kinematical quantity has been the next higher order derivative, namely the deceleration parameter $q$. Attempts to build up a dark energy model with $q$ as the starting point have been made by Gong and Wang~\cite{gong2007prd} and by Ting \etal~\cite{wang}. The deceleration parameter $q$ at various epoch can be estimated today with the help of observational quantities like luminosity of supernovae and the corresponding redshift, and is found to be evolving as well. For a recent work on this estimation, we refer to the work of vanPutten~\cite{vanPutten}. Hence the next higher order derivative of the scale factor, the jerk parameter $j$, is of importance now and should play a significant role in the game of reconstruction through kinematical quantities. Some work in this direction has been initiated by Luongo~\cite{luongo} and by Rapetti~\cite{rapetti}. Zhai \etal~\cite{zz}, starting from a parametric ansatz for $j$ such that the present (at $z=0$) value of $j$ is -1, reconstructed quite a few dark energy models. The motivation behind the choice of $j=-1$ at $z=0$ is the fact that it mimics the standard \lcdm model of the present acceleration of the Universe. That the jerk parameter should be instrumental for the reconstruction through kinematical quantities was categorically indicated a long time back by Alam \etal~\cite{alam}.

In a very recent work, Mukherjee and Banerjee~\cite{mukherjee2016prd} relaxed the requirement that $j(z=0)=-1$, and reconstructed several dark energy models. This is more general in the sense that the present value of $j$ is not controlled by hand. This  work also involves diverse data sets for the estimation of the model parameters as opposed to the work by Zhai \etal~\cite{zz} where only Observational Hubble Data (OHD) and the Union 2.1 Supernovae data were employed. In a later work, Mukherjee and Banerjee~\cite{mukherjee2017cqg} looked at the possible interaction between the cold dark matter and the dark energy, with the help of a model reconstructed through the jerk parameter, with an assumption that the jerk is varying very slowly, and can be approximated as a constant.

The reconstructed models, with a good choice of parameters, can fit well with various observational data. But in order to provide a useful description of the evolution, a model should also be able to describe certain other things. One crucial aspect is certainly a consistency with a growing mode of density perturbations, without which the formation of structures cannot be explained. The motivation of the present work is to investigate the density perturbations and to check whether the models reconstructed via $j$ can give rise to a growing mode of such perturbations. There is hardly any work on the matter perturbations in reconstructed models, except that of Hikage, Koyama and Heavens~\cite{kage}, where the perturbation of a model reconstructed from BAO data has been discussed. There is, however, some work on the reconstruction of perturbation itself~\cite{gonzal1, gonzal2, hunt, alam2009apj}.

This chapter deals with the density perturbation of models reconstructed from an ansatz on the jerk parameter. For a varying jerk, we pick up the models from reference~\cite{mukherjee2016prd}, as that is more general. We also consider interacting models, where the jerk is very slowly varying, as in reference~\cite{mukherjee2017cqg}. For detail discussion on the models considered here, we refer to~\cite{mukherjee2017}.

It should be realised that the perturbations of the model based on Einstein equations and that of a kinematically reconstructed model could well be different even for the same energy budget of the Universe. This is for the simple reason that in the former, one has more independent equations and thus contributions to perturbation from other sectors, such as the velocity perturbation could be manifest. 

%%%%%%%%%%%%%%%%%%%%%%%%%%%%%%%%%%%%%%%%%%%%%%%%%%%%%
\section{Background} \label{sec3:sec2}
A spatially flat, homogeneous and isotropic Universe is given by the metric
\begin{equation}\label{eq3:metric}
ds^2=-d t ^2+a^2(t)\gamma_{ij} d x^i dx^j,
\end{equation}
where $\gamma^i_{j}=\delta^i_j$ is the metric in the constant time hypersurface and $a(t)$ is the scale factor. The kinematic quantities of our interest are
\begin{enumerate}
\item[(i)] the fractional first order time derivative of the scale factor $a$, the well known \emph{Hubble parameter}, $H=\frac{\dot{a}}{a}$;
\item[(ii)] the second order derivative of $a$, defined in a dimensionless way, the  \emph{deceleration parameter}, $q=-\frac{\ddot{a}/a}{\dot{a}^2/a^2}$;
\item[(iii)] the third order derivative of $a$, again defined in a dimensionless way, the \emph{jerk parameter} $j(t)=-\frac{1}{a H^3}\left(\frac{d^3 a}{d t^3}\right)$.
\end{enumerate}

We pick up the convention in which $j$ is defined with a negative sign so that we can make use of the results from references~\cite{zz,mukherjee2016prd,mukherjee2017cqg} without any modification. In terms of redshift, $z=\frac{a_0}{a}-1$, ($a_0$ is the present value of the scale factor, taken to be unity throughout the calculation), the jerk parameter takes the form
\begin{equation} \label{eq3:jerk}
j(z) = -1 + \left(1+z \right) \frac{\left(E^{2}\right)'}{E^{2}} -\frac{1}{2}\left(1+z \right)^2 \frac{\left(E^{2}\right)''}{E^{2}},
\end{equation}
where $E(z)=\frac{H(z)}{H_0}$, $H_0$ being the present value of the Hubble parameter and prime denotes a differentiation  with respect to redshift, $z$.

%%%%%%%%%%%%%%%%%%%%%%%%%%%%%%%%%%%%%%%%%%%%%%%%%%%%%
\section{Standard Distribution Of Matter}\label{sec3:stanmat}
Although we shall be working with Eq.\ (\ref{eq3:jerk}), pretending that we do not know anything about the matter distribution in the universe, we should be able to identify the terms with the corresponding standard matter distribution consisting of a dark matter and a dark energy at some stage.

The energy-momentum tensor of a perfect fluid distribution is
\begin{equation} \label{eq3:stress}
T^{(m)}{}_{\mu \nu}= (p_m + \rho_m)u_\mu u_\nu + p_m g_{\mu \nu},
\end{equation}
where $\rho_m$ is the fluid density, $p_m$ is the pressure and $u_\mu$ is the comoving $4$-velocity, $u_{\mu} = (1,0,0,0)$. For a cold dark matter (CDM) $p_m=0$.

The contribution to the density and pressure from the dark energy are $\rho_{de}$ and $p_{de}$ respectively. The dark energy is also assumed to mimic a fluid distribution, and the corresponding energy momentum tensor is the same as Eq.\ (\ref{eq3:stress}) with $\rho_{m}$ and $p_m$ being replaced by $\rho_{de}$ and $p_{de}$ respectively. The equation of state (EoS) parameter of the dark energy component is $w_{de} = \frac{p_{de}}{\rho_{de}}$.

If an interaction between the dark matter and the dark energy is allowed, they do not conserve individually, and one can write the rate of transfer of energy in terms of an interaction term $Q$ so that 
\begin{eqnarray} 
\dot{\rho}_m + 3 H \rho_m &=& Q , \label{eq3:econ1}\\
\dot{\rho}_{de} + 3 H \left(1+w_{de}\right) \rho_{de} &=& -Q \label{eq3:econ2}.
\end{eqnarray}
Thus the above two equations combine to give the total conservation equation, which would follow from the standard Einstein equations. In the absence of any interaction, both the components conserve individually, indicating $Q=0$. 

%%%%%%%%%%%%%%%%%%%%%%%%%%%%%%%%%%%%%%%%%%%%%%%%%%%%%
\section{Interacting Model} \label{sec3:sec3}
If we now consider the jerk to be a very slowly varying function of $z$ so that one can consider it to be a constant for the purpose of integration~\cite{mukherjee2017cqg}, the Eq.\ (\ref{eq3:jerk}) can be integrated to yield 

\begin{equation} \label{eq3:hubble1}
E^{2}(z) =  A \left(1+z\right)^{\frac{3+\sqrt{9-8(1+j)}}{2}} +  \left(1-A\right) \left(1+z\right)^{\frac{3-\sqrt{9-8(1+j)}}{2}} .
\end{equation}
Here $A$ is a constant of integration. From definition, $E^{2}(z=0)$ has to be unity, so that the second constant of integration is chosen to be $(1-A)$ in order to satisfy the condition. From Eq.\ (\ref{eq3:hubble1}), one can easily see that for $j=-1$, which corresponds to a \lcdm model, the first term redshifts as a pressureless fluid and the second term corresponds to a constant. With this identification, the first term is easily picked up as the matter density parameter $\Omega_m$ and the second term as the dark energy density parameter $\Omega_{de}$ which reduces to a cosmological constant for $j=-1$. So one actually recovers the $G^{0}_{0}$ component of Einstein equations in terms of the dimensionless quantities as,
\begin{equation} \label{eq3:hubble2}
E^{2} = \Omega_m + \Omega_{de},
\end{equation}
where 
\begin{eqnarray}
\Omega_m &=& A \left( 1+z \right)^{\frac{3+\sqrt{9-8(1+j)}}{2}}, \label{eq3:omega1}\\
\mbox{and} \hspace{2mm} \Omega_{de}&=&\left(1-A\right) \left(1+z\right)^{\frac{3-\sqrt{9-8(1+j)}}{2}}. \label{eq3:omega2}
\end{eqnarray}
For values of $j$ other that $-1$, the second term is clearly an evolving dark energy rather than a constant. The sector identified to be the dark matter does not redshift as $(1+z)^3$ in this case. However, one can still identify that with the standard pressureless matter but has to allow an interaction amongst the dark sector as shown in reference~\cite{mukherjee2017cqg}.  The interaction, $Q$ between the two components can be expressed in terms of $z$ as
\begin{equation} \label{eq3:int}
Q(z) = A \rho_c \left(\frac{3-\sqrt{9-8(1+j)}}{2}\right) \left(1+z\right)^\frac{3+\sqrt{9-8(1+j)}}{2} H(z).
\end{equation}
The EoS of dark energy in terms of $z$ is obtained as
\begin{equation}\label{eq3:eos}
\begin{split}
w_{de}(z)= - \left(\frac{3+\sqrt{9-8(1+j)}}{6}\right)& \\
 -\left(\frac{A}{1-A}\right)& \left(\frac{3-\sqrt{9-8(1+j)}}{6}\right) \left(1+z\right)^{\sqrt{9-8(1+j)}} .
\end{split}
\end{equation}

Now small perturbations of densities and interaction are considered in the form $\bar{\rho}_m= \rho_m + \delta \rho_m$, $\bar{\rho}_{de}=\rho_{de}+\delta \rho_{de}$ and $\bar{Q}= Q +\delta Q$ and the resulting metric perturbation as , $\bar{H}=H+\delta H$ in Eqs.\ (\ref{eq3:econ1}), (\ref{eq3:econ2}) and (\ref{eq3:hubble2}). Expanding upto the first order the equations are obtained respectively as
\begin{eqnarray}
&-H \left(1+z\right) \delta \rho_m' + 3 H \delta \rho_m + 3 \delta H \rho_m = \delta Q , \label{eq3:pert1}\\
&-H \left(1+z\right) \delta \rho_{de}' + 3 H \left(1+w_{de}\right) \delta \rho_{de} + 3 \delta H \left(1+w_{de}\right) \rho_{de} =-\delta Q ,\\
&2 H \delta H = \delta \rho_m +\delta \rho_{de} \label{eq3:pert2}.
\end{eqnarray}
Using Eqs.\ (\ref{eq3:int}) and (\ref{eq3:eos}) in (\ref{eq3:pert1})-(\ref{eq3:pert2}), a second order differential equation for the density contrast of the dark matter $\delta=\frac{\delta \rho_m}{\rho_m}$ is obtained as

\begin{equation} \label{eq3:eq1}
\begin{split}
&2 \left( 1+z\right)^2 H^2 \rho_m \delta''(z)\\
-&\frac{\delta'(z)}{A \beta  \rho_{c} \left( 1+z\right)^{\alpha }-3 \rho_{m}}\left(\left( 1+z\right) \left(-4 \left( 1+z\right) H \rho_{m} H' \left(A \beta  \rho_{c} \left( 1+z\right)^{\alpha }-3 \rho_{m}\right)\right.\right.\\
+&\left.\left.2 H^2 \left(\rho_{m} \left(3 A \beta  \rho_{c} \left( 1+z\right)^{\alpha }w_{de}+A \left(\alpha +5\right) \beta  \rho_{c} \left( 1+z\right)^{\alpha }+3 \left( 1+z\right) \rho_{m}'\right)\right.\right.\right.\\
-&\left.\left.\left.2 A \beta  \rho_{c} \left( 1+z\right)^{\alpha +1} \rho_{m}'-3 \rho_{m}^2 \left(3 w_{de}+5\right)\right)\right.\right.\\
+&\left.\left.3 \rho_{m} \left(\rho_{de} \left(w_{de}+1\right)+\rho_{m}\right) \left(A \beta  \rho_{c} \left( 1+z\right)^{\alpha }-3 \rho_{m}\right)\right)\right)\\
-& \frac{\delta(z)}{A \beta  \rho_c \left( 1+z\right)^{\alpha }-3 \rho_m}\left(-4 \left( 1+z\right) H H' \left(\left( 1+z\right) \rho_m'-3 \rho_m\right)\right.\\
&\left. \left(A \beta  \rho_c \left( 1+z\right)^{\alpha }-3 \rho_m\right)-2 H^2 \left(3 \rho_m \left(3 w_{de}\right.\right.\right.\\
&\left.\left.\left. \left(A \beta  \rho_c \left( 1+z\right)^{\alpha }+\left( 1+z\right) \rho_m'\right)\right.\right.\right.\\
+& \left.\left.\left.3 A \beta  \rho_c \left( 1+z\right)^{\alpha }+A \alpha  \beta  \rho_c \left( 1+z\right)^{\alpha }-\left( 1+z\right)^2\rho_m''+2 \left( 1+z\right) \rho_m'\right)\right. \right.\\
+& \left. \left.\left( 1+z\right) \left( A \beta  \rho_c \left( 1+z\right)^{\alpha +1} \rho_m''\right.\right.\right.\\
&\left.\left.\left.-A \beta  \rho_c \left( 1+z\right)^{\alpha } \left(\alpha +3 w_{de}+5\right) \rho_m'+3 \left( 1+z\right) \rho_m'^2\right)\right. \right.\\
-& \left. \left. 27 \rho_m^2 \left(w_{de}+1\right)\right)+3 \left(A \beta  \rho_c \left( 1+z\right)^{\alpha }-3 \rho_m(z)\right) \left(-\rho_m\right.\right.\\
&\left.\left. \left(-A \beta  \rho_c \left( 1+z\right)^{\alpha } w_{de}-\left( 1+z\right) \rho_m'+3 \rho_m \left(w_{de}+1\right)\right)\right.\right.\\
-&\left.\left.\rho_{de} \left(w_{de}+1\right) \left(3 \rho_m-\left( 1+z\right) \rho_m'\right)\right)\right)=0,
\end{split}
\end{equation}
where $\alpha=\frac{3+\sqrt{9-8(1+j)}}{2}$ and $\beta=\frac{3-\sqrt{9-8(1+j)}}{2}$.

\begin{figure}[!h]
  \centering
\includegraphics[width=1\linewidth]{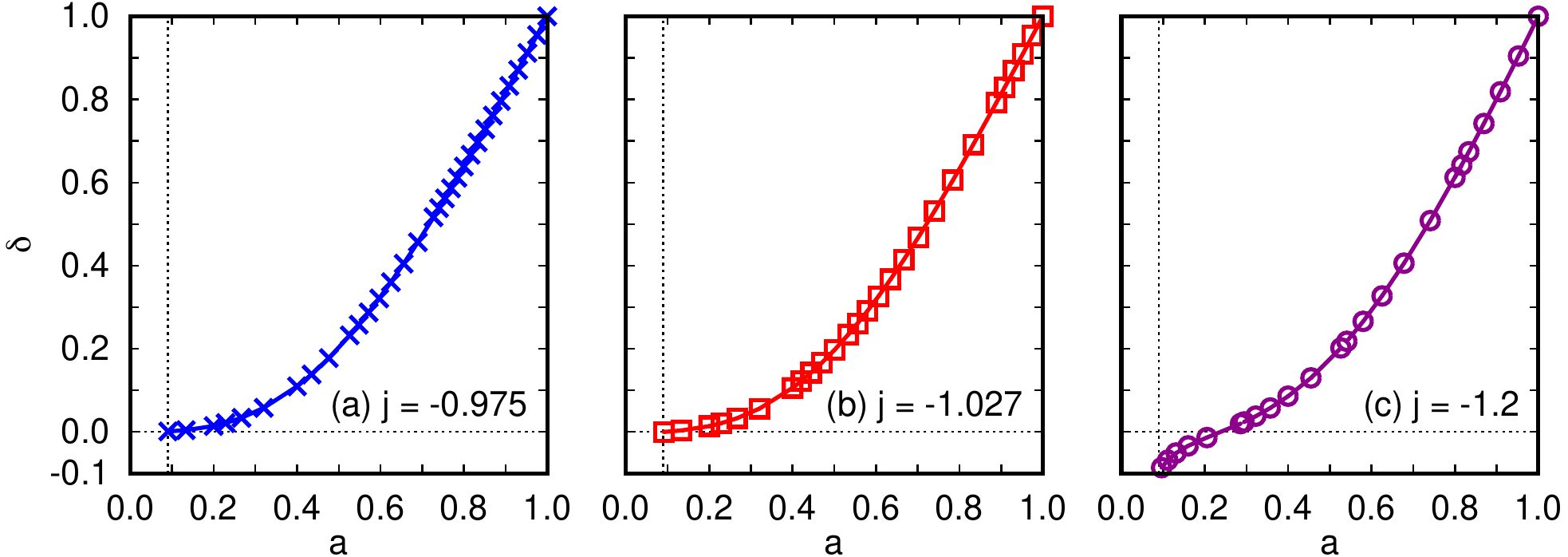}
\caption{Plot of $\delta$ against $z$ for $j=-0.975$ , $-1.027$ and $-1.20$. The vertical dotted line corresponds to $x=0.09$ and the horizontal dotted line corresponds to the zero crossing line.}\label{im3:fig1}
\end{figure}

The differential Eq.\ (\ref{eq3:eq1}) gives the evolution of the density contrast $\delta$ with $z$. This equation is solved numerically from $z=1100$ to the present epoch $z=0$ with standard initial conditions as given by Cembranos  \etal~\cite{cembranos2013prd} and Mehrabi  \etal~\cite{mehrabi2015prd} like $\delta(z=1100)=0.001$ and $\delta'(z=1100)=0.$ The value of the constant $A$, which is actually $\Omega_{m0}$ is taken as $0.286$~\cite{mukherjee2017cqg}. It deserves mention that the value is not taken from any particular observation, it is rather the best fit value as found by the statistical analysis given in reference~\cite{mukherjee2017cqg}. In order to get a qualitative picture of the perturbation, we scale $H_0$ to unity. The estimates are in gravitational units, where $G=1$. The evolution of $\delta$ is investigated for three different values of $j$ namely, $-1.027$, $-0.975$ and $-1.2$. The best fit value is $j=-1.027$, and the other two are  roughly the two extremes of the $2 \sigma$ confidence region as given in~\cite{mukherjee2017cqg}. Although the initial conditions are taken for $z=1100$, we show the plots of $\delta$ between $a=1$ (i.e.\ $z=0$) and $a=0.09$ (i.e.\ $z=10$) so as to have a closer look into the late time behaviour. The plots for the whole domain will give a much poorer resolution. The qualitative behaviour is shown in figure (\ref{im3:fig1}). In order to have tractable plots, we have normalised the value of $\delta$ by that at $a=1$. It is quite clearly seen from the plots that close to $a=1$, all the plots are linear.

One can see from Fig.\ \ref{im3:fig1} that for less that the best fit value of $j$, the density contrast changes sign with the evolution. However, if $j> - 1.027$, the best fit value, the matter perturbation has a monotonic growing mode. Thus even within the $2\sigma$ confidence level, the interacting model has a problem.
\begin{figure}[!h]
  \centering
\includegraphics[width=0.6\textwidth]{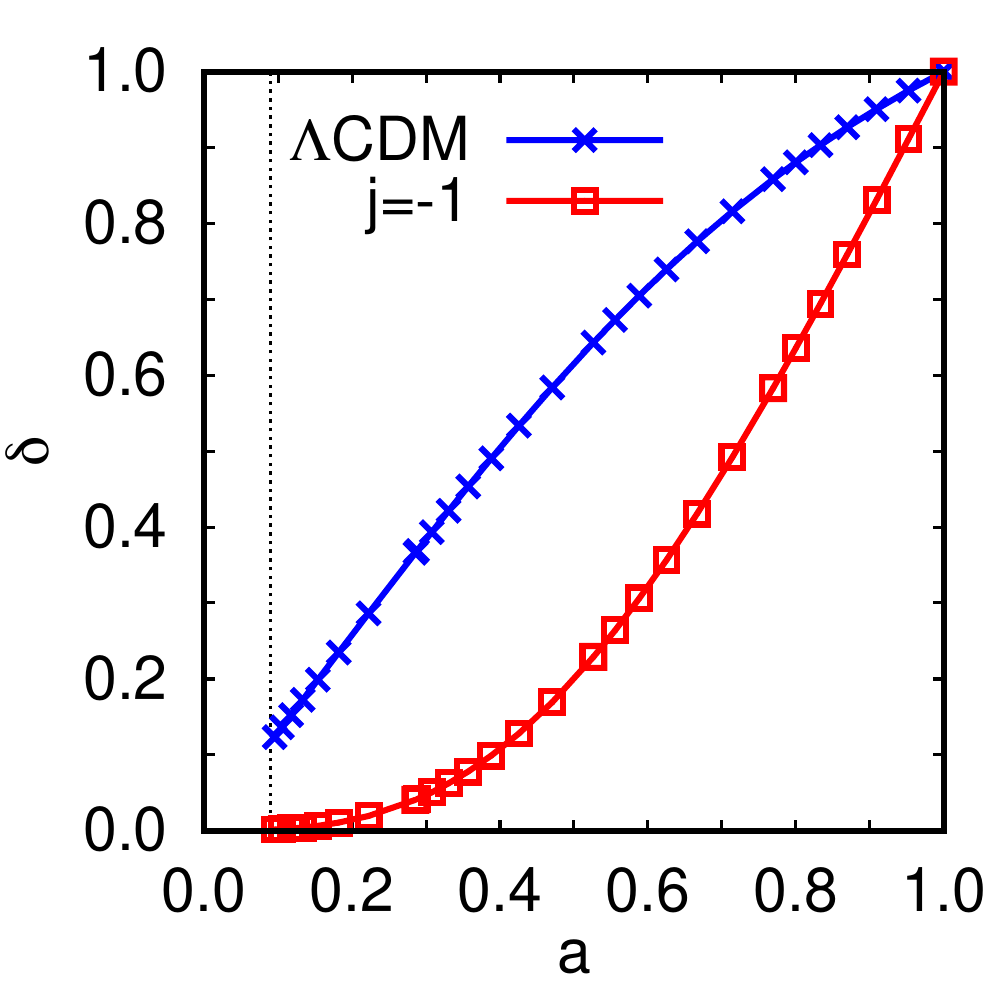}
\caption{Plot of $\delta$ against $z$ for \lcdm and $j=-1$. The vertical dotted line corresponds to $x=0.09$.}\label{im3:fig2}
\end{figure}

In figure (\ref{im3:fig2}) we show the plots of the standard \lcdm model. Vale and Lemos~\cite{vale2001mnras} gave the evolution equation of $\delta$ for \lcdm as
\begin{equation}\label{eq3:lcdm}
(z+1)^2 \delta_{\Lambda}^{\prime \prime}(z) + (z+1) \left(-1+\frac{(z+1) H^{\prime}(z)}{H(z)}\right) \delta_{\Lambda}^{\prime}(z)+\left(\frac{\Lambda }{2 H(z)^2}-\frac{3}{2}\right) \delta_{\Lambda}(z)=0.
\end{equation}
We numerically integrate this with the boundary conditions used in the present work and plot the result in figure (\ref{im3:fig2}) in a blue crossed line. The one in a red squared line is the plot of $\delta$ against $a$ from our numerical analysis of Eq.\ (\ref{eq3:eq1}) with $j=-1$ which is supposed to correspond to a \lcdm.  The plots are not really coincident even though the same initial conditions are used to integrate the Eqs.\ (\ref{eq3:eq1}) and (\ref{eq3:lcdm}). Both the plots are linear close to $a=1$, but definitely with a different slope. The difference is due to the following reason. The present model has only one equation, namely the Eq.\ (\ref{eq3:hubble1}), which is used for the perturbation. A standard \lcdm model, on the other hand, has an additional equation, that for $\dot{H}$. So the model from the reconstruction of jerk indeed mimics the \lcdm kinematically, but they are not really identical. Perturbations of the models strongly indicate that.

\section{Non-Interacting Model}\label{sec3:sec4}
We now relax the requirement of a very slowly varying jerk and allow the jerk to be a function of $z$. But the interaction in the dark sector is switched off, such that both DM and DE have their own conservation as in reference~\cite{mukherjee2016prd}.

The general form of $j$, as suggested by Mukherjee and Banerjee~\cite{mukherjee2016prd} is
\begin{equation} \label{eq3:jerkdef}
j(z)= -1 +j_1\frac{f(z)}{E^{2}(z)},
\end{equation}
where $f$ is an analytic function of $z$ and $j_1$ is parameter to be determined by observational data. The four different forms of $f(z)$ from~\cite{mukherjee2016prd} are
\begin{eqnarray}
\mbox{Case I:} \hspace{1cm} j(z)&=&-1+j_1\frac{1}{E^{2}(z)},\\ \label{eq3:casei}
\mbox{Case II:}\hspace{1cm} j(z)&=&-1+j_1\frac{\left(1+z\right)}{E^{2}(z)},\\
\mbox{Case III:}\hspace{1cm} j(z)&=& -1+j_1\frac{\left(1+z\right)^2}{E^{2}(z)},\\
\mbox{Case IV:}\hspace{1cm} j(z)&=&-1+j_1\frac{1}{\left(1+z\right)E^{2}(z)}. \label{eq3:caseiv}
\end{eqnarray}

The second order differential Eq.\ (\ref{eq3:jerk}) can be integrated using these expressions for $j$ to get four different cases as
\begin{eqnarray}
\mbox{Case I:} \hspace{1cm} E^{2}(z)&=& c_1 \left(1+z\right)^3+c_2 +\frac{2}{3} j_1 \log\left(1+z\right), \label{eq3:hubble3}\\
\mbox{Case II:}\hspace{1cm} E^{2}(z)&=&c_1 \left(1+z\right)^3+c_2 +j_1 \left(1+z\right),\\
\mbox{Case III:}\hspace{1cm} E^{2}(z)&=& c_1 \left(1+z\right)^3+c_2 +j_1 \left(1+z\right)^2,\\
\mbox{Case IV:}\hspace{1cm} E^{2}(z)&=&c_1 \left(1+z\right)^3+c_2 +j_1 \frac{1}{2\left(1+z\right)} \label{eq3:hubble3a},
\end{eqnarray}
where $c_1$ and $c_2$ are integration constants, which can be evaluated using initial data. The constants $c_1$, $c_2$ and the model parameter $j_1$ are connected by the fact that $h(z=0) = 1$ from its definition. The values of $j_1$ and $ c_1$ are used from~\cite{mukherjee2016prd} as given in table (\ref{tab2:title1}).

\begin{table}[!h]
\begin{center}
\caption{Values of the constants $c_1$ and $j_1$.} \label{tab2:title1} 
\begin{tabular}{ccc}
\hline \hline
\rule[-1ex]{0pt}{2.5ex}Cases &  \hspace{15ex}value of $c_1$ & \hspace{15ex} value of $j_1$ \\ 
\hline
\rule[-1ex]{0pt}{2.5ex}Case I &  \hspace{15ex}$0.2985$ &  \hspace{15ex}$0.078$ \\ 
\rule[-1ex]{0pt}{2.5ex}Case II &  \hspace{15ex}$0.299$ &  \hspace{15ex}$0.045$ \\ 
\rule[-1ex]{0pt}{2.5ex}Case III &  \hspace{15ex}$0.30$ &  \hspace{15ex}$0.017$ \\ 
\rule[-1ex]{0pt}{2.5ex}Case IV &  \hspace{15ex}$0.298$ &  \hspace{15ex}$0.112$ \\ 
\hline \hline
\end{tabular}
\end{center}
\end{table}

As the left hand side, of all the Eqs.\ (\ref{eq3:hubble3})-(\ref{eq3:hubble3a}), is the square of the Hubble parameter scaled by its present value, it is easy to pick up the first term in each equation as $\Omega_m$, the density parameter of the cold dark matter which does not interact with the other components of matter, as it redshifts as $(1+z)^3$. Also, the constant of integration $c_1$ can thus be identified with $\Omega_{m,0}$, the value of the density parameter at $z=0$. The rest of the right hand side of Eqs.\ (\ref{eq3:hubble3})-(\ref{eq3:hubble3a}) is thus picked up as the net $\Omega_{de}$. Thus the evolution of $\Omega_{de}$, in the four cases, will look like 
\begin{eqnarray}
\mbox{Case I:} \hspace{1cm} \Omega_{de}(z)&=& 1-c_1 +\frac{2}{3} j_1 \log\left(1+z\right), \\
\mbox{Case II:}\hspace{1cm} \Omega_{de}(z)&=&1-j_1-c_1 +j_1 \left(1+z\right),\\
\mbox{Case III:}\hspace{1cm} \Omega_{de}(z)&=& 1-j_1-c_1 +j_1 \left(1+z\right)^2,\\
\mbox{Case IV:}\hspace{1cm} \Omega_{de}(z)&=&1-\frac{j_1}{2}-c_1 +j_1 \frac{1}{2\left(1+z\right)}.
\end{eqnarray}

The corresponding expressions for $w_{de}$, the equation of state parameter of the dark energy, are used as given in~\cite{mukherjee2016prd}.
\begin{eqnarray}
\mbox{Case I:} \hspace{1cm} w_{de}(z)&=& -1 + \frac{\frac{2}{9} j_1}{\frac{2}{3} j_1 \log \left(1+z\right)+\left( 1-c_1\right)} , \\
\mbox{Case II:}\hspace{1cm} w_{de}(z)&=&-1  + \frac{\frac{1}{3} j_1\left(1+z\right)}{j_1 \left(1+z\right)+\left( 1-c_1-j_1\right)},\\
\mbox{Case III:}\hspace{1cm} w_{de}(z)&=& -1  + \frac{\frac{2}{3} j_1 \left(1+z\right)^2}{j_1 \left(1+z\right)^2+\left( 1-c_1-j_1\right)},\\
\mbox{Case IV:}\hspace{1cm} w_{de}(z)&=&-1+ \frac{\frac{j_1}{6\left(1+z\right)} }{\frac{-j_1}{2\left(1+z\right)} +\left( 1-c_1+\frac{1}{2}j_1 \right)}.
\end{eqnarray}

Considering a small perturbation, as discussed in the Section (\ref{sec3:sec3}),  Eqs.\ (\ref{eq3:econ1}), (\ref{eq3:econ2}) and (\ref{eq3:hubble2}) are combined to obtain the differential equation for the density contrast $\frac{\delta \rho_m}{\rho_m}= \delta$ as
\begin{equation} \label{eq3:eq2}
\begin{split}
& 2\left( 1+z\right)^2  H^2 \rho_m^2 \delta''(z) + \left( -\left( 1+z\right)  \left( 3 \rho_m^2 \left( \rho_m +\rho_{de} \left( 1+ w_{de} \right) \right)\right. \right. \\
-&\left.\left. 4 \left( 1+z\right) H H' \rho_m^2 +2 H^2 \rho_m \left(\rho_m  \left( 5+3 w_{de} \right) -\left( 1+z\right) \rho_m' \right) \right) \right) \delta'(z)\\ 
+&\left(4 \left( 1+z\right) H H' \rho_m \left( -3 \rho_m + \left( 1+z\right)\rho_m'\right) \right. \\
- &\left.  3 \rho_m \left( -\rho_{de} \left( 1+ w_{de} \right) \left( 3 \rho_m -\left( 1+z\right) \rho_m'\right) + \rho_m \left(-3 \rho_m \left( 1+ w_{de} \right) + \left( 1+z\right) \rho_m'\right) \right)  \right.
\end{split}
\end{equation}
\begin{equation*} \label{eq3:eq2}
\begin{split}
+& \left.  2 H^2 \left( 9 \rho_m^2 \left( 1+ w_{de} \right) \right. - \left. \left( 1+z\right)^2 \rho_m'^2 \right. \right. \\
+& \left. \left. \left( 1+z\right) \rho_m \left( - \left( 2+3 w_{de} \right) \rho_m' + \left( 1+z\right) \rho_m'' \right) \right) \right) \delta(z)=0.
\end{split}
\end{equation*}
Equations (\ref{eq3:hubble3})-(\ref{eq3:hubble3a}) explicitly show that the first term redshifts as $(1+z)^{3}$, it is easily picked up as the contribution from the pressureless dark matter which conserves by itself. So we have to fix $Q = 0$ in Eqs.\ (\ref{eq3:econ1}) and (\ref{eq3:econ2}), allowing both the components to have their own conservation. 

Equation (\ref{eq3:eq2}) is the dynamical equation for the density contrast $\delta$ against the redshift $z$. The said equation is solved numerically for each of the four ansatz mentioned in Eqs.\ (\ref{eq3:casei})--(\ref{eq3:caseiv}) as the unperturbed background.

\begin{figure}[!h]
\centering
\includegraphics[width=\linewidth]{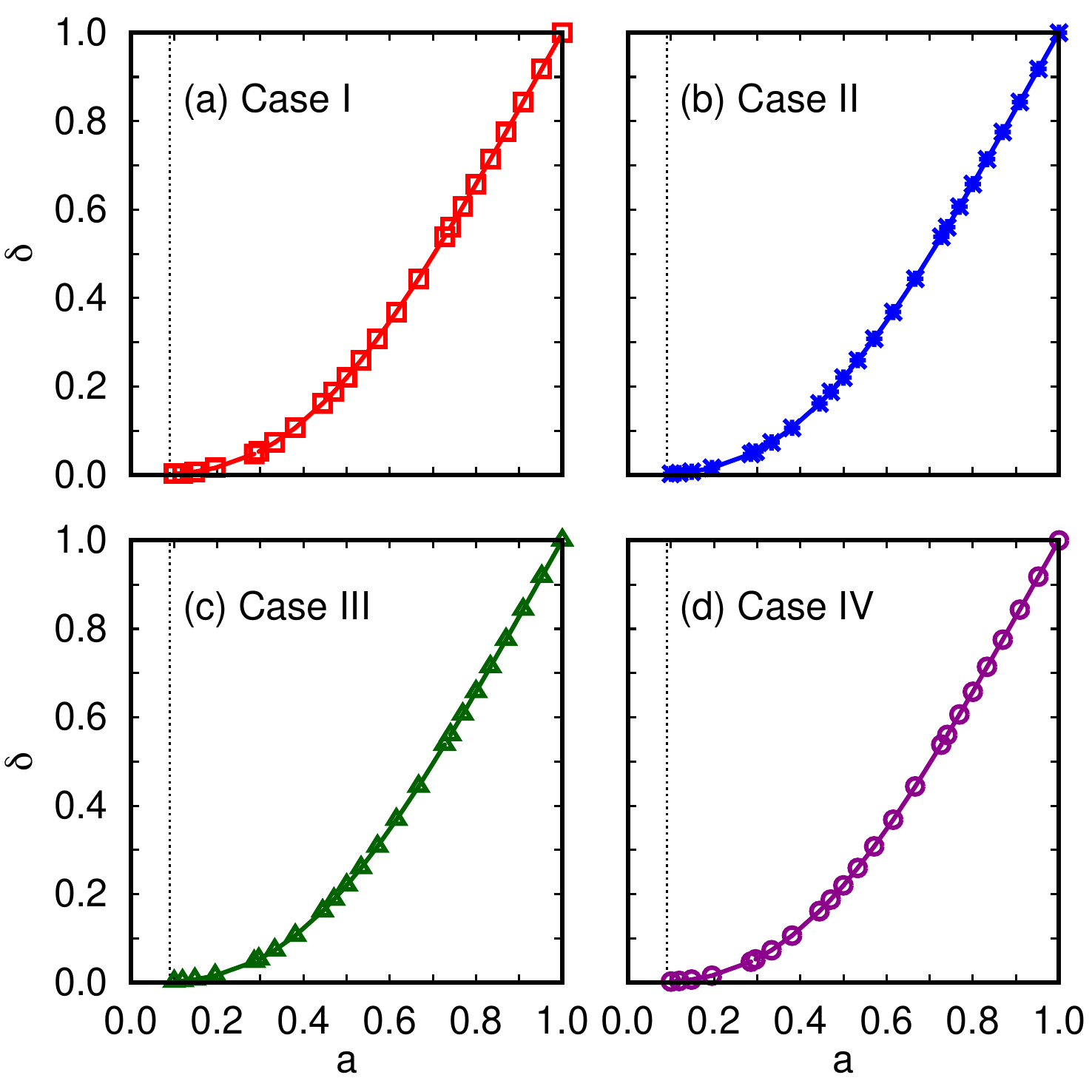}
\caption{Plot of $\delta$ against $z$ for different cases of varying jerk parameter. The vertical dotted line corresponds to $x=0.09$.}\label{im3:fig3}
\end{figure}
Figure (\ref{im3:fig3}) shows the plots of $\delta$ against $a$. All of them appear to be qualitatively similar, and they also possess the required nature of the perturbation, i.e.\ grow with the evolution (increase with $a$). Like the interacting case presented in Section (\ref{sec3:sec3}), here also the initial conditions are chosen at $z=1100$ as given by Cembranos  \etal~\cite{cembranos2013prd} and Mehrabi  \etal~\cite{mehrabi2015prd}, but the plots are given between $a=0.09$ (corresponding to $z=10$) and $a=1$ for the sake of better resolution.

%%%%%%%%%%%%%%%%%%%%%%%%%%%%%%%%%%%%%%%%%%%%%%%%%%%%%
\section{Summary And Discussion}\label{sec3:sum}
Albeit there is a proliferation of dark energy models reconstructed from the observational data, their suitability in connection with the structure formation, i.e., the possibility of the growth of density fluctuation, has very rarely been dealt with. The present work is an attempt towards looking at the growth of density perturbation in models reconstructed from the jerk parameter, the kinematical quantity gaining relevance recently. The perturbation equations are linearised in the fluctuations. The second order differential equations are solved numerically to plot the density contrast $\delta$ against the redshift $z$.

In both the examples of an interacting dark energy and a non-interacting one, the values of the parameters, though identified with some physically relevant quantities like the density parameter, are not taken from observational results, but rather from the best fit values as given in the reconstruction of the respective models. However, the values appear to be not too different from the recent observations like the Planck mission. 

It appears quite conclusively that the models allowing interaction in the dark sector fail to yield the required behaviour of $\delta$ even within the $2\sigma$ confidence level. The discrepancy is observed close $a=0.09$, i.e., roughly $z=10$. Non-interacting models, however, produce quite a suitable environment for the structure formation, at least qualitatively. The density contrast indeed has growing modes during the later time. So the non-interacting models appear to be favoured so far as the structure formation is concerned.

% Chapter 1
\chapter{Density Perturbation In An Interacting Holographic Dark Energy Model}% Main chapter title
\label{chap4:epjp}
\blfootnote{The work presented in this chapter is based on ``Density Perturbation In An Interacting Holographic Dark Energy Model'', \textbf{\authorname} and Narayan Banerjee, Eur.\ Phys.\ J. Plus, \textbf{135}, 779 (2020)}
\chaptermark{Density Perturbation In Interacting HDE}

%%%%%%%%%%%%%%%%%%%%%%%%%%%%%%%%%%%%%%%%%%%%%%%%%%%%%
\section{Introduction}\label{sec4:intro}
One of the most talked about forms of dark energy is the so-called holographic dark energy (HDE) based on the holographic principle in quantum gravity theory~\cite{susskind}. Based on the holographic principle, one of the characteristic features of the HDE is the long distance cut-off, called the infra-red (IR) cut-off~\cite{cohen}. In the context of cosmology, this cut-off is not uniquely specified but rather realised in various ways. One of the natural choices is the \emph{Hubble radius}~\cite{xu2009jcap}, but as shown by Hsu~\cite{hsu2004}, it does not provide the recent acceleration. In this context, Zimdahl and Pav\'{o}n in~\cite{pavon2006aip,zimdahl2007cqg} showed that allowing a non-gravitational interaction in the dark sector of the Universe not only solves this problem but also alleviate the nagging coincidence problem~\cite{pavon2005plb, hu}. Other possibilities are the \emph{particle horizo}n as suggested by Fischler and Susskind~\cite{fischler1998} and Bousso~\cite{bousso1999} and the \emph{future event horizon} as suggested by Li~\cite{li2004plb} and Huang and Li~\cite{huang2004}. A more recent choice for the cut-off scale is the \emph{Ricci scalar curvature} used by Gao \etal~\cite{gao2009prd}, Feng~\cite{feng2008, feng2008plb}, to name a few.  Although the HDE model is quite lucrative in many ways, it cannot avoid the phantom Universe~\cite{xin2005ijmpd,xu2009jcap}. One way of preventing the future ``big-rip'' singularity is to allow some phenomenological interaction between the dark matter and dark energy as shown by Wang \etal~\cite{wang141, wang357}. Thereafter the interacting HDE model has been studied extensively in~\cite{mohseni231,sen-diego, zhang26, wang1, diego-sen} and references therein. As there is a host of observational data in cosmology, almost all kinds of holographic dark energy models can now be tested against observations~\cite{campo, wu, mli2009jcap, su, ankan, dagostino2019, yun2013, li2013jcap, zhang2015, feng2018}. 

In this chapter, we will show how the evolution of perturbation of dark matter (DM) and holographic dark energy (HDE), are affected in the presence of an interaction between them. In this work, the characteristic infra-red (IR) cut-off is considered to be the future event horizon~\cite{li2004plb} as this ensures the recent acceleration even in absence of interaction. The evolution of the matter density perturbation can provide some knowledge about the components affecting it, in this case, the HDE. Since HDE is an evolving component of the Universe, it will have fluctuations like that of the DM and hence will not only affect the growth of matter perturbations~\cite{abramo2009prd} but also may cluster on its own~\cite{mehrabi2015mnras, batista2013jcap}. The inclusion of interaction leads to a brief transient oscillatory period for the density contrast for both HDE and DM. This existence of growing mode with a transient oscillatory behaviour is an interesting feature.
%\newpage
%%%%%%%%%%%%%%%%%%%%%%%%%%%%%%%%%%%%%%%%%%%%%%%%%%%%%
%%%%%%%%%%%%%%%%%%%%%%%%%%%%%%%%%%%%%%%%%%%%%%%%%%%%%

%%%%%%%%%%%%%%%%%%%%%%%%%%%%%%%%%%%%%%%%%%%%%%%%%%%%%
\section{The Background} \label{sec4:backgrnd}
We consider a spatially flat, homogeneous and isotropic Universe, described by the FLRW metric in conformal time as given in Eq.\ (\ref{eq1:metric-cart}),
\begin{equation}\label{eq4:metric}
ds^2= a^2(\tau)\paren*{- d \tau ^2+\gamma_{ij}\,d x^i dx^j}.\nonumber
\end{equation}
It is assumed that the Universe is filled with a perfect fluid dominated by a pressureless (cold) dark matter (CDM) and holographic dark energy (HDE). The energy densities and pressure are such that  $\rho=\rho_m+\rho_{de}$ and $p=p_{de}$ respectively. Subscript `\textit{m}' denotes the contribution of the CDM while `\textit{de}' denotes that of the HDE. There is an interaction between the two components of the Universe, CDM and HDE, hence a transfer of energy between the two.  The total energy balance equation 
\begin{equation}
 \rho^\prime + 3 \cH (\rho + p) = 0 \label{eq4:cons},
\end{equation}
is thus divided into two equations, 
\begin{eqnarray}
\rho^\prime_m+ 3 \cH \rho_m &=& aQ ,\label{eq4:cons1}\\
\rho^\prime_{de}+ 3 \cH \left(1+w_{de}\right) \rho_{de}&=& -aQ \label{eq4:cons2},
\end{eqnarray}
where $Q$ is the rate of energy density transfer, $w_{de}= \frac{p_{de}}{\rho_{de}}$ is the equation of state (EoS)\nomenclature{EoS}{Equation of State} parameter for the HDE. It is clear that equations (\ref{eq4:cons1}) and (\ref{eq4:cons2}) together give the conservation equation for the net matter content. The non-interacting scenario can be recovered simply by setting $Q=0$. If $Q>0$ energy is transferred from dark energy to dark matter and vice versa. There is, however, no compelling observational binding to take this interaction into account, but as the two sectors, DM and DE evolve together, this interaction adds to the generality of the model.

The expression for the energy density of HDE is
\begin{equation}
\rho_{de}=3 C^2 M_{P}^2 L^2, \label{eq4:def}
\end{equation}
where $3C^2$ is a numerical constant introduced for convenience, $M_{P}=\sqrt{\frac{1}{\kappa}}$ is the reduced Plank mass, $L$ is the characteristic length scale of the Universe which provides the IR cut-off of $\rho_{de}$. In the present work this cut-off is chosen as the future event horizon as suggested by Li~\cite{li2004plb}, 
\begin{equation}
L = a \int_t ^\infty\frac{d \tilde{t}}{a} = a \int_a^\infty \frac{d \tilde{a}}{H \tilde{a}^2}~,
\end{equation}
where $H$ is the Hubble parameter in cosmic time $t$. It has already been mentioned in the introduction that this is by no means the unique choice as the infra-red cut-off. 

The energy-momentum tensor of the fluid `A' (which stands for either `\textit{m}' or `\textit{de}') is $T^\mu_{\left(A\right) \nu}$ and is given by 
\begin{equation} \label{eq4:stressA}
T^\mu_{\left(A\right) \nu}= \left(\rho_A + p_A \right)u^\mu_{\left(A\right)} u_{\left(A\right)\nu} +  p_A  \delta^\mu{}_\nu.
\end{equation}
As discussed in Section (\ref{sec2:multi-fluid}), the source term for the interaction has the form
\begin{equation} \label{Q-def}
Q^\mu_m=\frac{1}{a} \left(Q_m, \vec{0}\right) = \frac{1}{a} \left(Q, \vec{0}\right)= \frac{1}{a} \left(-Q_{de}, \vec{0}\right) = -Q^\mu_{de}.
\end{equation}
%%%%%%%%%%%%%%%%%%%%%%%%%%%%%%%%%%%%%%%%%%%%%%%%%%%%%
%%%%%%%%%%%%%%%%%%%%%%%%%%%%%%%%%%%%%%%%%%%%%%%%%%%%%

%%%%%%%%%%%%%%%%%%%%%%%%%%%%%%%%%%%%%%%%%%%%%%%%%%%%%
\section{Interacting Holographic Dark Energy} \label{sec4:int-hde}
The evolution of the dimensionless HDE density parameter $\Omega_{de}=\frac{\rho_{de}}{\rho_c}$ where $\rho_c= 3 H_0^2 M_{P}^2$ is the critical density of the Universe, and the dimensionless Hubble parameter $E$, in the presence of an interaction are governed by the simultaneous differential equations~\cite{zhang2012jcap}
\begin{equation} \label{eq4:int-omega}
\frac{d \Omega_{de}}{d z} =-\frac{2\Omega_{de}\left( 1 -\Omega_{de} \right)}{1+z} \left( \frac{1}{2}+ \sqrt{\frac{\Omega_{de}}{C^2}}-\frac{\Omega_I}{2\left(1-\Omega_{de}\right)}\right),
\end{equation}
\begin{equation} \label{eq4:int-hubble}
\frac{1}{E}\frac{d E}{d z} = -\frac{ \Omega_{de}}{1+z} \left( \frac{1}{2}+\sqrt{\frac{\Omega_{de}}{C^2}} +\frac{\Omega_I- 3}{2 \Omega_{de}} \right),
\end{equation}
where $E =\frac{H}{H_0}$ is the Hubble parameter scaled by its present value $H_0$. The evolution equations 
(\ref{eq4:int-omega}) and (\ref{eq4:int-hubble}) are given in terms of the cosmic redshift $z$, which is a dimensionless quantity and is related to the scale factor $a$ as $z = \frac{a_0}{a} -1$, $a_0$ being the present value of the scale factor (taken to be unity). These two equations are obtained following the usual steps (also shown in~\cite{zhang2012jcap}).

The EoS parameter of DE, $w_{de}$, is an intrinsic characteristic of DE. From the system of equations, given by Einstein's equations and the conservation equations, $w_{de}$ imposes a constraint on $Q$ (see~\cite{zhang2012jcap}) as
\begin{equation} \label{eq4:int-eos}
w_{de}=-\frac{1}{3}-\frac{2}{3}\sqrt{\frac{\Omega_{de}}{C^2}}-\frac{\Omega_I}{3 \Omega_{de}}~,
\end{equation}
where $\Omega_I=\frac{Q}{H  \rho_c}$ is the interaction term expressed in a dimensionless form. 
 
To study the effect of interaction, we need to take a specific form of the interaction term $Q$. Models with interaction term $Q$ proportional to either $\rho_m$ or $\rho_{de}$ or any combination of them have been studied extensively in literature~\cite{zhang2012jcap, valiviita2008jcap, feng, miaoli09, yli2014prd, valiviita2015jcap, wang2016rpp, acosta2014prd, funo, clemson2012prd, bohmer2008prd}. It should be noted that there is no theoretical or observational compulsion for any one of these choices. In the present work we consider $Q \propto \rho_{de}$. We have taken the covariant form of the source term $Q^{\mu}_{m}\left(\tau\right) $ as
\begin{equation} \label{eq4:interaction}
Q^{\mu}_{m} = - Q^{\mu}_{de} =  \frac{\beta \cH \rho_{de} u^{\mu}_{de}}{a}~,
\end{equation}
where $\beta$ is the coupling constant whose magnitude determines the strength of the interaction rate and $u_{\left(de\right)\mu}= -a \delta^0_\mu$ is the comoving $4$-velocity of DE. Here we consider the Hubble parameter $\cH$ to be a global variable without any perturbation. When $\beta <0$, it is clear from equations (\ref{eq4:cons1}) and (\ref{eq4:cons2}) that DM redshifts faster than $a^{-3}$ while DE redshifts at a slower rate. This is physically problematic as more of the DM is expected to be transferred to the DE budget in the late time, rather than in the beginning. For $\beta>0$, this problem is avoided. As shown by Feng and Zhang in~\cite{feng}, for an HDE model, this form of interaction is favoured by geometrical data. We consider $\beta$ to be a free parameter. Using $u^{\mu}_{de} = \frac{1}{a} \delta^{\mu}_{0}$ and equation (\ref{Q-def}) in equation (\ref{eq4:interaction}) the interaction term $Q$ is obtained as
\begin{equation}
Q= \frac{\beta \cH \rho_{de}}{a}~.
\end{equation}
Since dark energy dominates at the present epoch, we assume $w_{de} < -\frac{1}{3}$. As the motivation of the present work is to investigate the perturbation for a model without a big rip singularity, we restrict $w_{de} \geqslant -1$. For the non-interacting case $\left(\beta=0\right)$, it is clear from (\ref{eq4:int-eos}) that for $w_{de} \rightarrow -1$ at $z=0$, $C \rightarrow \sqrt{\Omega_{de0}}$, $\Omega_{de0}$ being the value of the dark energy density parameter at the present epoch. The value of $\Omega_{de0}$ is taken from the \Planck 2018 data~\cite{planck2018cp} and is close to $0.6834$  which yields $C= 0.8267\simeq 0.83 $. In the presence of interaction, $C$ and $\beta$ have a correlation. The big rip singularity can be avoided if the interaction rate, $\beta$ lies between 
\begin{equation} \label{eq4:range}
-2 \sqrt{\frac{\Omega_{de0}}{C^2}} < \beta \leqslant 2- 2 \sqrt{\frac{\Omega_{de0}}{C^2}}~.
\end{equation}
The numerical values of $C$ and $\beta$ can be further constrained from other physical quantities like the deceleration parameter, $q$. For the IHDE\nomenclature{IHDE}{Interacting Holographic Dark Energy} model, $q$ depends on the parameters $C$ and $\beta$. In the subsequent part of this section, we will investigate the effect of interaction on the different physical quantities and try to constrain the parameter space for $C$ and $\beta$. 
\begin{figure}[!h]
  \centering
\includegraphics[width=\linewidth]{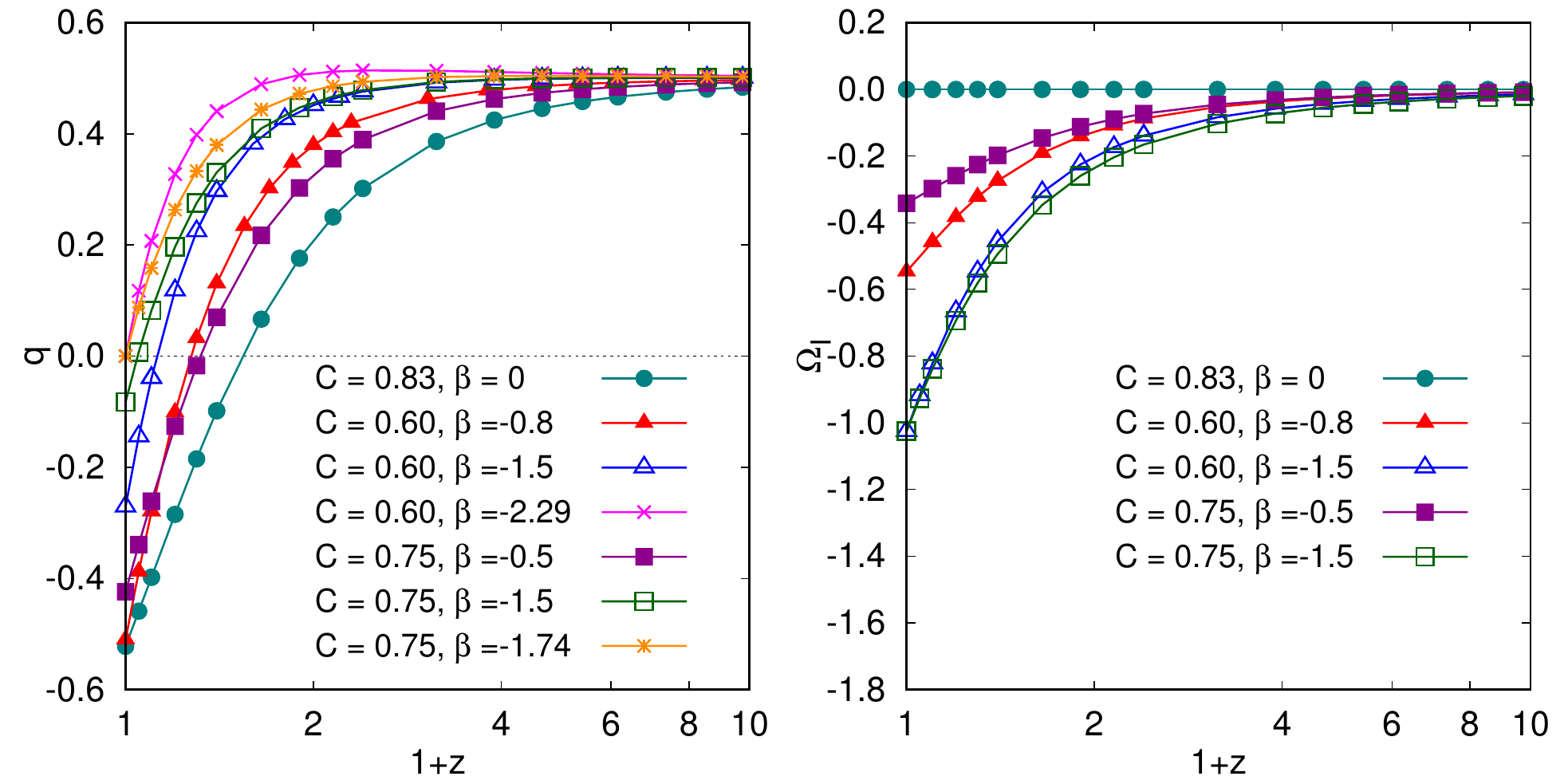}
\caption{(a) shows plot of $q$ against $\left(1+z\right)$ and (b) shows the variation of  $\Omega_I$ against $\left( 1+z\right)$ in logarithmic scale for different values of $C$ and $\beta$. The lines with solid circles corresponds to $C = 0.83$ and $\beta = 0$. The lines with triangles correspond to the interacting models with $C =  0.6$, $\beta = -0.8$ (solid) and $\beta = -1.5$ (hollow). The lines with squares correspond to the interacting models with $C =  0.75$, $\beta = -0.5$ (solid) and $\beta = -1.5$ (hollow). In figure (a) the lines with crosses ($C = 0.6$) and stars ($C = 0.75$) corresponds to the values of $\beta$ for which there is no acceleration at present.}\label{im4:qz}
\end{figure}
 
Figure (\ref{im4:qz}a) shows the variation of the deceleration parameter $q$ with $\left(1+z\right)$ in logarithmic scale for different sets of $C$ and $\beta$. In all the cases, $q$ increases with $z$ and approaches $0.5$ at higher redshifts. As seen from Fig.\ (\ref{im4:qz}a), for a fixed value of $C$ ($0.6$ and $0.75$), the smaller the value of $\beta$ ($-2.0$ and $-1.5$ respectively), the more recent is the transition from decelerated to accelerated Universe. For higher values of the coupling constant $\beta$ ($-0.8$ and $-0.5$), $q$ is nearly equal to $-0.5$ at present. Clearly, smaller values of $\beta (<-1.5)$ will give decelerated Universe at present. Thus future event horizon as IR cut-off does not necessarily ensure accelerated Universe in the presence of an interaction. For non-interacting case, acceleration comes naturally as pointed out by Li~\cite{li2004plb}. This figure brings out some new features, such as it puts a limit on the strength of interaction. For achieving an accelerated model,  $\beta$ should be greater than $-1.741$ (for $C = 0.75$) or greater than $-2.292$ (for $C = 0.6$).

Figure (\ref{im4:qz}b) shows the variation of the dimensionless interaction term $\Omega_{I}$ with with $\left(1+z\right)$ in logarithmic scale. For any pair of $C$ and $ \beta \left( \neq 0\right)$ the interaction term is nearly zero at higher redshifts and increases in magnitude with decrease in $z$. For $\beta< 0$, from the definition of $\Omega_I$, we can see $Q<0$, which means energy is transferred from DM to DE. Thermodynamically energy transfer should be from DE to DM following Le Ch\^{a}telier--Braun principle as shown by Pav\'{o}n and Wang~\cite{pavon2009grg}. In case of an HDE model a negative $\beta$ (DM $\rightarrow$ DE) is slightly favoured by the data as shown by Zhang \etal~\cite{zhang2012jcap}.

%%%%%%%%%%%%%%%%%%%%%%%%%%%%%%%%%%%%%%%%%%%%%%%%%%%%%
\section{The Perturbations} \label{sec4:pert}
In what follows, a scalar perturbation of the metric in longitudinal gauge is considered. The perturbed metric as given in Section (\ref{sec2:gaugechoices}), takes the form
\begin{equation} \label{eq4:metric2}
ds^2= a^2\left[-\left(1+2\Phi\right)d \tau^2 +\left(1-2\Psi \right)dx^i dx^j\right].\nonumber
\end{equation}
It is assumed that there is no anisotropic stress, hence longitudinal gauge becomes identical to Newtonian gauge, as discussed in Section (\ref{sec2:gaugechoices}), making $\Phi=\Psi$. 
Writing the perturbation in Fourier modes, the $(00)$, $(0i) \equiv (i0)$ and $(ij)$ components of the Einstein field equations up to the first order in perturbation will read as~\cite{mukhanov1992pr},
%%%%%%%%%%%%%%%%%%%
\begin{eqnarray}
&2\left[-3 \cH \left( \cH \Phi +\Phi^\prime \right) -k^2 \Phi \right]=\frac{3\cH^2}{\rho}\delta \rho, &  \label{eq4:g002}\\
&2 k^2\left( \cH\Phi +\Phi^\prime\right)=-\frac{3\cH^2}{\rho}\left(p+\rho\right)\theta, & \label{eq4:g0j2}\\
& \hspace{-0.8cm}-2\left[ \left(2\cH^\prime +\cH^2\right) \Phi + 3\cH\Phi^\prime +\Phi^{\prime \prime} \right] \delta^i{}_j=-\frac{3\cH^2}{\rho}\delta p \delta^i{}_j. &\label{eq4:gij2}
\end{eqnarray}
%%%%%%%%%%%%%%%%%%%
The temporal and spatial parts of the first order in perturbation of the energy balance equation of the  fluid `A'~\cite{valiviita2008jcap} are
\begin{eqnarray}
\begin{split}
\delta \rho^\prime_A + \left( \rho_A + p_A\right)\left(\theta_A -3 \Phi^\prime \right)& + 3 \cH \left( \delta\rho_A +\delta p_A\right)\\
 =&aQ_A\Phi +a \delta Q_A , \label{e1}
\end{split}\\
\begin{split}
\left[\theta_A\left( \rho_A + p_A \right)\right]^\prime  &+ 4 \cH \theta_A\left( \rho_A  + p_A \right)- k^2 \delta p_A\\
-k^2& \Phi\left( \rho_A  + p_A \right)=-k^2 a f_A+aQ_A \theta. \label{m1}
\end{split}
\end{eqnarray}
 
For an adiabatic perturbation in interacting fluids, the pressure perturbation $\delta p_A$ depends on $\delta \rho_A$ as well as on the interaction term $Q_A$ as~\cite{waynehu1998apj, valiviita2008jcap, bean2004prd} 
\begin{equation}\label{eq4:pert-p}
\delta p_A=c_{s,A}^2 \delta \rho_A+\left(c_{s,A}^2-c_{a,A}^2\right)\left[3 \cH\left(1+w_A\right)\rho_A -a Q_A\right]\frac{\theta_A}{k^2},
\end{equation}
where 
\begin{equation}\label{eq4:ca2}
c_{a,A}^2=\frac{p_A^\prime}{\rho_A^\prime}=w_A+\frac{w_A^\prime}{\rho_A^\prime/p_A^\prime}
\end{equation}
is the adiabatic speed of sound of fluid `A' and $c_{s,A}^2$ is the effective speed of sound of fluid `A', defined as
\begin{equation}\label{eq4:cs2}
c_{s,A}^2=\frac{\delta p_{A}}{\delta \rho_{A}} \bigg \rvert_{rest,A},
\end{equation}
i.e.\ the ratio of pressure fluctuation to density fluctuation in the rest frame of  fluid `A'.
%%%%%%%%%%%%%%%%%%%%%%%%%%%%%%%%%%%%%%%%%%%%%%%%%%%%%
%%%%%%%%%%%%%%%%%%%%%%%%%%%%%%%%%%%%%%%%%%%%%%%%%%%%%
\section{Evolution Of The Density Contrasts} \label{sec4:evolution}
We shall now frame the differential equations for density contrasts for both DM and DE that determine their evolution with redshift. For that, we need to know the perturbation in the interaction term also. From equations (\ref{eq2:Q-pert-def}) and (\ref{eq2:pert-condition}), it follows that
\begin{eqnarray}
\delta Q_{m}= -\delta Q_{de}  = \frac{\beta \cH \delta \rho_{de} }{a}, \label{eq4:pert-int1}  \\
 f_{m} = - f_{de} = \frac{\beta \cH \rho_{de}\left(\theta -\theta_{de}\right)}{a k^2}. \label{eq4:pert-int2}
\end{eqnarray}

The density contrasts of CDM and IHDE are $\delta_m =\frac{\delta \rho_m}{\rho_m}$ and $\delta_{de} =\frac{\delta \rho_{de}}{\rho_{de}}$ respectively. Using (\ref{eq4:pert-int1}) and (\ref{eq4:pert-int2}), the equations (\ref{e1}) and (\ref{m1}) for CDM and IHDE can be written respectively as 
\begin{eqnarray}
\delta_{m}^\prime+\theta_m-3 \Phi^\prime &=& \beta \cH\frac{\rho_{de}}{\rho_m}\left(\Phi-\delta_m +\delta_{de}\right), \label{eq4:e2m}\\
\theta_m^\prime+\cH\theta_m-k^2 \Phi&=& -\beta \cH\frac{\rho_{de}}{\rho_m}\left(\theta_m -\theta_{de}\right), \label{eq4:m2m}
\end{eqnarray}
%%%%%%%%%%%%%%%%%%%
\begin{equation}
\begin{split}
\delta_{de}^\prime +&3 \cH\left(c_{s,de}^{2}-w_{de}\right)\delta_{de}+\left(1+w_{de}\right)\left(\theta_{de}-3\Phi^\prime\right)\\
+&3\cH\left[3 \cH \left(1+w_{de}\right)\left(c_{s,de}^{2}-w_{de}\right)\right]\frac{\theta_{de}}{k^2} +3\cH w_{de}^\prime\frac{\theta_{de}}{k^2} \\
=&-\beta \cH\left[\Phi+3 \cH\left(c_{s,de}^{2}-w_{de}\right)\frac{\theta_{de}}{k^2}\right], \label{eq4:e2de}
\end{split}
\end{equation}
\begin{equation}
\theta_{de}^\prime+\cH\left(1-3c_{s,de}^{2}\right)\theta_{de}-k^2\Phi-\frac{k^2\delta_{de}c_{s,de}^{2} }{\left(1+w_{de}\right)}=\frac{\beta \cH}{\left(1+w_{de}\right)}\left(-c_{s,de}^{2}\theta_{de}\right). \label{eq4:m2de}
\end{equation}
%%%%%%%%%%%%%%%%%%%

Eliminating $\theta_m$ from equations (\ref{eq4:e2m}), (\ref{eq4:m2m}) and $\theta_{de}$ from equations (\ref{eq4:e2de}), (\ref{eq4:m2de}),  the coupled differential equations for CDM and IHDE are obtained respectively in terms of redshift as
%%%%%%%%%%%%%%%%%%%
\begin{equation}
\begin{split}
\mathbf{C^{(m)}_1} \frac{\partial^2 \delta_m}{\partial z^2}+ \mathbf{C^{(m)}_2} \frac{\partial\delta_m}{\partial z}& +\mathbf{C^{(m)}_3} \delta_m + \mathbf{C^{(m)}_4} \frac{\partial^2 \delta_{de}}{\partial z^2}+\mathbf{C^{(m)}_5} \frac{\partial \delta_{de}}{\partial z}+ \mathbf{C^{(m)}_6} \delta_{de}\\
&+ \mathbf{C^{(m)}_7} \frac{\partial^2 \Phi}{\partial z^2}+ \mathbf{C^{(m)}_8} \frac{\partial \Phi}{\partial z} + \mathbf{C^{(m)}_9} \Phi  = 0~, \label{eq4:finaldm}
\end{split}
\end{equation}
\begin{equation}
\begin{split}
\mathbf{C^{(de)}_1} \frac{\partial^2 \delta_{de}}{\partial z^2}+ \mathbf{C^{(de)}_2} \frac{\partial\delta_{de}}{\partial z} &+\mathbf{C^{(de)}_3} \delta_{de} + \mathbf{C^{(de)}_4} \frac{\partial^2 \delta_{m}}{\partial z^2}+\mathbf{C^{(de)}_5} \frac{\partial \delta_{m}}{\partial z} + \mathbf{C^{(de)}_6}\delta_m \\
&+\mathbf{C^{(de)}_7} \frac{\partial^2 \Phi}{\partial z^2}+ \mathbf{C^{(de)}_8} \frac{\partial \Phi}{\partial z} + \mathbf{C^{(de)}_9} \Phi  = 0~. \label{eq4:finalde}
\end{split}
\end{equation}
%%%%%%%%%%%%%%%%%%%
%
The coefficients $\mathbf{C^{(m)}_i}$ and $\mathbf{C^{(de)}_i}$ where $\mathbf{i}=\mathbf{1}, \mathbf{2},\ldots,\mathbf{9}$ are given in the Appendix \ref{appen:epjp}. The coefficients $\mathbf{C^{(de)}_4}$, $\mathbf{C^{(de)}_5}$ and $\mathbf{C^{(de)}_6}$ are zero in equation (\ref{eq4:finalde}) indicating that the evolution of DE is not directly affected by DM fluctuation but the converse is not true. The coefficients $\mathbf{C^{(m)}_j}$ and $\mathbf{C^{(de)}_j}$ where $\mathbf{j}=\mathbf{7}, \mathbf{8},\mathbf{9}$ are non zero in both the equations (\ref{eq4:finaldm}) and (\ref{eq4:finalde}) which implies that the potential $\Phi$ will affect the evolution of both DM and DE density contrasts. The evolution of $\Phi$ is governed by the equation (\ref{eq4:g002}) and is not approximated by the Poisson equation. The equations (\ref{eq4:finaldm}) and (\ref{eq4:finalde}) along with the equation (\ref{eq4:g002}) are solved numerically to find the evolution of the density contrasts of the CDM and IHDE. In order to do that, in the matter-dominated era, i.e. from the initial redshift $z_{in} = 1100$, it is assumed that $\Phi\left(z_{in}\right)=\mbox{constant}=\phi_0$ and $\Phi^\prime\left(z_{in}\right)=0$. It is also assumed that $\Omega_{m}\left(z_{in}\right) >> \Omega_{de}\left(z_{in}\right)$ so that the term with the ratio $\frac{\Omega_{de}\left(z_{in}\right)}{\Omega_m\left(z_{in}\right)}$ can be neglected for $\delta_m\left(z_{in}\right)$ only. As discussed in~\cite{abramo2009prd}, the initial conditions for $\delta_m$, $\delta^\prime_m$, $\delta_{de}$ and $\delta^\prime_{de}$ are
%%%%%%%%%%%%%%%%%%%
\begin{eqnarray}
\delta_{mi} = \delta_m\left(z_{in}\right) &=& -2 \phi_0\left[1+\frac{\left(1+z_{in}\right)^2 k_{in}^2}{3 H_{in}^2}\right],\\
\delta_{mi}^\prime =\frac{d \delta_m}{d z}\bigg \rvert_{\substack{z=z_{in}}} &= &-4 \phi_0 \frac{\left(1+z_{in}\right) k_{in}^2}{3 H_{in}^2} \left[1-\frac{3}{2}\left (1+\frac{w_{dei} \Omega_{dei}}{\Omega_{mi}}\right )\right],\\
\delta_{dei} = \delta_{de}\left(z_{in}\right) &=& \frac{\delta_{mi}}{3-\beta \frac{\Omega_{dei}}{\Omega_{mi}}}\lbrace 3\left(1+w_{dei}\right)+\beta \rbrace,\\
\delta_{dei}^\prime =\frac{d \delta_{de}}{d z}\bigg \rvert_{\substack{z=z_{in}}} &=&\frac{3 \left(\frac{d w_{de}}{d z}\right)\Big \rvert_{\substack{z=z_{in}}}}{3-\beta \frac{\Omega_{dei}}{\Omega_{mi}}}+\frac{\delta_{mi}^\prime}{3-\beta \frac{\Omega_{dei}}{\Omega_{mi}}}\lbrace 3\left(1+w_{dei}\right)+\beta\rbrace \nonumber\\
&+&\delta_{mi}\lbrace 3\left(1+w_{dei}\right)+\beta\rbrace \frac{\beta \left[ \frac{d}{d z}\left(\frac{\Omega_{de}}{\Omega_m}\right)\right]_{\substack{z=z_{in}}}}{\left(3-\beta \frac{\Omega_{dei}}{\Omega_{mi}}\right)^2},
\end{eqnarray}
%%%%%%%%%%%%%%%%%%%
where $H_{in}=H\left(z_{in}\right)$, $w_{dei}=w_{de}\left(z_{in}\right)$, $\Omega_{mi}=\Omega_m\left(z_{in}\right)$, $\Omega_{dei}=\Omega_{de}\left(z_{in}\right)$ and $k_{in}$ is the mode entering the Hubble horizon at $z_{in}$. The value of $k_{in}$ is taken as $\left(1+z\right)^{-1}H_{in}$. The numerical values for $H_{in}$, $w_{dei}$, $\Omega_{mi}$ and $\Omega_{dei}$ are obtained from the solutions of the equations (\ref{eq4:int-omega}), (\ref{eq4:int-hubble}) and (\ref{eq4:int-eos}). The value of $\phi_0$ is given by hand.

For the Fourier mode, $k$ in equations (\ref{eq4:finaldm}), (\ref{eq4:finalde}) and (\ref{eq4:g002}), the domain considered is in the linear regime given by the galaxy power spectrum~\cite{percival07}
\begin{equation} \label{eq4:k-linear}
0.01 h \hskip1ex \mbox{Mpc}^{-1} \lesssim k \lesssim 0.2 h \hskip1ex \mbox{Mpc}^{-1},
\end{equation}
where $h = \frac{H_0}{100 \hskip1ex \footnotesize{\mbox{km s}^{-1} \mbox{Mpc}^{-1}}}$ is the dimensionless Hubble parameter at the present epoch. The value of $H_0 = 67.27$ is taken from the \Planck 2018 data~\cite{planck2018cp}. For $k > 0.2 h \hskip1ex \mbox{Mpc}^{-1}$ (smaller scales), non-linear effects become prominent whereas for $k < 0.01 h \hskip1ex \mbox{Mpc}^{-1}$ observations are not very accurate. So we consider $k=0.1 h \hskip1ex \mbox{Mpc}^{-1}$ following~\cite{mehrabi2015mnras}. For our calculation we have considered $\phi_0 =10^{-5}$ and $c_{s,de}^2=0$. 

\begin{figure}[!h]
  \centering
\includegraphics[width=\linewidth]{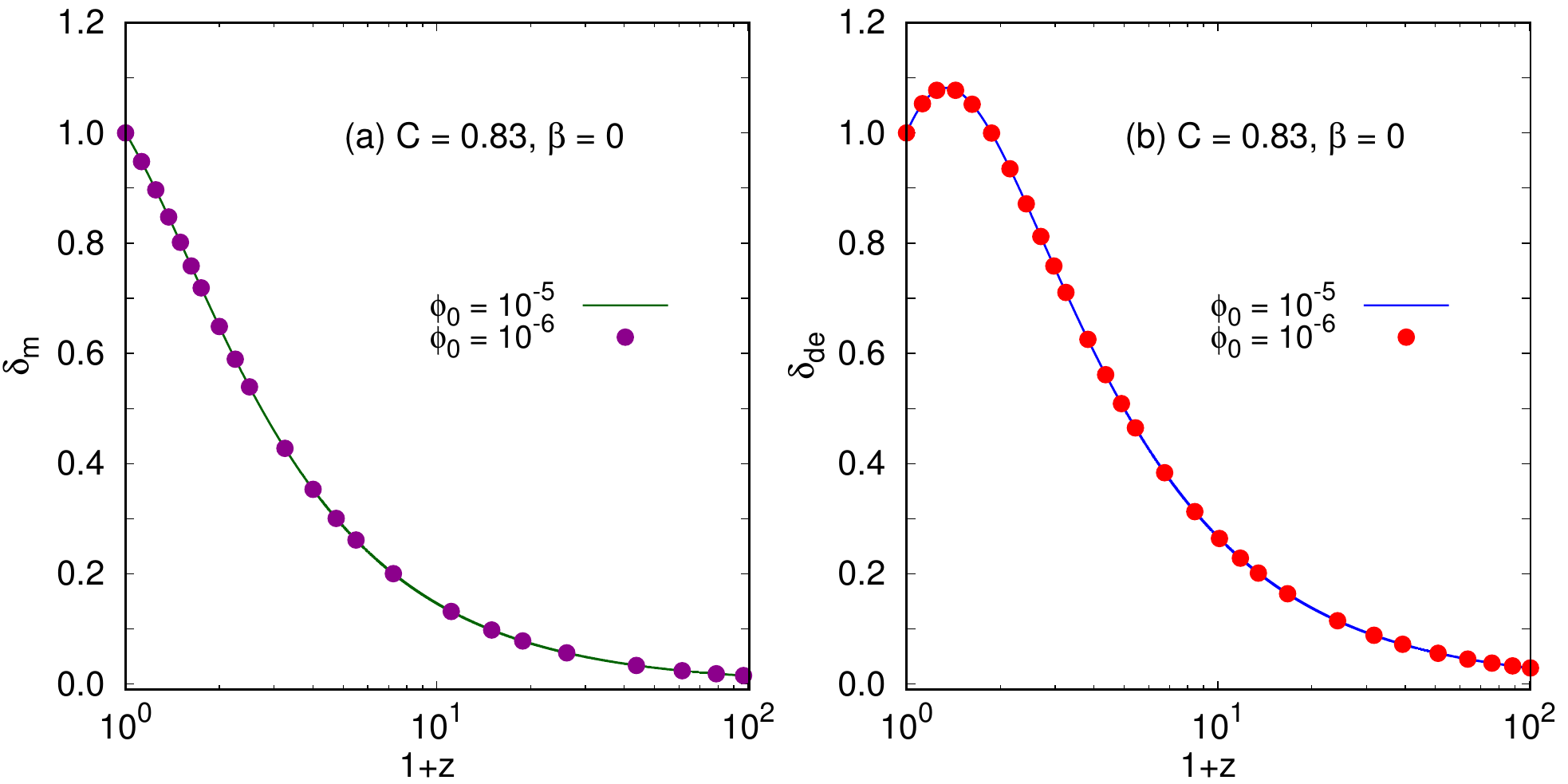}
\caption{(a) shows plot of $\delta_m$ against $\left(1+z\right)$ and (b) shows the plot of $\delta_{de}$ against $\left(1+z\right)$ in logarithmic scale for $C =0.83$ and $\beta = 0$. The line shows the variation of $\delta_m$ and $\delta_{de}$ for the initial condition, $\phi_0 = 10^{-5}$ and the solid circles represent the same corresponding to the initial condition, $\phi_0 = 10^{-6}$~.}\label{im4:figmul0183}
  \centering
\includegraphics[width=\linewidth]{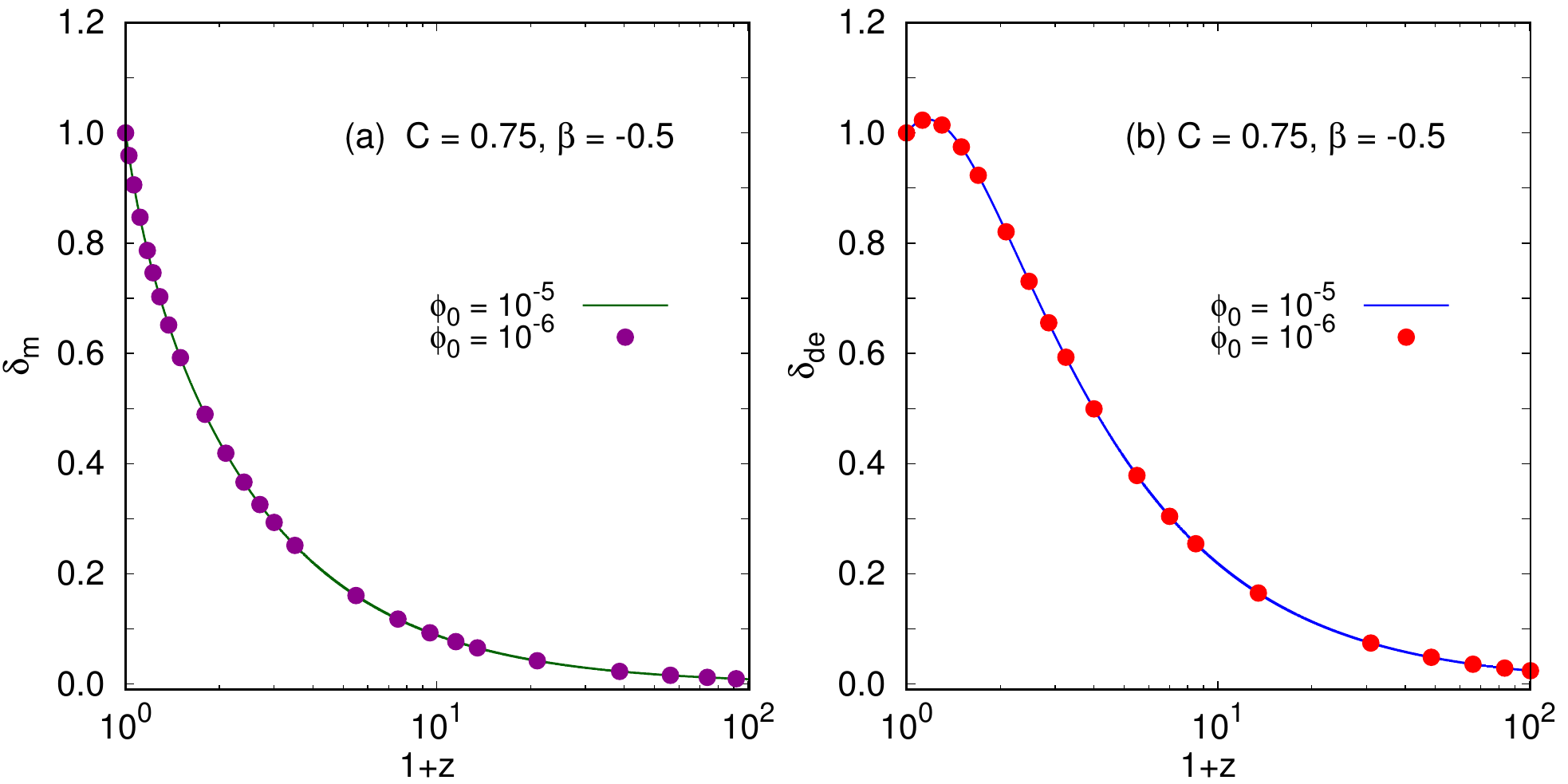}
\caption{(a) shows plot of $\delta_m$ against $\left(1+z\right)$ and (b) shows the plot of $\delta_{de}$ against $\left(1+z\right)$ in logarithmic scale for $C =0.75$ and $\beta =-0.5$. The line shows the variation of $\delta_m$ and $\delta_{de}$ for the initial condition, $\phi_0 = 10^{-5}$ and the solid circles represent the same corresponding to the initial condition, $\phi_0 = 10^{-6}$~.}\label{im4:figmul0175}
\end{figure}

\begin{figure}[!h]
  \centering
\includegraphics[width=\linewidth]{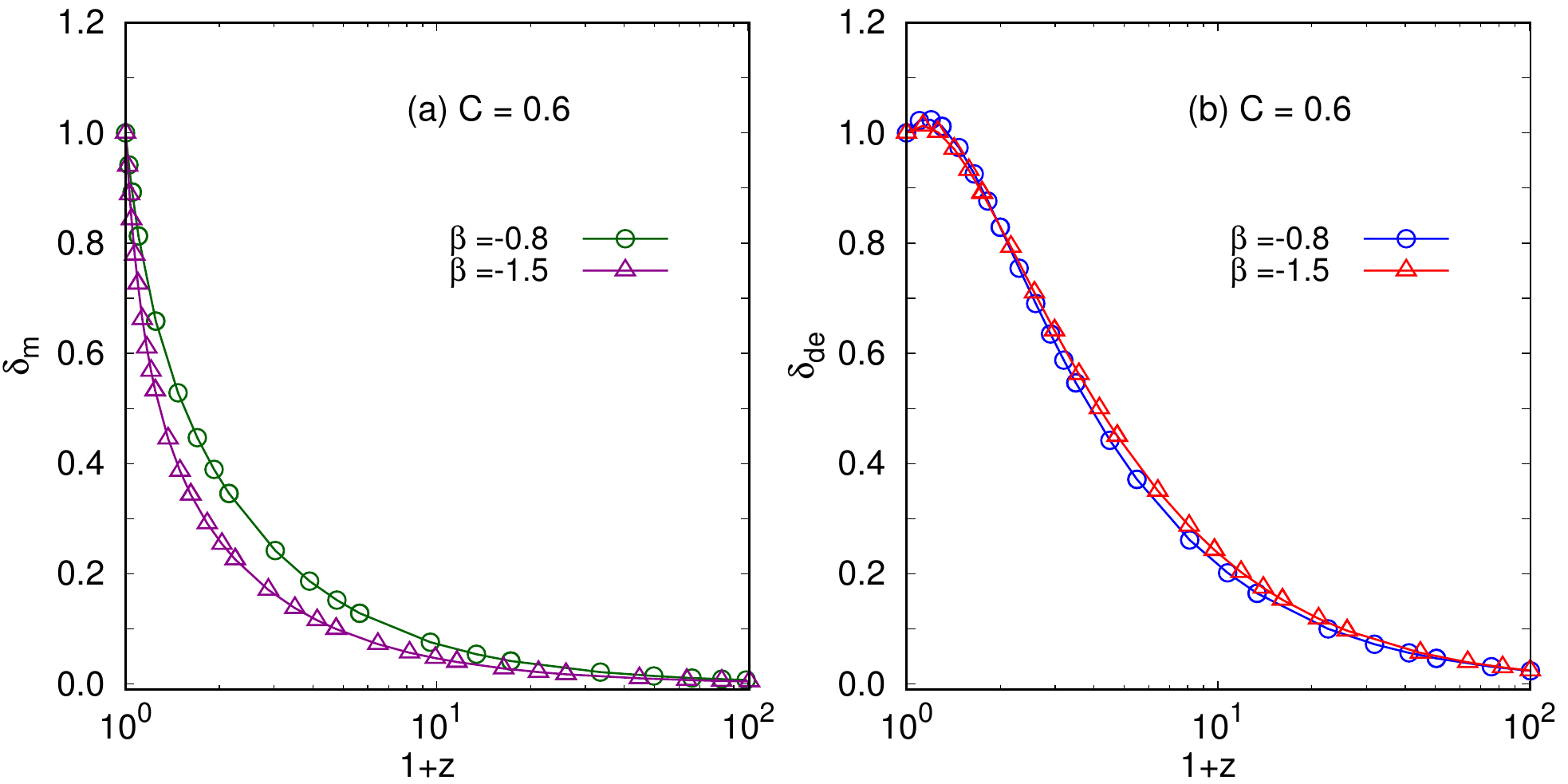}
\caption{(a) shows plot of $\delta_m$ against $\left(1+z\right)$ and (b) shows the plot of $\delta_{de}$ against $\left(1+z\right)$ in logarithmic scale for $C =0.6$ and two different values of $\beta$. The line with circles is for $\beta=-0.8$ and the line with triangles is for $\beta = -1.5$~.}\label{im4:fig016all}
  \centering
\includegraphics[width=\linewidth]{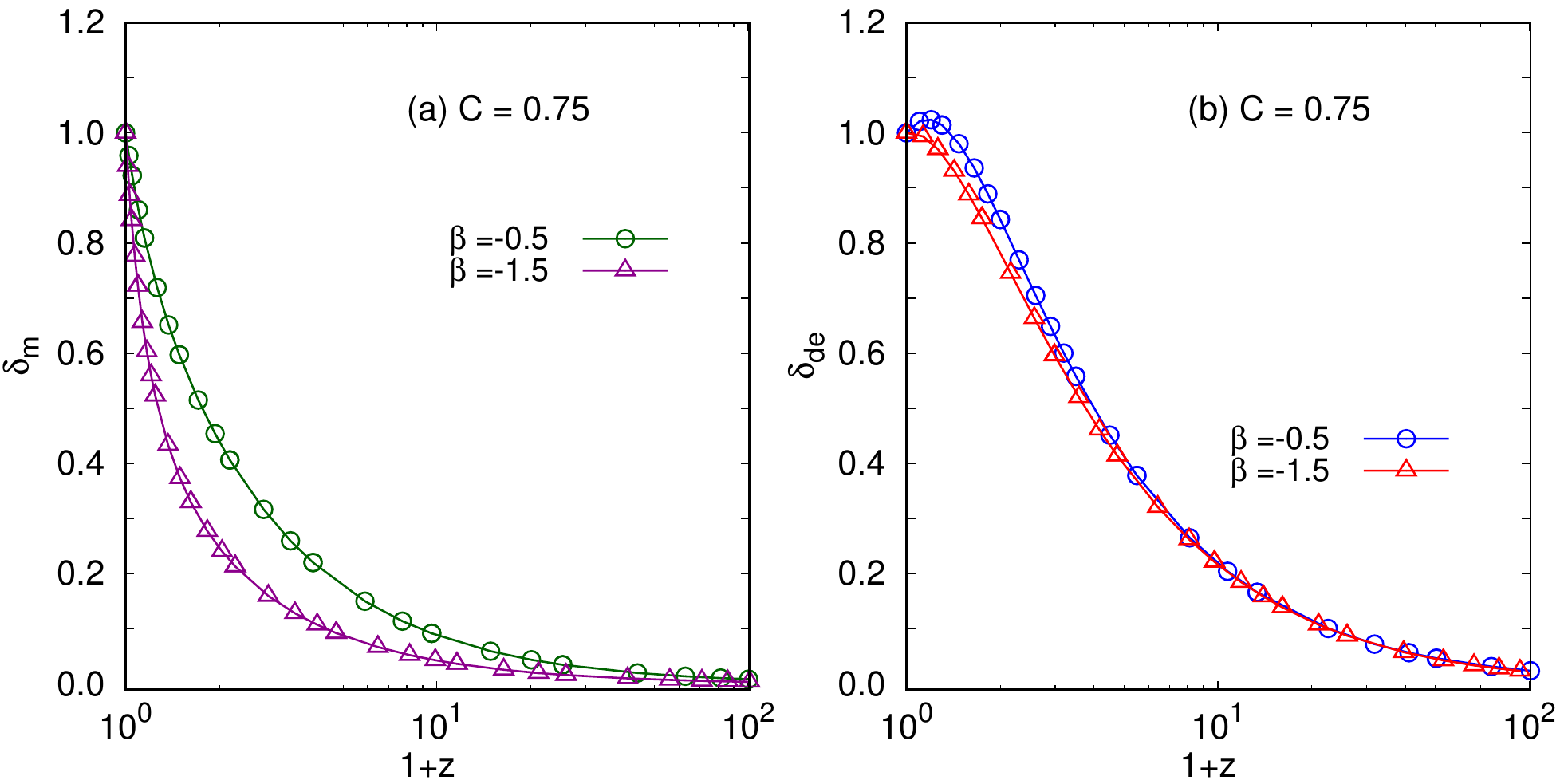}
\caption{(a) shows plot of $\delta_m$ against $\left(1+z\right)$ and (b) shows the plot of $\delta_{de}$ against $\left(1+z\right)$ in logarithmic scale for $C =0.75$ and two different values of $\beta$. The line with circles is for $\beta=-0.5$ and the line with triangles is for $\beta = -1.5$~.}\label{im4:fig0175all}
\end{figure}

\begin{figure}[!h]
  \centering
    \includegraphics[width=\linewidth]{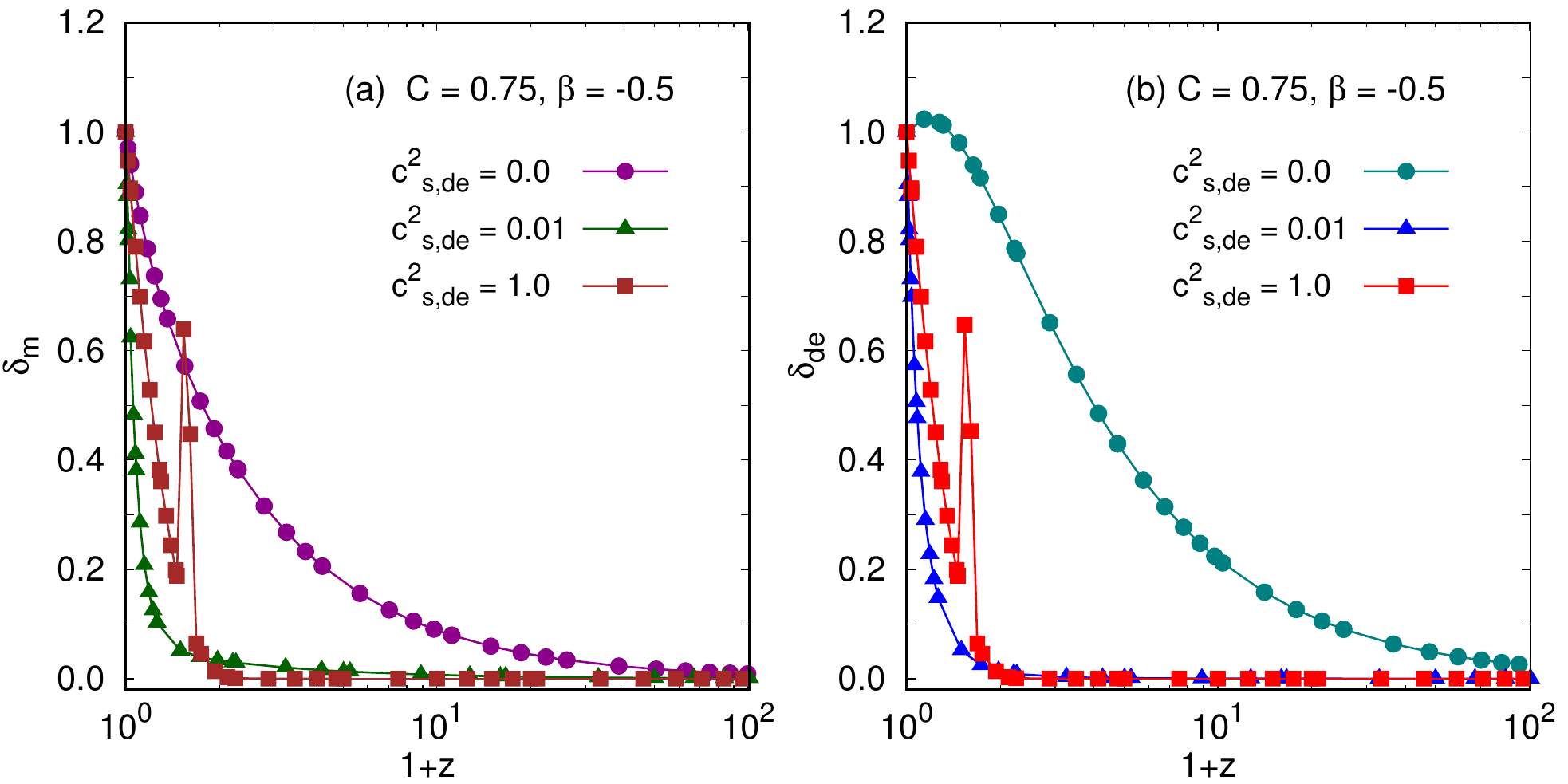}
\caption{(a) shows plot of $\delta_m$ against $\left(1+z\right)$ and (b) shows the plot of $\delta_{de}$ against $\left(1+z\right)$ in logarithmic scale for different values of $c_{s,de}^2$ with $C =0.75$ and $\beta=-0.5 $. The line with solid circles corresponds to $c_{s,de}^2 =0$, the line with solid triangles corresponds to $c_{s,de}^2 =0.01$ and the line with solid squares corresponds to $c_{s,de}^2 =1.0 $~.} \label{im4:figcs}
\end{figure}

The density contrast of DM ($\delta_m$) has over density (positive solution) while that of DE ($\delta_{de}$) has under density (negative solution). All the Figs.\ (\ref{im4:figmul0183}) - (\ref{im4:figcs}) are shown in logarithmic scale from $z=0$ to $z=100$ with $C=0.83$ and $\beta=0$ for the non-interacting case. For the interacting case, we have chosen $C=0.75$ and $\beta = -0.5$. In all the figures, the density contrast is scaled by its present value. For $C=0.83$ and $\beta=0$, the present values of dark matter and dark energy fluctuations are $\delta_{c}\paren*{z=0} = 0.71068$ and $\delta_{de}\paren*{z=0}=-0.24166$ respectively, while for $C=0.75$ and $\beta = -0.5$ the values are $\delta_{c}\paren*{z=0} = 0.81901$ and $\delta_{de}\paren*{z=0}=-0.23467$ respectively. To study the effect of interaction in the growth of the density contrasts we have considered different sets $C$ and $\beta$.
%%%%%%%%%%%%%%%%%%%

Figure (\ref{im4:figmul0183}) shows the variation of $\delta_m$ and $\delta_{de}$ with $\left(1+z\right)$ in logarithmic scale for the non-interacting case for $\phi_0 = 10^{-5}$ and $\phi_0 = 10^{-6}$. When scaled by their respective present value, the nature of the growth of $\delta_m$ and $\delta_{de}$ is hardly sensitive to the value of $\phi_0$. This is clear from Figs.\ (\ref{im4:figmul0183}a) and (\ref{im4:figmul0183}b). Figure (\ref{im4:figmul0175}) shows the same for the interacting case with $C=0.75$ and $\beta =-0.5$. One can clearly see from Figs.\ (\ref{im4:figmul0183}a) and (\ref{im4:figmul0175}a) that the interaction ($\beta \ne 0$) makes the slopes different.   For the variation of $\delta_{de}$, we see that it first grows up to a maximum and then decreases to unity at $z=0$. The position as well as the height of this peak is different in the Figs.\ (\ref{im4:figmul0183}b) and (\ref{im4:figmul0175}b). The presence of an interaction has decreased the height of the maximum and made the growth a little flat.

Figures (\ref{im4:fig016all}a) and (\ref{im4:fig0175all}a) show the variation of $\delta_m$ with $\left(1+z\right)$ in logarithmic scale for the same value of $C$ but different values of $\beta$. For $C=0.6$, as $\beta$ decreases from $-0.8$ to $-2.0$ the curve tends to become steeper and the growth becomes faster. This behaviour is similar in both the cases of $C=0.6$ and $C=0.75$.  For $\delta_{de}$ (Figs.\ (\ref{im4:fig016all}b) and (\ref{im4:fig0175all}b)), the change in slope for smaller $\beta$ is more prominent in smaller $C$ value. In Fig.\ (\ref{im4:fig016all}b), for $C=0.6$, $\delta_{de}$ for $\beta = -1.5$, changes faster than that for $\beta=-0.8$. Similarly in Fig.\ (\ref{im4:fig0175all}b), for $C=0.75$, $\delta_{de}$ for $\beta=-1.5$ changes faster than that for $\beta=-0.5$. The change in the direction of the growth rate takes place after the Universe starts accelerating and has a correlation with the deceleration parameter $q$ changing its sign. For Fig.\ (\ref{im4:fig016all}b),  the maximum of ${\delta}_{de}$ for $\beta = -1.5$ is at a slightly lower redshift than that for $\beta =-0.8$, and for Fig.\ (\ref{im4:fig0175all}b), the maximum of ${\delta}_{de}$ for $\beta =-1.5$ is at a lower redshift than that for $\beta =-0.5$. If $\beta$ is decreased below $-1.5$, no such correlation is seen.
\begin{figure}[!h]
  \centering
    \includegraphics[width=\linewidth]{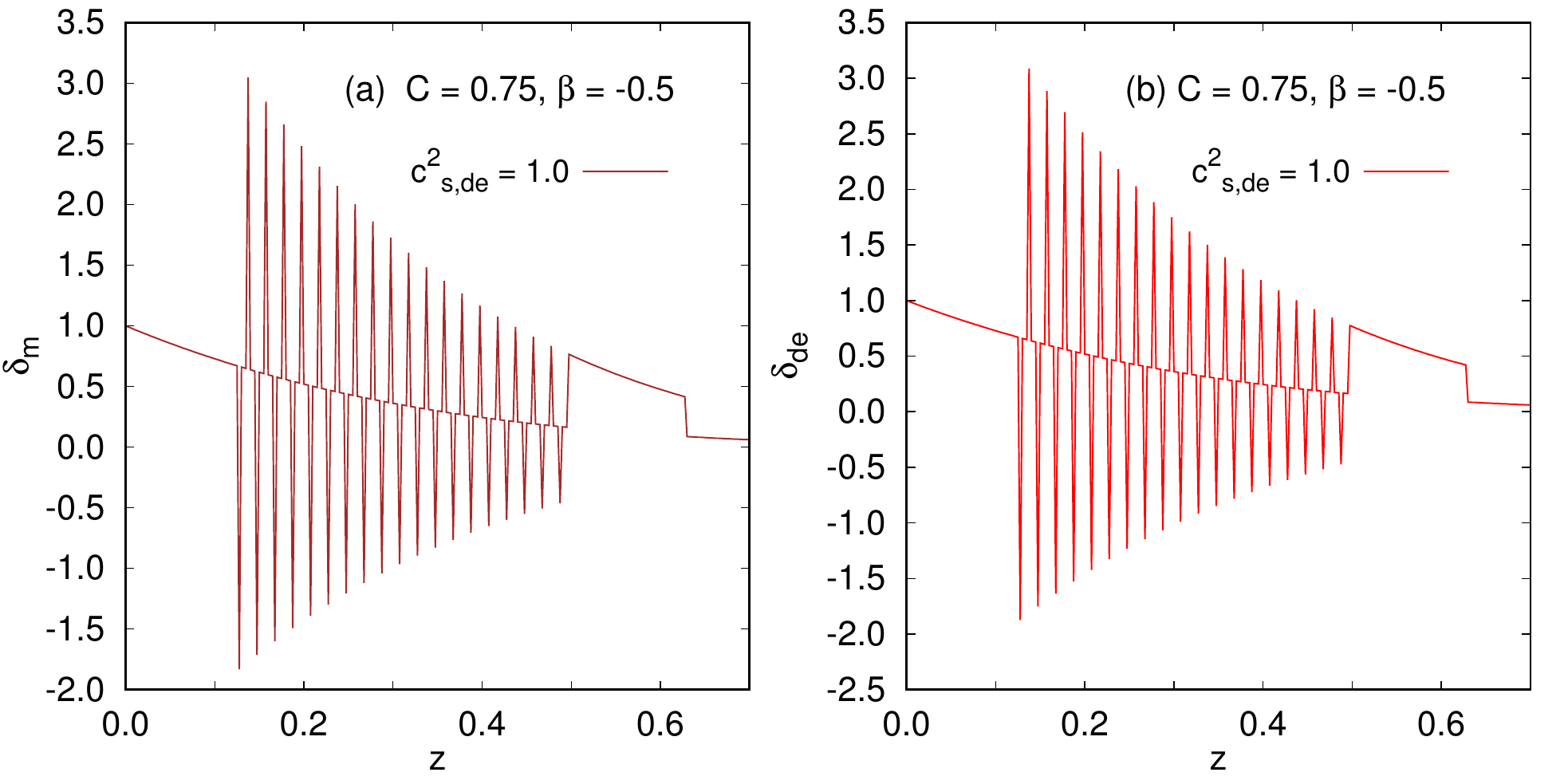}
\caption{(a) shows plot of $\delta_m$ against $z$ and (b) shows the plot of $\delta_{de}$ against $z$ from $z=0$ to $z=0.7$ for $c_{s,de}^2=1.0$ with $C =0.75$ and $\beta=-0.5$~.} \label{im4:figcs1}
 \includegraphics[width=\linewidth]{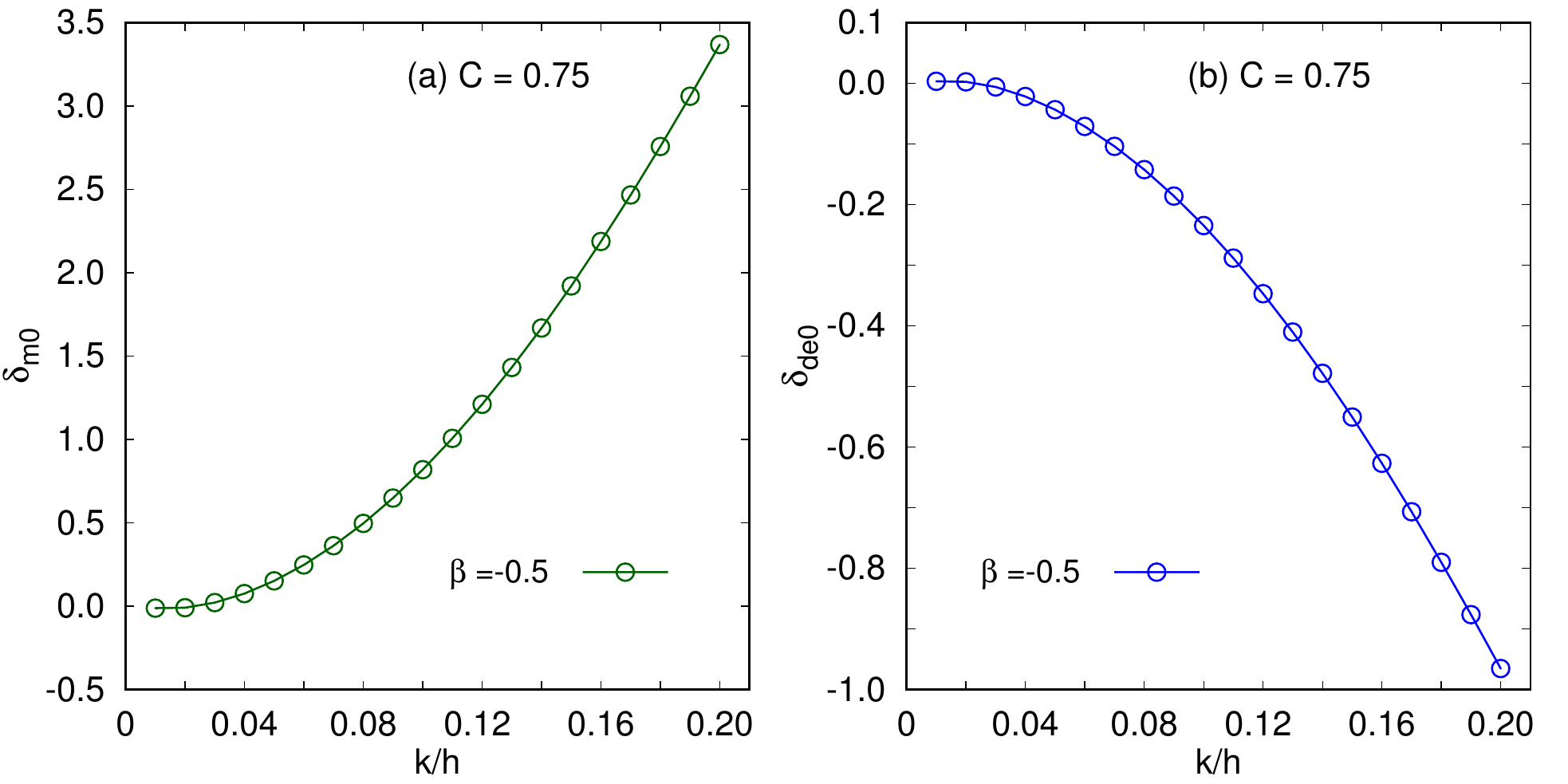}
\caption{(a) shows plot of $\delta_{m0}$ against $k$ and (b) shows the plot of $\delta_{de0}$ against $k$ at $z=0$ for $C =0.75$ and $\beta=-0.5$~.} \label{im4:fig01k}
\end{figure}

Figure (\ref{im4:figcs}) shows the variation of the density contrasts $\delta_m$ and $\delta_{de}$ for the different values of the effective speed of sound of dark energy perturbation, $c_{s,de}^2$. From the expression of $\delta p_{de}$ Eq.\ (\ref{eq4:pert-p}), we can see that the pressure perturbation not only depends on the product $c_{s,de}^2 \delta \rho_{de}$ but also on the background quantities like $w_{de}$, $ \rho_{de}$, $Q$ as well as the velocity perturbation through $\theta_{de}$. With $c_{s,de}^2=0$, the effect of the velocity perturbation present in the second term is prominent. The presence of the interaction $Q$ actually decreases this effect (Figs.\ (\ref{im4:figmul0183}b) and (\ref{im4:figmul0175}b)). In presence of the $c_{s,de}^2$ (i.e.\ any non-zero value), the effect of $\delta \rho_{de}$ comes into play. When zoomed into smaller redshift region ($z=0.1275$ to $z=0.4875$), rapid oscillations are observed (Fig.\ (\ref{im4:figcs1})). 

Figure \ref{im4:fig01k} shows the variation of $\delta_m$ and $\delta_{de}$ with $k/h$ at $z=0$.  In Fig.\ (\ref{im4:fig01k}a), for a given value of $C$ and $\beta$, as $k/h$ increases $\delta_{m0}$ increases --- the growth rate of DM over densities increase for smaller scales entering the horizon. Though the increase is not linear for $k/h$ less than $ \sim 0.1h$, but when scaled by $\delta_{m0}$, the growth rates of $\delta_m$ for different $k$-modes are independent of the $k$-modes. For Fig.\ (\ref{im4:fig01k}b), negative values of $\delta_{de}$ increases with larger scales. The change in slope in this case is also not linear for modes smaller than $\sim 0.1h$ and the growth of $\delta_{de}/\delta_{de0}$ for different modes are identical.
%%%%%%%%%%%%%%%%%%%%%%%%%%%%%%%%%%%%%%%%%%%%%%%%%%%%%
%%%%%%%%%%%%%%%%%%%%%%%%%%%%%%%%%%%%%%%%%%%%%%%%%%%%%

%%%%%%%%%%%%%%%%%%%%%%%%%%%%%%%%%%%%%%%%%%%%%%%%%%%%%
\section{Summary And Discussion} \label{sec4:summary}
The primary motivation of the present work is to study the effect of interaction on density perturbation in the dark sector of the Universe in a Holographic dark energy model. Among various possibilities, we have chosen the future event horizon as the IR cut-off for the HDE model for which the Universe can accelerate even in the absence of an interaction. For the interaction between the DM and DE, we have chosen the interaction term to be proportional to the dark energy density $\rho_{de}$. The interaction term is of the form $Q=\frac{\beta \cH\rho_{de}}{a}$, in which the dependence on cosmic time comes through the global expansion rate, the Hubble parameter $\cH$ and the scale factor $a$. The coupling constant $\beta$ determines the strength of the interaction as well as the direction of the energy flow. No interaction between DE and DM is characterised by $\beta=0$. We restricted the model parameters $C$ and $\beta$ in such a way that at present the DE EoS parameter, $w_{de}$ is sufficiently negative to produce the late time acceleration but can avoid the ``phantom menace'' ($w_{de}< -1$).

The set of coupled second order differential equations for the density contrasts for the CDM ($\delta_m$) as well as the HDE ($\delta_{de}$) were obtained in the Newtonian gauge. Adiabatic initial conditions were used with the assumption that the DE density parameter is small compared to the DM density parameter ($\frac{\Omega_{de}}{\Omega_m} \ll 1$) at the onset of the matter dominated epoch ($z=1100$) and has no influence on the estimation of $\delta_m$ at $z=1100$. We solved the differential equations numerically from $z=0$ to $z=1100$ and found that $\delta_m$ is increasing in the positive direction, whereas $\delta_{de}$ is increasing in the negative direction. However, as $\delta_m$ and $\delta_{de}$ are always scaled by their respective present values, this difference in signature is not reflected in the plots. 

A small negative value of $\beta$ indicates that dark matter decays into dark energy and the interaction in the dark sector, if any, has to be small. 

The effect of effective sound speed of DE, $c_{s,de}^2$ on density perturbation was also looked at. The absence of an interaction, with $w_{de} =-1$ and $c_{s,de}^2=0$ indicates no pressure perturbation in DE and the DE density perturbation is expected to grow like that of DM. In the present HDE model, the scenario is entirely different; even in the absence of interaction and $c_{s,de}^2$, the pressure perturbation remains non-vanishing as $w_{de}$ is now a varying function of redshift, $z$. The pressure perturbation is then governed by the velocity perturbation through $\theta_{de}$ and the background quantities $\rho_{de}$, $w_{de}$ and $Q_{de}$.

The DE density contrast, $\delta_{de}$ also grows almost in a similar fashion like its DM counterpart, $\delta_m$, for most of the evolution after the radiation dominated era, but right at the present moment is actually decaying after hitting a maximum in the recent past. This is true even in the absence of interaction (characterised by $\beta=0$). This maximum is found to occur after the onset of the present accelerated phase of expansion. The height of the maximum is related to the strength of the interaction, $\beta$. On decreasing the strength (smaller magnitude) of the interaction, the position of maximum shifts to lower redshifts and the height decreases. This feature is observed for the zero value of the effective sound speed, $c_{s,de}^2$(Fig.\ \ref{im4:figcs}b).

When $c_{s,de}^2=0$ the first part in the expression for $\delta p_{de}$ (see equation (\ref{eq4:pert-p})) is zero and from the second part, we can say that $\delta_{de}$ reaching a maximum is characterised by $\theta_{de}$ whereas $Q_{de}$ actually suppresses this feature (Figs.\ \ref{im4:figmul0183} b and \ref{im4:figmul0175}b). When $c_{s,de}^2 \ne 0$, the contribution from the first part ($c_{s,de}^2 \delta \rho_{de}$) results in the steep rise in $\delta_{de}$ at lower values of $z$ (Fig.\ \ref{im4:figcs}b). This apparently is engineered by $\theta_{de}$. Except for the peak in the growth rate, similar features are also seen for $\delta_{m}$ (Fig.\ \ref{im4:figcs}a).
For $c_{s,de}^2=1$, a rapid short-lived oscillations in the DE density contrast is found between $z=0.1275$ to $z=0.4875$. These oscillations are characteristic of $c_{s,de}^2=1$ and $\beta \ne 0$ (Fig.\ \ref{im4:figcs1}). The oscillations in $\ddm$ are present only in presence of DE perturbations; in absence of DE perturbations (smooth DE), oscillations in $\ddm$ would vanish.

For an interaction in dark sector, without a holographic bound, there is an instability in the perturbation~\cite{valiviita2008jcap}. This does not appear in the present case where there is a holographic bound.

Thus the present investigation has some new inputs leading to quite new and intriguing features. The new physical input at the outset is certainly the introduction of an interaction between DM and DE in the study of density contrasts. In the techniques and approximations, writing down the full relativistic perturbation equations a priori is new. Even in the most recent and general treatment~\cite{mehrabi2015prd}, $\Phi^\prime$ is neglected for the estimations, but in the present work, its contribution is also respected. The appearance of a peak $\delta_{de}$ for $c^{2}_{s,de} =0$ is a completely new feature observed, which is not due to the interaction, as it is there even for $\beta =0$~. So this is due to the inclusion of $\Phi^\prime$ in the estimation. For $c^{2}_{s,de} =1$, no growth for $\delta_{de}$ is normally observed. The recent work of Batista and Pace~\cite{batista2013jcap} shows an almost negligible growth. The present work shows a very steep growth for small $z$. This is there both in the presence and absence of interaction. For an interacting scenario, there is also a short-lived oscillatory period in the growing mode of $\delta_{de}$ (Figs.\ (\ref{im4:figcs1}a) and (\ref{im4:figcs1}b)).

We presented the calculations with $\phi_0=10^{-5}$ and $k=0.1h$, these results are insensitive to changes in $\phi_0$ and $k$ mode entering the horizon.

% Chapter 5
\chapter{Perturbations In A Scalar Field Model With Virtues Of \lcdm}\label{chap5:jcap}\blfootnote{The work presented in this chapter is based on ``Perturbations In A Scalar Field Model With Virtues Of \lcdm'', \textbf{\authorname} and Narayan Banerjee, JCAP \textbf{04} 060 (2021)} % Main chapter title
\chaptermark{Perturbations In A Scalar Field Model}

%%%%%%%%%%%%%%%%%%%%%%%%%%%%%%%%%%%%%%%%%%%%%%%%%%%%%
\section{Introduction}\label{sec5:intro}
Over the last decade, the availability of high precision data from various surveys has suggested that EoS parameter of dark energy $w=-1.03\pm 0.03$ within the $95\%$ confidence level~\cite{planck2018cp}, consistent with a cosmological constant. One is tempted to conclude that the cosmological constant as dark energy with cold dark matter (\lcdm) is by far the most suitable model that describes the evolution of the Universe at the present epoch. But the \lcdm model is plagued with problems like the fine-tuning problem~\cite{sahni2002cqg} and the coincidence problem~\cite{steinhardt2003jstor,velten2014epjc}. The fine-tuning problem is that the initial conditions are needed to be set to an exact value so that the cosmological constant term dominates at the current epoch. The coincidence problem is related to the question why the energy densities of dark matter and dark energy are of the same order of magnitude at the present epoch. These problems in the \lcdm model has forced us to look for other candidates that can drive the acceleration. A scalar field rolling down a slowly varying potential not only gives rise to acceleration but also alleviates the cosmological coincidence problem. Such a scalar field, dubbed as `quintessence', has been studied extensively in the literature~\cite{peebles1988apjl, ratra1988prd, ferreira1997prl, copeland1998prd,ferreira1998prd, wetterich1988npb, efstathiou1999mnras, kim1999jhep, zlatev1999prl, steinhardt1999prd, brax1999plb, brax2000prd1,barreiro2000prd, sahni2000prd, albrecht2000prl, dodelson2000prl, wang2000apj, sen2002plb, amendola2006prd, dimopoulos2017jcap, mishra2017jcap, nandan2018prd}. 

The `tracking' models as discussed in Section \ref{sec1:sfm} can resolve the fine-tuning problem and coincidence problem but cannot give rise to the acceleration with $w=-1$. The values of $w$ that can be obtained are $w=-0.6$~\cite{wang2000apj}, $w=-0.8$~\cite{efstathiou1999mnras}, $w\approx-0.82$~\cite{brax2000prd1}, $w<-0.8$~\cite{zlatev1999prl,barreiro2000prd} to mention a few. Thus, the quintessence models do not appear to be a complete solution. One should therefore look for a model that will have the virtues of both the \lcdm and a quintessence but will be devoid of the flaws. A `tracking' quintessence model with an inverse power law potential, however, was shown to be consistent with observational data sets~\cite{zhai2017apj,park2018apj,ooba2019ass,park2020prd}. The inverse power law potential with a dynamical dark energy gives $w = - 1.03 \pm 0.07$ at $z=0$ within the $68.27\%$ confidence limit~\cite{ooba2019ass}, in agreement with the recent observations. However, the present model is different from that discussed in~\cite{efstathiou1999mnras, zlatev1999prl, barreiro2000prd, wang2000apj, brax2000prd1, zhai2017apj, park2018apj, bag2018jcap, ooba2019ass, park2020prd} and yields $w=-1$ at the present epoch independent of the model parameters.

To construct a model without the problem of fixing the initial conditions, the ``scaling'' potentials or the ``tracking'' potentials are among the natural choices. However, the scaling solution does not drive an acceleration, whereas the best-known tracking potentials cannot give the observationally preferred value of $w \simeq -1$. In the present work, we introduce a scalar field model with a potential such that it will have an accelerated expansion with an equation of state at the present epoch similar to that given by \lcdm and the current dark energy density comparable to that of dark matter independent of the initial conditions. We engineer the model such that the scalar field $\vphi$ is subdominant as a tracking dark energy at early times and start dominating as a cosmological constant in the recent past driving the acceleration. The presence of a scalar field from early times will have its imprints on the growth of perturbations and hence on the large scale structures of the Universe. The scalar field will evolve through the history of the Universe, and unlike \lcdm, will have fluctuations similar to the other matter components. These fluctuations will affect the formation of structures~\cite{abramo2009prd} and can also cluster on their own~\cite{mehrabi2015mnras, batista2013jcap}. Thus, structure formation will help break the degeneracy between the \lcdm model and our scalar field model ($\vphi$CDM). This work aims to investigate the perturbation in such a dark energy model and look for the distinguishing features from the standard \lcdm model. The present work is not an attempt to constrain the model parameters with the observational datasets but rather to bring out the characteristic features of the model by solving the perturbation equations. It must be mentioned that the motivation of this work is not to unify inflation and dark energy and we will consider the evolution of the $\vphi$CDM long after the completion of inflation. 

%%%%%%%%%%%%%%%%%%%%%%%%%%%%%%%%%%%%%%%%%%%%%%%%%%%%%
\section{The Scalar Field Model} \label{sec5:scf}
We consider a homogeneous and isotropic Universe with spatially flat constant time hypersurface, described by the well-known FLRW metric in conformal time as,
\begin{equation}\label{eq5:metric}
ds^2= a^2(\tau)\paren*{- d \tau ^2+\delta_{ij}\,d x^i dx^j}.
\end{equation}
The Universe is filled with non-interacting fluids, namely photons ($\gamma$), neutrinos ($\nu$), baryons ($b$), cold dark matter ($c$) and a scalar field ($\vphi$) with a potential $\Vp$ acting as dark energy. 
The energy density and pressure of each component are respectively $\rho_{i}$ and $p_{i}$, where $i = \gamma, \nu, b, c, \vphi$. The equation of state (EoS) parameter is given as $w_{i} = \frac{p_{i}}{\rho_{i}}$. For the photons and neutrinos, $w_{\gamma}=w_{\nu}=1/3$\,, for baryons and CDM, $w_{b} = w_{c} = 0$\,. For the scalar field, $\rphi = \frac{1}{2 a^{2}}\vphi^{\prime \,2} +\Vp$ and $\pphi = \frac{1}{2 a^{2}}\vphi^{\prime \,2} -\Vp$ and the EoS parameter is given by 
\begin{eqnarray}
\wphi = \frac{\pphi}{\rphi}~=~ \frac{\frac{1}{2 a^{2}}\vphi^{\prime \,2} -\Vp}{\frac{1}{2 a^{2}}\vphi^{\prime \,2} +\Vp}~=~ 1-\frac{2 \,\Vp}{\rphi}~. \label{eq5:wphi}
\end{eqnarray}
The equation of equation of the scalar field is written as
\begin{equation} \label{eq5:kg1}
\vphi^{\prime \prime} + 2 \cH \vphi^{\prime} + a^{2} \frac{d V}{d \vphi}= 0~.
\end{equation}
Here, $\cH(\tau)= \frac{a^\prime}{a}$ is the conformal Hubble parameter and  prime $(^\prime)$ denotes the derivative with respect to the conformal time, $\tau$.

\begin{figure}[!h]
        \centering
\includegraphics[width=0.6\linewidth]{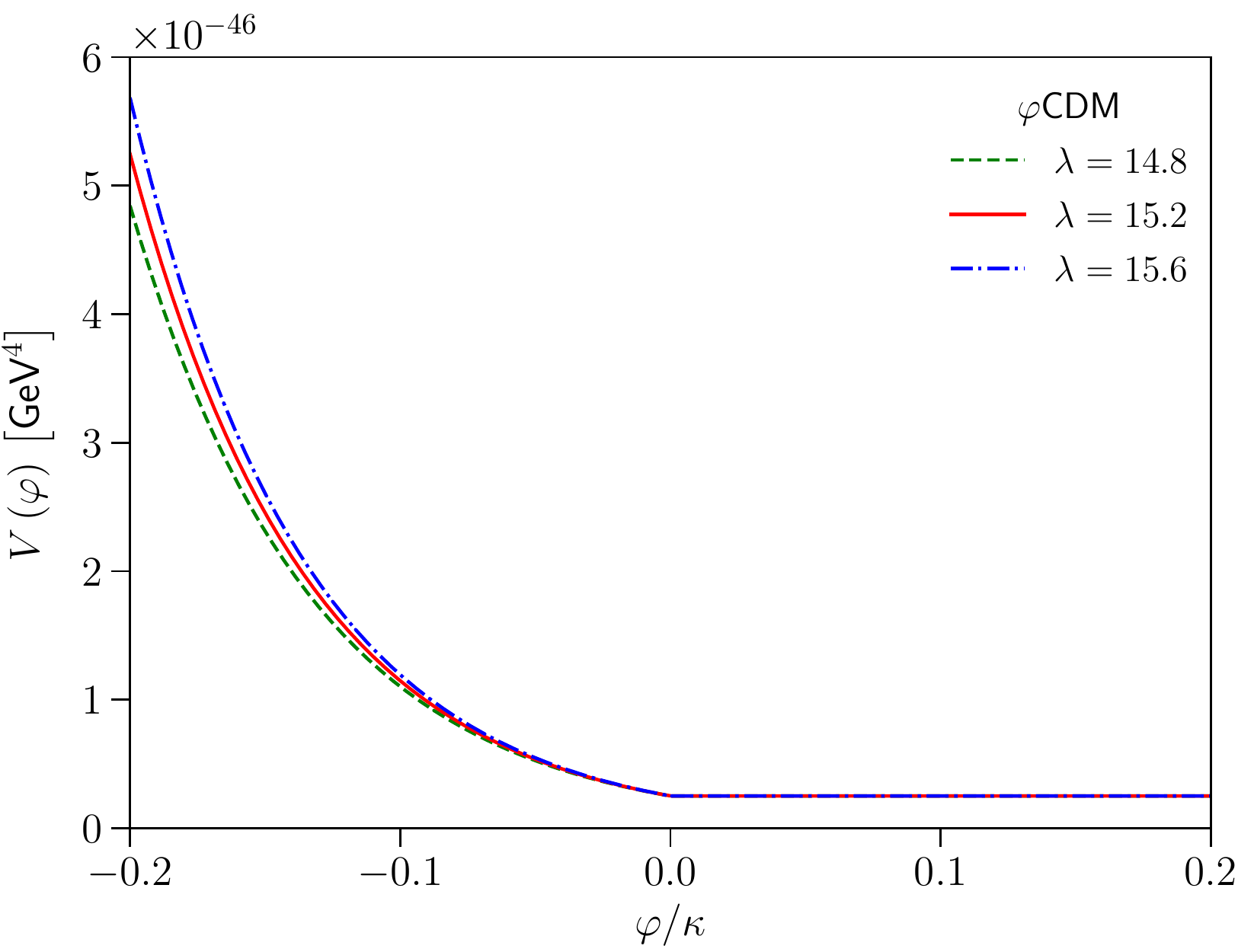}
\caption{Plot of the potential $\Vp$ in units of $\mbox{Gev}^{4}$ against $\vphi/\kappa$ with $V_{0} = 2.510 \times 10^{-47}~\mbox{Gev}^{4}$ and $\lambda = 14.8$ (dashed line), $\lambda = 15.2$ (solid line) and $\lambda = 15.6$ (dashed-dot line). Changing $V_{0}$ will change $\Omega_{\vphi\,0}$.}\label{im5:pot}
\end{figure}

We construct the potential such that the scalar field behaves as a quintessence field in the past and a cosmological constant at the present epoch. We consider the potential as the sum of an exponential potential and a constant potential, shown in Fig.\ (\ref{im5:pot}). The potential is written as,
\begin{equation}
\Vp = V_{0}\, e^{-\lambda \kappa \vphi} \Theta\paren*{-\vphi} + V_{0}\, \Theta\paren*{\vphi}~, \label{eq5:pot}
\end{equation}
where $V_{0}$ is a constant and $\Theta\paren*{\vphi}$ is the Heaviside theta defined as
\begin{equation}
\Theta\paren*{\vphi} = \begin{cases}
0 & \vphi < 0, \\
1 & \vphi \geq 0.
\end{cases}
\end{equation}
The potential given in Eq.\ (\ref{eq5:pot}) is continuous. In the exponential part, the scalar field tracks the evolution of the dominant background fluid with $\wphi = w_{D}$ and $\Ophi = 3\paren*{1+w_{D}}/\lambda^{2}$ with the condition $\lambda^{2} > 3\paren*{1+w_{D}}$, $w_{D}$ being the EoS parameter of the background fluid and $\Ophi$ is the energy density parameter defined as $\frac{\rho_{\vphi}}{3\,H^2 /\kappa}$. Here, $H$ is the Hubble parameter defined with respect to the cosmic time $t$. This attractor solution is called the ``scaling solution'', introduced by Ratra \& Peebles in~\cite{ratra1988prd} (see also~\cite{ferreira1997prl, copeland1998prd, ferreira1998prd}). The scalar field then leaves the scaling regime and enters the constant potential regime. The constant part of the potential arrests the fall of the scalar field and it starts to slow-roll and eventually dominate the energy density of the Universe as the cosmological constant. This drives an accelerated expansion at a late time with $\wphi = -1$. 

\begin{figure*}[!h]
\centering
\includegraphics[width=\textwidth]{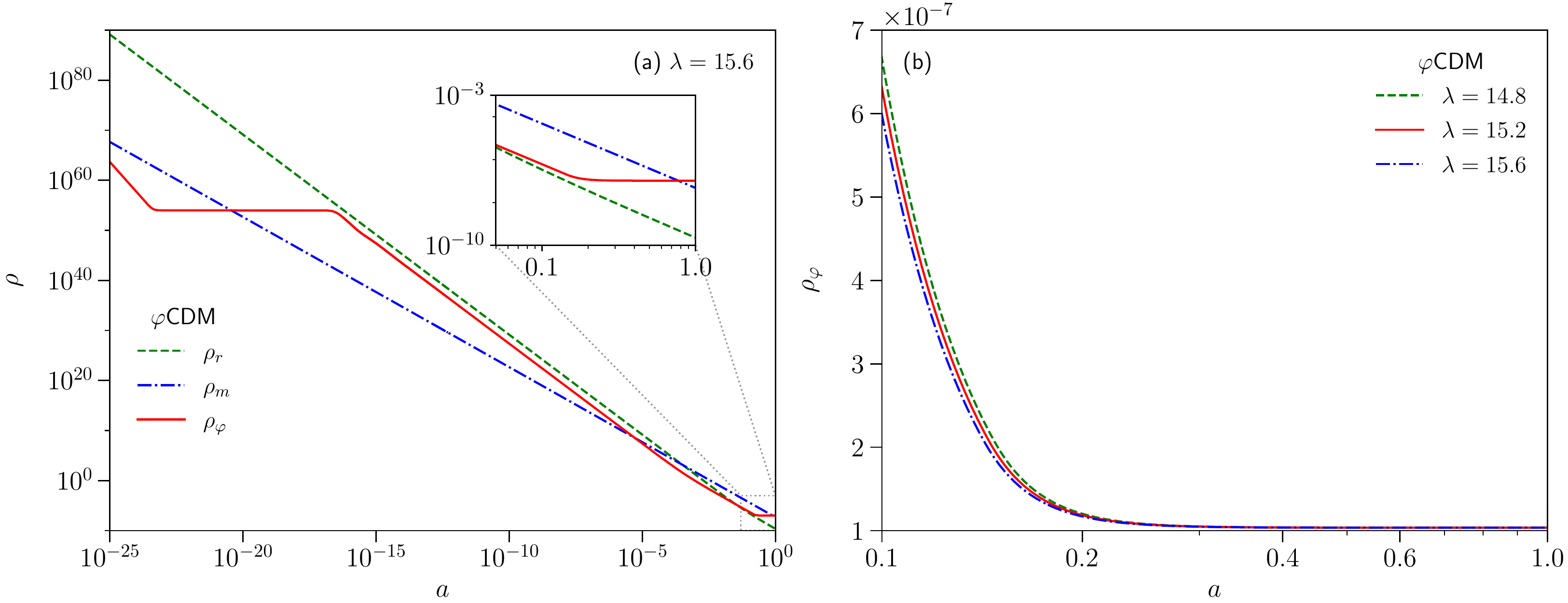}
\caption{(a) Plot of energy density $\rho$ against scale factor $a$ in logarithmic scale where the role of dark energy is played by a scalar field ($\vphi$) in presence of photons ($\gamma$), neutrinos ($\nu$), baryons ($b$), cold dark matter ($c$). For simplicity, only radiation $(r \equiv \gamma+\nu)$ and matter $(m \equiv b+c)$ are shown along with $\vphi$, labelling the model as $\vphi$CDM. The late time evolution of $\rho$ is enlarged in the inset. Only $\lambda =15.6$ is considered here. (b) Plot of $\rphi$ against scale factor $a$ shows that the evolution of $\rho_{\vphi}$ at late time is same for different values of the model parameter $\lambda$ for a fixed value of $V_{0}$.}\label{im5:bck1}
\end{figure*}

Fig.\ (\ref{im5:bck1}a) shows the variation of the energy density of radiation, $\rho_{r}$ $(r \equiv \gamma+\nu)$, matter, $\rho_{m}$ $(m \equiv b+c)$ and scalar field, $\rphi$, with the scale factor $a$ in logarithmic scale. Before reaching the tracking mode, the scalar field evolves through different regimes as shown in Fig.\ (\ref{im5:bck1}a). We integrate the Klein-Gordon equation (\ref{eq5:kg1}) numerically, using the potential (\ref{eq5:pot}), starting from $a=10^{-25}$ and consider that the initial $\rphi$ is smaller than $\rho_{r}$ and $\rho_{m}$ at that epoch. As $\vphi$ rolls down the potential, $E_{K} \gg E_{P}$ suppressing $\frac{d V}{d \vphi}$ relative to the first two terms in Eq.\ (\ref{eq5:kg1}). $E_{K}$ redshifts as $a^{-6}$ while $E_{P}$ remains constant and $\rphi $ is dominated by $E_{K}$. This kinetic-dominated regime is followed by the potential-dominated regime, where $\vphi$ rolls down very slowly making $\vphi^{\prime \prime}$ inconsequential. In this regime $\rphi$ is determined by $\Vp$ and becomes flat as $\vphi$ hardly evolves. When the dominant background, $\rho_{r}$ in this case, reaches the flat attractor value, they start evolving together. Thereafter, $\rphi$ tracks $\rho_{r}$ and subsequently $\rho_{m}$, depending on which dominates the background as discussed by Ratra \& Peebles~\cite{ratra1988prd}. For a single component background along with the scalar field, such as pure radiation and pure matter, similar results can be obtained analytically as well~\cite{ratra1988prd,steinhardt1999prd,brax2000prd2}. Later, when the constant potential $V_{0}$ takes over, the scalar field energy density, $\rphi$ behaves like the cosmological constant. It should be noted that this transition is instantaneous as it is implemented by a step function. Figure (\ref{im5:bck1}b) confirms that the cosmological constant like behaviour is ensured for any value of the parameter $\lambda$ for a given value of $V_{0}$.
 
\begin{figure*}[!h]
        \centering
            \subfloat{\includegraphics[width=.5\linewidth]{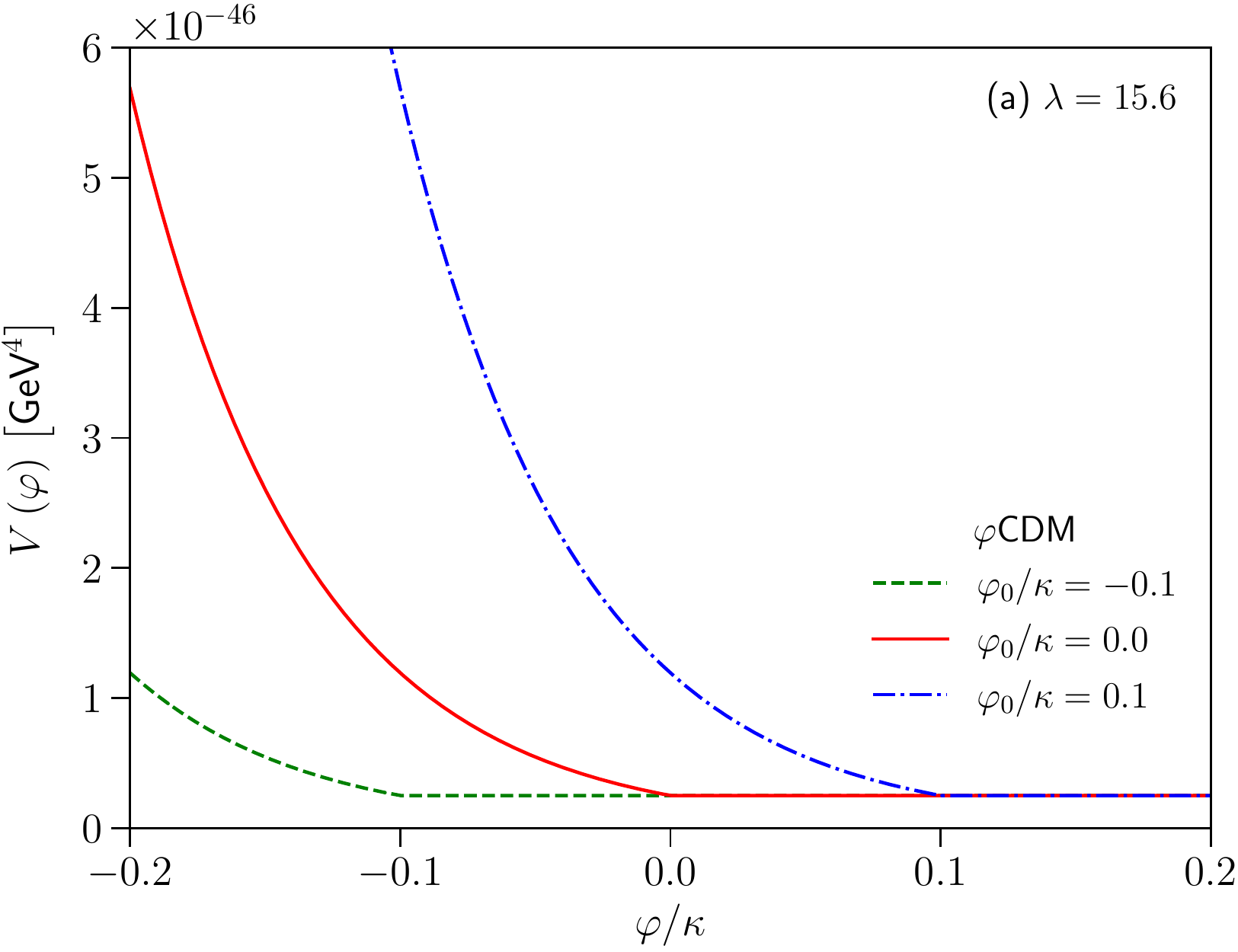}}\hfill
            \subfloat{\includegraphics[width=.5\linewidth]{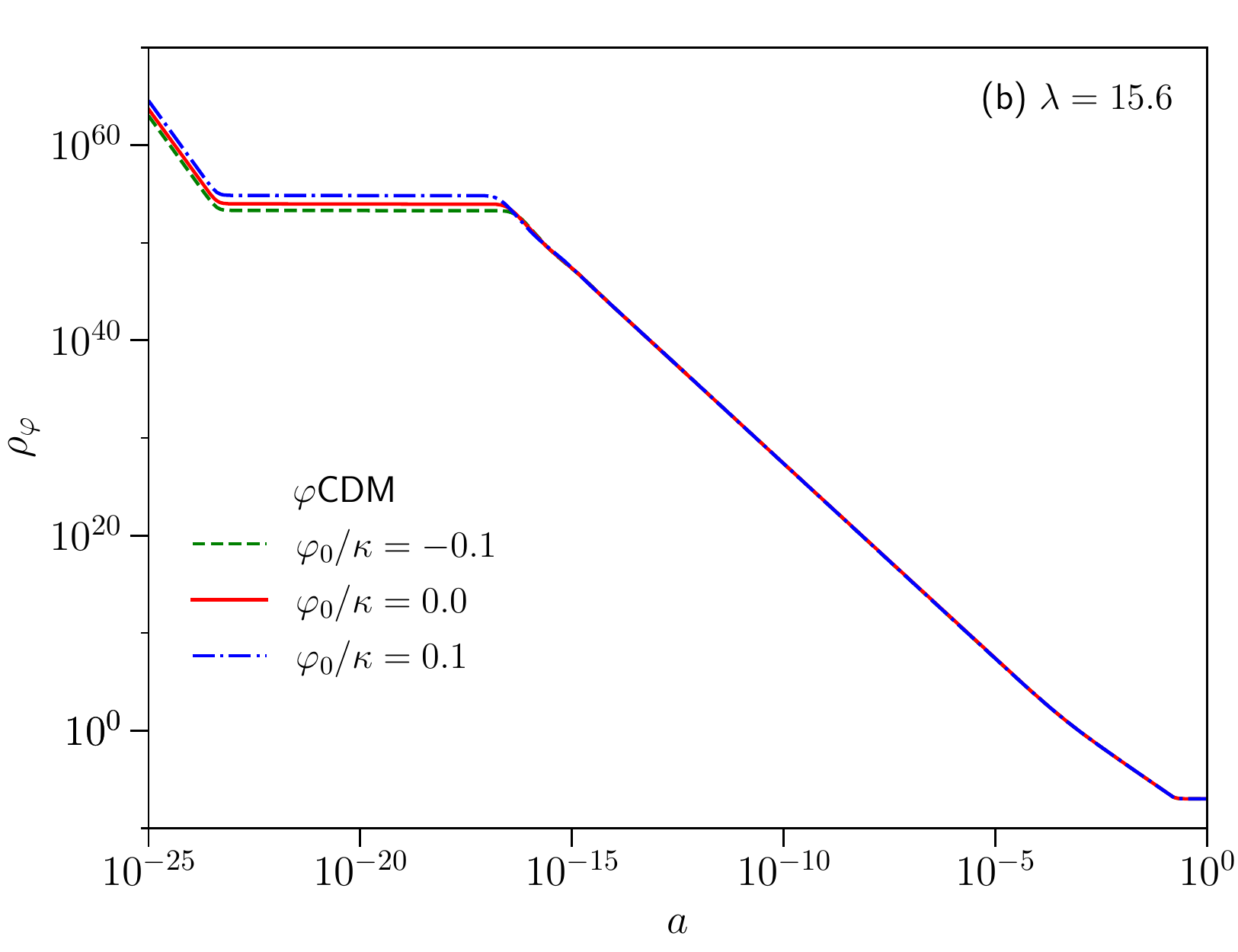}}\hfill
        \caption{(a) Plot of the potential $\Vp$ in units of $\mbox{Gev}^{4}$ against $\vphi/\kappa$ with $V_{0} = 2.510 \times 10^{-47}~\mbox{Gev}^{4}$, $\lambda = 15.6$ and transition at $\vphi_{0}/\kappa = -0.1$ (dashed line), $\vphi_{0}/\kappa = 0.0$ (solid line) and $\vphi_{0}/\kappa = 0.1$ (dashed-dot line). Changing $V_{0}$ will change $\Omega_{\vphi\,0}$. (b) Plot of $\rphi$ against scale factor $a$ shows that the evolution of $\rho_{\vphi}$ is different only at early times for different values of transition point $\vphi_{0}$ for fixed values of $V_{0}$ and $\lambda$.}\label{im5:bck4}
\end{figure*}

The advantages of this potential (\ref{eq5:pot}) are that at late time $\wphi =-1$, irrespective of the model parameters, $\lambda$ and $V_{0}$, or initial conditions and that the fraction of dark energy density present today, $\Omega_{\vphi0}$ depends on the height of the slow-roll region, $V_{0}$. It deserves mention that $V_{0}$ is not a free parameter but is fixed by the other model parameters like the $\Omega_b h^2$, $\Omega_c h^2$, $H_{0}$, such that $\Omega_{\vphi0}$ matches the observed value of $\Omega_{\Lambda}$ ($\sim 0.6847$)~\cite{planck2018cp}. The $\Theta$ function switches off the effect of the exponential potential in the constant potential part so that the scalar field is dominated completely by the $E_{P}$ after leaving the scaling region. It deserves mention that the value of $\vphi$ for the transition from exponential to constant potential is a free parameter. The effect in the evolution of the scalar field due to the change in the transition value can be seen from Fig.\ (\ref{im5:bck4}), where we consider three examples. When the transition point is shifted from zero to $\vphi_{0}$, the exponential part of the potential, Eq.\ (\ref{eq5:pot}) changes as $V_{0}\,e^{-\lambda \kappa \paren*{\vphi-\vphi_{0}}}$ to accommodate for the continuity of $\Vp$. The change in the steepness of the exponential potential changes the evolution of the scalar field before it behaves as a tracking field. Once it starts to track the dominant background components, its evolution remains unaffected by the change in the transition point, $\vphi_{0}$. So without loss of much of generality, we define our potential with $\vphi_{0}=0$.

The constraint on the parameter $\lambda$ comes from Big Bang Nucleosynthesis (BBN) condition~\cite{wetterich1988npb,copeland1998prd,ferreira1998prd},
\begin{equation}
\Ophi\paren*{a\sim 10^{-10}} \lesssim 0.09.
\end{equation}
It should be noted that in all the subsequent discussion, the scale factor, $a$, is scaled so that its present value, $a_{0}=1$. Considering $V_{0} = 2.510 \times 10^{-47}~\mbox{Gev}^{4}$ and $\lambda = 15.6$ with the parameter values listed in Table \ref{tab5:bck} gives $\Ophi\paren*{a\sim 10^{-10}} = 0.01642$ and $\Ophi\paren*{a = 1} = 0.6840$. The parameter values listed in Table \ref{tab5:bck}, are taken from the latest data release in 2018 of the \Planck collaboration~\cite{planck2018cp} (\Planck 2018, henceforth) and are based on the fiducial spatially flat \lcdm model. For our calculation we have considered $\vphi_{i} = -\frac{8.99}{\kappa}$ at $a_{i} = 10^{-25}$. It turns out that $\vphi = 0$ at $a = 0.14237$ (for the chosen values of $V_{0}$ and $\lambda$), where the potential changes its role from a scaling potential to effectively a cosmological constant. The dimensionless density parameter, $\Omega_i$ is given by $\frac{\rho_i}{3\,H^2 /\kappa}$ where the suffix $i$ stands for the $i$-th component. The dimensionless Hubble parameter at the present epoch is defined as $h = \frac{H_0}{100 \hskip1ex \footnotesize{\mbox{km s}^{-1} \mbox{Mpc}^{-1}}}$. Figure (\ref{im5:bck1}) is obtained by solving the Klein-Gordon Eq.\ (\ref{eq5:kg1}) numerically with the potential (\ref{eq5:pot}) using these parameter values. For the study of detailed dynamics of the scalar field during tracking region we refer to~\cite{ratra1988prd, ferreira1997prl, copeland1998prd, ferreira1998prd, steinhardt1999prd, brax2000prd2}.
\begin{table}[!h]
\begin{center}
\caption{\label{tab5:bck}
Values of background parameters from the \Planck 2018 collaboration.
}
\begin{adjustbox}{width=0.7\textwidth}
\begin{tabular}{cc}
\hline \hline
Parameter& \hspace{24ex} Value\\
\hline
\rule[-1ex]{0pt}{2.5ex}$\Omega_b h^2$ & \hspace{24ex} $0.0223828$ \\
\rule[-1ex]{0pt}{2.5ex}$\Omega_c h^2$ & \hspace{24ex} $0.1201075$ \\
\rule[-1ex]{0pt}{2.5ex}$H_0 \left[ \mbox{km s}^{-1} \mpci \right]$ & \hspace{24ex} $67.32117$ \\
\hline
\hline
\end{tabular}
\end{adjustbox}
\end{center}
\end{table}

\begin{figure*}[!h]
\centering
\includegraphics[width=\textwidth]{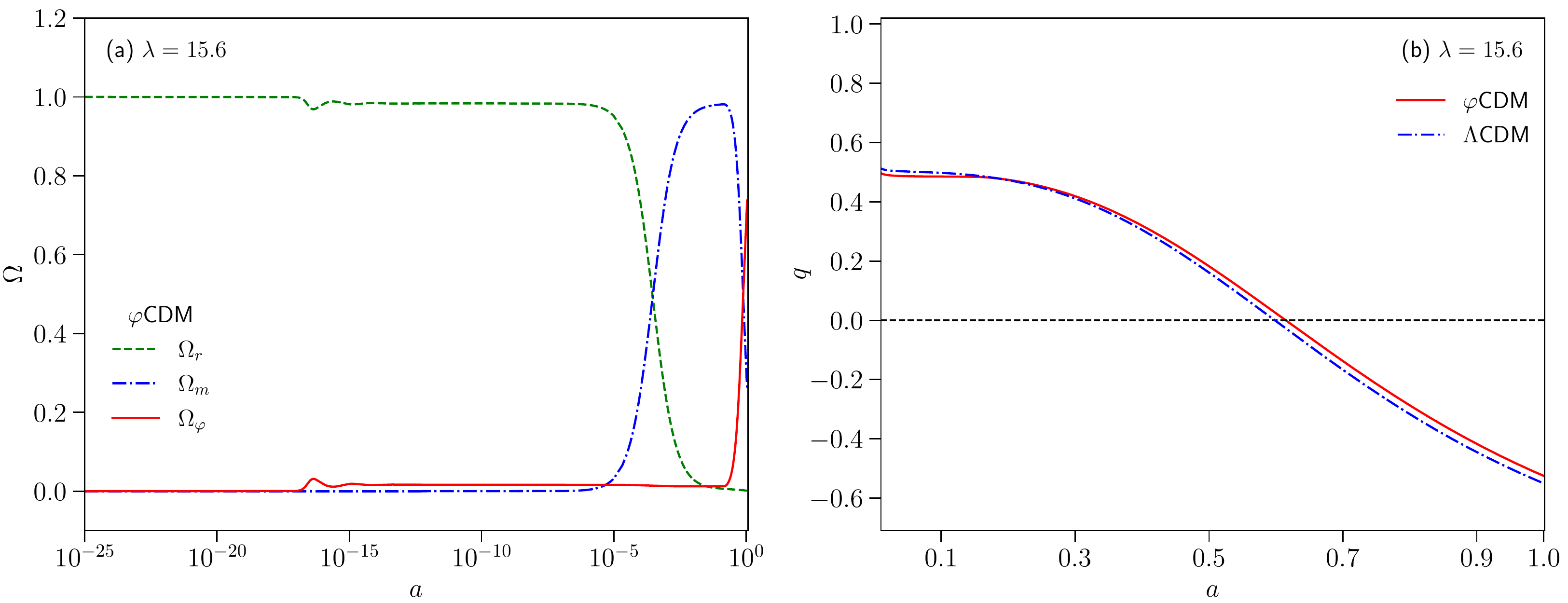}
\caption{(a) Plot of density parameter $\Omega$ against scale factor $a$ in logarithmic scale. For simplicity, only radiation $(r \equiv \gamma+\nu)$ and matter $(m \equiv b+c)$ are shown along with $\vphi$. (b) Plot of deceleration parameter $q$ against scale factor $a$ for $\vphi$CDM (solid line) and \lcdm (dashed-dot line). Only $\lambda =15.6$ is considered here.}\label{im5:om}
\end{figure*}

The evolution of the energy density parameters, $\Omega \paren*{\equiv \Omega_{i}}$ of radiation $(r \equiv \gamma+\nu)$, matter $(m \equiv b+c)$ and scalar field $(\vphi)$ with the scale factor $a$, in logarithmic scale, are shown in Fig.\ (\ref{im5:om}a) and that of the deceleration parameter $q = -\paren*{\frac{a\, a^{\prime \prime}}{a^{\prime \,2}}-1}$ with $a$ in Fig.\ (\ref{im5:om}b) for $\lambda=15.6$. Figure (\ref{im5:om}) shows that the evolution dynamics of the Universe is different from the \lcdm model even though $\wphi=-1$ at the present epoch. The two models are qualitatively very similar, but not really overlapping. For the scalar field model, henceforth called $\vphi$CDM, the accelerated expansion starts at a little higher value of $a$ compared to the \lcdm model.

%%%%%%%%%%%%%%%%%%%%%%%%%%%%%%%%%%%%%%%%%%%%%%%%%%%%%
\section{The Perturbations}\label{sec5:per}
The scalar field model given by Eq.\ (\ref{eq5:pot}) can have fluctuations and thereby affect the evolution of perturbations of other components. The scalar perturbation equations in synchronous gauge are considered in the present work and the differential equations are solved using the suitably modified version of the publicly available Boltzmann code \camb\footnote{Available at: \href{https://camb.info}{https://camb.info}}~\cite{lewis1999bs}. To study the dependence of the fluctuations on the model parameters, we used different values of $\lambda$ keeping $V_{0}$ constant (varying $V_{0}$ will change $\Omega_{\vphi\,0}$).
%%%%%%%%%%%%%%%%%%%%%%%%%%%%%%%%%%%%%%%%%%%%%%%%%%%%%

\subsection{Effect On Density Perturbation}\label{sec5:dp}

The scalar perturbation of the FLRW metric takes the form as given in Eq.\ (\ref{eq2:sc-metric2})  
\begin{equation} \label{eq5:metric2}
\begin{split}
ds^2=a^2\paren*{\tau} & \left\{ -\paren*{1+2\phi}d\tau^2+2\,\partial_iB \,d\tau \,dx^i +\left[\paren*{1-2\psi}\delta_{ij}+2\partial_i\partial_jE\right]dx^idx^j \right\},
\end{split}
\end{equation}
where $\phi, \psi, B, E$ are gauge-dependent functions of both space and time. In synchronous gauge, $\phi=B=0$, $\psi=\eta$ and $k^2 E=-\msh/2-3\eta$, where $\eta$ and $\msh$ are the synchronous gauge fields defined in the Fourier space and $k$ is the wavenumber~\cite{ma1995apj}.
The perturbation equations in the matter sector in the Fourier space are
\begin{eqnarray}
\delta_{i}^\prime+ k \,v_{i} +\frac{\msh^\prime}{2} &=& 0,\label{e2dm} \\ 
v_{i}^\prime+\cH v_{i}&=& 0, \label{m2dm}
\end{eqnarray}
where $\delta_{i} = \delta \rho_{i}/\rho_{i}$ is the density contrast and $v_{i}$ is the peculiar velocity of $i$-th $\paren*{i= b,c}$ fluid. Assuming there is no momentum transfer in CDM frame, $v_{c}$ is set to zero. For the details of this set of equations, we refer to the refernces~\cite{kodama1984ptps,mukhanov1992pr, ma1995apj, malik2003prd}. The perturbation $\delta \vphi$ in the scalar field has the equation of motion~\cite{cembranos2016jhep}
\begin{equation} \label{eq5:kg2}
\delta \vphi^{\prime \prime} + 2 \cH \delta \vphi^{\prime} + k^{2}\delta \vphi +a^{2}~\frac{d^{2} V}{d \vphi^{2}} \delta \vphi + \frac{1}{2} \vphi^{\prime} \msh^{\prime}= 0,
\end{equation}
in the Fourier space with wavenumber $k$. The perturbation in energy density $\delta \rphi$ and pressure $\delta \pphi$ are given as
\begin{eqnarray}
\delta \rphi &=& -\delta T_{0 \paren*{\vphi}}^{0}~=~ \frac{\vphi^{\prime} \delta \vphi^{\prime}}{a^{2}}+\delta \vphi \frac{d V}{d \vphi} , \label{eq5:pe2}\\
\delta T_{0\paren*{\vphi}}^{j} &=& - \frac{ \im k_{j}\, \vphi^{\prime}\, \delta \vphi}{a^{2}}, \hspace{0.6cm} \im \equiv \sqrt{-1} \label{eq5:pv2}\\
\delta \pphi \delta^{i}_{j}&=& \delta T_{j\paren*{\vphi}}^{i}~=~ \paren*{\frac{\vphi^{\prime} \delta \vphi^{\prime}}{a^{2}}-\delta \vphi \frac{d V}{d \vphi}}\delta^{i}_{j}\label{eq5:pp2},
\end{eqnarray}
when expanded in the Fourier space. Here $\delta T^{\mu}_{\nu\paren*{\vphi}}$ is the perturbed stress-energy tensor of the scalar field.

For an adiabatically expanding Universe, the square of sound speed is $\csf = \pphi^{\prime}/\rphi^{\prime}$. Using the Klein-Gordon Eq.\ (\ref{eq5:kg1}), the square of adiabatic sound speed~\cite{martin1998prd,brax2000prd2} for the scalar field reads as
\begin{equation}
\csf = -\frac{1}{3}-\frac{2 \vphi^{\prime \prime}}{3 \cH \vphi^{\prime}} = 1+\frac{2 a^{2}}{3 \cH \vphi^{\prime}} \frac{d V}{d \vphi}.\label{eq5:sound}
\end{equation}
In order to solve the perturbation Eq.\ (\ref{eq5:kg2}), the second derivative of the potential is written in terms of the square of sound speed, $\csf$ as
\begin{equation}
\frac{d^{2} V}{d \vphi^{2}} = \frac{3}{2} \frac{\cH^{2}}{a^{2}}\left[\frac{c_{s,\vphi}^{2 \,\,\prime}}{\cH}-\frac{1}{2}\paren*{\csf-1}\paren*{3 \csf+5}+\frac{\cH^{\prime}}{\cH}\paren*{\csf-1}\right].\label{eq5:v2}
\end{equation}
The square of sound speed, $\csf$ is constant in the different phases of evolution, e.g.\ in the scaling regime $\csf = \wphi = w_{D}$ and in the slow-roll regime $\csf = 1$. We shall henceforth take it to be described by Eq.\ (\ref{eq5:sound}) but neglect its derivative, $c_{s,\vphi}^{2 \,\,\prime}$~\cite{brax2000prd2} in Eq.\ (\ref{eq5:v2}). The perturbation Eqs.\ (\ref{e2dm}) and (\ref{m2dm}) are solved along with Eqs.\ (\ref{eq5:kg2}), (\ref{eq5:pe2}) and (\ref{eq5:pv2}) with adiabatic initial conditions and $k = \left[1.0, 0.1, 0.01 \right] h$ $\mpci$ using \camb.

%%%%%%%%%%%%%%%%%%%%%%%%%%%%%%%%%%%%%%%%%%%%%%%%%%%%%
\begin{figure*} [!h]
 \centering
\includegraphics[width=0.95\textwidth]{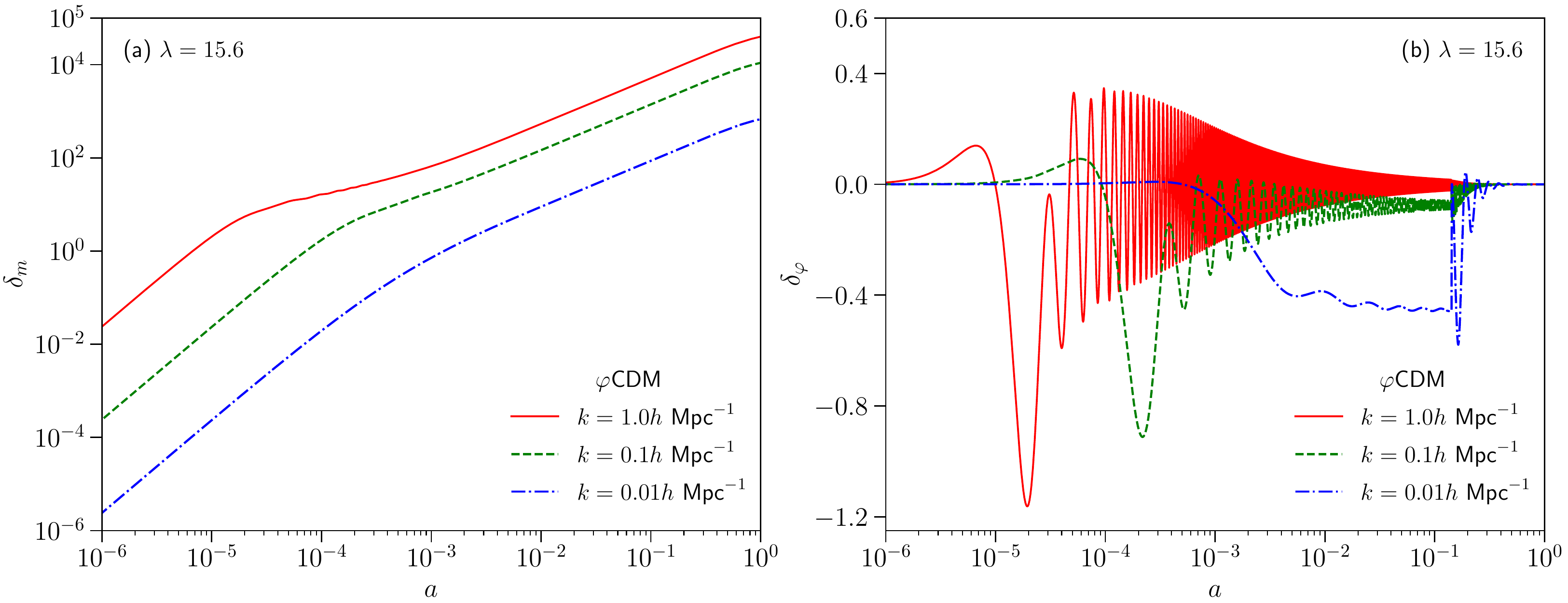}
\caption{(a) Plot of the matter density contrast $\delta_{m}$ against $a$. Both the axes are in logarithmic scale. (b) Plot of scalar field density contrast $\delta_{\vphi}$ against $a$. In (b), only $a$ is in logarithmic scale. The solid line represents $k = 1.0\,h$ $\mpci$, dashed line represents $k = 0.1\,h$ $\mpci$ and dashed-dot line represents $k = 0.01\,h$ $\mpci$ with $\lambda = 15.6$.}\label{im5:delta2}
\end{figure*}

\begin{figure}[!h]
        \centering
\includegraphics[width=0.6\linewidth]{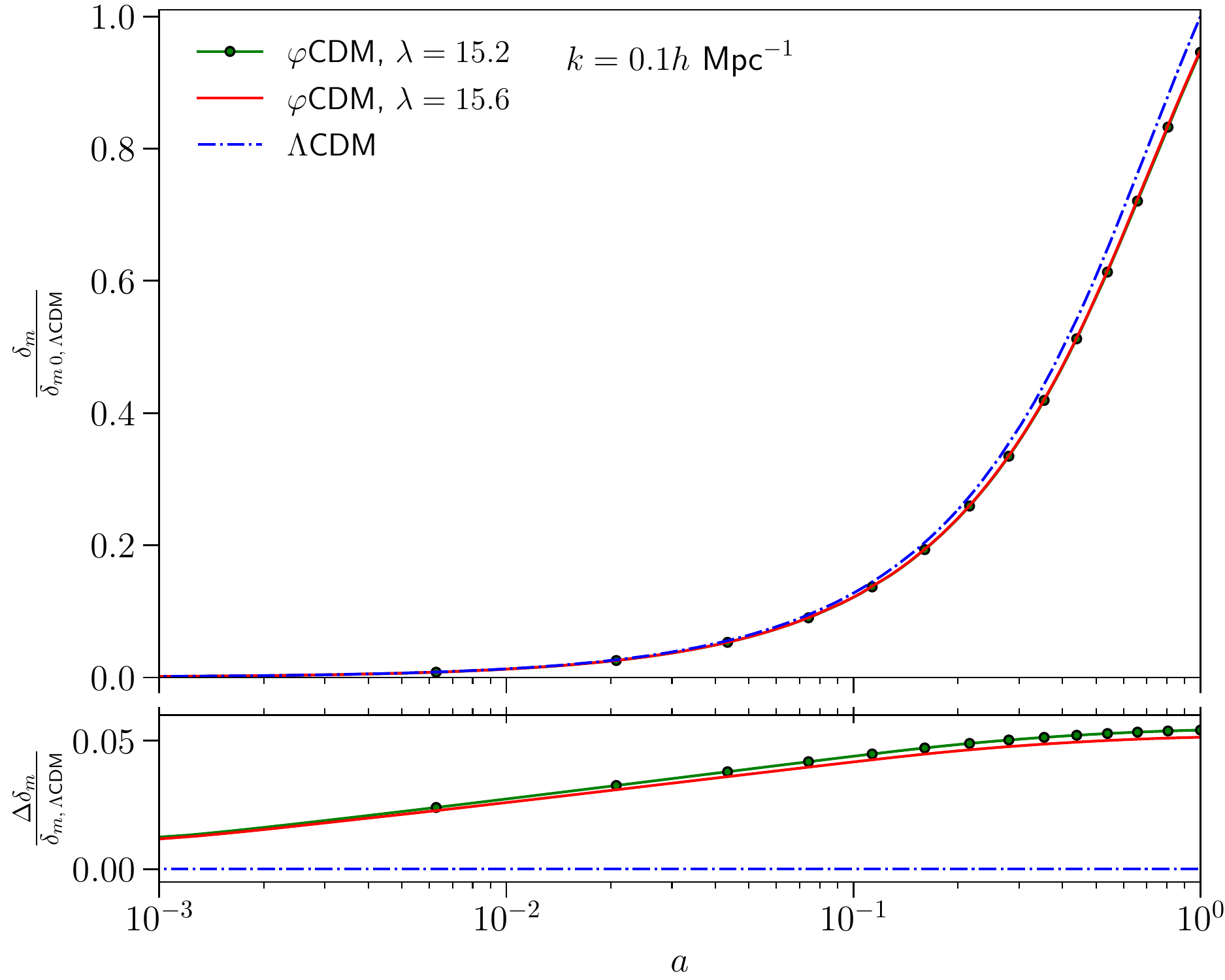}
\caption{\textbf{Upper Panel} : Plot of the matter density contrast $\frac{\delta_{m}}{\delta_{m\,0,\,\scriptsize{\Lambda\text{CDM}}}}$ against $a$ in logarithmic scale for $\vphi$CDM with $\lambda = 15.2$ (solid line with solid circles) and $\lambda = 15.6$ (solid line) and \lcdm (dashed-dot line) for $k = 0.1\,h$ $\mpci$. The difference in the growth of $\delta_{m}$ for $\vphi$CDM and \lcdm is prominent in the recent past, so the plot starts from $a =10^{-3}$. \textbf{Lower Panel} : Plot of the fractional growth rate relative to the \lcdm model. The fractional growth rate is defined as $\frac{\Delta \delta_{m}}{\delta_{m,\,\scriptsize{\Lambda\text{CDM}}}} = \paren*{1- \frac{\delta_{m,\,\scriptsize{\vphi\text{CDM}}}}{\delta_{m,\,\scriptsize{\Lambda\text{CDM}}}}}$.}\label{im5:delta6}
\end{figure}

Figure (\ref{im5:delta2}a) shows the variation of the density contrast, $\delta_{m} = \delta \rho_{m}/\rho_{m}$ for the cold dark matter ($c$) together with the baryonic matter ($b$) and Fig.\ (\ref{im5:delta2}b) shows the variation of the density contrast $\delta_{\vphi} = \delta\rho_{\vphi}/\rho_{\vphi}$ of the scalar field against $a$ in logarithmic scale for $k = \left[1.0, 0.1, 0.01 \right] h$ $\mpci$. In the matter dominated era, the modes of $\delta_{m}$ grow in a very similar fashion. The modes of $\delta_{\vphi}$ oscillate rapidly with decreasing amplitude after entering the horizon. Figure (\ref{im5:delta6}) shows the evolution of the matter density contrast $\delta_{m}$, for $\vphi$CDM and \lcdm. For a better comparison, $\delta_{m}$ for both the models have been scaled by $\delta_{m0} = \delta_{m}\paren*{a=1}$ of \lcdm. It can be seen that there is a difference in the growth of $\delta_{m}$ in the two models ($\vphi$CDM and \lcdm). To distinguish the effect of the parameter, $\lambda$, of the present potential with the \lcdm model, we have shown the fractional matter density contrast, $\frac{\Delta \delta_{m}}{\delta_{m,\,\scriptsize{\Lambda\text{CDM}}}} = \paren*{1- \frac{\delta_{m,\,\scriptsize{\vphi\text{CDM}}}}{\delta_{m,\,\scriptsize{\Lambda\text{CDM}}}}}$ in the lower panel of Fig.\ (\ref{im5:delta6}). It is clearly seen that, $\delta_{m}$ for $\lambda=15.2$ takes a slightly smaller value compared to that of $\delta_{m}$ for $\lambda=15.6$. Thus, the growth of matter density fluctuation decreases with decrease in the parameter, $\lambda$.

%%%%%%%%%%%%%%%%%%%%%%%%%%%%%%%%%%%%%%%%%%%%%%%%%%%%%
\subsection{Effect On CMB Temperature, Matter Power Spectra And $\fsg$}\label{sec5:ps}
\begin{figure}[!h]
 \centering
\includegraphics[width=\linewidth]{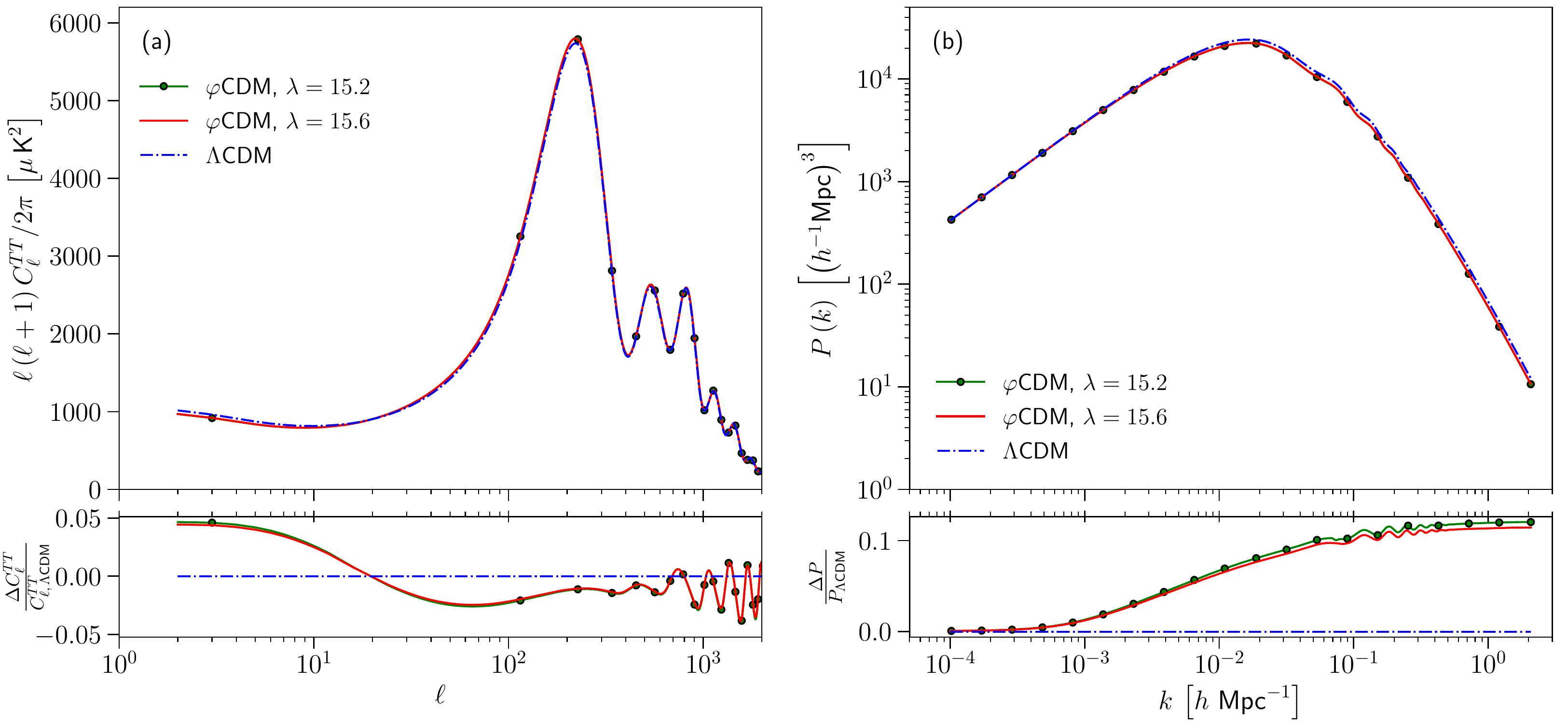}
\caption{\textbf{Upper Panel} : (a) Plot of CMB temperature power spectrum in units of $\mu \mbox{K}^{2}$ with the multipole index $\ell$ in logarithmic scale. (b) Plot of matter power spectrum $P\paren*{k}$ in units of $\paren*{h^{-1}\mpc}^{3}$ with wavenumber $k$ in units of $h\,\mpci$. Both the axes are in logarithmic scales in (b). \textbf{Lower Panel} : Plot of fractional change in the temperature spectrum, $\frac{\Delta C_{\ell}^{TT}}{C_{\ell,\, \scriptsize{\Lambda\text{CDM}}}^{TT}} = \paren*{1- \frac{C_{\ell,\, \scriptsize{\vphi\text{CDM}}}^{TT}}{C_{\ell,\, \scriptsize{\Lambda\text{CDM}}}^{TT}}}$ and the fractional change in matter power spectrum, $\frac{\Delta P}{P_{\scriptsize{\Lambda\text{CDM}}}} = \paren*{1- \frac{P_{\scriptsize{\vphi\text{CDM}}}}{P_{\scriptsize{\Lambda\text{CDM}}}}}$. For both panels, the solid line with solid circles represents $\vphi$CDM with $\lambda = 15.2$ and solid line represents $\vphi$CDM with $\lambda = 15.6$ while the dashed-dot line is for \lcdm at $a=1$.}\label{im5:mt}
\end{figure}

For more insight into the effect of the scalar field $\vphi$ on different physical quantities, we look at the CMB temperature spectrum, matter power spectrum and $\fsg$. The CMB temperature power spectrum is given as
\begin{equation}
C_{\ell}^{TT} = \frac{2}{k} \int k^{2} d k \,P_{\zeta}\paren*{k} \Delta^{2}_{T\ell}\paren*{k},
\end{equation}
where $P_{\zeta}\paren*{k}$ is the primordial power spectrum, $\Delta_{T\ell}\paren*{k}$ is the temperature transfer function, $\ell$ is the multipole index and $T$ stands for temperature. For the detail calculation of the CMB spectrum we refer to~\cite{hu1995apj,seljak1996apj}. The matter power spectrum is given as
\begin{equation} \label{power}
P\left(k,a\right)= A_s \,k^{n_s} T^2\left(k\right) D^2\left(a\right),
\end{equation}
where $A_{s}$ is the normalising constant, $n_{s}$ is the spectral index, $T\left(k\right)$ is the matter transfer function and $D\left(a\right)=\frac{\delta_m\left(a\right)}{\delta_m\left(a=1\right)}$ is the normalised density contrast. For the detailed method of calculation we refer to the monograph by Dodelson~\cite{dodelson2003}. $C_{\ell}^{TT}$ and $P\paren*{k,a}$ are computed numerically using \camb. The values $A_{s} = 2.100549 \times 10^{-9}$ and $n_{s} = 0.9660499$ are taken from the \Planck 2018 data~\cite{planck2018cp}, and hence, depend on the fiducial \lcdm model. Figure (\ref{im5:mt}a) shows that the CMB temperature power spectra, $C_{\ell}^{TT}$, are almost independent of the values of the model parameter, $\lambda$. For clarity of the plots only two values of $\lambda$ are given. The presence of the scalar field $\vphi$ decreases the matter content of the Universe slightly during matter domination making the amplitude of first two peaks of the CMB spectra marginally higher than that in the \lcdm model. The scalar field also lowers the low-$\ell$ CMB spectrum through the integrated Sachs-Wolfe (ISW) effect. These features are clear from the lower panel of Fig.\ (\ref{im5:mt}a), which shows the fractional change ($=\Delta C_{\ell}^{TT}/C_{\ell,\, \scriptsize{\Lambda\text{CDM}}}^{TT}$) in $C_{\ell}^{TT}$ of the $\vphi$CDM models relative to the \lcdm model; a smaller $\lambda$ produces slightly lower low-$\ell$ modes. A lesser amount of matter leads to a marginally lower matter power spectrum at small scales (Fig.\ (\ref{im5:mt}b)), which is clear from the positive fractional change in matter power spectrum, $\Delta P/P_{\scriptsize{\Lambda\text{CDM}}}$, relative to the \lcdm model (lower panel). Both these figures are for the present epoch.
 
To differentiate the $\vphi$CDM and \lcdm decisively, we have studied the linear growth rate, 
\begin{equation} \label{eq5:growth_rate}
f\paren*{a}= \frac{d \ln \delta_{m}}{d \ln a}~=~ \frac{a}{\delta_{m}\paren*{a}}\frac{d \delta_{m}}{d \,a}~.
\end{equation}
Observationally the growth rate is measured using the perturbation of the galaxy density $\delta_{g}$, which is related to the matter density perturbations $\delta_{m}$ as $\delta_{g} = b \delta_{m}$, where $b \in \left[1,3\right]$ is the bias parameter. The estimate of the growth rate $f$ is sensitive to the bias parameter, and thus not very reliable. A more dependable observational quantity is the product $f\paren*{a}\se\paren*{a}$~\cite{percival2009mnras}, where $\se\paren*{a}$ 
is the root-mean-square (rms) fluctuations of the linear density field within the sphere of radius $R = 8 h^{-1}\, \mpc$. The rms mass fluctuation can be written as $\se\paren*{a} = \se\paren*{1}\frac{\delta_{m}\paren*{a}}{\delta_{m}\paren*{1}}$, where $\se\paren*{1}$ is the value at $a=1$ (Table \ref{tab5:s8}), calculated by integrating the matter power spectrum over all the values of the wavenumber $k$ using \camb. Thus, the combination becomes
\begin{equation} \label{eq5:fsigma}
\fsg\paren*{a} \equiv f\paren*{a}\se\paren*{a} = \se\paren*{1}\frac{a}{\delta_{m}\paren*{1}}\frac{d \delta_{m}}{d \,a}~.
\end{equation}

Since $\fsg$ measurements provide a tighter constraint on the cosmological parameters, it will give a better insight into the growth of the density perturbations. We have studied the variation of $f$ and $\fsg$ with redshift $z$ for three different values of $\lambda$. Redshift $z$ is related to the scale factor $a$ as $z = \paren*{\frac{a_0}{a} -1}$, $a_0$ being the present value. The linear growth rate $f$ and $\fsg$ are independent of the wavenumber $k$ for low redshift. As the $\fsg$ analysis is valid for $z \in \left[0,2\right]$, the redshift from $z=0$ to $z=2$ are considered here.
\begin{table}[h!]
\begin{center}
\caption{\label{tab5:s8}
Values of $\se$ at $a = 1$ for the $\vphi$CDM and \lcdm models.
}
\begin{adjustbox}{width=0.7\textwidth}
\begin{tabular}{ccc}
\hline
\hline
Model& \hspace{15ex}$\lambda$&\hspace{15ex} $\se$ \\
\hline
\rule[-1ex]{0pt}{2.5ex}&\hspace{15ex}$14.8$ &\hspace{15ex} $0.7638$ \\
\rule[-1ex]{0pt}{2.5ex}$\vphi$CDM&\hspace{15ex}$15.2$ &\hspace{15ex} $0.7664$ \\
\rule[-1ex]{0pt}{2.5ex}&\hspace{15ex}$15.6$ &\hspace{15ex} $0.7687$ \\
\rule[-1ex]{0pt}{2.5ex}\lcdm&\hspace{15ex} --- &\hspace{15ex} $0.8123$ \\
\hline
\hline
\end{tabular}
\end{adjustbox}
\end{center}
\end{table}

\begin{figure*}[!h]
 \centering
\includegraphics[width=\textwidth]{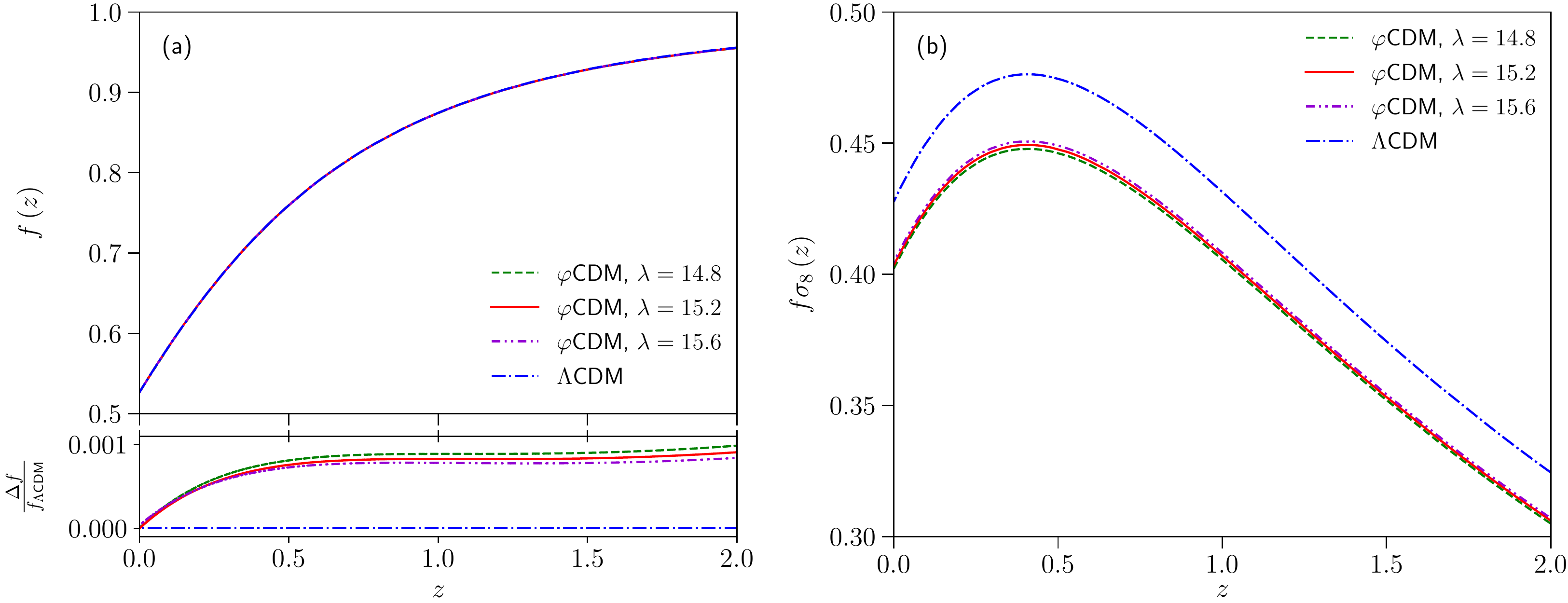}
\caption{(a) Plot of \textbf{Upper Panel} : linear growth rate $f$ and \textbf{Lower Panel} : fractional growth rate, $\frac{\Delta f}{f_{\scriptsize{\Lambda\text{CDM}}}}  = \paren*{1-\frac{f_{\scriptsize{\vphi\text{CDM}}}}{f_{\scriptsize{\Lambda\text{CDM}}}}}$ relative to the \lcdm model. (b) Plot of $\fsg$ against redshift $z$. For all the plots, the dashed line represents $\vphi$CDM with $\lambda = 14.8$, solid line represents $\lambda = 15.2$ and dashed-dot-dot represents $\lambda = 15.6$ while the dashed-dot line is for \lcdm.}\label{im5:fs}
\end{figure*}

The linear growth rate $f$ is almost same for all the models at low redshift (Fig.\ (\ref{im5:fs}a)). The little change in the growth rate, $f$ due to change in $\lambda$ is visible in the fractional change in growth rate, $\Delta f/f_{\scriptsize{\Lambda\text{CDM}}}$, relative to the \lcdm model (lower panel of Fig.\ (\ref{im5:fs}a)). The difference in matter power spectrum is manifested in its amplitude $\se$ as given in Table (\ref{tab5:s8}) and in $\fsg$ in Fig.\ (\ref{im5:fs}b). It is interesting to note that there is a substantial difference in $\fsg$ for $\vphi$CDM and \lcdm which is not there in the CMB temperature and matter power spectra. Thus, a low $\fsg$ can be said to be the characteristic distinguishing feature of the present $\vphi$CDM model from the \lcdm model. It must be noted that $\lambda$ is chosen in such a way that is compatible with the age of the Universe which is around $13.797 \pm 0.023$ Giga years according to the recent \Planck 2018 data~\cite{planck2018cp}. 

%%%%%%%%%%%%%%%%%%%%%%%%%%%%%%%%%%%%%%%%%%%%%%%%%%%%%
\section{Summary And Discussion}\label{sec5:sum}
In the present work, we have introduced a scalar field model that will retain the virtues of the \lcdm model without its shortcomings. We have investigated the perturbation in such a dynamical dark energy model that will alleviate the initial condition problem associated with the cosmological constant and attain an EoS parameter $\wphi = -1$ at the present epoch. At early times the scalar field energy density tracks the dominant component of the background fluid and later on starts to roll sufficiently slowly to drive the accelerated expansion of the Universe. A scalar field with an exponential potential at early epoch and a constant potential at late time, connected by Heaviside $\Theta$ functions (Eq.\ \ref{eq5:pot}), appears to serve the purpose. That $\wphi = -1$ for the present epoch is independent of the choice of the model parameters and the present dark energy density parameter, $\Omega_{\vphi\,0}$ is dependent on the height of the constant potential, $V_{0}$.

We have worked out a detail perturbation analysis to differentiate the scalar field model ($\vphi$CDM) with the \lcdm model. The linearised scalar perturbations of the FLRW metric in synchronous gauge are studied using our modified \camb. The growth of matter density contrast, $\delta_{m}$ is similar to the \lcdm model and is smaller for the smaller value of $\lambda$. The linear growth rate $f$, which is the logarithmic derivative of $\delta_{m}$ with respect to $a$ is same for both the models. The presence of the scalar field slightly decreases the matter content of the Universe during the evolutionary history. This decrease in matter content is manifested in the matter power spectrum and even more clearly in the evolution of the $\fsg$. Thus, $\fsg$ helps in breaking the degeneracy between the present $\vphi$CDM model and the standard \lcdm. Another interesting result is that the decrease in the rate of clustering decreases the variance of the linear matter perturbation, $\se$. As seen from Table (\ref{tab5:s8}), the $\se$ obtained here is more towards the side of the value obtained from the galaxy cluster counts using thermal Sunyaev-Zel'dovich (tSZ) signature~\cite{planck2018cp,zubeldia2019mnras}, $\se = 0.77^{+0.04}_{-0.03} $ rather than the value obtained from \Planck spectrum~\cite{planck2018cp}, $\se = 0.811 \pm 0.006$. It must be mentioned here that as shown in~\cite{banerjee2020prdl,eoin2019plb}, quintessence-CDM model also prefer lower $H_{0}$ compared to the standard \lcdm model. A detailed study of the parameter space is required to confirm if this model can solve the $\se$ tension and prefer a lower value of $H_{0}$. Such an analysis is outside the scope of the present work and will be considered in a separate work. It has already been mentioned that the values of the background parameters ($\Omega_b h^2$, $\Omega_c h^2$, $H_{0}$, $A_{s}$ and $n_{s}$) are fixed to the mean values obtained by the \Planck 2018 collaboration~\cite{planck2018cp}, for the fiducial spatially flat \lcdm model.  We have used them as an illustrative example in the absence of any parameter values obtained by constraining the present model with the observational data sets.

It can be said quite conclusively that this scalar field model resolves the initial condition problem, produces late-time acceleration with $\wphi = -1$ as predicted by the recent data as well as decreases rms mass fluctuation $\se$. This model is also successful in the context of the structure formation in the Universe. The model looks to be promising, but it has to be tested against observational datasets, and compared with \lcdm and other competing models in connection with the evidence criteria.

% Chapter 1
\chapter{Differentiating Interaction In The Dark Sector With Perturbation}% Main chapter title
\label{chap6:prd}\blfootnote{The work presented in this chapter is based on ``Differentiating Dark Interactions With Perturbation'', \textbf{\authorname}, Phys.\ Rev.\ D \textbf{103}, 123547 (2021) }
\chaptermark{Differentiating Interaction With Perturbation}

\section{Introduction}\label{sec6:intro}
The fact that the dark matter and dark energy have energy densities of the same order of magnitude opens the possibility that there is an energy exchange between the two. Interactions between dark matter and dark energy in various dark energy models have been studied and tested against observations extensively~\cite{billyard2000prd, pavon2004jcap, amendola2004jcap, curbelo2006cqg, gonzalez2006cqg,guo2007prd,olivares2008prd, bohmer2008prd, quercellini2008prd, bean2008prd2, quartin2008jcap, he2008jcap, chimento2010prd, amendola2012prd, pettorino2012prd, salvatelli2014prl, yang2014prd1, wang2014aa, caprini2016jcap, nunes2016prd, mukherjee2017cqg, yang2017prd2, pan2018mnras, yang2018prd, yang2018jcap, visinelli2019prd, vagnozzi2020mnras}. For detailed reviews on interacting dark matter-dark energy models, we refer to~\cite{bamba2012ass, bolotin2015ijmpd, wang2016rpp}.

The presence of a coupling in the dark sector may not be ruled out \emph{a priori}~\cite{billyard2000prd, pavon2004jcap, amendola2004jcap, curbelo2006cqg, gonzalez2006cqg, guo2007prd, olivares2008prd, bohmer2008prd, quercellini2008prd, bean2008prd2, quartin2008jcap, he2008jcap, caldera2009jcap, chimento2010prd, amendola2012prd, pettorino2012prd, chimento2012prd2, chimento2013prd, salvatelli2014prl, yang2014prd1, wang2014aa, caprini2016jcap, nunes2016prd, mukherjee2017cqg, yang2017prd2, pan2018mnras, yang2018prd, yang2018mnras, bruck2017prd}. It naturally raises the question whether the interaction was there from the beginning of the Universe and exists through its evolution or is a recent phenomenon, or it was entirely an early phenomenon and not at all present today. A modification of the phenomenological interaction term by an evolving coupling parameter instead of its being a constant, may answer this question. A constant coupling parameter indicates the interaction is present throughout the evolution of the Universe~\cite{guo2007prd, yang2017prd1}. In this work, we have considered the coupling parameter to be evolving with the scale factor. Interaction with an evolving coupling parameter is not studied much in literature and warrants a detailed analysis. Rosenfeld~\cite{rosenfeld2007prd} and Yang \etal~\cite{yang2019du} have considered the dynamical coupling parameter, but the motivation as well as the analytical form of the parameter used in the present work are different.

There is no theoretically preferred form of the phenomenological interaction term. In this work, two possible scenarios are considered --- (a) the presence of interaction is significant during the late time but not at early time and (b) the presence of interaction is significant in the early times but not at late time. The rate of energy transfer is considered to be proportional to the dark energy density. The dynamical coupling parameter will affect the evolution of the dark matter and hence have its imprints in the growth of perturbations. Thus the presence of dynamical interaction can give rise to new features in structure formation. The motivation of the present work is to investigate the effects of interaction on clustering of matter perturbation, understand the evolution of the interaction using perturbation and test the models against observational datasets. 

We tested the interacting models with different observational datasets like the cosmic microwave background (CMB)~\cite{planck2018cp}, baryon acoustic oscillation (BAO)~\cite{beutler2011mnras, ross2015mnras, alam2017sdss3}, Type Ia Supernovae (SNe Ia)~\cite{scolnic2018apj} data and their different combinations. For a complete understanding of the effect of interaction on structure formation, it is necessary to consider the effect of the large scale structure (LSS) information on the cosmological constraints. In the present work, we have considered the redshift-space distortions (RSD) data~\cite{kaiser1987mnras} as the LSS data. Combining the RSD data with CMB, BAO and Supernovae data is expected to break the degeneracy between the different interacting models with similar background evolution as well as provide a tight constraint on the interaction parameter.

The LSS data, which includes Planck Sunyaev-Zel'dovich survey~\cite{planck2013szc}, Canada France Hawaii Telescope Lensing Survey (CFHTLens)~\cite{kilbinger2013mnras,heymans2013mnras}, South Pole Telescope (SPT)~\cite{schaffer2011apj,vanengelen2012apj}, RSD survey, are in disagreement with CMB observations for the root-mean-square mass fluctuation in sphere with radius $8 h^{-1}\, \mpc$, (called $\se$) and hence for the matter density parameter $\Omega_{m}$ and the Hubble parameter $H_{0}$~\cite{pourtsidou2016prd, vandebruck2018prd, mohanty2018jaa, an2018jcap, martinelli2019mnras, lambiase2019epjc, eoin2019plb, banerjee2020prdl}. The LSS observations prefer lower values of $\se$ and $\Omega_{m}$ and a higher value of $H_{0}$ compared to the CMB results. Many attempts have been made to settle the disagreement between the two datasets~\cite{gomez2017epl, sakr2018aa, kazantzidis2018prd, gomez2018mnras, ooba2019ass, park2020prd}. Some more of the notable work with RSD data are~\cite{wang2014prd, yang2014prd1, yang2014prd2, costa2017jcap, nesseris2017prd, akhlaghi2018mnras, sagredo2018prd, skara2020prd, borges2020prd}.

It must be mentioned here that the model with constant coupling parameter has been tested rigorously against different observational datasets and priors ranges~\cite{martinelli2019mnras, divalentino2020pdu, vagnozzi2020prd} to name a few. In this work, we used different datasets and different prior ranges and an ``evolving'' coupling parameter in the interaction term. Moreover, we considered an evolving dark energy with EoS given by the Chevallier-Polarski-Linder (CPL) parametrisation. However, the present work is not an attempt to alleviate the $\se$ or $H_{0}$ tensions but to understand the evolution of the interaction using perturbation and test the models against observational datasets.
%
%%%%%%%%%%%%%%%%%%%%%%%%%%%%%%%%%%%%%%%%%%%%%%%%%%%%%%

%%%%%%%%%%%%%%%%%%%%%%%%%%%%%%%%%%%%%%%%%%%%%%%%%%%%%%
\section{Interacting Dark Matter-Dark Energy Fluid} \label{sec6:bckgrnd}
The Universe is considered to be described by a spatially flat, homogeneous and isotropic FLRW metric,
\begin{equation}\label{eq6:metric}
ds^2= a^2(\tau)\paren*{- d \tau ^2+\delta_{ij} d x^i dx^j}.
\end{equation}
The Universe is filled with five components of matter, all formally represented as perfect fluids --- photons ($\gamma$), neutrinos ($\nu$), baryons ($b$), cold dark matter ($c$) and dark energy ($de$). We assume that there is an energy transfer only in the dark sector of the Universe such that the conservation equations are
\begin{eqnarray}
\rho^\prime_c+ 3 \cH \rho_c &=& -aQ\,,\label{eq6:con1}\\
\rho^\prime_{de}+ 3 \cH \paren*{1+\wde} \rde&=& aQ. \label{eq6:con2}
\end{eqnarray}
A prime indicates differentiation with respect to the conformal time $\tau$.  The pressure, $p_{c} = 0$ for cold dark matter. The other three fluids --- photons ($\gamma$), neutrinos ($\nu$) and baryons ($b$) conserve independently and hence, have no energy transfer among them. Their conservation equations are written as
\begin{equation}
\rho^\prime_{A}+ 3 \cH \paren*{1+w_{A}} \rA = 0\, \label{eq6:con3},
\end{equation}
where $w_{A} = \pA/\rA$ is the equation of state parameter (EoS) of the $A$-th fluid and $A = \gamma, \nu, b$. For photons and neutrinos, the EoS parameter is $w_{\gamma} = w_{\nu} = 1/3$, for baryons and cold dark matter, the EoS parameter is $w_{b} = w_{c} = 0$ and for dark energy, the EoS parameter is $\wde = \pde/\rde$. 
%%%%%%%%%%%%%%%%%%%%%%%%%%%%%%%%%%%%%%%%%%%%%%%%%%%%%
%%%%%%%%%%%%%%%%%%%%%%%%%%%%%%%%%%%%%%%%%%%%%%%%%%%%%
%
In Eqs.\ (\ref{eq6:con1}) and (\ref{eq6:con2}), $Q$ gives the rate of energy transfer between the two fluids. If $Q<0$, energy is transferred from dark energy to dark matter (DE $\rightarrow$ DM) and if $Q>0$, energy is transferred from dark matter to dark energy (DM $\rightarrow$ DE). When $Q>0$, dark matter redshifts faster than $a^{-3}$ and when $Q<0$, dark matter redshifts slower than $a^{-3}$. The dark energy evolution depends on the difference $\wde-\frac{a Q}{3 \cH \rde}$. Thus, the interaction manifests itself by changing the scale factor dependence of the dark matter as well as dark energy. 
There are different forms of the choice of the phenomenological interaction term $Q$, the models with $Q$ proportional to either $\rdc$ or $\rde$ or any combination of them are among the more popular choices,~\cite{bohmer2008prd,clemson2012prd,acosta2014prd,yang2014prd1,yang2018prd} to mention a few. It must be mentioned here that there is no particular theoretical compulsion for any of these choices. We have taken the covariant form of the source term such that it is proportional to the dark energy density ($Q^{\mu} \propto \rde$) and is written in terms of DM 4-velocity $u^{\mu}_{c}$ as
\begin{equation} \label{eq6:inter}
Q^{\mu} =   \frac{\cH \rde \, u^{\mu}_{c} \, \bela}{a}\,.
\end{equation}
Here, $\bela$ is the coupling parameter evolving with the scale factor, $a$.  The coupling parameter determines the strength of interaction and direction of energy flow; $\beta = 0 $ indicates that there is no coupling in the dark sector. In this work, we considered two possible scenarios,
\begin{description}
\item[Model L\,] If the coupling was not significant in the early Universe ($a=0$) and is felt only at the recent epoch. 
\item[Model E\,] If the interaction is predominantly an early phenomenon and is insignificant now ($a=1$). 
\end{description}
We compared the models with the Universe with a constant interaction parameter ({\bf Model C}). The ansatz chosen for the models are simple analytic functions of $a$ which are well-behaved in the region $a \in \left[0,1\right]$.
\begin{subequations}
\begin{eqnarray}
\mbox{\hypertarget{MI}{\bf Model L}\,:} \hspace{1cm} \bela &=& \beta_{0}\paren*{\frac{2\,a}{1+a}}, \label{eq6:b1}\\
\mbox{\hypertarget{MII}{\bf Model E}\,:} \hspace{1cm} \bela &=& \beta_{0}\paren*{\frac{1-a}{1+a}},\label{eq6:b2}\\
\mbox{\hypertarget{MIII}{\bf Model C}\,:} \hspace{1cm} \bela &=& \beta_{0}. \label{eq6:b3}
\end{eqnarray}
\end{subequations}
The terms in parenthesis in the Eqs.\ (\ref{eq6:b1}) and (\ref{eq6:b2}) are positive definite for the domain of $a$ under consideration and hence the direction of energy flow is determined by the signature of the constant $\beta_{0}$.

It is considered in this work that the DE has a dynamical EoS parameter given by the well-known Chevallier-Polarski-Linder (CPL) parametrisation~\cite{chevallier2001ijmpd, linder2003prl} as
\begin{equation}
\wde = w_0 + w_{1} \paren*{1-a} \, ,\label{eq6:w-cpl}
\end{equation}
where $w_{0}$ and $w_{1}$ are constants. A dimensionless interaction term is defined as $\Omega_{I} = \frac{Q}{3 H^{3}/\kappa}$ and the dimensionless density parameter of matter (baryonic matter  and cold dark matter (DM), denoted as `$m\paren*{=b+c}$') and dark energy (DE) are defined as $\Omega_{m}=\frac{\rdm}{3\,H^2 /\kappa}$ and $\Omega_{de}=\frac{\rde}{3\,H^2 /\kappa}$ respectively. Similarly, energy density parameter for radiation is $\Omega_{r}=\frac{\rho_{r}}{3\,H^2 /\kappa}$. Here $H$ is the Hubble parameter defined with respect to the cosmic time $t$ and the dimensionless Hubble parameter at the present epoch is defined as $h = \frac{H_0}{100 \hskip1ex \footnotesize{\mbox{km s}^{-1} \mbox{Mpc}^{-1}}}$. The parameter values used in this work are listed in table \ref{tab5:bck}, where the values are taken from the latest 2018 data release of the \Planck collaboration~\cite{planck2018cp} (\Planck 2018, henceforth).

As shown by Pav{\'o}n and Wang~\cite{pavon2009grg}, energy transfer from dark energy to dark matter (DE $\rightarrow$ DM) is thermodynamically favoured following the Le Ch{\^a}telier-Braun principle. Observational data, on the other hand, prefer energy transfer from dark matter to dark energy (DM $\rightarrow$ DE)~\cite{zhang2012jcap, yang2018mnras, yang2018prd, yang2018jcap}. It must be noted that though the parameters $\beta_{0}$ and $\wde$ are in principle independent, they largely affect the perturbation evolutions and hence are correlated in parameter space of perturbation constraints. It had been shown in~\cite{valiviita2008jcap, he2009plb, majerotto2010mnras} that gravitational instabilities arise for constant $\wde\simeq -1$ due the interaction term in non-adiabatic pressure perturbations of dark energy.  The early time instabilities in the evolution of dark energy perturbation~\cite{valiviita2008jcap, he2009plb, gavela2009jcap, jackson2009prd, caldera2009prd, chongchitnan2009prd, xia2009prd, gavela2010jcap, clemson2012prd, mehrabi2015mnras} depend on the parameters $\beta_{0}$ and $\paren*{1+\wde}$ via a ratio called the doom factor, given as
\begin{equation}
d \equiv -\frac{aQ}{3\cH\rde\paren*{1+\wde}}.
\end{equation}
To avoid early time instabilities, $d$ must be negative semi-definite ($d \le 0$)~\cite{gavela2009jcap}, ensuring that $\beta_{0}$ and $\paren*{1+\wde}$ have the same sign. Thus stable perturbations can be achieved with either energy flow from dark matter to dark energy ($\beta_{0}>0$) and non-phantom or quintessence EoS ($\paren*{1+\wde}>0$) or energy flow from dark energy to dark matter ($\beta_{0}<0$) and phantom EoS ($\paren*{1+\wde}<0$). 

\begin{figure}[!htbp]
        \centering
            \subfloat{\includegraphics[width=.5\linewidth]{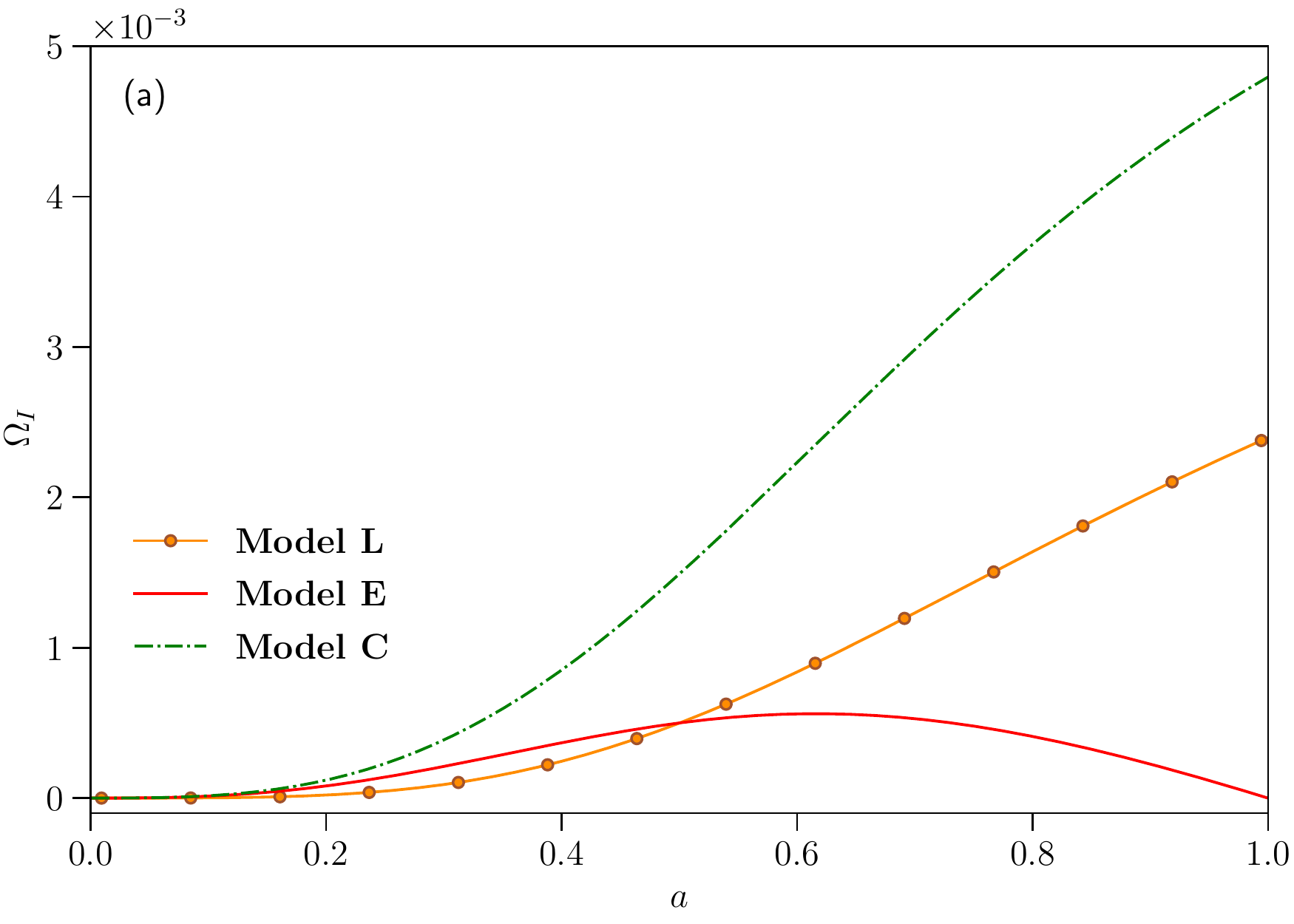}}\hfill
            \subfloat{\includegraphics[width=.5\linewidth]{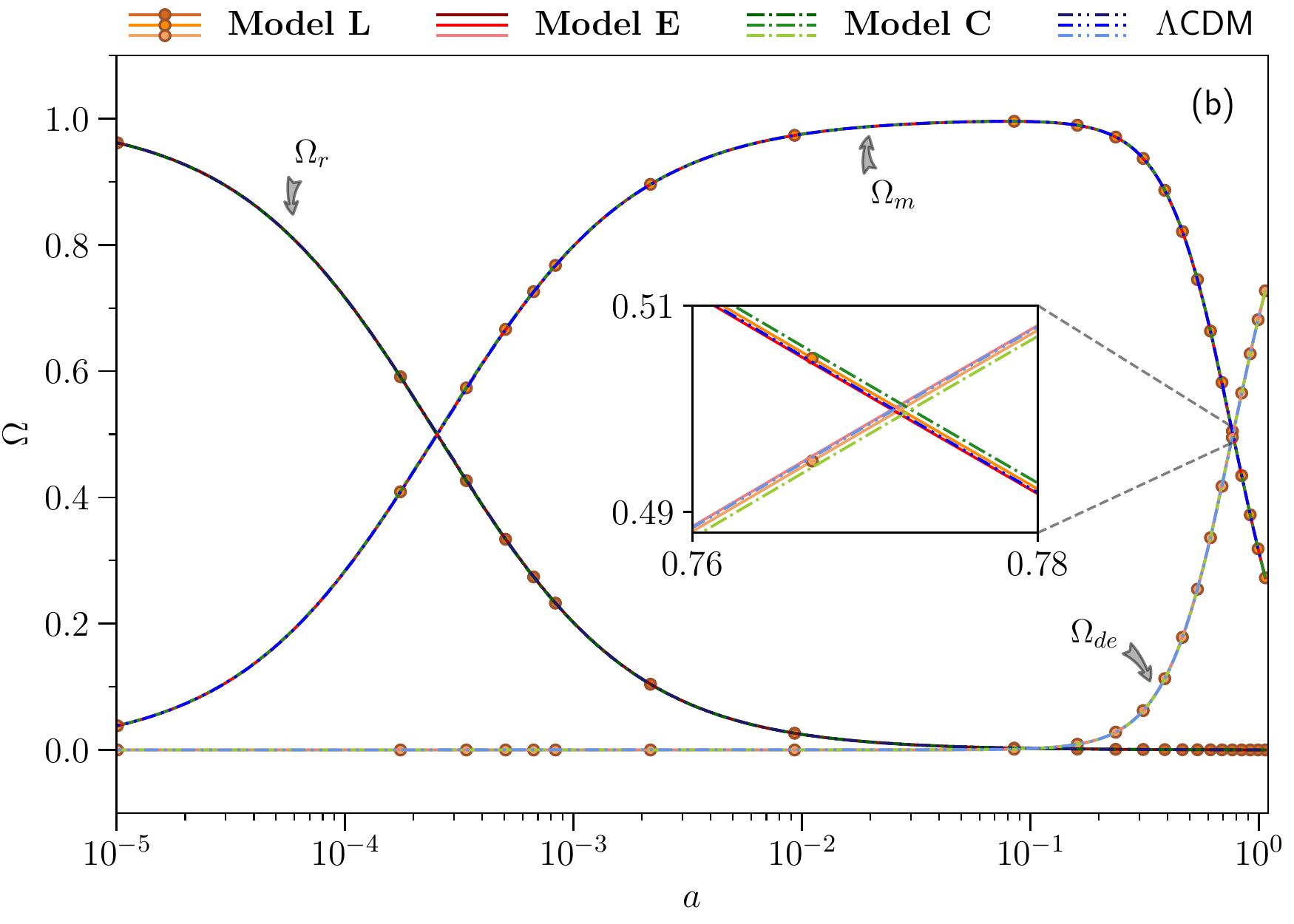}}\hfill
        \caption{Plot of (a) the dimensionless interaction parameter $\Omega_{I}$ and (b) density parameter $\Omega$ against scale factor $a$. The x-axis in Fig.\ (b) is in logarithmic scale The solid line with solid circles represents {\bf Model L}, solid line represents {\bf Model E} and dashed-dot line represents {\bf Model C} while the dashed-dot-dot line is for \lcdm. The inset shows the zoomed-in portion for the region $a= 0.76$ to $a = 0.78$.}\label{im6:bck2}
\end{figure}

In this section and the next (Section \ref{sec6:pert}), we have considered the energy flow from dark matter to dark energy and $\beta_{0}$ to be positive and hence $\wde>-1$. We have chosen the magnitude of $\beta_{0}$ to be small consistent with the observational results given in~\cite{yang2014prd1, yang2017prd1, yang2017prd2, pan2018mnras, vagnozzi2020mnras}. The particular value used here, $\beta_{0}=0.007$, is an example chosen such that no instability in the dark energy perturbation arises. For the background and perturbation analyses (Section \ref{sec6:pert}), we have chosen the example values of the parameter, $w_{0}$ and $w_{1}$ in $\wde$ (Eq.\ (\ref{eq6:w-cpl})) as
\begin{equation}\label{eq6:w}
w_{0} = -0.9995, ~~w_{1} = 0.005.
\end{equation}
The chosen values of the parameters $w_{0}$ and $w_{1}$ also ensure that $w_{de} \sim -1$ at $a=1$. It must be mentioned that, EoS parameter in the quintessence region is considered solely to avoid DE models with a  future ``big-rip'' singularity associated with phantom EoS parameter. Several instances of interacting DE models with $\wde <-1$ are found in the literature~\cite{valiviita2008jcap, gavela2009jcap, divalentino2017prd1, divalentino2017prd2, yang2017prd1, yang2017prd2, pan2018mnras, yang2018prd, yang2018jcap}. Figure (\ref{im6:bck2}a) shows the evolution of $\Omega_{I}$ with scale factor $a$ for \modellate, \modelearly and \modelcons. In Fig.\ (\ref{im6:bck2}a) the direction of energy flow is from dark matter to dark energy and the magnitude of $\Omega_{I}$ is the rate of energy transfer. The variation of density parameters of radiation ($\Omega_{r}$), dark matter together with baryons ($\Omega_{m}$) and dark energy ($\Omega_{de}$) with scale factor $a$ in logarithmic scale is shown in Fig.\ (\ref{im6:bck2}b) for the three models and the \lcdm model. It is clear from Figs.\ (\ref{im6:bck2}a) and (\ref{im6:bck2}b) that the effect of interaction will be very small in its contribution to the density parameters, $\Omega_{A}$, where $A = r,\,m,\,de$. In Figs.\ (\ref{im6:bck2}a) the y-axis is scaled by $10^{-3}$.

%%%%%%%%%%%%%%%%%%%%%%%%%%%%%%%%%%%%%%%%%%%%%%%%%%%%%%
%%%%%%%%%%%%%%%%%%%%%%%%%%%%%%%%%%%%%%%%%%%%%%%%%%%%%%
\section{Evolution Of Perturbations}\label{sec6:pert}
The perturbed FLRW metric in a general gauge takes the form~\cite{kodama1984ptps, mukhanov1992pr, ma1995apj, malik2003prd} 
\begin{equation} \label{eq6:metric2}
\begin{split}
ds^2=a^2\paren*{\tau} & \left\{ -\paren*{1+2\phi}d\tau^2+2\,\partial_iB \,d\tau \,dx^i +\right. \\
& \left.\left[\paren*{1-2\psi}\delta_{ij}+2\partial_i\partial_jE\right]dx^idx^j \right\},
\end{split}
\end{equation}
where $\phi, \psi, B, E$ are gauge-dependant scalar functions of space and time. In presence of interaction, the covariant form of the energy-momentum conservation equation will be
\begin{equation} \label{eq6:condition}
T^{\,\mu \nu}_{\paren*{A}; \nu} =Q^{\,\mu}_{\paren*{A}} \,, \hspace{0.3cm} \mbox{where}  \hspace{0.2cm} \sum_{A}Q^{\mu}_{\paren*{A}} =0 ~.
\end{equation}
The energy-momentum transfer function for the fluid `$A$', $Q^{\,\mu}_{\paren*{A}}$, can be split into the energy transfer rate, $Q_{\paren*{A}}$ and the momentum transfer rate, $F^{\mu}_{\paren*{A}}$, relative to the total $4$-velocity as~\cite{valiviita2008jcap, majerotto2010mnras, clemson2012prd}
\begin{equation} \label{Q-pert-def}
Q^{\,\mu}_{\paren*{A}} = Q_{\paren*{A}} u^{\,\mu} + F^{\,\mu}_{\paren*{A}}\,, \hspace{0.5cm} u_{\mu} F^{\,\mu}_{\paren*{A}} =0\,, \hspace{0.5cm} F^{\,\mu}_{\paren*{A}}= a^{-1}\, \paren*{ 0,\partial^{i}\, f_A}.
\end{equation}
Writing the total 4-velocity, $u^{\,\mu}$, in terms of the total peculiar velocity, $v$ as 
\begin{equation}\label{eq6:v4}
u^{\,\mu} = a^{-1}\paren*{1-\phi, v^{i}},
\end{equation}
the temporal and spatial components of the 4-energy-momentum transfer rate can be written as
\begin{eqnarray}
Q^{0}_{\paren*{A}}&=&a^{-1}\,\left[Q_A(1-\phi)+\delta Q_A\right], \\
\hspace{0.5cm} \mbox{and} \hspace{0.5cm} Q^{i}_{\paren*{A}} &=&a^{-1}\,\left[Q_A\,v^{i}+ \partial^{i}\,f_A\right]
\label{eq6:Q-component}
\end{eqnarray}
respectively, where $\delta Q_A$ is the perturbation in the energy transfer rate and $f_A$ is the momentum transfer potential.

The perturbed conservation equations of the fluid `$A$' in the Fourier space are written as
\begin{eqnarray}
\begin{split}
\delta \rho^\prime_A  -3\paren*{\rA+\pA}\psi'+ &k\paren*{\rA+\pA}\paren*{\vA+E^{\prime}} + \\3 \cH \,\paren*{\delta \rA+\delta \pA} =&~aQ_A\phi +a \delta Q_A , \label{eq6:e1}
\end{split}\\
\begin{split}
\left[\paren*{\rA+\pA}\paren*{\vA+B}\right]'+ &4\cH \paren*{\rA+\pA}\paren*{\vA+B}-\\
k\paren*{\rA+\pA}\phi-k\,\delta \pA &=aQ_A\paren*{v+B}-a\, kf_A ~. \label{eq6:m1}
\end{split}
\end{eqnarray}
In Eqs.\ (\ref{eq6:e1}) and (\ref{eq6:m1}), $\delta \rA$ is the perturbation in the energy density, $\delta \pA$ is the perturbation in pressure, $u^{\,\mu}_{A} = a^{-1}\paren*{1-\phi, \vA^{i}}$ is the 4-velocity with peculiar velocity $\vA$ of the fluid `$A$' and $k$ is the wavenumber. For an adiabatic perturbation, the pressure perturbation in presence of interaction is
\begin{equation}\label{eq6:pert-p}
\delta \pA=c_{s,\,A}^2 \delta \rA+\paren*{c_{s,\,A}^2-c_{a,\,A}^2}\left[3 \cH\paren*{1+w_A}\rA -a Q_A\right]\frac{\vA}{k},
\end{equation}
where $c_{a,\,A}^2=\frac{\pA^\prime}{\rA^\prime}$ is the square of adiabatic sound speed and $c_{s,\,A}^2=\frac{\delta \pA}{\delta \rA}$ is the square of effective sound speed in the rest frame of $A$-th fluid. 

The dynamical coupling parameter $\beta_{0}$ defined in Eq.\ (\ref{eq6:inter}) in the previous section is considered to be not affected by  perturbation. This assumption is valid for the EoS parameter defined in Eq.\ (\ref{eq6:w-cpl}) and the Hubble parameter, $\cH$. These perturbation equations are solved along with the perturbation equations~\cite{kodama1984ptps, mukhanov1992pr, ma1995apj} of the radiation, neutrino and baryon using the publicly available Boltzmann code \camb\footnote{Available at: \href{https://camb.info}{https://camb.info}}~\cite{lewis1999bs} after suitably modifying it. 

Using (\ref{eq6:v4}), Eq.\ (\ref{eq6:inter}) can be conveniently written as
\begin{equation} \label{eq6:q1}
Q= \frac{\cH \rde\, \beta\paren*{a}}{a}.
\end{equation}
Defining the density contrasts of the dark matter and dark energy as $\ddc = \delta\rdc/\rdc$ and $\dde = \delta\rde/\rde$ respectively and using Eqs.\ (\ref{eq6:pert-p}) and (\ref{eq6:q1}), the perturbation Eqs.\ (\ref{eq6:e1}) and (\ref{eq6:m1}) are written in synchronous gauge~\cite{ma1995apj} ($\phi=B=0$, $\psi=\eta$ and $k^2\,E=-\msh/2-3\eta$, where $\eta$ and $\msh$ are synchronous gauge fields in the Fourier space) as
\begin{eqnarray}
\ddc^\prime+ k v_{c} +\frac{\msh^\prime}{2} &=& \cH \bela \frac{\rde}{\rdc}\paren*{\ddc-\dde}, \\ \label{eq6:e2dm}
v_{c}^\prime+\cH v_{c}&=& 0~, \label{eq6:m2dm}
\end{eqnarray}
\begin{equation}
\begin{split}
\dde^\prime +3  \cH & \paren*{\cde-\wde}\dde+\paren*{1+\wde}\paren*{k \vde+\frac{\msh^\prime}{2}}\\
+3 \cH & \left[3 \cH \paren*{1+\wde}\paren*{\cde-\wde}\right]\frac{\vde}{k} +3\cH \wde^\prime\frac{\vde}{k} \\
=\, 3 \cH^{2} & \bela  \paren*{\cde-\wde}\frac{\vde}{k}, \label{eq6:e2de}
\end{split}
\end{equation}
\begin{equation}
\vde^\prime+\cH\paren*{1-3\cde}\vde-\frac{k\,\dde\,\cde }{\paren*{1+\wde}}=\frac{\cH\, \bela}{\paren*{1+\wde}}\, \left[\vdc-\paren*{1+\cde}\vde\right]. \label{eq6:m2de}
\end{equation}

The coupled differential Eqs.\ (\ref{eq6:e2dm})-(\ref{eq6:m2de}) are solved with $k = 0.1\,h$ $\mpci$ and the adiabatic initial conditions using \camb. Using the gauge-invariant quantity~\cite{malik2003prd, malik2009pr, he2009plb, chongchitnan2009prd, xia2009prd} $\zeta_{A} = \paren*{-\psi - \cH \frac{\delta \rho_{A}}{\rho_{A}^{\prime}}}$ and relative entropy perturbation $S_{AB} = 3\paren*{\zeta_{A}-\zeta_{B}}$, the adiabatic initial conditions for $\ddc$, $\dde$ in presence of interaction are obtained respectively as 
\begin{subequations}
\begin{eqnarray}
\delta_{ci} &=& \left[3 +\frac{\rde}{\rdc}\bela\right] \frac{\delta_{\gamma}}{3\paren*{1+w_{\gamma}}}, \label{eq6:initial-m}\\
\delta_{dei} &=& \left[3\,\paren*{1+\wde} - \bela \right] \frac{\delta_{\gamma}}{3\paren*{1+w_{\gamma}}}, \label{eq6:initial-de}
\end{eqnarray}
\end{subequations}
Here, $\delta_{\gamma}$ is the density fluctuation of photons. As can be seen from Eq.\ (\ref{eq6:m2dm}), there is no momentum transfer in the DM frame, hence initial value for $v_{c}$ is set to zero ($v_{ci} =0$)~\cite{bean2008prd2, chongchitnan2009prd, xia2009prd, valiviita2008jcap}. The initial value for the dark energy velocity, $\vde$ is assumed to be same as the initial photon velocity, $v_{dei} = v_{\gamma\,i}$.
To avoid the instability in dark energy perturbations due to the the propagation speed of pressure perturbations, we have set $\cde = 1$~\cite{waynehu1998apj, bean2004prd, gordon2004prd, afshordi2005prd, valiviita2008jcap}.

%%%%%%%%%%%%%%%%%%%%%%%%%%%%%%%%%%%%%%%%%%%%%%%%%%%%%
\begin{figure}[!h]
  \centering
\includegraphics[width=\textwidth]{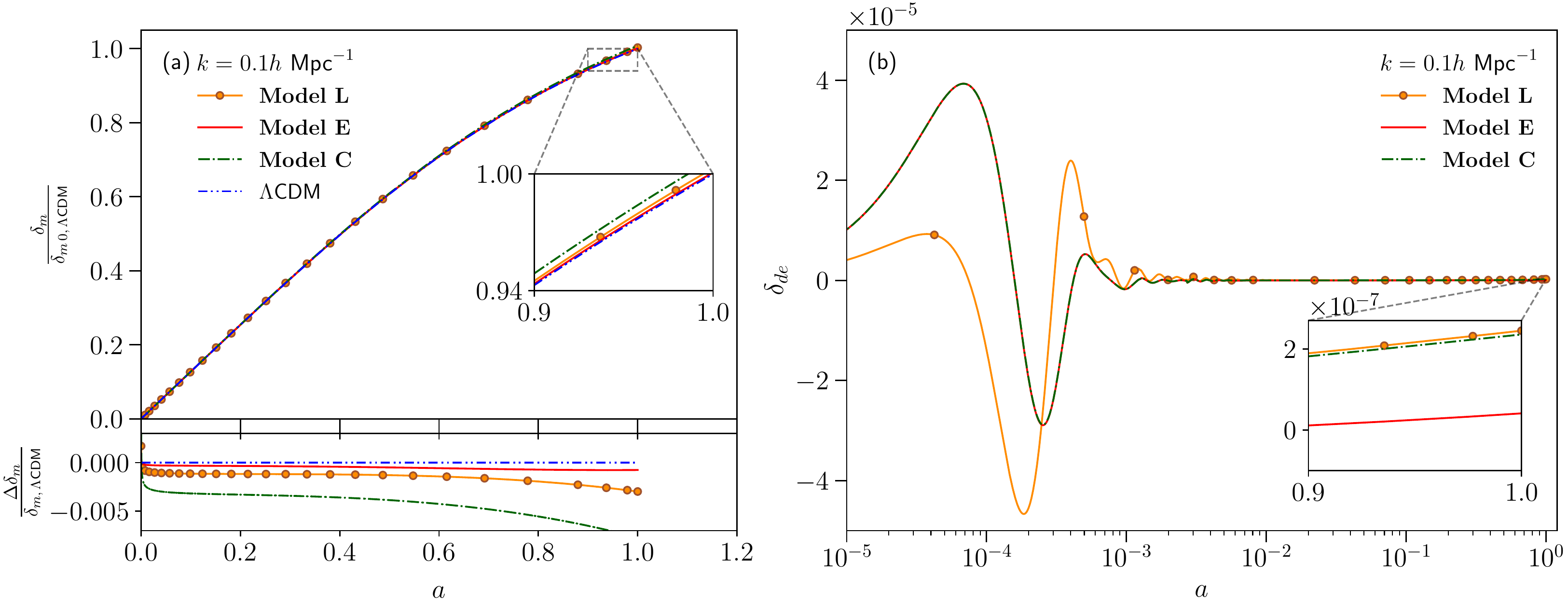}
\caption{(a) Plot of \textbf{Upper Panel} : the matter density contrast $\frac{\delta_{m}}{\delta_{m\,0,\,\scriptsize{\Lambda\text{CDM}}}}$ and \textbf{Lower Panel} : fractional growth rate is defined as $\frac{\Delta \delta_{m}}{\delta_{m,\,\scriptsize{\Lambda\text{CDM}}}} = \paren*{1- \frac{\delta_{m}}{\delta_{m,\,\scriptsize{\Lambda\text{CDM}}}}}$ relative to the \lcdm model against $a$. The origin on the x-axis represents $10^{-5}$. (b) Plot of the dark energy density fluctuation, $\dde$ against $a$ in logarithmic scale for $k = 0.1\,h$ $\mpci$. The solid line with solid circles represents {\bf Model L}, solid line represents {\bf Model E} and dashed-dot line represents {\bf Model C} while the dashed-dot-dot line is for \lcdm. The inset shows the zoomed-in portion from $a = 0.9$ to $a = 1.0$.}\label{im6:delta1}
\end{figure}

Figure (\ref{im6:delta1}a) shows the variation of the density contrast, $\ddm = \delta \rdm/\rdm$ for the cold dark matter ($c$) taken together and the baryonic matter ($b$) against $a$ for \modellate, \modelearly and \modelcons along with the \lcdm model. For a better comparison with the \lcdm model, $\ddm$ is scaled by $\delta_{m0} = \ddm\paren*{a=1}$ of \lcdm\footnote{The origin on the x-axis is actually $10^{-5}$}. As can be seen from the Fig.\ (\ref{im6:delta1}a), the growth of density fluctuation $\ddm$ is similar in all the model at early times. The effect of interaction comes into play at late time. The late-time growth of $\ddm$ (inset of (\ref{im6:delta1}a)) shows that \modelearly agrees well with the \lcdm model, whereas \modellate and \modelcons grow to a little higher value. Figure (\ref{im6:delta1}b) shows the variation of the dark energy density contrast $\dde$ for \modellate, \modelearly and \modelcons. At early time, $\dde$ oscillates and then decays to very small values. In \modelcons, the early time evolution of $\dde$ is similar to \modelearly while the late time evolution is similar to \modellate. To understand the differences among the three models and the \lcdm model, we have shown the fractional matter density contrast, $\frac{\Delta \delta_{m}}{\delta_{m,\,\scriptsize{\Lambda\text{CDM}}}} = \paren*{1- \frac{\delta_{m}}{\delta_{m,\,\scriptsize{\Lambda\text{CDM}}}}}$ in the lower panel of Fig.\ (\ref{im6:delta1}a). It is clearly seen that, $\delta_{m}$ for \modelearly evolves close to the \lcdm model. 

%%%%%%%%%%%%%%%%%%%%%%%%%%%%%%%%%%%%%%%%%%%%%%%%%%%%%%

%%%%%%%%%%%%%%%%%%%%%%%%%%%%%%%%%%%%%%%%%%%%%%%%%%%%%%
\subsection{Effect On CMB Temperature, Matter Power Spectrum And $\fsg$}\label{sec6:result}
It is necessary to have an insight into other physical quantities like the CMB temperature spectrum, matter power spectrum and the logarithmic growth of matter perturbation, to differentiate the interacting models. The CMB temperature power spectrum is given as
\begin{equation}
C_{\ell}^{TT} = \frac{2}{k} \int k^{2} d k \,P_{\zeta}\paren*{k} \Delta^{2}_{T\ell}\paren*{k},
\end{equation}
where $\ell$ is the multipole index, $P_{\zeta}\paren*{k}$ is the primordial power spectrum, $\Delta_{T\ell}\paren*{k}$ is the temperature transfer function and $T$ represents the temperature. For a detailed analysis on the CMB spectrum we refer to~\cite{hu1995apj,seljak1996apj}. The matter power spectrum is written as
\begin{equation} \label{eq6:power}
P\left(k,a\right)= A_s \,k^{n_s} T^2\left(k\right) D^2\left(a\right),
\end{equation}
where $A_{s}$ is the scalar primordial power spectrum amplitude, $n_{s}$ is the spectral index, $T\left(k\right)$ is the matter transfer function and $D\left(a\right)=\frac{\ddm\left(a\right)}{\ddm\left(a=1\right)}$ is the normalised density contrast. For a detailed description we refer to~\cite{dodelson2003}. Both $C_{\ell}^{TT}$ and $P\paren*{k,a}$ are computed numerically using \camb. The values of power spectrum amplitude, $A_{s} = 2.100549 \times 10^{-9}$ and spectral index, $n_{s} = 0.9660499$ are taken from \Planck 2018 data~\cite{planck2018cp}.

\begin{figure}[!htbp]
\centering
\includegraphics[width=\textwidth]{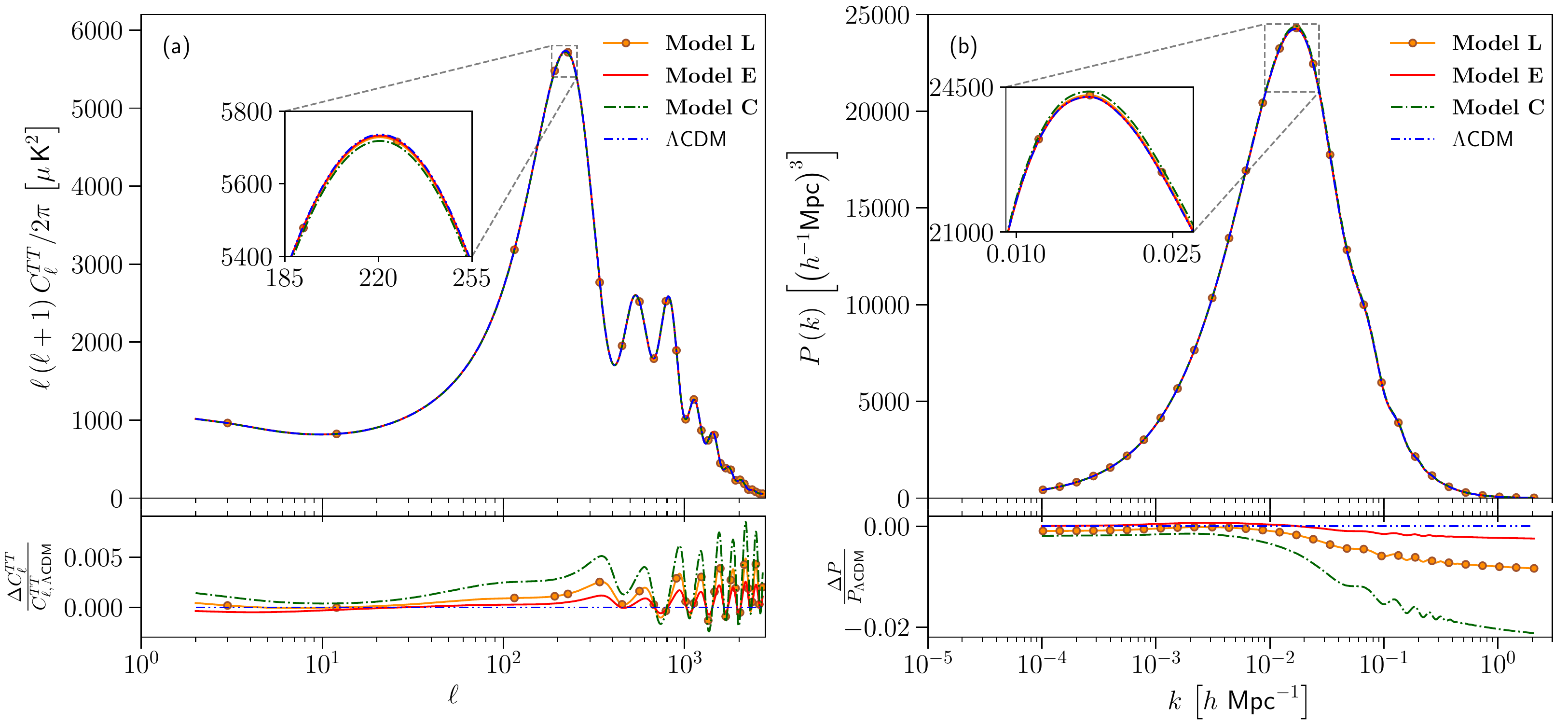}
\caption{\textbf{Upper Panel} : (a) Plot of CMB temperature power spectrum in units of $\mu \mbox{K}^{2}$ with the multipole index $\ell$ in logarithmic scale. (b) Plot of matter power spectrum $P\paren*{k}$ in units of $\paren*{h^{-1}\mpc}^{3}$ with wavenumber $k$ in units of $h\,\mpci$. \textbf{Lower Panel} : Plot of fractional change in the temperature spectrum, $\frac{\Delta C_{\ell}^{TT}}{C_{\ell,\, \scriptsize{\Lambda\text{CDM}}}^{TT}} = \paren*{1- \frac{C_{\ell}^{TT}}{C_{\ell,\, \scriptsize{\Lambda\text{CDM}}}^{TT}}}$ and the fractional change in matter power spectrum, $\frac{\Delta P}{P_{\scriptsize{\Lambda\text{CDM}}}} = \paren*{1- \frac{P}{P_{\scriptsize{\Lambda\text{CDM}}}}}$. For both the panels, the solid line with solid circles represents {\bf Model L}, solid line represents {\bf Model E} and dashed-dot line represents {\bf Model C} while the dashed-dot-dot line is for \lcdm at $a=1$. The inset shows the zoomed-in versions of the peaks.}\label{im6:cl}
\end{figure}
Figure (\ref{im6:cl}) shows the temperature and matter power spectrum for \modellate, \modelearly, \modelcons and \lcdm at $a=1$. In \modellate and \modelcons, more matter content results in lower amplitude of the first peak of the CMB spectrum compared to the \lcdm model. The lower panel of Fig.\ (\ref{im6:cl}a), shows the fractional change ($=\Delta C_{\ell}^{TT}/C_{\ell,\, \scriptsize{\Lambda\text{CDM}}}^{TT}$) in $C_{\ell}^{TT}$. It is seen from the lower panel of Fig.\ (\ref{im6:cl}a), that the low-$\ell$ modes of \modelearly increases through the integrated Sachs-Wolfe (ISW) effect. More matter content also increases the matter power spectrum compared to the \lcdm model. The deviations from the \lcdm model are prominent for the smaller modes. These features are clear from the lower panel of Fig.\ (\ref{im6:cl}b), which shows the  fractional change in matter power spectrum, $\Delta P/P_{\scriptsize{\Lambda\text{CDM}}}$ of the interacting models relative to the \lcdm model.

The presence of the interaction modifies the logarithmic growth rate which helps in differentiating between the models even better. The growth rate is the logarithmic derivative of the density fluctuation of matter (baryon and CDM) and is written as
\begin{equation}
f\paren*{a}= \frac{d \ln \ddm}{d \ln a} ~=~ a \frac{d}{d\,a}\paren*{\frac{\delta \rdm}{\rdm}}.
\end{equation}
Since, $\delta \rdm = \paren*{\delta_{c}\rdc + \delta_{b}\rho_{b}}$, $\delta_{b}$ being the baryon density fluctuation, in presence of interaction the growth rate~\cite{costa2017jcap} will be
\begin{equation} \label{eq6:growth_rate}
f\paren*{a} = a \paren*{\frac{\delta_{c,\,a} \,\, \rdc + \delta_{b,\,a} \,\,\rho_{b}}{\ddm \, \rdm} - \frac{a Q\, \ddc}{\ddm \,\rdm} - \frac{a Q}{\rdm}},
\end{equation}
where `$_{,\,a}$' denotes the derivative with respect to the scale factor $a$ and $Q$ is given by Eq.\ (\ref{eq6:q1}). It must be noted that the last two terms involving interaction $Q$ is introduced in Eq.\ (\ref{eq6:growth_rate}) via the evolution of $\rdc$ (Eqs.\ (\ref{eq6:con1})). We have calculated the growth rate, $f$ for the different models using \camb.

Observationally the galaxy density fluctuation, $\delta_{g}$ is measured, which in turn gives the matter density fluctuation, $\ddm$ as $\delta_{g} = b \ddm$, where $b \in \left[1,3\right]$ is the bias parameter. This $\ddm$ is used to calculate the logarithmic growth rate, $f$. Thus, $f$ is sensitive to $b$ and is not a very reliable quantity. A more dependable observational quantity is defined as the product $f\paren*{a}\se\paren*{a}$~\cite{percival2009mnras}, where $\se\paren*{a}$ is the root-mean-square (rms) mass fluctuations within the sphere of radius $R = 8 h^{-1}\, \mpc$. 
The rms linear density fluctuation is also written as $\se\paren*{a} = \se\paren*{1}\frac{\ddm\paren*{a}}{\ddm\paren*{1}}$, where $\se\paren*{1}$ and $\ddm\paren*{1}$ are the values at $a=1$, and $f$ and $\se\paren*{1}$ for the different models are obtained from Eq.\ (\ref{eq6:growth_rate}) and (\ref{eq6:sigma2}) using our modified version of \camb. The combination $\fsg$ is written as
\begin{equation} \label{eq6:fsigma}
\fsg\paren*{a} \equiv f\paren*{a}\se\paren*{a} = \se\paren*{1}\frac{a}{\ddm\paren*{1}}\frac{d \ddm}{d \,a}.
\end{equation}

\begin{figure}[!htbp]
\centering
\includegraphics[width=\textwidth]{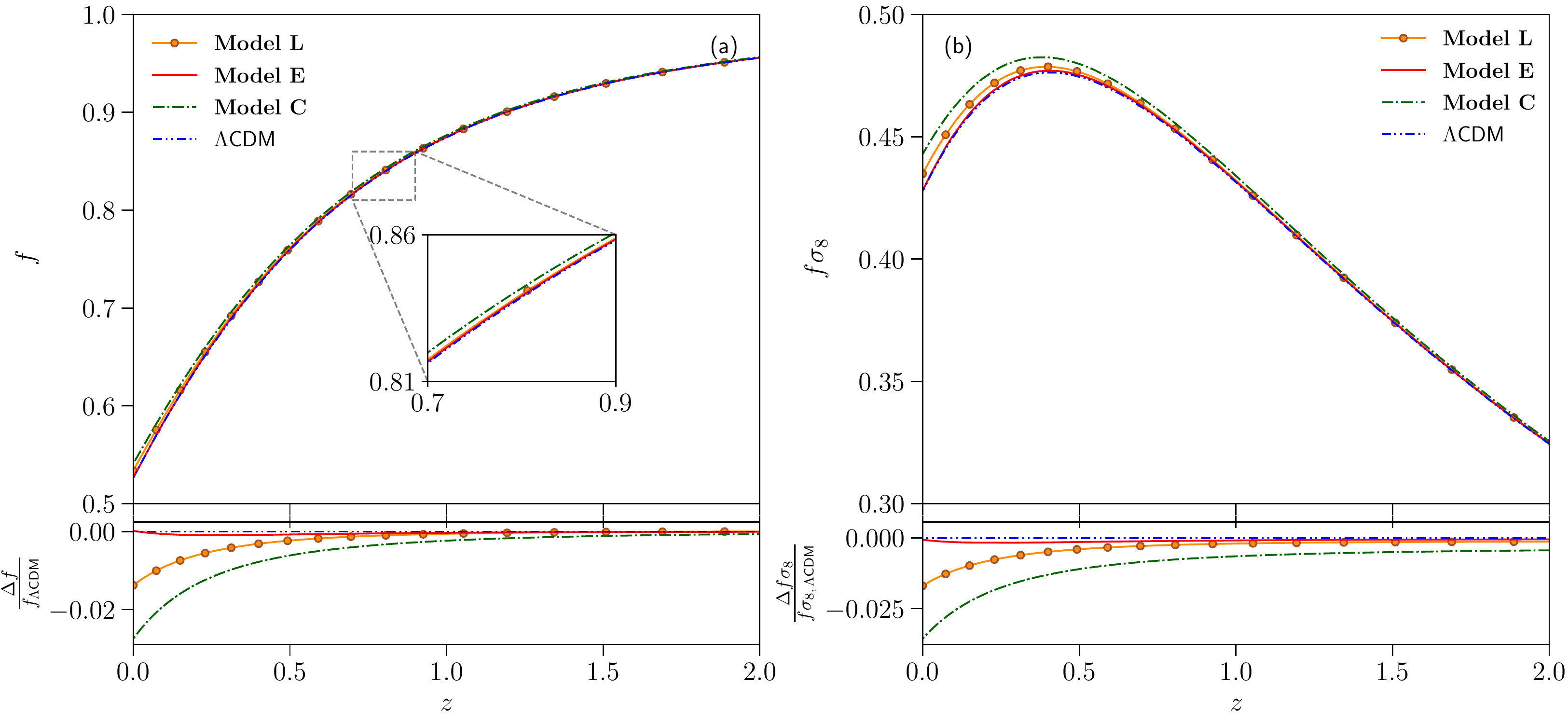}
\caption{\textbf{Upper Panel} : Plot of (a) linear growth rate $f$ and (b) $\fsg$ against redshift $z$. The inset shows the zoomed-in portion from $z = 0.7$ to $z = 0.9$. \textbf{Lower Panel} : Plot of fractional change in the temperature spectrum, $\frac{\Delta f}{f_{\scriptsize{\Lambda\text{CDM}}}} = \paren*{1- \frac{f}{f_{\scriptsize{\Lambda\text{CDM}}}}}$ and the fractional change in matter power spectrum, $\frac{\Delta \fsg}{f\sigma_{8,\,\scriptsize{\Lambda\text{CDM}}}} = \paren*{1- \frac{\fsg}{f\sigma_{8,\,\scriptsize{\Lambda\text{CDM}}}}}$. For both the panels, the solid line with solid circles represents {\bf Model L}, solid line represents {\bf Model E} and dashed-dot line represents {\bf Model C} while the dashed-dot-dot line is for \lcdm.}\label{im6:fs}
\end{figure}
The logarithmic growth rates, $f$ and $\fsg$ are independent of the wavenumber $k$ for smaller redshift, $z$, so only the domain $z=0$ to $z=2$ is considered here. The redshift, $z$ and the scale factor $a$ are related as $z = \frac{a_0}{a} -1$, $a_0$ being the present value (taken to be unity). The difference in the models is magnified in the $f$ and $\fsg$ analysis. As can be seen from Fig.\ (\ref{im6:fs}), growth rates ($f$) and $\fsg$ are different for the different models in the recent past. The differences due to the evolution of interaction are seen in $f$ and $\fsg$. Both \modellate and \modelcons have slightly higher values of $f$ and $\fsg$ at $z=0$, compared to the value obtained from the \lcdm model. For \modelearly and the \lcdm model, the values of $f$ and $\fsg$ are same at $z=0$. When the energy transfer rates were different in the recent past, \modelearly had a slightly larger value of $f$ and $\fsg$ (compared to the \lcdm model) when the interaction was non-zero. The fractional changes in growth rate ($\Delta f/f_{\scriptsize{\Lambda\text{CDM}}}$) and $\fsg$ ($\Delta \fsg/f\sigma_{8,\,\scriptsize{\Lambda\text{CDM}}}$) of the interacting models relative to the \lcdm model are shown in the lower panels. The difference among the three models is distinctly seen.
%%%%%%%%%%%%%%%%%%%%%%%%%%%%%%%%%%%%%%%%%%%%%%%%%%%%%%

%%%%%%%%%%%%%%%%%%%%%%%%%%%%%%%%%%%%%%%%%%%%%%%%%%%%%%
\section{Observational Constraints}\label{sec6:mcmc}
In this section, \modellate, \modelearly and \modelcons are tested against observational datasets like the CMB, BAO, Supernovae and RSD data by using the Markov Chain Monte Carlo (MCMC) analysis of the publicly available, efficient MCMC simulator \cosmomc\footnote{Available at: \href{https://cosmologist.info/cosmomc/}{https://cosmologist.info/cosmomc/}}\nomenclature{CosmoMC}{Cosmological MonteCarlo}~\cite{lewis2013hha,lewis2002ah}. The datasets and the methodology are discussed in the Appendix \ref{appen:a}. The datasets are used to constrain the nine-dimensional parameter space given as 
\begin{equation}\label{eq6:parameter}
P \equiv \lbrace \Omega_{b} h^2, \Omega_{c} h^2, 100\theta_{MC}, \tau, \beta_0, w_0, w_{1}, \ln\paren*{10^{10} A_s}, n_{s}\rbrace,
\end{equation}
where $\Omega_b h^2$ is the baryon density, $\Omega_c h^2$ is the cold dark matter density, $\theta_{MC}$ is the angular acoustic scale, $\tau$ is the optical depth, $\beta_{0}$, $w_{0}$ and $w_{1}$ are the free model parameters, $A_{s}$ is the scalar primordial power spectrum amplitude and $n_{s}$ is the scalar spectral index. The parameter space, $P$, for all the three models, is explored for the flat prior ranges given in Table \ref{tab6:prior}. We allowed the prior of $\beta_{0}$ to cross the zero and set the prior of $w_{0}$ and $w_{1}$ such that such that $\wde$ is always in the quintessence region.
\begin{table}[!h]
\begin{center}
\caption{Prior ranges of nine independent parameters used in the \cosmomc analysis.}\label{tab6:prior}
%\centering
\begin{adjustbox}{width=0.7\textwidth}
\begin{tabular}{cc}
\hline \hline
\rule[-1ex]{0pt}{2.5ex}Parameter &  \hspace{24ex} Prior\\
\hline
\rule[-1ex]{0pt}{2.5ex}$\Omega_{b} h^2$&  \hspace{24ex} $\left[0.005, 0.1\right]$ \\
\rule[-1ex]{0pt}{2.5ex}$\Omega_{c} h^2$&  \hspace{24ex} $\left[0.001, 0.99\right]$\\
\rule[-1ex]{0pt}{2.5ex}$100\theta_{MC}$&  \hspace{24ex} $\left[0.5, 10\right]$\\
\rule[-1ex]{0pt}{2.5ex}$\tau$&  \hspace{24ex} $\left[0.01, 0.8\right]$\\
\rule[-1ex]{0pt}{2.5ex}$\beta_0$&  \hspace{24ex} $\left[-1.0, 1.0\right]$\\
\rule[-1ex]{0pt}{2.5ex}$w_0$&  \hspace{24ex} $\left[-0.9999, -0.3333\right]$\\
\rule[-1ex]{0pt}{2.5ex}$w_{1}$&  \hspace{24ex} $\left[0.005, 1.0\right]$\\
\rule[-1ex]{0pt}{2.5ex}$\ln\paren*{10^{10} A_s}$&  \hspace{24ex} $\left[1.61, 3.91\right]$\\
\rule[-1ex]{0pt}{2.5ex}$n_{s}$&  \hspace{24ex} $\left[0.8, 1.2\right]$\\
\hline \hline
\end{tabular}
\end{adjustbox}
\end{center}
\end{table}

\subsection{Model L: $\bela = \beta_{0}\paren*{\frac{2\,a}{1+a}}$}

For \modellate, the marginalised values with errors at $1\sigma$ ($68\%$ confidence level) of the nine free parameters and three derived parameters, $H_{0}$, $\Omega_{m}$ and $\se$, are listed in Table \ref{tab6:mean-1}. Henceforth, the 1D marginalised values given in the tables will be referred to as mean values. The correlations between the model parameters ($\beta_{0}$, $w_{0}$, $w_{1}$) and the derived parameters ($H_{0}$, $\Omega_{m}$, $\se$) and their marginalised contours are shown in Fig.\ \ref{im6:tri1}. The contours contain $1\sigma$ region ($68\%$ confidence level) and $2\sigma$ region ($95\%$ confidence level). When only the \Planck data is considered, the mean value of the coupling parameter, $\beta_{0}(=0.00788^{+0.00815+0.0158}_{-0.00815-0.0162})$, is positive with zero in the $1\sigma$ region indicating energy transfers from DM to DE. The parameters $w_{0} (<-0.909< -0.800 )$ and $w_{1} (< 0.174 < 0.365)$ remain unconstrained even in the $2\sigma$ region. For other parameters, the mean values are compared with their \lcdm counterparts from the \Planck estimation~\cite{planck2018cp}. The Hubble expansion rate, $H_{0}$, is obtained at a value lower than $67.36\pm 0.54$ in $\mbox{km s}^{-1} \mbox{Mpc}^{-1}$, as obtained for the \lcdm model~\cite{planck2018cp}. Though the mean value is lower than that obtained from the \Planck estimate, the presence of high error bars results in $3.5\sigma$ tension with the local measurement as $H_{0} = 74.03\pm1.42$ $\mbox{km s}^{-1} \mbox{Mpc}^{-1}$. The value of the late time clustering amplitude ($\se$) is skewed towards the value, $\se = 0.77^{+0.04}_{-0.03} $, as obtained by the galaxy cluster counts using thermal Sunyaev-Zel'dovich (tSZ) signature~\cite{planck2018cp,zubeldia2019mnras}. Thus, \Planck data alone alleviates the $\se$ tension in the \modellate. Figure \ref{im6:tri1} highlights the positive correlation between $H_{0}$ and $\se$ and strong negative correlations of $\Omega_{m}$ with $H_{0}$ and $\se$. The parameter $w_{0}$ has negative correlations with $w_{1}$, $H_{0}$ and $\se$ and positive correlation with $\Omega_{m}$. The coupling parameter ($\beta_{0}$) is uncorrelated to the others. 

Addition of the BAO to the \Planck data, increases the value of $\beta_{0}$ to $0.00814^{+0.00755+0.0146}_{-0.00755-0.0151}$ with zero outside the $1\sigma$ region. The \Planck and BAO combination cannot constrain the parameters $w_{0}$ and $w_{1}$. The mean value of the Hubble parameter increases considerably but is still smaller than the corresponding value for \lcdm, $H_{0} = 67.66\pm 0.42$ in $\mbox{km s}^{-1} \mbox{Mpc}^{-1}$ in the $1\sigma$ region. The considerable decrease in error bar increased the $H_{0}$ tension to $\sim 4\sigma$. The values of $\Omega_{m}$ decreases and $\se$ increases and are higher than the \lcdm counterpart ($\Omega_{m} = 0.3111\pm 0.0056$ and $\se = 0.8102\pm 0.006$) in the $1\sigma$ region. Thus, addition of the BAO data to the \Planck data restores the $\se$ tension ($\sim 0.79\sigma$) in \modellate. The combination also lowers the error regions substantially. 

Interestingly, addition of $\fsg$ to the \Planck data changes the parameter mean values in the similar fashion like the \Planck and BAO combination but the error bars become higher. This is also clear from Fig.\ \ref{im6:tri1}. The mean value of $\beta_{0} (=0.00752^{+0.00757+0.0145}_{-0.00757-0.0151})$ is smaller the \Planck and BAO combination. Clearly, addition of the $\fsg$ data restores the $\se$ tension in \modellate.

Addition of the BAO and Pantheon to the \Planck data, increases the mean value of $\beta_{0} (=0.00859^{+0.00745+0.0145}_{-0.00745-0.0148} )$ with zero in the $2\sigma$ region. The parameters $w_{0}$ and $w_{1}$ still remain unconstrained. The combination increases the $H_{0}$ mean value but is still slightly smaller than the fiducial \lcdm value. The central value of $\Omega_{m}$ at the present epoch remains slightly larger whereas $\se$ remains slightly smaller than the \lcdm case. Clearly, the $\se$ tension is restored.

Combining $\fsg$ data with \Planck, BAO and Pantheon lowers the mean values of both $\Omega_{m}$ and $\se$ but increases the value of $H_{0}$ compared to the baseline \Planck values~\cite{planck2018cp}. Thus, addition of all the datasets worsen the $H_{0}$ tension ($\sim 4.2\sigma$) and the $\se$ tension ($\sim 0.87\sigma$). The mean value of $\beta_{0} (=0.00818^{+0.00731+0.0142}_{-0.00731-0.0146})$ decreases slightly with zero in the $2\sigma$ region. Although the error bars on $w_{0}$ and $w_{1}$ become small, they still remain unconstrained.

Combination of all the datasets significantly reduced the error bars. The parameters, $\beta_{0}$ and $w_{1}$ become very weakly correlated with other parameters. However, the correlations among the rest of the parameters remain unchanged. 
%%Table I
\begin{center}
\begin{table*}[!h]
\centering
\caption{Observational constraints on the nine dependent model parameters with three derived parameters separated by a horizontal line and the error bars correspond to $68\%$ confidence level for {\bf Model L}, using different observational datasets.}
\label{tab6:mean-1}
\begin{adjustbox}{width=\textwidth}
\begin{tabular} { l  c c c c c}
\hline\hline \noalign{\vskip 2pt}
 {\normalsize Parameter} &  {\normalsize \Planck} &  {\normalsize \Planck \dataplus $f\sigma_{8}$} &  {\normalsize \Planck \dataplus BAO} &  {\normalsize \thead{\Planck \\ \dataplus BAO \dataplus Pantheon}} &  {\normalsize \thead{\Planck \dataplus BAO \\ \dataplus Pantheon \dataplus $f\sigma_{8}$}}\\
\hline \noalign{\vskip 2pt}
\rule[-1.5ex]{0pt}{2.7ex}{\boldmath$\Omega_b h^2   $} & $0.022362\pm 0.000168      $ & $0.022483\pm 0.000163      $ & $0.022487\pm 0.000156      $ & $0.022500\pm 0.000154      $ & $0.022542\pm 0.000152      $\\
\rule[-1.5ex]{0pt}{2.7ex}{\boldmath$\Omega_c h^2   $} & $0.12005\pm 0.00129        $ & $0.11853\pm 0.00117        $ & $0.11848\pm 0.00102        $ & $0.118381\pm 0.000977      $ & $0.117838\pm 0.000927      $\\
\rule[-1.5ex]{0pt}{2.7ex}{\boldmath$100\theta_{MC} $} & $1.040773\pm 0.000326      $ & $1.040938\pm 0.000316      $ & $1.040951\pm 0.000315      $ & $1.040954\pm 0.000316      $ & $1.041007\pm 0.000313      $\\
\rule[-1.5ex]{0pt}{2.7ex}{\boldmath$\tau           $} & $0.05475\pm 0.00773        $ & $0.05641^{+0.00705}_{-0.00794}$ & $0.05732^{+0.00701}_{-0.00787}$ & $0.05707^{+0.00691}_{-0.00777}$ & $0.05791\pm 0.00760        $\\
\rule[-1.5ex]{0pt}{2.7ex}{\boldmath$\beta_0        $} & $0.00788\pm 0.00815        $ & $0.00752\pm 0.00757        $ & $0.00814\pm 0.00755        $ & $0.00859\pm 0.00745        $ & $0.00818\pm 0.00731        $\\
\rule[-1.5ex]{0pt}{2.7ex}{\boldmath$w_0            $} & $< -0.909                  $ & $< -0.976                  $ & $< -0.968                  $ & $< -0.980                  $ & $< -0.985                  $\\
\rule[-1.5ex]{0pt}{2.7ex}{\boldmath$w_{1}          $} & $< 0.174                   $ & $< 0.0672                  $ & $< 0.0707                  $ & $< 0.0623                  $ & $< 0.0500                  $\\
\rule[-1.5ex]{0pt}{2.7ex}{\boldmath${\rm{ln}}(10^{10} A_s)$} & $3.0489\pm 0.0149          $ & $3.0489\pm 0.0147          $ & $3.0512\pm 0.0148          $ & $3.0507\pm 0.0145          $ & $3.0512\pm 0.0146          $\\
\rule[-1.5ex]{0pt}{2.7ex}{\boldmath$n_s            $} & $0.96330\pm 0.00444        $ & $0.96672\pm 0.00436        $ & $0.96674\pm 0.00413        $ & $0.96690\pm 0.00418        $ & $0.96818\pm 0.00412        $\\
\hline \noalign{\vskip 2pt}
\rule[-1.5ex]{0pt}{2.7ex}$H_0 \left[\mbox{km s}^{-1} \mbox{Mpc}^{-1}\right]                       $ & $63.98^{+2.45}_{-1.47}     $ & $66.93^{+1.04}_{-0.719}    $ & $66.770^{+0.792}_{-0.602}  $ & $67.179^{+0.579}_{-0.521}  $ & $67.596\pm 0.524           $\\
\rule[-1.5ex]{0pt}{2.7ex}$\Omega_m                  $ & $0.3507^{+0.0157}_{-0.0292}$ & $0.31643^{+0.00846}_{-0.0116}$ & $0.31778^{+0.00681}_{-0.00832}$ & $0.31368\pm 0.00632        $ & $0.30871\pm 0.00598        $\\
\rule[-1.5ex]{0pt}{2.7ex}$\sigma_8                  $ & $0.7825^{+0.0228}_{-0.0141}$ & $0.8027^{+0.0102}_{-0.00836}$ & $0.8020^{+0.0104}_{-0.00892}$ & $0.80541\pm 0.00862        $ & $0.80560\pm 0.00803        $\\
\hline\hline
\end{tabular}
\end{adjustbox}
\end{table*}
%\end{center}
%%Figure I
%\begin{center}
\begin{figure}[!h]
        \centering
         \includegraphics[width=0.7\linewidth]{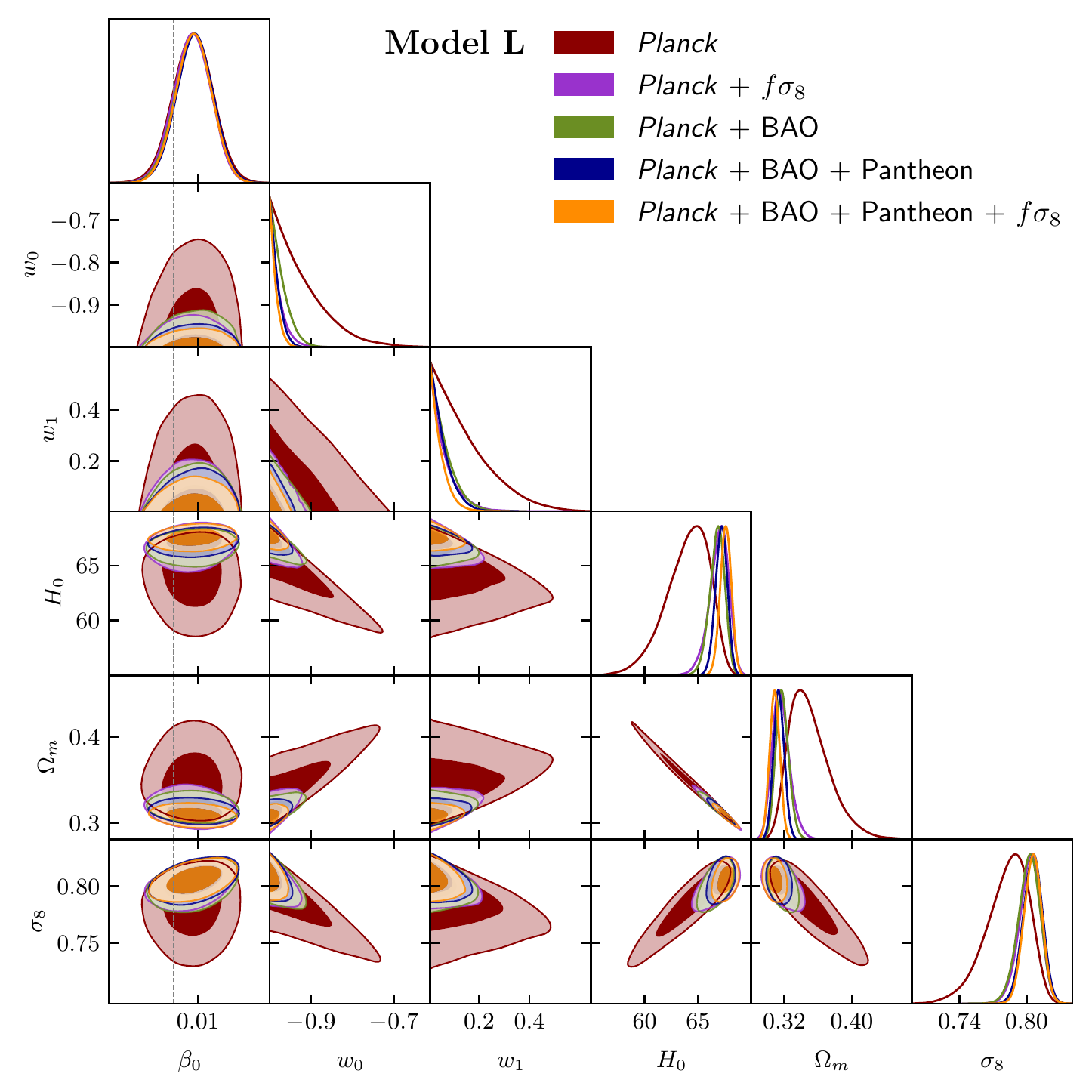}
         \caption{Plot of 1-dimensional marginalised posterior distributions and 2-dimensional marginalised constraint contours on the parameters of {\bf Model L} containing $68\%$ and $95\%$ probability. The dashed line represents the $\beta_{0}=0$ value.}\label{im6:tri1}
\end{figure}
\end{center}

\subsection{Model E: $\bela = \beta_{0}\paren*{\frac{1-a}{1+a}}$}

For \modelearly, the mean values with $1\sigma$ errors of the nine free parameters along with the three derived parameters, $H_{0}$, $\Omega_{m}$ and $\se$, are given in Table \ref{tab6:mean-2}. The correlations of the model parameters ($\beta_{0}$, $w_{0}$, $w_{1}$) with the derived parameters ($H_{0}$, $\Omega_{m}$, $\se$) and their marginalised contours are shown in Fig.\ \ref{im6:tri2}. When only \Planck data is considered, the mean value of $\beta_{0}(=0.0339^{+0.0372+0.0724}_{-0.0372-0.0746})$ is large compared to that in \modellate, though $\beta_{0}=0$ remains within the $1\sigma$ region. The CPL parameters, $w_{0} (< -0.914< -0.809)$ and $w_{1} (< 0.168 < 0.355)$, remain unconstrained even within the $2\sigma$ region. The values of $H_{0}$ is greater and that of $\se$ is slightly grater whereas $\Omega_{m}$ is slightly lower than those in \modellate. Similar to \modellate, the discrepancy in the value of $H_{0}$ with the local measurement is at $3.5\sigma$. In \modelearly also, the \Planck data alleviates the $\se$ tension. The distinguishing feature of \modelearly is that the mean value of $\beta_{0}$ is greater than that obtained in \modellate for all the dataset combinations. 

Addition of the BAO to the \Planck data, increases the mean value of $\beta_{0}(=0.0432^{+0.0376+0.0733}_{-0.0376-0.0744})$ with zero allowed in the $2\sigma$ region. The parameters $w_{0}$ and $w_{1}$ remain unconstrained. The mean value of $H_{0}$ increases considerably but is still smaller than the corresponding value for \lcdm. The values of $\Omega_{m}$ decreases and $\se$ increases and are higher than the \lcdm counterpart. The addition of BAO data to \Planck data restores the $H_{0}$ ($\sim 4\sigma$) and $\se$ ($\sim 0.77\sigma$) tensions in \modelearly. The combination also lowers the error bars considerably.

Similar to \modellate, addition of $\fsg$ to the \Planck data changes the parameter mean values like the \Planck \dataplus BAO combination but the error bars still remain a little higher. This is also clear from Fig.\ \ref{im6:tri2}. The mean value of $\beta_{0} (=0.0395^{+0.0381+0.0735}_{-0.0381-0.0750})$ is slightly smaller than the \Planck \dataplus BAO combination. The $H_{0}$ and $\se$ tensions are restored on addition of $\fsg$ to the \Planck data.

Combining BAO and Pantheon with \Planck data increases the mean value of $\beta_{0} (=0.0448^{+0.0377+0.0738}_{-0.0377-0.0733})$ and $\beta_{0} =0$ is within the $2\sigma$ region. The \Planck \dataplus BAO \dataplus Pantheon results in a very small change in the mean values of the parameters along with reduced error bars. The mean values of $H_{0}$ and $\se$ increase and $\Omega_{m}$ decreases relative to the \Planck \dataplus BAO combination. Again, the $\se$ tensions are not alleviated.

Addition of $\fsg$ to the combination \Planck \dataplus BAO \dataplus Pantheon, increases the mean value of $H_{0}$ slightly and decreases the mean value of $\Omega_{m}$ very slightly keeping $\se$ almost unchanged. The mean value of $\beta_{0} (=0.0446^{+0.0370+0.0724}_{-0.0370-0.0726})$ decreases slightly with zero in the $2\sigma$ region. Clearly, the addition of datasets do not improve the $H_{0}$ and $\se$ tension in \modelearly. Addition of the datasets significantly reduces the error bars. The correlations between the parameters for \modelearly remain same as in \modellate. 
%%Table II
\begin{center}
\begin{table*}[!h]
\centering
\caption{Observational constraints on the nine dependent model parameters with three derived parameters separated by a horizontal line and the error bars correspond to $68\%$ confidence level for {\bf Model E}, using different observational datasets.}
\label{tab6:mean-2}
\begin{adjustbox}{width=\textwidth}
\begin{tabular} { l  c c c c c}
\hline\hline \noalign{\vskip 2pt}
 {\normalsize Parameter} &  {\normalsize \Planck} &  {\normalsize \Planck \dataplus $f\sigma_{8}$} &  {\normalsize \Planck \dataplus BAO} &  {\normalsize \thead{\Planck \\ \dataplus BAO \dataplus Pantheon}} &  {\normalsize \thead{\Planck \dataplus BAO \\ \dataplus Pantheon \dataplus $f\sigma_{8}$}}\\
\hline \noalign{\vskip 2pt}
\rule[-1.5ex]{0pt}{2.7ex}{\boldmath$\Omega_b h^2   $} & $0.022358\pm 0.000165      $ & $0.022490\pm 0.000162      $ & $0.022489\pm 0.000156      $ & $0.022500\pm 0.000152      $ & $0.022546\pm 0.000151      $\\
\rule[-1.5ex]{0pt}{2.7ex}{\boldmath$\Omega_c h^2   $} & $0.12008\pm 0.00126        $ & $0.11848\pm 0.00117        $ & $0.11850\pm 0.00101        $ & $0.118405\pm 0.000970      $ & $0.117845\pm 0.000909      $\\
\rule[-1.5ex]{0pt}{2.7ex}{\boldmath$100\theta_{MC} $} & $1.040769\pm 0.000324      $ & $1.040941\pm 0.000318      $ & $1.040941\pm 0.000313      $ & $1.040945\pm 0.000315      $ & $1.040999\pm 0.000313      $\\
\rule[-1.5ex]{0pt}{2.7ex}{\boldmath$\tau           $} & $0.05466^{+0.00699}_{-0.00779}$ & $0.05630^{+0.00703}_{-0.00797}$ & $0.05704^{+0.00704}_{-0.00792}$ & $0.05697\pm 0.00749        $ & $0.05778^{+0.00700}_{-0.00790}$\\
\rule[-1.5ex]{0pt}{2.7ex}{\boldmath$\beta_0        $} & $0.0339\pm 0.0372          $ & $0.0395\pm 0.0381          $ & $0.0432\pm 0.0376          $ & $0.0448\pm 0.0377          $ & $0.0446\pm 0.0370          $\\
\rule[-1.5ex]{0pt}{2.7ex}{\boldmath$w_0            $} & $< -0.914                  $ & $< -0.977                  $ & $< -0.969                  $ & $< -0.981                  $ & $< -0.985                  $\\
\rule[-1.5ex]{0pt}{2.7ex}{\boldmath$w_{1}          $} & $< 0.168                   $ & $< 0.0645                  $ & $< 0.0707                  $ & $< 0.0604                  $ & $< 0.0489                  $\\
\rule[-1.5ex]{0pt}{2.7ex}{\boldmath${\rm{ln}}(10^{10} A_s)$} & $3.0486\pm 0.0147          $ & $3.0488\pm 0.0148          $ & $3.0509\pm 0.0148          $ & $3.0507\pm 0.0144          $ & $3.0511\pm 0.0146          $\\
\rule[-1.5ex]{0pt}{2.7ex}{\boldmath$n_s            $} & $0.96315\pm 0.00453        $ & $0.96681\pm 0.00434        $ & $0.96652\pm 0.00419        $ & $0.96672\pm 0.00418        $ & $0.96802\pm 0.00404        $\\
\hline \noalign{\vskip 2pt}
\rule[-1.5ex]{0pt}{2.7ex}$H_0 \left[\mbox{km s}^{-1} \mbox{Mpc}^{-1}\right]                       $ & $64.12^{+2.40}_{-1.39}     $ & $67.00^{+1.02}_{-0.702}    $ & $66.787^{+0.775}_{-0.600}  $ & $67.200^{+0.577}_{-0.516}  $ & $67.631\pm 0.516           $\\
\rule[-1.5ex]{0pt}{2.7ex}$\Omega_m                  $ & $0.3492^{+0.0149}_{-0.0282}$ & $0.31569^{+0.00834}_{-0.0114}$ & $0.31765^{+0.00678}_{-0.00812}$ & $0.31353^{+0.00590}_{-0.00658}$ & $0.30842\pm 0.00588        $\\
\rule[-1.5ex]{0pt}{2.7ex}$\sigma_8                  $ & $0.7836^{+0.0221}_{-0.0138}$ & $0.80265^{+0.00992}_{-0.00800}$ & $0.8019^{+0.0102}_{-0.00866}$ & $0.80539\pm 0.00830        $ & $0.80573\pm 0.00774        $\\
\hline\hline
\end{tabular}
\end{adjustbox}
\end{table*}
%\end{center}
%%Figure II
%\begin{center}
\begin{figure}[!h]
        \centering
         \includegraphics[width=0.7\linewidth]{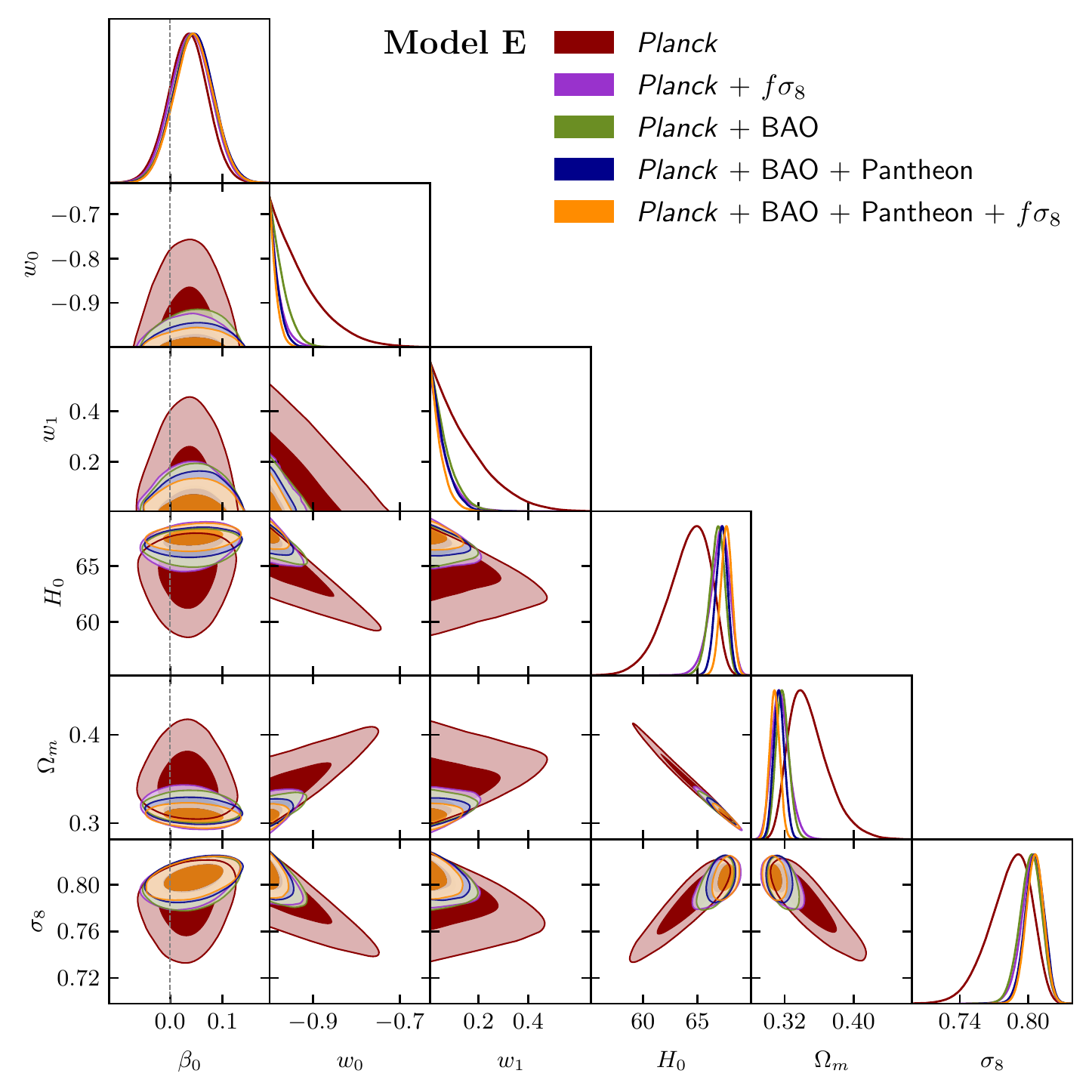}
         \caption{Plot of 1-dimensional marginalised posterior distributions and 2-dimensional marginalised constraint contours on the parameters of {\bf Model E} containing $68\%$ and $95\%$ probability. The dashed line represents the $\beta_{0}=0$ value.}\label{im6:tri2}
\end{figure}
\end{center}

\subsection{Model C: $\bela = \beta_{0}$}

The mean values of the parameters with $1\sigma$ errors for \modelcons are given in Table \ref{tab6:mean-3}. In Table \ref{tab6:mean-3}, the mean values and $1\sigma$ errors of the three derived parameters, $H_{0}$, $\Omega_{m}$ and $\se$, are also quoted. The correlations between the model parameters ($\beta_{0}$, $w_{0}$, $w_{1}$) and the derived parameters ($H_{0}$, $\Omega_{m}$, $\se$) along with their marginalised contours are shown in Fig.\ \ref{im6:tri3}. The parameter values of \modelcons are very close to those of \modellate and they respond to the datasets in the similar fashion as well. Similar to \modellate and \modelearly, $H_{0}$ tension is at $\sim4\sigma$ and only the \Planck data alleviates the $\se$ tension in \modelcons whereas consideration of other datasets restore the tension. The main difference is that the mean values of $\beta_{0}$ in \modelcons is slightly smaller than that in \modellate. These features are clearly seen from Table \ref{tab6:mean-3}.
%%Table III
\begin{center}
\begin{table}[!h]
\centering
\caption{Observational constraints on the nine dependent model parameters with three derived parameters separated by a horizontal line and the error bars correspond to $68\%$ confidence level for {\bf Model C}, using different observational datasets.}
\label{tab6:mean-3}
\begin{adjustbox}{width=\textwidth}
\begin{tabular} { l  c c c c c}
\hline\hline \noalign{\vskip 2pt}
 {\normalsize Parameter} &  {\normalsize \Planck} &  {\normalsize \Planck \dataplus $f\sigma_{8}$} &  {\normalsize \Planck \dataplus BAO} &  {\normalsize \thead{\Planck \\ \dataplus BAO \dataplus Pantheon}} &  {\normalsize \thead{\Planck \dataplus BAO \\ \dataplus Pantheon \dataplus $f\sigma_{8}$}}\\
\hline \noalign{\vskip 2pt}
\rule[-1.5ex]{0pt}{2.7ex}{\boldmath$\Omega_b h^2   $} & $0.022358\pm 0.000164      $ & $0.022482\pm 0.000164      $ & $0.022487\pm 0.000156      $ & $0.022499\pm 0.000151      $ & $0.022545\pm 0.000152      $\\
\rule[-1.5ex]{0pt}{2.7ex}{\boldmath$\Omega_c h^2   $} & $0.12007\pm 0.00128        $ & $0.11854\pm 0.00118        $ & $0.11849\pm 0.00100        $ & $0.118388\pm 0.000977      $ & $0.117824\pm 0.000935      $\\
\rule[-1.5ex]{0pt}{2.7ex}{\boldmath$100\theta_{MC} $} & $1.040772\pm 0.000322      $ & $1.040939\pm 0.000321      $ & $1.040947\pm 0.000314      $ & $1.040954\pm 0.000312      $ & $1.041011\pm 0.000311      $\\
\rule[-1.5ex]{0pt}{2.7ex}{\boldmath$\tau           $} & $0.05491\pm 0.00757        $ & $0.05638^{+0.00708}_{-0.00788}$ & $0.05718^{+0.00685}_{-0.00788}$ & $0.05730\pm 0.00751        $ & $0.05800^{+0.00707}_{-0.00790}$\\
\rule[-1.5ex]{0pt}{2.7ex}{\boldmath$\beta_0        $} & $0.00624\pm 0.00673        $ & $0.00621\pm 0.00626        $ & $0.00696\pm 0.00629        $ & $0.00708\pm 0.00631        $ & $0.00693\pm 0.00615        $\\
\rule[-1.5ex]{0pt}{2.7ex}{\boldmath$w_0            $} & $< -0.907                  $ & $< -0.977                  $ & $< -0.969                  $ & $< -0.981                  $ & $< -0.985                  $\\
\rule[-1.5ex]{0pt}{2.7ex}{\boldmath$w_{1}          $} & $< 0.174                   $ & $< 0.0681                  $ & $< 0.0728                  $ & $< 0.0610                  $ & $< 0.0511                  $\\
\rule[-1.5ex]{0pt}{2.7ex}{\boldmath${\rm{ln}}(10^{10} A_s)$} & $3.0493\pm 0.0146          $ & $3.0489\pm 0.0147          $ & $3.0511\pm 0.0146          $ & $3.0511\pm 0.0144          $ & $3.0514\pm 0.0147          $\\
\rule[-1.5ex]{0pt}{2.7ex}{\boldmath$n_s            $} & $0.96331\pm 0.00444        $ & $0.96670\pm 0.00435        $ & $0.96670\pm 0.00416        $ & $0.96695\pm 0.00415        $ & $0.96823\pm 0.00414        $\\
\hline \noalign{\vskip 2pt}
\rule[-1.5ex]{0pt}{2.7ex}$H_0 \left[\mbox{km s}^{-1} \mbox{Mpc}^{-1}\right]                      $ & $63.93^{+2.51}_{-1.44}     $ & $66.93^{+1.03}_{-0.714}    $ & $66.759^{+0.795}_{-0.594}  $ & $67.187^{+0.581}_{-0.516}  $ & $67.606^{+0.552}_{-0.493}  $\\
\rule[-1.5ex]{0pt}{2.7ex}$\Omega_m                  $ & $0.3513^{+0.0153}_{-0.0299}$ & $0.31643^{+0.00843}_{-0.0115}$ & $0.31789^{+0.00671}_{-0.00832}$ & $0.31361\pm 0.00631        $ & $0.30860\pm 0.00605        $\\
\rule[-1.5ex]{0pt}{2.7ex}$\sigma_8                  $ & $0.7821^{+0.0232}_{-0.0140}$ & $0.8026^{+0.0100}_{-0.00823}$ & $0.8020^{+0.0104}_{-0.00879}$ & $0.80558\pm 0.00856        $ & $0.80564\pm 0.00807        $\\
\hline\hline
\end{tabular}
\end{adjustbox}
\end{table}
%\end{center}
%%Figure III
%\begin{center}
\begin{figure}[!h]
        \centering
         \includegraphics[width=0.7\linewidth]{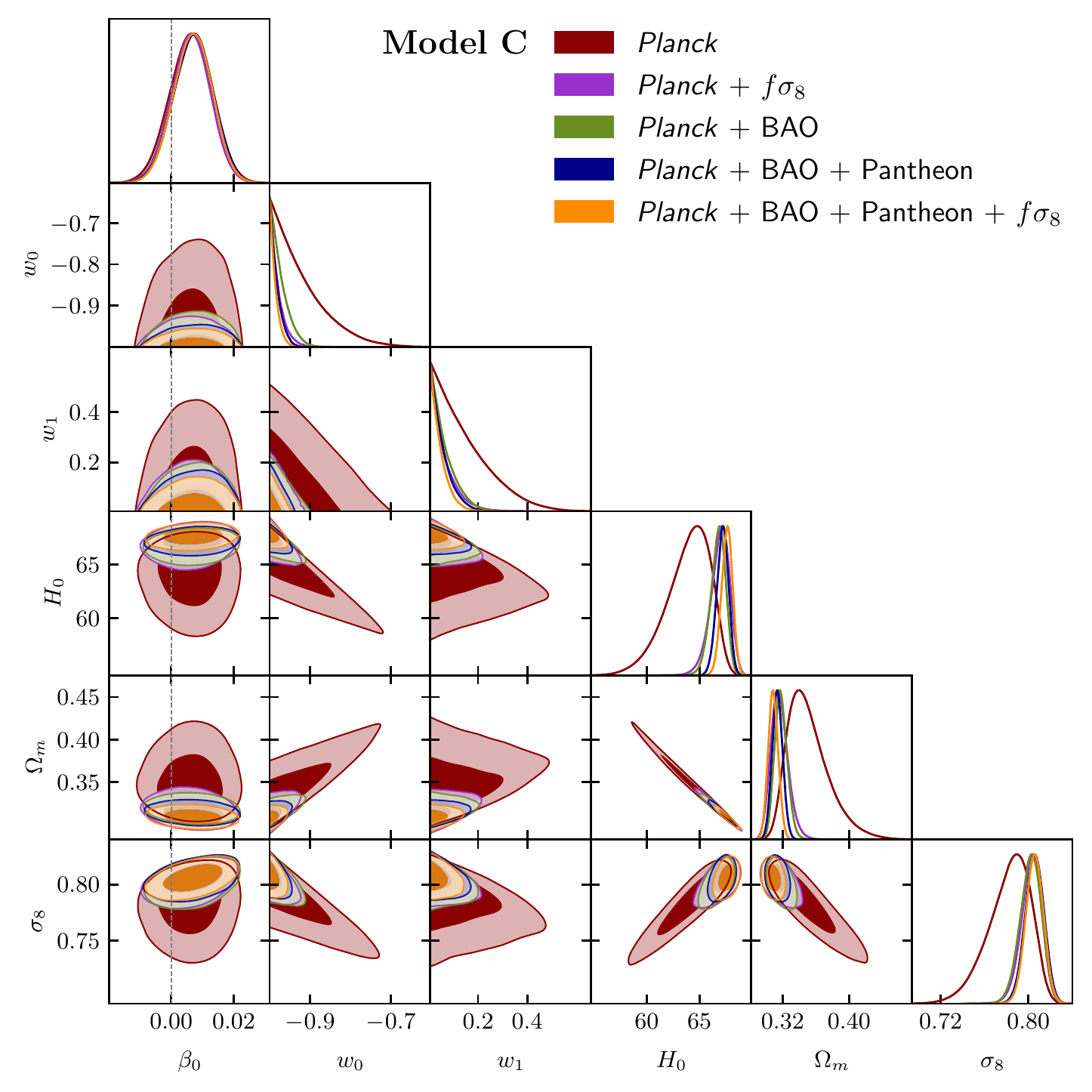}
         \caption{Plot of 1-dimensional marginalised posterior distributions and 2-dimensional marginalised constraint contours on the parameters of {\bf Model C} containing $68\%$ and $95\%$ probability. The dashed line represents the $\beta_{0}=0$ value.}\label{im6:tri3}
\end{figure}
\end{center}

%%%%%%%%%%%%%%%%%%%%%%%%%%%%%%%%%%%%%%%%%%%%%%%%%%%%%%
%%%%%%%%%%%%%%%%%%%%%%%%%%%%%%%%%%%%%%%%%%%%%%%%%%%%%%
\section{Bayesian Evidence}\label{sec6:evi}
Finally, we aim to investigate which one of \modellate, \modelearly and \modelcons is statistically favoured by the observational data. Hence, the Bayesian evidence or more precisely, the logarithm of the Bayes factor, $\ln B_{ij}$ given in Eq.\ (\ref{eq6:bayesfac}), for each of the three models is calculated. Here, $i$ corresponds to \modellate, \modelearly and \modelcons for each of the the dataset combination, $j$ corresponds to the reference model, $M_{j}$. The details on the Bayes factor, $B_{ij}$, are discussed in Appendix \ref{appen:b}. The fiducial \lcdm model is considered to be the reference model, and therefore, a negative value ($\ln B_{ij} < 0$) indicates a preference for the \lcdm model. The logarithmic Bayes factor, $\ln B_{ij}$, is calculated directly from the MCMC chains using the publicly available cosmological package \mcevidence\footnote{Available on GitHub: \href{https://github.com/yabebalFantaye/MCEvidence}{https://github.com/yabebalFantaye/MCEvidence}}~\cite{heavens2017prl,heavens2017ax}. The computed values of $\ln B_{ij}$ for \modellate, \modelearly and \modelcons are summarised in Table \ref{tab6:evidence}. From Table \ref{tab6:evidence}, it is clear that the \lcdm model is preferred over the interacting models by all the dataset combinations. However, the motive is to assess if there is any observationally preferable evolution stage when the interaction is significant. As can be seen from the relative differences of $\abs{\ln B_{ij}}$ (values corresponding to column $\Delta \ln B_{ij}$) in Table \ref{tab6:evidence}), when compared with \modelcons, \modellate is strongly disfavoured while \modelearly is weakly disfavoured by observational data over \modelcons.
\begin{table}[!htbp]
\begin{center}
\caption{The values of $\ln B_{ij}$, where $j$ is the \lcdm model and $i$ is the interacting model. A negative sign indicates $M_{j}$ is favoured over $M_{i}$. The $\abs{\ln B_{ij}}$ values are compared with Table \ref{tab6:bij-values}. The column $\Delta \ln B_{ij}$ corresponds to the comparison of {\bf Model L} and {\bf Model E} with {\bf Model C}.
  }\label{tab6:evidence}
\begin{adjustbox}{width=0.7\textwidth}
\begin{tabular}{ccccc}
\hline \hline
\rule[-1ex]{0pt}{2.5ex}Model& Dataset&\hspace{0ex} ~~$\ln B_{ij}$&\hspace{0.5ex}~~~$\Delta \ln B_{ij}$\\
    \hline
    \multirow{5}{*}{\rule[-1ex]{0pt}{2.5ex}{\bf Model L}}& \Planck &\hspace{0ex} $-8.843$&\hspace{0.5ex} $-2.244$ \\
    \rule[-1ex]{0pt}{2.5ex}& \Planck \dataplus $f\sigma_{8}$ &\hspace{0ex} $-11.410$&\hspace{0.5ex} $-2.245$\\
    \rule[-1ex]{0pt}{2.5ex}& \Planck \dataplus BAO &\hspace{0ex} $-10.610$&\hspace{0.5ex} $-2.187$ \\
    \rule[-1ex]{0pt}{2.5ex}& \Planck \dataplus BAO \dataplus Pantheon &\hspace{0ex} $-11.354$&\hspace{0.5ex} $-2.104$\\
    \rule[-1ex]{0pt}{2.5ex}& \Planck \dataplus BAO \dataplus Pantheon \dataplus $f\sigma_{8}$ &\hspace{0ex}$-11.977$&\hspace{0.5ex} $-2.328$ \\
    \hline
    \multirow{5}{*}{\rule[-1ex]{0pt}{2.5ex}{\bf Model E}}& \Planck &\hspace{0ex} $-7.233$  &\hspace{0.5ex} $-0.633$\\
    \rule[-1ex]{0pt}{2.5ex}& \Planck \dataplus $f\sigma_{8}$ &\hspace{0ex} $-9.730$&\hspace{0.5ex} $-0.566$\\
    \rule[-1ex]{0pt}{2.5ex}& \Planck \dataplus BAO &\hspace{0ex} $-9.047$&\hspace{0.5ex} $-0.624$ \\
    \rule[-1ex]{0pt}{2.5ex}& \Planck \dataplus BAO \dataplus Pantheon &\hspace{0ex} $ -9.733$&\hspace{0.5ex} $-0.483$ \\
    \rule[-1ex]{0pt}{2.5ex}& \Planck \dataplus BAO \dataplus Pantheon \dataplus $f\sigma_{8}$ &\hspace{0ex} $-10.192$&\hspace{0.5ex} $-0.542$\\
    \hline
    \multirow{5}{*}{\rule[-1ex]{0pt}{2.5ex}{\bf Model C}}& \Planck &\hspace{0ex} $-6.599$ &\hspace{0.5ex} 0.0\\
    \rule[-1ex]{0pt}{2.5ex}& \Planck \dataplus $f\sigma_{8}$ &\hspace{0ex} $-9.164$&\hspace{0.5ex} 0.0 \\
    \rule[-1ex]{0pt}{2.5ex}&  \Planck \dataplus BAO &\hspace{0ex} $-8.423$&\hspace{0.5ex} 0.0\\
    \rule[-1ex]{0pt}{2.5ex}& \Planck \dataplus BAO \dataplus Pantheon &\hspace{0ex} $-9.250$&\hspace{0.5ex} 0.0\\
    \rule[-1ex]{0pt}{2.5ex}& \Planck \dataplus BAO \dataplus Pantheon \dataplus $f\sigma_{8}$ &\hspace{0ex} $-9.650$&\hspace{0.5ex} 0.0\\
\hline
\hline
\end{tabular}
\end{adjustbox}
\end{center}
\end{table}\newpage

%
%%%%%%%%%%%%%%%%%%%%%%%%%%%%%%%%%%%%%%%%%%%%%%%%%%%%%%
%%%%%%%%%%%%%%%%%%%%%%%%%%%%%%%%%%%%%%%%%%%%%%%%%%%%%%
\section{Summary And Discussion} \label{sec6:sum}
The present work deals with the matter perturbations in a cosmological model where the dark energy has an interaction with the dark matter. We investigate the possibility whether the coupling parameter between the two dark components can evolve. We have considered two new examples, (a) the interaction is a recent phenomenon (\modellate; Eq.\ (\ref{eq6:b1})), and (b) the interaction is an early phenomenon (\modelearly; Eq.\ (\ref{eq6:b2})) and compared them with the normally talked about model where the coupling is a constant (\modelcons; Eq.\ (\ref{eq6:b3})), in the context of density perturbations. The results are compared with the standard \lcdm model as well. The rate of energy transfer is proportional to the dark energy density, $\rde$ and energy flows from dark matter to dark energy. The interaction term is given by Eq.\ (\ref{eq6:q1}). We have also considered the dark energy to have a dynamical EoS parameter, $\wde$ being given by the CPL parametrisation (Eq.\ (\ref{eq6:w-cpl})). 

We have worked out a detailed perturbation analysis of the models in the synchronous gauge and compared them with each other. The background dynamics of the three interacting models are almost the same, which is evident from the smallness of the coupling parameter and the domination of dark energy at late times. The signature of the presence of interaction at different epochs for different couplings are noticeable in the perturbation analysis.

In all the three interacting models, the fractional density perturbation of dark matter is marginally higher than that in a \lcdm model, indicating more clumping of matter. From the CMB temperature spectrum, matter power spectrum and the evolution of growth rate, we note that the presence of interaction for a brief period in the evolutionary history (\modelearly), makes the Universe behave like the \lcdm model with a slightly higher value of $\fsg$ at the epoch when the interaction prevails. The first part of the present work shows that \modelearly behaves in a closely similar fashion as the \lcdm model and leads to the conclusion that \modelearly performs better than \modellate and \modelcons in describing the evolutionary history of the Universe. 

To determine further the evolution stage when the interaction is significant, we have tested the interacting models with the observational datasets. We have tested \modelearly, \modellate and \modelcons against the recent observational datasets like CMB, BAO, Pantheon and RSD with the standard six parameters of \lcdm model along with the three model parameters, $\beta_{0}$, $w_{0}$ and $w_{1}$. We have obtained the mean value of the coupling parameter, $\beta_{0}$ to be positive, indicating an energy flow from dark matter to dark energy. When only CMB data is used, $\beta_{0}=0$ lies within the $1\sigma$ error region while when different combinations of the datasets are used, $\beta_{0}=0$ lies outside the $1\sigma$ error region. The priors of $w_{0}$ and $w_{1}$ are set such that $\wde$ remains in the quintessence region. Hence, $w_{0}$ and $w_{1}$ remain unconstrained. Moreover, for all the three interacting models, the $\se$ tension is alleviated when CMB data is used.  Though the estimated parameter values are prior dependent, it can be said conclusively that the CMB data and RSD data are in agreement when the interacting models are considered. Addition of other datasets restore the $\se$ tension in all the three interacting models. However, the tension in $H_{0}$ value persists for all the three interacting models.

From the Bayesian evidence analysis, we see that all the three interacting dark energy models are rejected by observational data when compared with the fiducial \lcdm model. However, a close scrutiny reveals that both \modelearly and \modelcons are favoured over \modellate. Though the Bayesian evidence analysis ever so slightly favours \modelcons over \modelearly, the difference is too small to choose a clear winner. Thus, to conclude from the results of the perturbation analysis and observational data we infer that the interaction, if present, is likely to be significant only at some early stage of evolution of the Universe.

% Chapter 7
\chapter{Conclusions}
% Main chapter title
\label{chap7}

\section{Summary And Discussion}\label{sec7:sum}
Various independent cosmological observational data like the Type Ia Supernovae (SNe Ia) measurements~\cite{riess1998anj, schmidt1998apj,perlmutter1999apj,scolnic2018apj}, cosmic microwave background (CMB)~\cite{eisenstein1998apj, planck2015cp, planck2018cp}, Particle Data Group~\cite{partphys2018prd}, large scale structure (LSS)~\cite{reid2010mnras, descol2019prl, alam2017sdss3, alam2020sdss4} has consolidated the fact that the Universe is expanding with an acceleration and $68\%$ of the energy content of the Universe of unknown nature is the reason for this acceleration. The cosmological constant $\Lambda$ as dominant component, named \emph{dark energy}, though observationally most favoured, is riddled with problems like cosmological constant and coincidence problem. Hence, other candidates as dark energy have been looked for, and no one of them seems to be a clear winner. Density perturbations may provide a way to distinguish between dark energy models. The motivation of the present thesis is to study linear density perturbations in different DE models. After a brief description of the perturbation equations in chapter \ref{chap2:gi-pert}, the next four chapters contain the perturbation evolution in different dark energy models.

Chapter \ref{chap3:grg} discusses density perturbations in dark energy models that are reconstructed from the kinematical quantity jerk. The models, already there in the literature~\cite{mukherjee2016prd, mukherjee2017cqg} are reconstructed independent of any prior assumption about the theory of gravity or the nature of the DE, and considers only a parametric ansatz for jerk parameter. The present work deals with the possibility whether these reconstructed models can successfully generate large scale structures from fluctuations. We considered two types of reconstructed models -- one in which the components of the dark sector conserves independently and the other in which there is an energy transfer between DM and DE. It can be said quite conclusively that the reconstructed models with no interaction among the components in the dark sector favourably produce large scale structures in the Universe.

In chapter \ref{chap4:epjp}, we have studied the effects of interaction in density perturbations in a Universe dominated by Holographic dark energy at the present epoch. We have considered the IR cut-off for the HDE to be the future event horizon and the rate of energy transfer proportional to DE. The DE density perturbations are found to grow in the absence of any effective sound speed of DE perturbation ($c_{s,de}^2\equiv \frac{\delta p_{de}}{\delta \rde}=0$).

In chapter \ref{chap5:jcap}, we have introduced a scalar field model that behaves like a cosmological constant at the present epoch but is devoid of the initial condition problem. The potential is so designed that at an early epoch, the scalar field starts rolling down an exponential potential and tracks the dominant energy component of the Universe until recently when it lands on a constant potential and rolls sufficiently slowly to produce the late time acceleration. The EoS parameter for the scalar field at the present epoch is $\wphi = -1$ and is independent of the model parameters. Though the background evolution in the recent past is similar to the \lcdm model, the growth of density perturbations is different from \lcdm. In the present work, we also studied the CMB temperature spectrum and the matter power spectrum and concluded that the power spectra are different from that of the \lcdm model. The striking characteristic of the scalar field model is that it reduces the rms mass fluctuation $\se$ substantially to a value closer to that obtained from the galaxy cluster counts, hinting that it might alleviate the $\se$ tension. 

Chapter \ref{chap6:prd} deals with the matter perturbations in a cosmological model where DE interacts with DM, and interaction between the two has an evolving coupling parameter. We have considered two possibilities --- coupling parameter dominates at the late time (\modellate), and coupling parameter dominates at the early epoch (\modelearly). We have assumed the DE to be a fluid with a dynamical EoS parameter described by the CPL parametrisation. We compared the two models with the well-known case of constant coupling parameter (\modelcons) and the \lcdm model. This work aims to study the effect of dynamical coupling parameter on the growth of density perturbations and different power spectra. It can be said conclusively that \modelearly is favoured over \modellate and \modelcons in describing the evolution of the perturbations in the Universe. The models are then tested against recent observational datasets using MCMC to obtain the mean values of the model parameters. From the Bayesian evidence analysis, we have shown that all the three interacting models perform worse than the fiducial \lcdm model. Amongst the interacting models, \modelearly and \modelcons are marginally favoured over \modellate. Thus, to conclude from the results of the perturbation analysis and observational data, we infer that the interaction, if at all present, is likely to be significant only at some early stage of the evolution of the Universe. 

%
%

%

%-------------------------------------------------------------------------------
%	THESIS CONTENT - APPENDICES
%-------------------------------------------------------------------------------
%\begin{appendices} % Cue to tell LaTeX that the following "chapters" are Appendices
\appendix % Cue to tell LaTeX that the following "chapters" are Appendices
%
% Appendix A
\chapter{Coefficients Of The Coupled Differential Equations}% Main appendix title
\label{appen:epjp}% For referencing this appendix elsewhere, use \ref{AppendixA}
\sloppy
The coefficients of equations (\ref{eq4:finaldm}) and (\ref{eq4:finalde}) are given below.
\begin{itemize}
\item[(i)] The coefficients of equation (\ref{eq4:finaldm}) are:
\begin{myequation}
\mathbf{C^{(m)}_1} = -H_0 E  ~,
\end{myequation}
%%%%%%%%%%%%%%%%%%%%%%%%%%%%%%%%%%%
\begin{myequation}
\mathbf{C^{(m)}_2} =
-E   ((2 \beta -1) H_0 \Omega_{de}  +H_0)/(z+1) (\Omega_{de}  -1)-H_0 E ' ~,
\end{myequation}
%%%%%%%%%%%%%%%%%%%%%%%%%%%%%%%%%%%%
%%%%%%%%%%%%%%%%%%%
\begin{myequation}
\begin{split}
\mathbf{C^{(m)}_3} =
&-\left.\left(H_0 \beta  \left((z+1) (\Omega_{de}  -1) \Omega_{de}   E '\right.\right.\right. +\\
&\left.\left.\left. E \left(2 \Omega_{de}  +(\beta -2) \Omega_{de}^2- (z+1) \Omega_{de} ' \right)\right)\right)\right/(z+1)^2 (\Omega_{de}  -1)^2~,
\end{split}
\end{myequation}
%%%%%%%%%%%%%%%%%%%%%%%%%%%%%%%%%%%
\begin{myequation}
\mathbf{C^{(m)}_4} = 0 ~,
\end{myequation}
%%%%%%%%%%%%%%%%%%%%%%%%%%%%%%%%%%%
\begin{myequation}
\begin{split}
\mathbf{C^{(m)}_5} = 
& \left(H_0 \beta  E   \Omega_{de}   \left(-9 H_0^2 w_{de}^2 E^2+\right.\right.
 w_{de}   \left(3 H_0^2 \left(-\beta +3 c_{s,de}^2-3\right) E^2+k^2 (z+1)^2\right)+\\
& \left.\left.\left.3 H_0^2 E^2 \left((\beta +3) c_{s,de}^2-(z+1) w_{de} ' \right)\right)\right)\right/(z+1 
 \Omega_{de}  -1\\
&  \left(-k^2 (z+1)^2-9 H_0^2 w_{de}^2 E^2+\right.\\
& w_{de}   \left(3 H_0^2 \left(-\beta +3 c_{s,de}^2-3\right) E^2+k^2 (z+1)^2\right)+\\
&
 \left.\left.3 H_0^2 E^2 \left((\beta +3) c_{s,de}^2-(z+1) w_{de} ' \right)\right)\right) ~,
\end{split}
\end{myequation}
%%%%%%%%%%%%%%%%%%%%%%%%%%%%%%%%%%%
\begin{myequation}
\begin{split}
\mathbf{C^{(m)}_6} = 
& (H_0 \beta  (E   \Omega_{de}  -
 E   (\Omega_{de}  -1) \Omega_{de}  -
 E   (\Omega_{de}^2+\beta  E   (\Omega_{de}^2+\\
 &
 \left.3 c_{s,de}^2 k^2 (z+1) E   (\Omega_{de}  -1) \Omega_{de}  \right/
 \left(k^2 (z+1)^2+\right.\\
& \left.9 H_0^2 w_{de}^2 E^2-\right.
 w_{de}   \left(3 H_0^2 \left(-\beta +3 c_{s,de}^2-3\right) E^2+k^2 (z+1)^2\right)+\\
& \left.3 H_0^2 E^2 \left((z+1) w_{de} ' -(\beta +3) c_{s,de}^2\right)\right)+\\
&
 \left.3 c_{s,de}^2 k^2 z (z+1) E   (\Omega_{de}  -1) \Omega_{de}  \right/
 \left(k^2 (z+1)^2+9 H_0^2 w_{de}^2 E^2-\right.\\
& w_{de}   \left(3 H_0^2 \left(-\beta +3 c_{s,de}^2-3\right) E^2+k^2 (z+1)^2\right)+\\
& \left.3 H_0^2 E^2 \left((z+1) w_{de} ' -(\beta +3) c_{s,de}^2\right)\right)-
 \left(3 k^2 (z+1) w_{de}   E  \right.
 (\Omega_{de}  -1) \Omega_{de}  )/\\
& \left(k^2 (z+1)^2+9 H_0^2 w_{de}^2 E^2-\right.
 w_{de}   \left(3 H_0^2 \left(-\beta +3 c_{s,de}^2-3\right) E^2+k^2 (z+1)^2\right)+\\
& \left.3 H_0^2 E^2 \left((z+1) w_{de} ' -(\beta +3) c_{s,de}^2\right)\right)-
 \left(3 k^2 z (z+1) w_{de}   E  \right.\\
& (\Omega_{de}  -1) \Omega_{de}  )/
 \left(k^2 (z+1)^2+9 H_0^2 w_{de}^2 E^2-\right.\\
 &
 w_{de}   \left(3 H_0^2 \left(-\beta +3 c_{s,de}^2-3\right) E^2+k^2 (z+1)^2\right)+\\
& \left.3 H_0^2 E^2 \left((z+1) w_{de} ' -(\beta +3) c_{s,de}^2\right)\right)+
 (z+1) (\Omega_{de}  -1) \Omega_{de}   E ' +\\
& (z+1) E   (\Omega_{de}  -1) \Omega_{de} ' -
 \left.\left.\left.(z+1) E   \Omega_{de}   \Omega_{de} ' \right)\right)\right/
 (z+1)^2 (\Omega_{de}  -1)^2 ~,
\end{split}
\end{myequation}
%%%%%%%%%%%%%%%%%%%%%%%%%%%%%%%%%%%
\begin{myequation}
\mathbf{C^{(m)}_7} = 3 H_0 E   ~,
\end{myequation}
%%%%%%%%%%%%%%%%%%%%%%%%%%%%%%%%%%%
\begin{myequation}
\begin{split}
\mathbf{C^{(m)}_8} = 
& \left(H_0 \left(3 \beta  k^2 E   \Omega_{de}  +6 \beta  k^2 z E  \right.\right.
 \Omega_{de}  +\\
 &3 \beta  k^2 z^2 E   \Omega_{de}  -
 3 \beta  k^2 w_{de}   E   \Omega_{de}  -
 6 \beta  k^2 z w_{de}   E   \Omega_{de}  -
 3 \beta  k^2 z^2 w_{de}   E   \Omega_{de}  -\\
& 3 E   \left(k^2 (z+1)^2+9 H_0^2 w_{de}^2 E^2-\right.
 w_{de}   \left(3 H_0^2 \left(-\beta +3 c_{s,de}^2-3\right) E^2+k^2 (z+1)^2\right)+\\
& \left.3 H_0^2 E^2 \left((z+1) w_{de} ' -(\beta +3) c_{s,de}^2\right)\right)+
 3 E   \Omega_{de}  
 \left(k^2 (z+1)^2+9 H_0^2 w_{de}^2 E^2-\right.\\
& w_{de}   \left(3 H_0^2 \left(-\beta +3 c_{s,de}^2-3\right) E^2+k^2 (z+1)^2\right)+\\
&
 \left.3 H_0^2 E^2 \left((z+1) w_{de} ' -(\beta +3) c_{s,de}^2\right)\right)-
 4 \beta  E   \Omega_{de}  \\
& \left(k^2 (z+1)^2\right.+\left.9 H_0^2 w_{de}^2 E^2-\right.
 w_{de}   \left(3 H_0^2 \left(-\beta +3 c_{s,de}^2-3\right) E^2+k^2 (z+1)^2\right)+\\
& \left.3 H_0^2 E^2 \left((z+1) w_{de} ' -(\beta +3) c_{s,de}^2\right)\right)+\\
&
 3 (z+1) (1-\Omega_{de}  )
 \left(k^2 (z+1)^2+9 H_0^2 w_{de}^2 E^2-\right.\\
& w_{de}   \left(3 H_0^2 \left(-\beta +3 c_{s,de}^2-3\right) E^2+k^2 (z+1)^2\right)+\\
&
 \left.3 H_0^2 E^2 \left((z+1) w_{de} ' -(\beta +3) c_{s,de}^2\right)\right) \\
& \left.\left.\left.E ' \right)\right)\right/((z+1) (1-\Omega_{de}  )
 \left(k^2 (z+1)^2+9 H_0^2 w_{de}^2 E^2-\right.\\
 &
 w_{de}   \left(3 H_0^2 \left(-\beta +3 c_{s,de}^2-3\right) E^2+k^2 (z+1)^2\right)+\\
& \left.\left.3 H_0^2 E^2 \left((z+1) w_{de} ' -(\beta +3) c_{s,de}^2\right)\right)\right) ~,
\end{split}
\end{myequation}
%%%%%%%%%%%%%%%%%%%%%%%%%%%%%%%%%%%
\begin{myequation}
\begin{split}
\mathbf{C^{(m)}_9} = 
& -\frac{k^2}{H_0 E  }+\frac{\beta  H_0 \Omega_{de}   E ' }{(z+1) (\Omega_{de}  -1)}+\\
&
 (H_0 \beta  E   (\Omega_{de}  -(\Omega_{de}  -1)
 \Omega_{de}  -(\Omega_{de}^2+\beta  (\Omega_{de}^2+
 \left.\beta  k^2 (z+1) (\Omega_{de}  -1) \Omega_{de}  \right/\\
& \left(k^2 (z+1)^2+9 H_0^2 w_{de}^2 E^2-\right.
 w_{de}   \left(3 H_0^2 \left(-\beta +3 c_{s,de}^2-3\right) E^2+k^2 (z+1)^2\right)+\\
& \left.3 H_0^2 E^2 \left((z+1) w_{de} ' -(\beta +3) c_{s,de}^2\right)\right)+\\
&
 \left.\beta  k^2 z (z+1) (\Omega_{de}  -1) \Omega_{de}  \right/
 \left(k^2 (z+1)^2+9 H_0^2 w_{de}^2 E^2-\right.\\
& w_{de}   \left(3 H_0^2 \left(-\beta +3 c_{s,de}^2-3\right) E^2+k^2 (z+1)^2\right)+\\
&
 \left.3 H_0^2 E^2 \left((z+1) w_{de} ' -(\beta +3) c_{s,de}^2\right)\right)+
 (z+1) (\Omega_{de}  -1) \Omega_{de} ' -\\
& \left.\left.\left.(z+1) \Omega_{de}   \Omega_{de} ' \right)\right)\right/
 (z+1)^2 (\Omega_{de}  -1)^2 ~.
\end{split}
\end{myequation}
%%%%%%%%%%%%%%%%%%%%%%%%%%%%%%%%%%%%%%%%%%%%%%%%%%%%%%%%%%%%%%%%%%%%%%%%%%%%%%%%%%%%%%%%%%%%%%%%%%%%%%%%%%%%%%%%%%%%%%%%%%%%%%%%%%%%%%%%%%%%%%%%%%%%%%%%%%%%%%%%%%%%%
\item[(ii)] The coefficients of equation (\ref{eq4:finalde}) are:
\begin{myequation}
\mathbf{C^{(de)}_1}=-H_0 E  ~,
\end{myequation}
%%%%%%%%%%%%%%%%%%%%%%%%%%%%%%%%%%%
\begin{myequation}
\begin{split}
\mathbf{C^{(de)}_2}=
& -\left(\left(H_0 \left(-k^2 (z+1)^2 E   \left(-1-\beta  c_{s,de}^2+\right.\right.\right.\right.
 (\Omega_{de}^2-3 (\Omega_{de}^3+(z+1) w_{de} ' +\\
& \left.w_{de}   \left((z+1) w_{de} ' +\beta  c_{s,de}^2+3\right)\right)+
 k^2 (z+1)^3 \left((\Omega_{de}^2-1\right) E ' +\\
& 3 H_0^2 (z+1) (w_{de}  +1) E^2
  \left((\beta +3) \left(-c_{s,de}^2\right)+\left(\beta -3 c_{s,de}^2+3\right) w_{de}  +\right.\\
& \left.(z+1) w_{de} ' +3 (\Omega_{de}^2\right) E ' +
 3 H_0^2 E^3 \left(3 c_{s,de}^2+\beta  c_{s,de}^2-3 \beta  c_{s,de}^4-\beta ^2 c_{s,de}^4+\right.\\
& 3 \left(-\beta +3 c_{s,de}^2-7\right) (\Omega_{de}^3-9 (\Omega_{de}^4+
 (z+1) \left(\beta +(\beta -3) c_{s,de}^2+3\right) w_{de} ' +\\
& (3 (z+1) \Omega_{de}^2 w_{de} ' -4 \beta +3 (2 \beta +7) c_{s,de}^2-15+
(z+1)^2 w_{de} ''  +\\
& w_{de}   \left(-\beta -3 \beta  c_{s,de}^4+\beta ^2 c_{s,de}^2+7 \beta  c_{s,de}^2+15 c_{s,de}^2-(z+1)-3\right.\\
& \left.\left.\left.\left.\left. \left(-\beta +3 c_{s,de}^2-6\right) w_{de} ' +(z+1)^2 w_{de} '' \right)\right)\right)\right)\right/\\
&
 \left(z+1 w_{de}  +1 \left(-9 H_0^2 (\Omega_{de}^2-k^2 (z+1)^2\right.\right.
 E^2+\\
& w_{de}   \left(3 H_0^2 \left(-\beta +3 c_{s,de}^2-3\right) E^2+k^2 (z+1)^2\right)+\\
&
 \left.\left.\left.3 H_0^2 E^2 \left((\beta +3) c_{s,de}^2-(z+1) w_{de} ' \right)\right)\right)\right) ~,
\end{split}
\end{myequation}
%%%%%%%%%%%%%%%%%%%%%%%%%%%%%%%%%%%
\begin{myequation}
\begin{split}
\mathbf{C^{(de)}_3}=
& \left(c_{s,de}^2 k^4 (z+1)^4-3 H_0^2 k^2 (z+1)^2 E^2\right.
 -\left(3 c_{s,de}^2-1\right) (z+1) w_{de} ' +\\
 &3 (\beta +1) c_{s,de}^4+2 c_{s,de}^2+ 3 c_{s,de}^2 H_0^2 k^2 (z+1)^3 E   E ' +\\
 &
 27 H_0^4 (z+1) (\Omega_{de}^4 E^3 E ' +
 9 c_{s,de}^2 H_0^4 (z+1) E^3 \left((\beta +3) c_{s,de}^2-(z+1) w_{de} ' \right)\\
& 3 H_0^2 (\Omega_{de}^3 E  +E ' 
 \left(9 H_0^2 E^3-\left(3 c_{s,de}^2+2\right) k^2 (z+1)^2 E  \right.
 (z+1) w_{de} ' +\\
 &2 \beta  c_{s,de}^2+k^2 (z+1)^3 E ' - \left.3 H_0^2 (z+1) \left(-\beta +6 c_{s,de}^2-6\right) E^2 E ' \right)+\\
&
 9 H_0^4 E^4 \left(2 \beta  (\beta +3) c_{s,de}^6-c_{s,de}^2 (z+1) \left(3 \beta  c_{s,de}^2+1\right)\right.\\
&  w_{de} ' +\left(c_{s,de}^2-1\right) (z+1)^2 \left(w_{de} ' \right)^2-
 \left.c_{s,de}^2 (z+1)^2 w_{de} '' \right)+\\
&(\Omega_{de}^2 
 \left(c_{s,de}^2 k^4 (z+1)^4+3 c_{s,de}^2 H_0^2 k^2 (z+1)^2 \left(-3 \beta +3 c_{s,de}^2+2\right) E^2-\right.\\
& 3 c_{s,de}^2 H_0^2 k^2 (z+1)^3 E   E ' +\\
&9 H_0^4 (z+1)
 E^3 \left((z+1) w_{de} ' +\beta +3 c_{s,de}^4-2 (\beta +6) c_{s,de}^2+3\right)\\
&  E ' -9 H_0^4 E^4 \left(2 \beta  c_{s,de}^2 \left(-\beta +6 c_{s,de}^2-3\right)+\right.\\
&
 \left.\left.\left(3 c_{s,de}^2-4\right) (z+1) w_{de} ' -(z+1)^2 w_{de} '' \right)\right)+\\
& w_{de}   \left(-3 H_0^2 k^2 (z+1)^2 E^2-2 c_{s,de}^2 k^4 (z+1)^4\right.\\
& \left(\left(c_{s,de}^2+1\right) (z+1) w_{de} ' -3 \beta  c_{s,de}^4-3 (\beta +1) c_{s,de}^2-2\right)-\\
&27 H_0^6 (z+1) E^3+k^2 (z+1)^3 E   E ' \\
& -\left(c_{s,de}^2-1\right) (z+1) w_{de} ' +(\beta +6) c_{s,de}^4-2 (\beta +3) c_{s,de}^2
 9 H_0^4 E^4+E ' \\
& \left(6 \beta  c_{s,de}^6-4 \beta  (\beta +3) c_{s,de}^4+(z+1) \left((3 \beta -4) c_{s,de}^2+1\right) w_{de} ' -\right.\\
& \left.\left.\left.\left.(z+1)^2 \left(w_{de} ' \right)^2-\left(c_{s,de}^2-1\right) (z+1)^2 w_{de} '' \right)\right)\right)\right/\\
&
 \left(H_0 (z+1)^2 (w_{de}  +1) E  \right.
 \left(k^2 (z+1)^2+9 H_0^2 (\Omega_{de}^2 E^2-\right.\\
& w_{de}   \left(3 H_0^2 \left(-\beta +3 c_{s,de}^2-3\right) E^2+k^2 (z+1)^2\right)+\\
&
 \left.\left.3 H_0^2 E^2 \left((z+1) w_{de} ' -(\beta +3) c_{s,de}^2\right)\right)\right) ~,
\end{split}
\end{myequation}
%%%%%%%%%%%%%%%%%%%%%%%%%%%%%%%%%%%
\begin{myequation}
\begin{split}
\mathbf{C^{(de)}_4}= \mathbf{C^{(de)}_5}=\mathbf{C^{(de)}_6}=0 ~,
\end{split}
\end{myequation}
%%%%%%%%%%%%%%%%%%%%%%%%%%%%%%%%%%%
\begin{myequation}
\begin{split}
\mathbf{C^{(de)}_7}=3 H_0 (w_{de}  -1) E~,
\end{split}
\end{myequation}
%%%%%%%%%%%%%%%%%%%%%%%%%%%%%%%%%%%
\begin{myequation}
\begin{split}
\mathbf{C^{(de)}_8}=
& \left(H_0 k^2 (z+1)^2 (w_{de}  -1)\right.
 \left(3+3 (\beta -3) c_{s,de}^2+\beta +\left(\beta -3 \beta  c_{s,de}^2\right) w_{de}  +\right.\\
& \left.\left(9 c_{s,de}^2-3\right) (\Omega_{de}^2\right) E  +
 3 H_0 k^2 (z+1)^3 (w_{de}  -1)^2 (w_{de}  +1) E ' +\\
& 9 H_0^3 (z+1) \left((\Omega_{de}^2-1\right) E^2
 \left((\beta +3) \left(-c_{s,de}^2\right)+\left(\beta -3 c_{s,de}^2+3\right) w_{de} \right.\\
 & +3 (\Omega_{de}^2+ \left.(z+1) w_{de} ' \right) E ' -\\
 &3 H_0^3 E^3
 \left(9 c_{s,de}^2+27 c_{s,de}^4-\beta ^2 c_{s,de}^2-3 \beta ^2 c_{s,de}^4+9 \left(3 c_{s,de}^2+1\right) (\Omega_{de}^4-\right.\\
& (z+1) \left(-4 \beta +27 c_{s,de}^2-9\right) w_{de} ' +3 (z+1)^2 \left(w_{de} ' \right)^2-\\
& 3 (\Omega_{de}^3 \left(3 (z+1) w_{de} ' -2 \beta +9 c_{s,de}^4-6 c_{s,de}^2-3\right)+\\
& w_{de}   \left(-9-18 c_{s,de}^2+27 c_{s,de}^4-6 \beta  c_{s,de}^2+\beta ^2+2 \beta ^2 c_{s,de}^2+\right.\\
& 3 \beta ^2 c_{s,de}^4-(z+1) \left(-4 \beta +6 (\beta +3) c_{s,de}^2-27\right) w_{de} ' +
 \left.3 (z+1)^2 \left(w_{de} ' \right)^2\right)+\\
& 3 w_{de} '' +6 z w_{de} '' +3 z^2 w_{de} '' +
 (\Omega_{de}^2 \left(-9-36 c_{s,de}^2-27 c_{s,de}^4+6 \beta \right.-\\
 &\left.6 \beta  c_{s,de}^2+\beta ^2-3 \beta ^2 c_{s,de}^2+\right.\\
& \left.\left.\left.\left.9 \left(c_{s,de}^2+1\right) (z+1) w_{de} ' -3 (z+1)^2 w_{de} '' \right)\right)\right)\right/\\
&
 \left(z+1 w_{de}  +1 \left(-k^2 (z+1)^2-\right.\right.\\
& 9 H_0^2 (\Omega_{de}^2 E^2+
 w_{de}   \left(3 H_0^2 \left(-\beta +3 c_{s,de}^2-3\right) E^2+k^2 (z+1)^2\right)+\\
 &
 \left.\left.3 H_0^2 E^2 \left((\beta +3) c_{s,de}^2-(z+1) w_{de} ' \right)\right)\right) ~,
\end{split}
\end{myequation}
%%%%%%%%%%%%%%%%%%%%%%%%%%%%%%%%%%%
\begin{myequation}
\begin{split}
\mathbf{C^{(de)}_9}=
& \left(81 H_0^4 (\Omega_{de}^5 E^4-9 (\Omega_{de}^4+k^4 (z+1)^4\right.
 3 H_0^4 \left(-2 \beta +6 c_{s,de}^2-9\right) E^4+\\
 &2 H_0^2 k^2 (z+1)^2 E^2+ H_0^2 k^2 (z+1)^2 E^2
 (\beta +6) (z+1) w_{de} ' -\\
 &2 \beta -\left(\beta ^2+3 \beta +18\right) c_{s,de}^2+
 \beta  H_0^2 k^2 (z+1)^3 E   E ' -\\
& 3 \beta  H_0^4 (z+1) E^3 \left((z+1) w_{de} ' -(\beta +3) c_{s,de}^2\right)
 E ' +\\
 &(\Omega_{de}^3 \left(k^4 (z+1)^4+\right. H_0^2 k^2 (z+1)^2 \left(-\beta +3 c_{s,de}^2-3\right) E^2+9 H_0^4 E^4\\
& 6 (z+1) w_{de} ' +\beta ^2+12 \beta +9 c_{s,de}^4-9 (\beta +6) c_{s,de}^2+27-
 \left.9 \beta  H_0^4 (z+1) E^3 E ' \right)-\\
& (\Omega_{de}^2 \left(H_0^2 k^2 (z+1)^2 E^2+k^4 (z+1)^4\right.
  6 (z+1) w_{de} ' -2 (\beta +9)-3 (\beta +6) c_{s,de}^2-\\
& 9 H_0^4 E^4 \left((\beta +3)^2+3 (\beta +9) c_{s,de}^4-2 \left(\beta ^2+9 \beta +27\right) c_{s,de}^2-\right.\\
&
 \left.6 \left(c_{s,de}^2-2\right) (z+1) w_{de} ' \right)+ \beta  H_0^2 k^2 (z+1)^3 E   E ' -\\
&
 \left.3 \beta  H_0^4 (z+1) \left(-\beta +3 c_{s,de}^2-6\right) E^3 E ' \right)+\\
& 3 H_0^4 E^4 \left(\left(\beta ^3+3 \beta ^2+9 \beta +27\right) c_{s,de}^4-\right.
 (z+1) \left(\beta  (\beta +4)+\left(\beta ^2+18\right) c_{s,de}^2\right) w_{de} ' +\\
& \left.3 (z+1)^2 \left(w_{de} ' \right)^2-\beta  (z+1)^2 w_{de} '' \right)-
 w_{de}   \left(k^4 (z+1)^4-H_0^2 k^2 (z+1)^2 E^2\right.\\
& \beta  (z+1) w_{de} ' +6 (\beta +3)+\left(\beta ^2-18\right) c_{s,de}^2+
 3 \beta  H_0^4 (z+1) E^3\\
& E '  \left((z+1) w_{de} ' +\beta -(\beta +6) c_{s,de}^2+3\right)-\\
& 3 H_0^4 E^4 \left(c_{s,de}^2 \left(-\beta ^3-6 \beta ^2-27 \beta +3 \left(\beta ^2+6 \beta +27\right) c_{s,de}^2-54\right)-\right.\\
& (z+1) \left(\beta ^2+4 \beta +36 c_{s,de}^2-18\right) w_{de} ' +
 \left.\left.\left.\left.3 (z+1)^2 \left(w_{de} ' \right)^2-\beta  (z+1)^2 w_{de} '' \right)\right)\right)\right/\\
& \left(H_0 (z+1)^2 (w_{de}  +1) E  \right.
 \left(k^2 (z+1)^2+9 H_0^2 (\Omega_{de}^2 E^2-\right.\\
 &
 w_{de}   \left(3 H_0^2 \left(-\beta +3 c_{s,de}^2-3\right) E^2+k^2 (z+1)^2\right)+\\
& \left.\left.3 H_0^2 E^2 \left((z+1) w_{de} ' -(\beta +3) c_{s,de}^2\right)\right)\right) ~.
\end{split}
\end{myequation}
%%%%%%%%%%%%%%%%%%%
\end{itemize}

% Appendix A
\chapter{Observational Data And Methodology}% Main appendix title
\label{appen:a}% For referencing this appendix elsewhere, use \ref{AppendixA}
Different observational datasets obtained from the publicly available cosmological probes have been used to constrain the parameters of the interacting models. The datasets used in this work are listed below.
\begin{description}
\item[CMB] 
We considered the cosmic microwave background (CMB) anisotropies data from the latest 2018 data release of the \Planck collaboration\footnote{Available at: \href{http://pla.esac.esa.int/pla/\#home}{ https://pla.esac.esa.int}}~\cite{planck2019cmb,planck2018cp}. The CMB likelihood consists of the low-$\ell$ temperature \commander likelihood, $C_{\ell}^{TT}$, the low-$\ell$ polarization \simall likelihood, $C_{\ell}^{EE}$, high-$\ell$ temperature-polarization likelihood, $C_{\ell}^{TE}$, high-$\ell$ combined TT, TE and EE \plik likelihood. The low-$\ell$ likelihoods span from $2\le \ell \le 29$ and the high-$\ell$ likelihoods consists of multipole values $\ell \ge 30$ and collectively make the combination \planckall. For CMB lensing data, the power spectrum of the lensing potential measured by \Planck collaboration is used. The \planckall, along with the lensing likelihood (\plancklensing) are denoted as `\Planck' in the results given in Sect.\ \ref{sec6:mcmc}. References~\cite{planck2019cmb,planck2018cp} provide a detailed study of the CMB likelihoods.
\item[BAO]
The photon-baryon fluid fluctuations in the early Universe leave their signatures as the acoustic peaks in the CMB anisotropies power spectrum. The anisotropies of baryon acoustic oscillations (BAO) provide tighter constraints on the cosmological parameters~\cite{eisenstein2005apj}. The BAO surveys measure the ratio, $D_{V}/r_{d}$ at different effective redshifts. The quantity $D_{V}$ is related to the comoving angular diameter $D_{M}$ and Hubble parameter $H$ as
\begin{equation}
D_{V}\paren*{z} = \left[D_{M}^{2}\paren*{z} \frac{c\,z}{H\paren*{z}}\right] ^{1/3},
\end{equation}
and $r_{d}$ refers to the comoving sound horizon at the end of baryon drag epoch. For the BAO data, three surveys are considered: the 6dF Galaxy Survey (6dFGS) measurements~\cite{beutler2011mnras} at redshift $z = 0.106$, the Main Galaxy Sample of Data Release $7$ of the Sloan Digital Sky Survey (SDSS-MGS)~\cite{ross2015mnras} at redshift $z = 0.15$ and the latest Data Release $12$ (DR12) of the Baryon Oscillation Spectroscopic Survey (BOSS) of the Sloan Digital Sky Survey (SDSS) III at redshifts $z = 0.38$, $0.51$ and $0.61$~\cite{alam2017sdss3}.
\item[Pantheon] 
We considered the latest `Pantheon' catalogue for the luminosity distance measurements of the Type Ia supernovae (SNe Ia)~\cite{scolnic2018apj}. The Pantheon sample is the compilation of 276 supernovae discovered by the Pan-STARRS1 Medium Deep Survey at $0.03 < z < 0.65$ and various low redshift and Hubble Space Telescope (HST) samples to give a total of 1048 supernovae data in the redshift range $0.01 < z < 2.3$.
\item[RSD]
Redshift-space distortion (RSD) is the cosmological effect where spatial galaxy maps produced by measuring distances from the spectroscopic redshift surveys show an anisotropic galaxy distribution. These galaxy anisotropies arise due to the galaxy recession velocities having components from both the Hubble flow and comoving peculiar velocities from the motions of galaxies and result in the anisotropies of the observed power spectrum. However, additional anisotropies in the power spectrum arise due to incorrect fiducial cosmology, $H\paren*{z}$ while converting the relative redshifts to comoving coordinates. The introduction of anisotropies due to incorrect fiducial cosmology is called Alcock-Paczy{\'n}ski (AP) effect~\cite{alcock1979nature}. The RSD surveys measure the matter peculiar velocities and provide the galaxy matter density perturbation, $\delta_{g}$~\cite{kaiser1987mnras}. As mentioned in Sect.\ \ref{sec6:result}, the combination $\fsg$ is the widely used quantity to study the growth rate of the matter density perturbation. In the present work, we considered the $\fsg$ data compilation by Nesseris \etal~\cite{nesseris2017prd}, Sagredo \etal~\cite{sagredo2018prd} and Skara and Perivolaropoulos~\cite{skara2020prd}. The surveys and the corresponding data points used in this work are shown in Table \ref{tab6:fs8data}, along with the corresponding fiducial cosmology used by the collaborations to convert redshift to distance in each case. The fiducial cosmology in Table \ref{tab6:fs8data} is used to correct the AP effect following Macaulay \etal~\cite{macaulay2013prl} as discussed in~\cite{sagredo2018prd,skara2020prd}. The RSD measurement is denoted as `$\fsg$' data in the results given in Sect.\ \ref{sec6:mcmc}.
\end{description}
The covariance matrices of the data from the WiggleZ~\cite{blake2012mnras} and the SDSS-IV~\cite{zhao2018mnras} surveys are given as
\begin{myequation} \label{eq6:wigglez}
\mbox{\textbf{\large{C}}}_{\mbox{\scriptsize{WiggleZ}}}=10^{-3}
\left(
\begin{array}{ccc}
 6.400 & 2.570 & 0.000 \\
 2.570 & 3.969 & 2.540 \\
 0.000 & 2.540 & 5.184 \\
\end{array}
\right),
\end{myequation}
and
\begin{myequation} \label{eq6:sdss}
\mbox{\textbf{\large{C}}}_{\mbox{\scriptsize{SDSS-IV}}}=10^{-2}
\left(
\begin{array}{cccc}
 3.098 & 0.892 & 0.329 & -0.021 \\
 0.892 & 0.980 & 0.436 & 0.076 \\
 0.329 & 0.436 & 0.490 & 0.350 \\
 -0.021 & 0.076 & 0.350 & 1.124 \\
\end{array}
\right)
\end{myequation}
respectively.
%Data Table
\begin{table}[!h]
\centering
\caption{
A compilation of $\fsg$ measurements with redshift $z$ and fiducial value of $\Omega_{m}$ from different surveys.
}\label{tab6:fs8data}
\begin{adjustbox}{width=1\textwidth}
\begin{tabular}{cccccc}
\hline
\hline
\rule[-1ex]{0pt}{2.ex}Survey &\hspace{5ex} $z$ &\hspace{5ex} $f\sigma_8(z)$ &\hspace{5ex} $\Omega_m$&\hspace{5ex} Refs. \\
\hline
\rule[-1ex]{0pt}{2.ex}6dFGS+SnIa &\hspace{5ex} $0.02$ &\hspace{5ex}  $0.428\pm 0.0465$ &  \hspace{5ex}$0.3$& \hspace{5ex}~\cite{huterer2017jcap} \\ 
\rule[-1ex]{0pt}{2.ex}SnIa+IRAS &\hspace{5ex} $0.02$&\hspace{5ex} $0.398 \pm 0.065$ &\hspace{5ex}$0.3$&\hspace{5ex}~\cite{turnbull2012mnras},~\cite{hudson2012ajl}\\
\rule[-1ex]{0pt}{2.ex}2MASS &\hspace{5ex} $0.02$&\hspace{5ex} $0.314 \pm 0.048$ &\hspace{5ex} $0.266$&\hspace{5ex}~\cite{davis2011mnras},~\cite{hudson2012ajl} \\
\rule[-1ex]{0pt}{2.ex}SDSS-veloc &\hspace{5ex} $0.10$ &\hspace{5ex} $0.370\pm 0.130$ &\hspace{5ex} $0.3$&\hspace{5ex}~\cite{feix2015prl}  & \\
\rule[-1ex]{0pt}{2.ex}SDSS-MGS &\hspace{5ex} $0.15$ &\hspace{5ex} $0.490\pm0.145$ &\hspace{5ex} $0.31$&\hspace{5ex}~\cite{howlett2015mnras} \\
\rule[-1ex]{0pt}{2.ex}2dFGRS &\hspace{5ex} $0.17$ &\hspace{5ex} $0.510\pm 0.060$ &\hspace{5ex} $0.3$&\hspace{5ex}~\cite{song2009jcap} \\
\rule[-1ex]{0pt}{2.ex}GAMA &\hspace{5ex} $0.18$ &\hspace{5ex} $0.360\pm 0.090$ &\hspace{5ex} $0.27$&\hspace{5ex}~\cite{blake2013mnras}\\
\rule[-1ex]{0pt}{2.ex}GAMA &\hspace{5ex} $0.38$ &\hspace{5ex} $0.440\pm 0.060$ &&\hspace{5ex}~\cite{blake2013mnras} \\
\rule[-1ex]{0pt}{2.ex}SDSS-LRG-200 &\hspace{5ex} $0.25$ &\hspace{5ex} $0.3512\pm 0.0583$ &\hspace{5ex} $0.25$&\hspace{5ex}~\cite{samushia2012mnras}\\
\rule[-1ex]{0pt}{2.ex}SDSS-LRG-200 &\hspace{5ex} $0.37$ &\hspace{5ex} $0.4602\pm 0.0378$ &&\hspace{5ex}~\cite{samushia2012mnras}\\
\rule[-1ex]{0pt}{2.ex}BOSS-LOWZ&\hspace{5ex} $0.32$ &\hspace{5ex} $0.384\pm 0.095$ &\hspace{5ex} $0.274$&\hspace{5ex}~\cite{sanchez2014mnras} \\
\rule[-1ex]{0pt}{2.ex}SDSS-CMASS &\hspace{5ex} $0.59$ &\hspace{5ex} $0.488\pm 0.060$ &\hspace{5ex} $0.307115$&\hspace{5ex}~\cite{chuang2016mnras}\\
\rule[-1ex]{0pt}{2.ex}WiggleZ &\hspace{5ex} $0.44$ &\hspace{5ex} $0.413\pm 0.080$ &\hspace{5ex} $0.27$&\hspace{5ex}~\cite{blake2012mnras} \\
\rule[-1ex]{0pt}{2.ex}WiggleZ &\hspace{5ex} $0.60$ &\hspace{5ex} $0.390\pm 0.063$ &\hspace{5ex} $\mathbf{C}_{ij}\rightarrow$ Eq.~(\ref{eq6:wigglez})&\hspace{5ex}~\cite{blake2012mnras} \\
\rule[-1ex]{0pt}{2.ex}WiggleZ &\hspace{5ex} $0.73$ &\hspace{5ex} $0.437\pm 0.072$ & &\hspace{5ex}~\cite{blake2012mnras}\\
\rule[-1ex]{0pt}{2.ex}VIPERS PDR-2&\hspace{5ex} $0.60$ &\hspace{5ex} $0.550\pm 0.120$ &\hspace{5ex} $0.3$&\hspace{5ex}~\cite{pezzotta2017aa} \\
\rule[-1ex]{0pt}{2.ex}VIPERS PDR-2&\hspace{5ex} $0.86$ &\hspace{5ex} $0.400\pm 0.110$ &&\hspace{5ex}~\cite{pezzotta2017aa}\\
\rule[-1ex]{0pt}{2.ex}FastSound&\hspace{5ex} $1.40$ &\hspace{5ex} $0.482\pm 0.116$ &\hspace{5ex} $0.27$&\hspace{5ex}~\cite{okumura2016pasj}\\
\rule[-1ex]{0pt}{2.ex}SDSS-IV&\hspace{5ex} $0.978$  &\hspace{5ex} $0.379 \pm 0.176$ &\hspace{5ex} $0.31$&\hspace{5ex}~\cite{zhao2018mnras}\\
\rule[-1ex]{0pt}{2.ex}SDSS-IV&\hspace{5ex} $1.23$ &\hspace{5ex} $ 0.385 \pm 0.099$ &\hspace{5ex} $\mathbf{C}_{ij}\rightarrow$ Eqn.~(\ref{eq6:sdss})&\hspace{5ex}~\cite{zhao2018mnras}\\
\rule[-1ex]{0pt}{2.ex}SDSS-IV&\hspace{5ex} $1.526$ &\hspace{5ex} $0.342 \pm 0.070$  &&\hspace{5ex}~\cite{zhao2018mnras}\\
\rule[-1ex]{0pt}{2.ex}SDSS-IV&\hspace{5ex} $1.944$ &\hspace{5ex} $0.364 \pm 0.106$  &&\hspace{5ex}~\cite{zhao2018mnras}\\
\rule[-1ex]{0pt}{2.ex}VIPERS PDR2 &\hspace{5ex} $0.60$ &\hspace{5ex} $0.49 \pm 0.12$ &\hspace{5ex} $0.31$&\hspace{5ex}~\cite{mohammad2018aa}\\
\rule[-1ex]{0pt}{2.ex}VIPERS PDR2 &\hspace{5ex} $0.86$ &\hspace{5ex} $0.46 \pm 0.09$ &&\hspace{5ex}~\cite{mohammad2018aa}\\
\rule[-1ex]{0pt}{2.ex}BOSS DR12 voids &\hspace{5ex} $0.57$ &\hspace{5ex} $0.501 \pm 0.051$ &\hspace{5ex} $0.307$&\hspace{5ex}~\cite{nadathur2019prd}\\
\rule[-1ex]{0pt}{2.ex}2MTF 6dFGSv &\hspace{5ex} $0.03$ &\hspace{5ex} $0.404 \pm 0.0815$ &\hspace{5ex} $0.3121$&\hspace{5ex}~\cite{qin2019mnras}\\
\rule[-1ex]{0pt}{2.ex}SDSS-IV &\hspace{5ex} $0.72$ &\hspace{5ex} $0.454 \pm 0.139$ &\hspace{5ex} $0.31$&\hspace{5ex}~\cite{icaza2019mnras}\\
\hline \hline
\end{tabular}

\end{adjustbox}
\end{table}

To compare the interacting model with the observational data, we calculated the likelihood as
\begin{equation}\label{eq6:chi-total}
\mathcal{L} \propto e^{-\chi^{2}/2}, \hspace{0.3cm} \mbox{where} \hspace{0.3cm} \chi^{2} = \chi_{\mbox{\scriptsize{CMB}}}^{2} +\chi_{\mbox{\scriptsize{BAO}}}^{2} + \chi_{\mbox{\scriptsize{Pantheon}}}^{2} +\chi_{\fsg}^{2}.
\end{equation}
The quantity $\chi^{2}$ for any dataset is calculated as 
\begin{equation}
\chi^{2}_{i} = V^i\, \mathbf{C}_{ij}^{-1}\,V^j,
\end{equation}
where the vector, $V^{i}$ is written as
\begin{equation}
V^{i} = \Theta_{i}^{\scriptsize{\mbox{obs}}} - \Theta^{\scriptsize{\mbox{th}}}\paren*{z_{i},P}
\end{equation}
with $\Theta$ being the physical quantity corresponding to the observational data (\Planck, BAO, Pantheon, $\fsg$) used, $z_{i}$ being the corresponding redshift, $\mathbf{C}^{-1}_{ij}$ is the corresponding inverse covariance matrix and $P$ is the parameter space. The posterior distribution (see Eqn.\ \ref{eq6:bayes-th}) is sampled using the Markov Chain Monte Carlo (MCMC) simulator through a suitably modified version of the publicly available code \cosmomc~\cite{lewis2013hha,lewis2002ah}. The statistical convergence of the MCMC chains for each model is set to satisfy the Gelman and Rubin criterion~\cite{gelman1992}, $R-1 \lesssim 0.01$. 

The correction for the Alcock-Paczy{\'n}ski effect is taken into account by the fiducial correction factor, $\mathcal{R}$~\cite{sagredo2018prd,skara2020prd} given as
\begin{equation} \label{eq6:ratio}
\mathcal{R}(z)=\frac{H(z) D_A\paren*{z}}{H^{\scriptsize{\mbox{fid}}}\paren*{z} D^{\scriptsize{\mbox{fid}}}_A\paren*{z}}
\end{equation}
where $H\paren*{z}$ is the Hubble parameter and $D_A\paren*{z}$ is the angular diameter distance  of the interacting models and that of the fiducial cosmology are denoted with superscript `fid'. The corrected vector, $V_{\fsg}^i(z,P)$ is corrected as
\begin{equation} \label{eq6:vector}
V_{\fsg}^i(z_{i},P) \equiv f\sigma_{8,i}^{\scriptsize{\mbox{obs}}} - \frac{f\sigma_{8}^{\scriptsize{\mbox{th}}}\paren*{z_{i},P}}{\mathcal{R}(z_{i})},
\end{equation}
where $f\sigma_{8,i}^{\scriptsize{\mbox{obs}}}$ is the $i$-th observed data point from Table \ref{tab6:fs8data}, $f\sigma_{8}^{\scriptsize{\mbox{th}}}\paren*{z_{i},P}$ is the theoretical prediction at the same redshift $z_{i}$ and $P$ is the parameter vector given by Eqn.\ \ref{eq6:parameter}. The corrected $\chi_{\fsg}^{2}$ is then written as
\begin{equation} \label{eq6:chi-fs8}
\chi_{\fsg}^{2}= V_{\fsg}^i\, \mathbf{C}_{ij,\fsg}^{-1}\,V_{\fsg}^j,
\end{equation}
where $\mathbf{C}_{ij,\fsg}^{-1}$ is the inverse of the covariance matrix, $\mathbf{C}_{ij,\fsg}$ of the $\fsg$ dataset given by 
%\begin{adjustbox}{width=0.5\textwidth}
\begin{myequation} \label{eq6:totalcov}
\mathbf{C}_{ij,\fsg}=
\left(
\begin{array}{cccccc}
 \sigma_1^2 & 0 & \cdots & 0 & \cdots & 0 \\
 0 & \sigma_2^2 & \cdots & 0 & \cdots & 0 \\
 \vdots &  \vdots &  \vdots &  \vdots &  \vdots &  0 \\
 0 & 0 & \cdots & \mathbf{C}_{\mbox{\scriptsize{WiggleZ}}} & \cdots & 0 \\
 0 & 0 & \cdots & 0 & \mathbf{C}_{\mbox{\scriptsize{SDSS-IV}}} & 0 \\
0 & 0 & \cdots & 0 & \cdots & \sigma_{\scriptsize{N}}^2\\
\end{array}
\right)
\end{myequation}
%\end{adjustbox}
where $N=27$ corresponds to total number of data points in Table \ref{tab6:fs8data}. Thus the covariance matrix, $\mathbf{C}_{ij,\fsg}$ is a $27 \times 27$ matrix with Eqns.\ (\ref{eq6:wigglez}) and (\ref{eq6:sdss}) at the positions of $\mathbf{C}_{\mbox{\scriptsize{WiggleZ}}}$ and $\mathbf{C}_{\mbox{\scriptsize{SDSS-IV}}}$ respectively and $\sigma_{i}$ is the error from Table \ref{tab6:fs8data}. To use the RSD measurements, we added a new likelihood module to the publicly available \cosmomc package to calculate the corrected $\chi_{\fsg}^{2}$. The results obtained by analysing the MCMC chains are explained in Sect.\ \ref{sec6:mcmc}.

% Appendix A
\chapter{Model Selection}% Main appendix title
\label{appen:b}% For referencing this appendix elsewhere, use \ref{AppendixA}
\emph{Bayesian evidence} is the Bayesian tool to compare models and is the integration of the likelihood over the multidimensional parameter space. Hence, it is also referred to as marginal likelihood. Using \emph{Bayes theorem}, the posterior probability distribution of a model, $M$ with parameters $\Theta$ for the given particular dataset $x$ is obtained as
\begin{equation}\label{eq6:bayes-th}
p\paren*{\Theta|x,M} = \frac{p\paren*{x|\Theta,M}\pi\paren*{\Theta|M}}{p\paren*{x|M}},
\end{equation}
where $p\paren*{x|\Theta,M}$ refers to the likelihood function, $\pi\paren*{\Theta|M}$ refers to the prior distribution and $p\paren*{x|M}$ refers to the Bayesian evidence. From Eqn.\ (\ref{eq6:bayes-th}), the evidence follows as the integral over the unnormalised posterior distribution,
\begin{equation}
E \equiv p\paren*{x|M} = \int d\Theta p\paren*{x|\Theta,M}\pi\paren*{\Theta|M}.
\end{equation}
To compare model $M_{i}$ with the reference model $M_{j}$, the ratio of the evidences, called the \emph{Bayes factor} is calculated.
\begin{equation}\label{eq6:bayesfac}
B_{ij} = \frac{p\paren*{x|M_{i}}}{p\paren*{x|M_{j}}}.
\end{equation}
The calculation of the multidimensional integral is undoubtedly computationally expensive. This problem is solved by the method developed by Heavens \etal~\cite{heavens2017prl, heavens2017ax}, where the Bayesian evidence is estimated directly from the MCMC chains generated by \cosmomc. This method for evidence estimation is publicly available in the form of \mcevidence. The \mcevidence package provides with the logarithm of the Bayes factor, $\ln B_{ij}$. The value of $\ln B_{ij}$ is then used to assess if model $M_{i}$ is preferred over model $M_{j}$ and if so, what is the strength of preference, by using the revised Jeffreys scale (Table \ref{tab6:bij-values}) by Kass and Raftery~\cite{kass1995jasa}. Thus, if $\ln B_{ij} > 0 $, model $M_{i}$ is preferred over model $M_{j}$.
\begin{table}[!htbp]
\begin{center}
\caption{Revised Jeffreys scale by Kass and Raftery to interpret the values of $\ln B_{ij}$ while comparing two models $M_{i}$ and $M_{j}$}\label{tab6:bij-values}
%\centering
\begin{adjustbox}{width=0.7\textwidth}
\begin{tabular}{cc}
\hline \hline
\rule[-1ex]{0pt}{2.5ex}$\ln B_{ij}$ &  \hspace{24ex} Strength\\
\hline
\rule[-1ex]{0pt}{2.5ex}$0 \le \ln B_{ij} <1$&  \hspace{24ex} Weak \\
\rule[-1ex]{0pt}{2.5ex}$1 \le \ln B_{ij} <3$&  \hspace{24ex} Definite/Positive\\
\rule[-1ex]{0pt}{2.5ex}$3\le \ln B_{ij} <5$&  \hspace{24ex} Strong \\
\rule[-1ex]{0pt}{2.5ex}$\ln B_{ij} \ge 5$&  \hspace{24ex} Very strong \\
\hline \hline
\end{tabular}
\end{adjustbox}
\end{center}
\end{table}%

The results of model comparison from the Bayesian evidence are discussed in Sect.\ \ref{sec6:mcmc}

%\end{appendices}

%-------------------------------------------------------------------------------
%	BIBLIOGRAPHY
%-------------------------------------------------------------------------------
\small
\addcontentsline{toc}{chapter}{Bibliography}
\bibliographystyle{JHEP}
%\bibliography{\myref} 
\bibliography{main} 

%-------------------------------------------------------------------------------

\end{document}